\documentclass[preprint]{elsarticle}  
\usepackage{graphicx}  
\usepackage{amssymb,amstext,amsmath,proof,stmaryrd,upgreek,amsthm}
\usepackage[]{pstricks}
\usepackage[all]{xy}
\usepackage{xspace}
\usepackage{url}
\usepackage{wrapfig}
\newif\ifignore 
\ignorefalse

\newcommand{\auxproof}[1]{
\ifignore\mbox{}\newline
\textbf{BEGIN: AUX-PROOF} \dotfill\newline
{#1}\mbox{}\newline
\textbf{END: AUX-PROOF}\dotfill\newline
\fi}

\xyoption{v2}
\xyoption{curve}
\xyoption{2cell}
\SelectTips{cm}{}  
\UseAllTwocells
\SilentMatrices
\def\labelstyle{\scriptstyle}
\def\twocellstyle{\scriptstyle}
\newdir{ >}{{}*!/-8pt/@{>}}
\newdir{|>}{%
!/1.6pt/@{|}*:(1,-.2)@^{>}*:(1,+.2)@_{>}}
\newdir{pb}{:(1,-1)@^{|-}}
  \def\pb#1{\save[]+<20 pt,0 pt>:a(#1)\ar@{pb{}}[]\restore}
\newcommand{\shifted}[3]{\save[]!<#1,#2>*{#3}\restore}

\newtheorem{mylemma}{Lemma}[section]
\newproof{mylemma*}{Lemma}
\newtheorem{myproposition}[mylemma]{Proposition}
\newproof{myproposition*}{Proposition}
\newtheorem{mytheorem}[mylemma]{Theorem}
\newtheorem{mycorollary}[mylemma]{Corollary}
\newdefinition{mydefinition}[mylemma]{Definition}
\newdefinition{myremark}[mylemma]{Remark}
\newdefinition{myassumption}[mylemma]{Assumption}
\newdefinition{mynotation}[mylemma]{Notation}
\newdefinition{myexample}[mylemma]{Example}
\newproof{myproof}{Proof}
\newcommand{\myqed}{\hspace*{\fill}{$\Box$}}

\DeclareMathOperator{\tr}{tr}

  \allowdisplaybreaks[1] 

\newcommand{\co}{\mathrel{\circ}}

\newcommand{\id}{\mathrm{id}}
\newcommand{\Sets}{\mathbf{Sets}}
\newcommand{\Rel}{\mathbf{Rel}}
\newcommand{\Pfn}{\mathbf{Pfn}}

\newcommand{\fdVect}{\mathbf{fdVect}}

\newcommand{\Kleisli}[1]{\mathcal{K}{\kern-.2ex}\ell(#1)}
 
\newcommand{\sem}[1]{\llbracket #1 \rrbracket} 
\newcommand{\semConst}[1]{\llbracket #1 \rrbracket_{\textrm{const}}} 
\newcommand{\semP}[1]{\llbracket #1 \rrbracket^{\mathrm{P}}} 
\newcommand{\fvsem}[1]{\llbracket #1 \rrbracket_{\mathrm{FV}}}

\newcommand{\C}{\mathbb{C}}

\newcommand{\Q}{\mathcal{Q}}

\newcommand{\place}{\mathop{\underline{\phantom{n}}}} 

\newcommand{\nat}{\mathbb{N}}
\newcommand{\real}{\mathbb{R}}
\newcommand{\op}{\mathop{\mathrm{op}}\nolimits}
\newcommand{\tuple}[1]{\langle#1\rangle}

\newcommand{\trace}{\mathsf{tr}}
\newcommand{\pow}{\mathcal P}

\newcommand{\dist}{\mathcal D}
\newcommand{\lift}{\mathcal{L}}
\newcommand{\iso}{\mathrel{\stackrel{
           \raisebox{.5ex}{$\scriptstyle\cong\,$}}{
           \raisebox{0ex}[0ex][0ex]{$\rightarrow$}}}}

\newcommand{\longto}{\longrightarrow}

\newcommand{\ev}{\mathsf{ev}}

\newcommand{\textlb}{
  \def\labelstyle{\textstyle}
  \def\twocellstyle{\textstyle}}

\def\compsign{\mathrel>\kern-2pt\joinrel>\kern-2pt\joinrel>}

\newcommand{\seq}[2]{{#1}_{1},\dotsc,{#1}_{#2}}

\newcommand{\x}{\mathtt{x}}

\newcommand{\defiff}{\stackrel{\text{def.}}{\Longleftrightarrow}}

\newcommand{\dar}{\ar@{..>}}
\newcommand{\lar}{\ar@{-}}

\newcommand{\Int}{\mathrm{Int}}
\newcommand{\PER}{\mathbf{PER}} 
\newcommand{\Kco}{\odot} 
\newcommand{\cmbt}[1]{\mathsf{#1}} 
\newcommand{\cmbtDelta}{\updelta} 
\newcommand{\limp}{\multimap} 
\newcommand{\bang}{\mathop{!}\nolimits} 
\newcommand{\relto}{\mathrel{\ooalign{\hfil\raisebox{.3pt}{$\shortmid$}\hfil\crcr$\rightarrow$}}}
\newcommand{\longrelto}{\mathrel{\ooalign{\hfil\raisebox{.3pt}{$\shortmid$}\hfil\crcr$\longrightarrow$}}}
\newcommand{\dirsum}{\prod} 
\newcommand{\bc}{\mathrm{bc}}
\newcommand{\punit}{\mathrm{I}} 
\newcommand{\ptensor}{\boxtimes} 
\newcommand{\qll}{\mathbf{q}\lambda_{\ell}}
\newcommand{\synt}[1]{\mathtt{#1}} 
\newcommand{\qbit}{\synt{qbit}}
\newcommand{\nqbit}{n\text{-}\synt{qbit}}

\newcommand{\mqbit}{m\text{-}\synt{qbit}}
\newcommand{\kqbit}{k\text{-}\synt{qbit}}
\newcommand{\Nqbit}{N\text{-}\synt{qbit}}
\newcommand{\zeroqbit}{0\text{-}\synt{qbit}}
\newcommand{\bit}{\synt{bit}}
\newcommand{\letcl}[2]{\synt{let}\,{#1}\,\synt{in}\,{#2}}
\newcommand{\meas}{\synt{meas}}
\newcommand{\new}{\synt{new}}
\newcommand{\qstate}{\synt{qstate}}
\newcommand{\DType}{\mathrm{DType}}
\newcommand{\injl}{\synt{inj}_{\ell}}
\newcommand{\injr}{\synt{inj}_{r}}
\newcommand{\matchcl}[2]{\synt{match}\,{#1}\,\synt{with}\,{#2}}
\newcommand{\letreccl}[2]{\synt{letrec}\,{#1}\,\synt{in}\,{#2}}

\newcommand{\cmp}{\synt{cmp}}
\newcommand{\subtp}{\mathrel{<:}}
\newcommand{\ttrue}{\mathtt{t{\kern-1.5pt}t}}
\newcommand{\ffalse}{\mathtt{f{\kern-1.5pt}f}}
\newcommand{\der}{\mathsf{der}}
\newcommand{\weak}{\mathsf{weak}}
\newcommand{\con}{\mathsf{con}}
\newcommand{\mult}{\mathrm{mult}}
\newcommand{\str}{\mathrm{str}}
\newcommand{\dtimes}{\mathbin{\dot{\mbox{$\times$}}}}
\newcommand{\fix}{\mathrm{fix}}
\newcommand{\Val}{\mathrm{Val}}
\newcommand{\ClTerm}{\mathrm{ClTerm}}
\newcommand{\EV}{\mathrm{EV}}
\newcommand{\EValt}{\overline{\mathrm{EV}}}

\newcommand{\oprred}{\curlyveedownarrow} 
\newcommand{\denred}{\Downarrow} 
\newcommand{\test}{\mathrm{test}}
\newcommand{\prb}{\mathrm{prob}}
\newcommand{\tree}{\mathrm{tree}}
\newcommand{\FV}{\mathrm{FV}}
\newcommand{\rest}[1]{|_{#1}}
\newcommand{\VdashP}{\Vdash_{\mathrm{P}}}

\newcommand{\DM}{\mathrm{DM}}  
\newcommand{\QO}{\mathrm{QO}}  
\newcommand{\IM}{\mathcal{I}} 
\newcommand{\Eop}{\mathcal{E}} 
\newcommand{\Fop}{\mathcal{F}} 
\newcommand{\bra}[1]{\langle #1 |}
\newcommand{\ket}[1]{| #1 \rangle}
\newcommand{\inpr}[2]{\langle #1 | #2 \rangle} 
\newcommand{\abs}[1]{\| #1 \|}
\newcommand{\trNorm}[1]{\| #1 \|_{\tr}}
\newcommand{\FrNorm}[1]{\| #1 \|_{\mathsf{Fr}}}

\newcommand{\Lle}{\sqsubseteq} 
\newcommand{\QOle}{\sqsubseteq_{\mathrm{QO}}} 
\newcommand{\retr}{\mathrel{\vartriangleleft}}
\newcommand{\Hoq}{\text{Hoq}}
\newcommand{\Fpbt}{F_{\mathrm{pbt}}}
\newcommand{\Bt}{\mathsf{Bt}}

\journal{Annals of Pure and Applied Logic}

\begin{document}



\begin{frontmatter}
 \title{Semantics of Higher-Order Quantum Computation via Geometry of
 Interaction\tnoteref{lics}}
 \tnotetext[lics]{An earlier version of this paper~\cite{HasuoH11} has been
 presented at the 
Twenty-Sixth Annual IEEE Symposium on
Logic in Computer Science (LICS 2011), 
21–24 June 2011, Toronto, Ontario, Canada.}
 \author[ih]{Ichiro Hasuo\fnref{ihack}\corref{cor1}}
 \ead{ichiro@is.s.u-tokyo.ac.jp}
 \ead[url]{http://www-mmm.is.s.u-tokyo.ac.jp/~ichiro}
 \fntext[ihack]{Supported by Grants-in-Aid 
 No.\ 24680001, 15K11984 \& 15KT0012, JSPS;
 by the
 JSPS-INRIA Bilateral Joint Research Project ``CRECOGI''; 
 and by Aihara
Innovative Mathematical Modelling Project, FIRST Program, JSPS/CSTP.}
 \cortext[cor1]{Corresponding author}
 \address[ih]{Department of Computer Science, The University of Tokyo,
 Hongo 7-3-1, Tokyo 113-8656, Japan}
 \author[nh]{Naohiko Hoshino\fnref{nhack}}
 \ead{naophiko@kurims.kyoto-u.ac.jp}
 \fntext[nhack]{Supported by
 Grants-in-Aid  No.\ 26730004 \& 15K11984, JSPS;
 and by the
 JSPS-INRIA Bilateral Joint Research Project ``CRECOGI''.
 }
 \ead[url]{http://www.kurims.kyoto-u.ac.jp/~naophiko}
 \address[nh]{Research Institute for Mathematical Sciences,
 Kyoto
 University,  Kitashirakawa-Oiwakecho, Kyoto 606-8502, Japan}

\begin{abstract}
  While much of the current study on quantum computation employs
 low-level formalisms such as quantum circuits, several high-level
 languages/calculi have been recently proposed aiming at structured
 quantum programming. The current work contributes to the semantical
 study of such languages by providing interaction-based semantics of a
 functional quantum programming language; the latter is, much like
 Selinger and Valiron's, based on linear
 lambda calculus and equipped with features like the $\bang$ modality
 and recursion.  The proposed denotational model is the first one that
 supports the full features of a quantum functional programming
 language; we  prove adequacy of our semantics. The construction of
 our model is by a series of existing techniques taken from
 the semantics of classical computation as well as from process theory.
 The most notable among them is Girard's \emph{Geometry of Interaction
 (GoI)}, categorically formulated by Abramsky, Haghverdi and Scott.  The
 mathematical genericity of these techniques---largely due to their
 categorical formulation---is exploited for our move from classical to
 quantum.
\end{abstract}


\end{frontmatter}









\auxproof{
To-do:
\begin{itemize}
 \item Cite~\cite{ArrighiD11,Malherbe10PhDThesis,DHondtP06,Grattage11}
 \item Categorical GoI goes back to~\cite{Abramsky96,AbramskyJ92}
 \item The language: HOQ
 \item map $\mapsto$ arrow
 \item ``define the'' $\mapsto$ ``define a''
 \item Language \& operational semantics first 
\begin{itemize}
 \item Language design
  \item Examples (Hoshino):
       \begin{itemize}
	\item Quantum teleportation
	\item (Recursive) coin flipping
	\item Grover's algorithm
       \end{itemize}
\end{itemize}
 \item Points to be made:
\begin{itemize}
 \item
   various GoI along two axis
\begin{itemize}
 \item 
  information on a token or on pipes (string/pipe/wire/bus)
 \item 
     Int-version or non-Int version. The former goes well with
       proof/interaction nets
\end{itemize}
 \item   Hilb (potentially infinite-dimensional) with $\otimes$ is not 
   a monoidal closed category, at least easily...
 \item Remark~\ref{remark:simplerOprSem}. It's hard to get $\bang$ that
       goes along with the Hilbert space tensor $\otimes$

\end{itemize}
 \item The issue of well-definedness of denotational semantics for a
       type judgment, what Valiron does in his PhD thesis is as follows
       (Chap.9-11). 
       \begin{itemize}
	\item First he introduces \emph{indexed terms} (as we do);
	\item then introduces \emph{neutral terms} for which
	      insensitivity to labels (or indices as he calls them) can
	      be easily established;
	\item and then uses normalization by evaluation.
       \end{itemize}
       Let's not bother too much on this point 
       (Valiron does not have sum
       types nor a recursion operator). We should be able to do with
       fewer labels than we had in
	      our lics paper, though.
 \item $\Q$ can be explained also in the analogy to matrices
 \item Comparison with [Pagani et al.]
       \begin{itemize}
	\item It's not fair for them to say that entanglement is missing 
	      in our calculus :)
	\item Their construction relies on \emph{quantitative semantics}
	      for linear logic. It is a variant of the method by Girard
	      (see also Hasegawa) and a general technique for
	      construction of linear logic models exploiting a certain
	      universal property of the Taylor expansion of the
	      exponential function
	\item Our GoI model has an operational flavor that can actually
	      lead
	      to a compiler. cf.\ GoS, Mackie
       \end{itemize}
 \item Comparison with Quipper
       \begin{itemize}
	\item Quipper: a language for generating quantum circuits
	\item Lacks full-fledged higher-order feature, or seamless
	      invocation of quantum operations. For the latter: Quipper
	      requires a constructed quantum circuit to be explicitly
	      invoked
	      by the \verb+run_generic+ (or
\verb+run_classical_generic+, \verb+run_clifford_generic+) command, and 
	      explicit use of dynamic lifting (distinction between
	      parameters and input)
	      
       \end{itemize}
\end{itemize}
}

\section{Introduction}\label{section:intro}
\subsection{Quantum Programming Languages}
Computation and communication using quantum data has attracted growing
attention. On the one hand, quantum computation provides a real
breakthrough in computing power---at least for certain applications---as
demonstrated by Shor's algorithm. On the other hand, quantum
communication realizes ``unconditional security'' e.g.\ via quantum key
distribution. Quantum communication is  being physically
realized and put into use.

The extensive research efforts on this new paradigm have identified some
challenges, too. On quantum computation, aside from a few striking ones
such as Shor's and quantum search algorithms, researchers are having a
hard time finding new ``useful'' algorithms. On quantum
communication, the counter-intuitive nature of quantum data becomes an
additional burden in the task of getting communication protocols
right---which has  proved extremely hard already with classical data.

 \emph{Structured programming} and \emph{mathematically formulated
semantics} are potentially useful tools against these
difficulties. Structured programming often leads to discovery of ingenious
algorithms; well-formulated semantics would provide a ground for proving
a communication protocol correct.  

In this direction, there have been proposed several high-level
languages tailored for quantum computation (see~\cite{Valiron13NGC} for
an excellent survey). 
Among the first ones is QCL~\cite{Oemer00} that is
imperative; the quantum IO monad~\cite{GreenA09} and its successor
Quipper~\cite{GreenLRSV13} are quantum extensions of Haskell that
facilitate generation of quantum circuits. Closely related to the latter two
is the one in~\cite{DelbecqueP08}, that is an (intuitionistic) $\lambda$-calculus with 
quantum stores.  

Another important
family---that is most strongly oriented towards mathematical
semantics---is those of \emph{quantum $\lambda$-calculi} that are very
often based
on \emph{linear $\lambda$-calculus}. While $\lambda$-calculus is a
prototype of functional programming languages and inherently supports
higher-order computation, linearity in a type system provides a useful means of
prohibiting duplication of quantum data (``no-cloning''). Examples of
such languages are found
in~\cite{Selinger04,SelingerV08,SelingerV09,DalLagoMZ11,DalLagoF11,Grattage11,vanTonder04}.

\subsection{Denotational Semantics of Quantum Programming Languages}
Models of linear logic (and hence of linear $\lambda$-calculus) have
been studied fairly well since 1990s; therefore denotational models for
the last family of quantum programming languages may well be based on
those well-studied models. Presence of quantum primitives---or more
precisely \emph{coexistence of ``quantum data, classical
control''}---poses unique challenges, however. It thus seems that
denotational semantics for quantum $\lambda$-calculi has attracted
research efforts, not only from those interested in quantum computation,
but also from the semantics community in general, since it offers unique
and interesting ``exercises'' to the semantical techniques developed over
many years, many of which are formulated in categorical terms and hence
are aimed at genericity.

Consider a quantum
$\lambda$-calculus that is essentially a linear $\lambda$-calculus with
quantum primitives. It is standard that \emph{compact closed categories}
provide models for the latter; so we are aiming a compact closed
category 1) with a quantum flavor, and 2) that allows interpretation of
the $\bang$ modality that is essential in duplicating classical
data. This turns out to be not easy at all. For example, the requirement
1) makes one hope that the category
$\mathbf{fdHilb}$ of finite-dimensional Hilbert spaces and linear maps
would work. This category however has no convenient ``infinity''
structure that
can be exploited for the requirement 2). Moving to the category
$\mathbf{Hilb}$
of possibly infinite-dimensional Hilbert spaces does not work either,
since it is not compact closed.

A few attempts have been made to address this
difficulty. In~\cite{SelingerV08} a categorical model is presented that is fully
abstract for the $\bang$-free fragment of a quantum $\lambda$-calculus
is presented. It relies on Selinger's category $\mathbf{Q}$
in~\cite{Selinger04}---it can be thought of as an extension of
$\mathbf{fdHilb}$ with non-duplicable classical information. The
works~\cite{Malherbe10PhDThesis,PaganiSV14} essentially take ``completions'' of
this model to accommodate the  $\bang$-modality: the
former~\cite{Malherbe10PhDThesis} uses presheaves and thus results in a
huge model; the construction in the latter~\cite{PaganiSV14} keeps a model in a tractable size by
the general semantical technique called \emph{quantitative
semantics}~\cite{Ehrhard05,DanosE11}. The difference between the two is
comparable to the one between Girard's normal functor
semantics~\cite{Girard88} (see also~\cite{Hasegawa02}) and quantitative semantics.


In this paper we take a different path towards a denotational model of a
quantum $\lambda$-calculus. Instead of starting from $\mathbf{fdHilb}$
(a purely quantum model) and completing it with structures suitable for
classical data, we start from a general family of models of
classical computation,\footnote{``Classical'' as opposed to ``quantum''; not as the
opposite of ``intuitionistic''.} and fix its parameter so that the
resulting instance accommodates quantum data too. The family of models
is the one given by Girard's \emph{geometry of interaction
(GoI)}~\cite{Girard89GoI}---more specifically its categorical
formulation by Abramsky, Haghverdi and Scott~\cite{AbramskyHS02}. GoI,
like game semantics~\cite{AbramskyJM00,HylandO00}, is an
interaction-based denotational semantics of (classical) computation that 
has a strong operational flavor, too. It thus possibly enables us to
extract a compiler from a denotational model, which is the case with
classical computation~\cite{Mackie95,GhicaS10,GhicaS11,GhicaSS11,MuroyaKHH14LOLA}.

\subsection{Contributions}
In this paper we introduce a  calculus $\Hoq$ and its denotational
model that supports the full features (including the $\bang$ modality
and recursion).  The language $\Hoq$ is almost the same as 
Selinger and Valiron's quantum $\lambda$-calculus~\cite{SelingerV09}---in particular we share their
principle of ``quantum data, classical control''---but is modified for a
better fit to our denotational model. We also define its operational
semantics and prove adequacy.

\begin{figure}[hbp]
 \begin{displaymath}\textlb
 \xymatrix@R=1.8em@C+2em{
 **[r]{\boxed{
\begin{array}{l}
\text{Monad $B$ for branching}
\end{array}}}
   \ar[d]^-{
\begin{array}{l}
 \text{\emph{Coalgebraic trace semantics}~\cite{HasuoJS07b,Jacobs10trace}}
\\
 \text{Take the Kleisli category $\Kleisli{B}$}
\end{array}
}
 &
 \\
 **[r]{\boxed{
\begin{array}{l}
 \text{Traced monoidal category $\C$}\\
 \text{($+$ other constructs
  $\to$ \emph{GoI situation}~\cite{AbramskyHS02})}
\end{array}
}}
   \ar[d]^-{
\begin{array}{l}
  \text{\emph{Categorical GoI}~\cite{AbramskyHS02}}
 \\
  \text{Take $\C(U,U)$}
\end{array}
}  
 &
 \\
 **[r]{\boxed{
\begin{array}{l}
   \text{Linear combinatory algebra $A$}
\end{array}
}}
   \ar[d]^{
\begin{array}{l}
  \text{\emph{Realizability}, e.g.~\cite{Longley94,AbramskyL05,Hoshino07}}
 \\
  \text{Take $\PER_{A}$}
\end{array}
}  
 &
 \\
 **[r]{\boxed{
\begin{array}{l}
 \text{Linear category}
\end{array}
}}
 } 
 \end{displaymath}
\caption{The construction of the model}
\label{figure:constructionOfModel}
\end{figure}
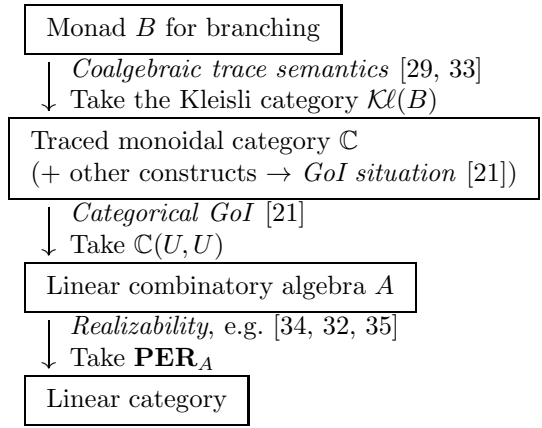

For the construction of the denotational model we employ a series of
existing techniques in theoretical computer science
(Figure~\ref{figure:constructionOfModel}). Namely: 1) a monad with an
order structure for modeling branching, used in the coalgebraic study of
state-based systems (e.g.\ in~\cite{HasuoJS07b}); 2) Girard's
\emph{Geometry of Interaction (GoI)}~\cite{Girard89GoI}, categorically
formulated by Abramsky, Haghverdi and Scott~\cite{AbramskyHS02},
providing interaction-based, game-like semantics for linear logic and
computation; 3) the \emph{realizability} technique that turns an
(untyped) combinatory algebra into a categorical model of a typed
calculus (in our case a \emph{linear
category}~\cite{Bierman95a,BentonW96}; 
the linear realizability technique is used e.g.\ in~\cite{AbramskyL05}); and 4) the
\emph{continuation-passing style (CPS)} semantics.  In each stage we
benefit from the fact that the relevant technique is formulated in the
language of category theory: the technique is originally for classical
computation but its genericity makes it applicable to quantum settings.

\subsection{Organization of the Paper} 
In~\S{}\ref{section:prelim} we fix the notations for quantum computation
and briefly review the semantical techniques used
later.
In~\S{}\ref{section:syntaxAndOprSem} we introduce our target language
$\Hoq$ and its operational semantics. The (subtle but important)
differences from its predecessor are discussed, too.
 In~\S{}\ref{section:quantumBranchingMonad} we introduce the
\emph{quantum branching monad} $\Q$ on $\Sets$; this is our choice for
the monad $B$ in Figure~\ref{figure:constructionOfModel}.  The resulting
linear category $\PER_{\Q}$ is described,
too. In~\S{}\ref{section:denotationalModel} we interpret $\Hoq$ in this
category; finally  in~\S{}\ref{section:adequacy} we  prove adequacy of
the denotational model.

Some details and proofs are deferred to appendices. They are found
in~\cite{HasuoH15APAPExtendedAtArxiv}.

\section{Preliminaries}\label{section:prelim}

We denote the syntactic equality by $\equiv$.

\subsection{Quantum Computation}\label{subsection:prelimQM}
We follow Kraus' formulation~\cite{Kraus83} of quantum mechanics, which
is by now standard and is used in e.g.~\cite{NielsenC00,Selinger04}.
For proofs and more detailed explanation, our principal
reference is the standard textbook~\cite[Chap.~2 \&
Chap.~8]{NielsenC00}.  

\begin{mynotation}
  $\IM_{m}$ denotes the $m\times m$
 identity matrix; $A^{\dagger}$ denotes a matrix $A$'s adjoint (i.e.\
 conjugate transpose).
\end{mynotation}

\subsubsection{Density Matrices}
 We motivate the formalism of \emph{density matrices} as one that
generalizes the state vector formalism. See~\cite[\S{}2.4]{NielsenC00}
for more details; for our developments later, it is crucial that we
allow density matrices $\rho$ such that the trace $\trace(\rho)$ is possibly less
than $1$.

A mathematical representation of a state of a quantum mechanical system
is standardly given by a normalized vector $\ket{v}$ with the norm
$\|\ket{v}\|=1$ in some Hilbert space $\mathcal{H}$. As is usual in the
context of quantum information and quantum computation, we will
be working exclusively with finite-dimensional systems
($\mathcal{H}\cong\C^{n}$ for some $n\in\nat$). 
As an example let us consider the following \emph{Bell state}:
 \begin{equation}\label{eq:stateVectorExample}
 \ket{\Phi^{+}}
\;=\;
 \frac{1}{\sqrt{2}}
 \bigl(\,
 \ket{00}+\ket{11}\,\bigr)
\;=\;
 \frac{1}{\sqrt{2}}
 \bigl(\,
 \ket{0_{1}0_{2}}+\ket{1_{1}1_{2}}\,\bigr)
\enspace.
\end{equation}
The vector $\ket{\Phi^{+}}\in\C^{4}$ is a state of a 2-qubit system; we shall sometimes use explicit subscripts $1,2$ as in $0_{2}$ above
 to designate which of the two qubits we are referring to.

We now consider the measurement of the first qubit with respect to the
 basis consisting of $\ket{0_{1}}$ and $\ket{1_{1}}$. The outcome is
 $\ket{0_{1}}$ or $\ket{1_{1}}$ with the same probabilities $1/2$; in
 each case the state vector gets \emph{reduced} and becomes
 $\ket{0_{1}0_{2}}$ or $\ket{1_{1}1_{2}}$, respectively. In other words,
 the result of the measurement is a probability distribution
 \begin{displaymath}
  \bigl[\,
   \ket{0_{1}0_{2}} \mapsto \frac{1}{2}
   \enspace,
  \quad
   \ket{1_{1}1_{2}} \mapsto \frac{1}{2}
  \,\bigr]
 \end{displaymath}
 over state
 vectors. Such is called an \emph{ensemble}.

\emph{Density matrices} generalize state vectors and also encompass
ensembles; in other words, they represent both \emph{pure} and
\emph{mixed} states. Given an ensemble
\begin{displaymath}
 \bigl[\,
 \ket{v_{i}}\mapsto p_{i}
\,\bigr]
 \quad\text{with}\quad
 p_{i}\in\real_{\ge 0} \quad\text{and}\quad
 \sum_{i}p_{i}\le 1\enspace,
\end{displaymath}
the corresponding density matrix is defined to be
\begin{displaymath}
 \sum_{i}p_{i}\ket{v_{i}}\bra{v_{i}}\enspace,
\end{displaymath}
where $\bra{v_{i}}=\ket{v_{i}}^{\dagger}$ as usual. For example, 
the Bell state  $\ket{\Phi^{+}}$ is represented by the density
matrix
 \begin{equation}\label{eq:DMExampleForBellState}
  \ket{\Phi^{+}}\bra{\Phi^{+}}
  \;=\;
  \frac{1}{2}\left(\footnotesize
 \begin{array}{cccc}
  1&0 &0 &1 \\
  0&0 &0 &0 \\
  0&0 &0 &0 \\
  1&0 &0 &1 
 \end{array}
\right)\enspace;
 \end{equation}
 the ensemble
 \begin{math}
  \bigl[\,
   \ket{0_{1}0_{2}} \mapsto \frac{1}{2}
   \enspace,
  \;
   \ket{1_{1}1_{2}} \mapsto \frac{1}{2}
  \,\bigr]
 \end{math}
 that results from the measurement above is represented by
 \begin{equation}\label{eq:DMExampleForEnsemble}
  \frac{1}{2}\left(\footnotesize
 \begin{array}{cccc}
  1&0 &0 &0 \\
  0&0 &0 &0 \\
  0&0 &0 &0 \\
  0&0 &0 &1 
 \end{array}
\right)\enspace.
 \end{equation}

\begin{mynotation}\label{notation:lexicoIndex}
 In~(\ref{eq:DMExampleForBellState}--\ref{eq:DMExampleForEnsemble})
 we followed the common \emph{lexicographic indexing} convention: the
 matrices are with respect to the basis vectors
 $\ket{00},\ket{01},\ket{10}$ and $\ket{11}$ in this order.  See
 e.g.~\cite[\S{}3.2]{Selinger04}. This convention will be used in the
 rest of the paper too.
\end{mynotation}

Here is an ``axiomatic'' definition of density matrix. Every density matrix
arises in the way described above from some ensemble;
see~\cite[Theorem~2.5]{NielsenC00}. 
\begin{mydefinition}[Density matrix]
\label{definition:densityMatrix}
An \emph{$m$-dimensional density matrix} is an $m\times m$ matrix
 $\rho\in\C^{m\times m}$
which is positive and satisfies
 $\trace(\rho)\in [0,1]$. Here $[0,1]$ denotes the unit interval.
The set of all $m$-dimensional density matrices is denoted by $\DM_{m}$.
\end{mydefinition}
\noindent Note that we allow density matrices with trace less than
$1$. This will be the case typically when ``some probability is
missing,'' such as when the original ensemble
$\bigl[\,\ket{v_{i}}\mapsto p_{i}\,\bigr]$ is such that
$\sum_{i}p_{i}<1$. This generality turns out to be very useful later
when we model classical control structures that depend on the outcome of
measurements.
One can also recall, as a related phenomenon in program
semantics, that the semantics of possibly \emph{diverging} probabilistic
program is given by a \emph{sub}distribution where the probabilities can
add
up to less than $1$. The missing probability is then that  for divergence.

We note that if a quantum system consists of $N$ qubits, then the system
is $2^{N}$-dimensional (we can take a basis that consists of 
$\ket{0_{1}0_{2}\dotsc 0_{N}}, \ket{0_{1}0_{2}\dotsc 1_{N}}, \dotsc,
\ket{1_{1}1_{2}\dotsc 1_{N}}$). In this case a density matrix that
represents a (pure or mixed) state will be $2^{N}\times 2^{N}$.



The following order is standard and used e.g.\
in~\cite{NielsenC00,Selinger04}.
\begin{mydefinition}[L\"{o}wner partial order]
\label{definition:LoewnerPartialOrder}
 The order $\Lle$ on the set $\DM_{m}$ of density matrices is defined by:
 $\rho\Lle \sigma$ if and only if $\sigma-\rho$ is a positive matrix.
\end{mydefinition}
A prototypical situation in which we have $\rho\sqsubseteq \sigma$ is when
\begin{itemize}
 \item 
 $\rho$ arises from an ensemble $\bigl[\,\ket{v_{i}}\mapsto
       p_{i}\,\bigr]_{i\in I}$;
 \item 
 $\sigma$ arises from an ensemble $\bigl[\,\ket{v_{i}}\mapsto
       q_{i}\,\bigr]_{i\in I}$; and
 \item $p_{i}\le q_{i}$ for each $i\in I$.
\end{itemize}
That is, when, in comparing $\rho$ and $\sigma$ thought of as ensembles,
$\rho$ has some components missing. 

The following fact is crucial in this work. 
 It is proved
 in~\cite[Proposition~3.6]{Selinger04} using a translation into quadratic
 forms; in \ref{appendix:CPOStrOfDensityMatrix} we present 
 another proof using matrix norms.
\begin{mylemma}\label{lem:LoewnerPartialOrderIsOrderAndCPO}
 The relation $\Lle$  in Definition~\ref{definition:LoewnerPartialOrder} is
 indeed a partial order. Moreover it is an $\omega$-CPO: 
 any increasing $\omega$-chain $\rho_{0}\Lle
 \rho_{1}\Lle\cdots$ in $\DM_{m}$ has the least upper bound.
 \myqed
\end{mylemma}


\subsubsection{Quantum Operations}
Built on top of the density matrix formalism, the notion of
\emph{quantum operation} captures the general concept of ``what we can
do to quantum systems,'' unifying preparation, unitary transformation and
measurement. See~\cite[Chap.~8]{NielsenC00} for details.

\begin{mydefinition}[Quantum operation, QO]
 \label{definition:quantumOperation}
 A \emph{quantum operation (QO)} from an $m$-dimensional system to an $n$-dimensional
 system is a mapping $\Eop:\DM_{m}\to \DM_{n}$ subject to the following
 axioms.
\begin{enumerate}
 \item (Trace condition)\; 
\begin{displaymath}
   \frac{\trace\bigl(\Eop(\rho)\bigr)}{\trace(\rho)}\in [0,1]
  \quad\text{for any $\rho\in
       \DM_{m}$ such that $\trace(\rho)>0$.}
\end{displaymath} 
\item\label{item:QODefLinearity} (Convex linearity) 
      Let
 $(\rho_{i})_{i\in I}$ be a family of $m$-dimensional density
       matrices; and  $(p_{i})_{i\in I}$ be a probability
      subdistribution (meaning $p_{i}\in\real_{\ge 0}$ and $\sum_{i}p_{i}\le 1$).
      Then:
       \begin{displaymath}
	\Eop\bigl(\,\sum_{i\in I}p_{i}\rho_{i}\,\bigr)
	\;=\;
	\sum_{i\in I}p_{i}\Eop(\rho_{i})\enspace.
       \end{displaymath}
     Here $I$ is a possibly infinite index set---we can assume that $I$ is at
     most countable since a discrete probability subdistribution
       $(p_{i})_{i\in I}$ necessarily has a countable support. From this
     and that the trace of $\rho_{i}$ is bounded by $1$,  it
     easily follows that the infinite sums on both sides are well-defined.
 \item (Complete positivity)  An arbitrary 
       ``extension'' of $\Eop$ of the form $\id_{k}\otimes \Eop: M_{k}\otimes M_{m}\to
       M_{k}\otimes M_{n}$ carries a positive matrix to a positive
       one.
       (Here $\id_{k}\colon M_{k}\to M_{k}$ is the identity function.)
       In particular, so does $\Eop$ itself.
\end{enumerate}
The set of QOs of the type $\DM_{m}\to \DM_{n}$ shall be
denoted by $\QO_{m,n}$.
\end{mydefinition}
 The definition slightly differs from the one
 in~\cite[\S{}8.2.4]{NielsenC00}. This difference---which is technically
 minor but conceptually important---is because we allow
 density matrices with trace less than $1$.


QO has two alternative definitions other than the above ``axiomatic''
one. One is by the \emph{operator-sum representation}
$\sum_{i}A_{i}(\place) A_{i}^{\dagger}$ and useful in concrete
calculations. This is presented below.
The other is ``physical'' and describes a QO as a
certain succession of operations to a system, namely: combining with an auxiliary
quantum state; a unitary transformation; and measurement.  
See~\cite[\S{}8.2]{NielsenC00} for further details.
\begin{myproposition}[Operator-sum representation]
\label{proposition:oprSumRepr}
 A mapping $\Eop:\DM_{m}\to \DM_{n}$ is a QO if and only if it can 
 be represented in the form
 \begin{equation}\label{eq:oprSumRepr}
  \Eop(\rho) = \sum_{i\in I} E^{(i)}\rho (E^{(i)})^{\dagger}\enspace,  
 \end{equation}
 where $I$ is a finite index set, $E^{(i)}$ is an $n\times m$ matrix for
 each $i$, and
\begin{displaymath}
  \sum_{i\in I}\bigl(E^{(i)}\bigr)^{\dagger}E^{(i)}
  \;\Lle\; \IM_{m}\enspace.
\end{displaymath} 
Here  the order $\Lle$ refers to the one in
 Definition~\ref{definition:LoewnerPartialOrder}.
\end{myproposition}
\begin{myproof}
See~\cite[\S{}8.2.4]{NielsenC00}. \myqed 
\end{myproof}
We call the right-hand side of~(\ref{eq:oprSumRepr}) an
\emph{operator-sum representation} of a QO $\Eop$. 
Given a QO $\Eop$, its operator-sum representation is not uniquely
determined. However:
\begin{mydefinition}[The matrix $M(\Eop)$]
 \label{definition:theMatrixMofE}
 For a QO $  \Eop = \sum_{i} E^{(i)}(\place) (E^{(i)})^{\dagger}$, we
 define an $m\times m$ matrix $M(\Eop)$ by
\begin{displaymath}
 M(\mathcal{E})\;:=\;\sum_{i}(E^{(i)})^{\dagger}E^{(i)}\enspace.
\end{displaymath}
\end{mydefinition}
\begin{mylemma}
 The matrix $M(\Eop)$ for a QO $\Eop$ does not depend on the choice of an operator-sum representation.
\end{mylemma}
\begin{myproof}
 There is only ``unitary freedom'' in the choice of an operator-sum
 representation~\cite[Theorem~8.2]{NielsenC00}: given two operator-sum
 representations
 \begin{displaymath}
 \Eop = \textstyle\sum_{i} E^{(i)}(\place) (E^{(i)})^{\dagger}
  = \textstyle\sum_{j} F^{(j)}(\place) (F^{(j)})^{\dagger}\enspace,
 \end{displaymath}
 there exists a unitary matrix $U=(u_{i,j})_{i,j}$ such that
 $E^{(i)}=\sum_{j}u_{i,j}F^{(j)}$.
 We have
 \begin{align*}
  \textstyle\sum_{i} (E^{(i)})^{\dagger}E^{(i)}
 &=
    \textstyle\sum_{i} 
   \bigl(
    \textstyle\sum_{j} u^{*}_{i,j}(F^{(j)})^{\dagger}
  \bigr)
   \bigl(
    \textstyle\sum_{k} u_{i,k}F^{(k)}
  \bigr)
 \\
 &=
    \textstyle\sum_{j,k} \,
  \bigl(\textstyle\sum_{i}u^{*}_{i,j}u_{i,k}\bigr)\,(F^{(j)})^{\dagger}F^{(k)}
 \\
 &=
    \textstyle\sum_{j} \,
  (F^{(j)})^{\dagger}F^{(j)}\enspace,
 \end{align*}
where the last equality is because 
$\textstyle\sum_{i}u^{*}_{i,j}u_{i,k}$ is the $(j,k)$-entry of
 $U^{\dagger}U = \IM$.
 \myqed
\end{myproof}

The following property is immediate.
\begin{mylemma}\label{lem:MPreservesSums}
The operation $M(\place)$ preserves sums.  More precisely,
 let $(\Eop_{i})_{i\in I}$ be a
 (at most countably infinite)
 family of quantum operations of
the same dimensions; assume that $\sum_{i}\Eop_{i}$ is again a quantum
 operation. Then
 \begin{math}
  M\bigl(\sum_{i}\Eop_{i}\bigr)
 =  \sum_{i} M(\Eop_{i})
 \end{math}. \myqed
\end{mylemma}

We exhibit some concrete QOs. Application of a unitary transformation
 $U$ to a quantum state (pure or mixed) that is represented by a density
 matrix $\rho$ corresponds to a QO
 \begin{displaymath}
  U(\place)U^{\dagger}
  \;:\;
  \rho
  \;\longmapsto\;
  U\rho U^{\dagger}\enspace.
 \end{displaymath}
 For illustration consider a special case where $\rho=\ket{v}\bra{v}$;
 the outcome is
 \begin{displaymath}
  U\ket{v}\bra{v}U^{\dagger}
  =
  U\ket{v}\bigl(\,\ket{v}\,\bigr)^{\dagger}U^{\dagger}
  =
  U\ket{v}\bigl(\,U\ket{v}\,\bigr)^{\dagger}\enspace,
 \end{displaymath}
 i.e.\ the density matrix that is induced by the state vector $U\ket{v}$.
 
 We explain measurement using a concrete example. Recall the Bell state
 $\ket{\Phi^{+}}$ in~(\ref{eq:stateVectorExample}) and the
 corresponding density matrix $\ket{\Phi^{+}}\bra{\Phi^{+}}$
 in~(\ref{eq:DMExampleForBellState}).  Consider now the
 measurement of the first qubit with respect to the basis consisting of
 $\ket{0_{1}}$ and $\ket{1_{1}}$.
The corresponding QO is
 \begin{equation}\label{eq:QOExampleForMeas}
  \bra{0_{1}}\place\ket{0_{1}} +
  \bra{1_{1}}\place\ket{1_{1}}\enspace,
 \end{equation}
 where, for example, $\ket{0_{1}}$ is concretely given by
 \begin{displaymath}
  \ket{0_{1}} =
  \ket{0}\otimes \IM_{2} =
  \left(\footnotesize
  \begin{array}{c}
   1\\ 0
  \end{array}
\right)
  \otimes 
  \left(\footnotesize
  \begin{array}{cc}
   1 &0 \\
   0&1 
  \end{array}
\right)
  =
  \left(\footnotesize
  \begin{array}{cc}
   1 &0 \\
   0&1 \\
   0&0 \\
   0&0
  \end{array}
\right)\enspace.
 \end{displaymath}
 Here we followed
 the lexicographic indexing
 convention (Notation~\ref{notation:lexicoIndex}).
 
Applying the measurement to the Bell state $\ket{\Phi^{+}}$---i.e.\
 applying the QO in~(\ref{eq:QOExampleForMeas}) to the density
 matrix in~(\ref{eq:DMExampleForBellState})---results in the
 following density matrix.
\begin{align*}
& \left(\footnotesize
 \begin{array}{cccc}
  1&0 &0 &0 \\
  0&1 &0 &0 
 \end{array}
\right)
   \ket{\Phi^{+}}\bra{\Phi^{+}}
  \left(\footnotesize
  \begin{array}{cc}
   1 &0 \\
   0&1 \\
   0&0 \\
   0&0
  \end{array}
\right)
+
 \left(\footnotesize
 \begin{array}{cccc}
  0&0 &1 &0 \\
  0&0 &0 &1
 \end{array}
\right)
  \ket{\Phi^{+}}\bra{\Phi^{+}}
  \left(\footnotesize
  \begin{array}{cc}
   0 &0 \\
   0&0 \\
   1&0 \\
   0&1
  \end{array}
\right)
\\
&=\;
\frac{1}{2}
\left(\footnotesize
\begin{array}{cc}
 1&0 \\
 0&0
\end{array}
\right)
+
\frac{1}{2}
\left(\footnotesize
\begin{array}{cc}
 0&0 \\
 0&1
\end{array}
\right)
\;=\;
\frac{1}{2}
\ket{0}\bra{0}
+
\frac{1}{2}
\ket{1}\bra{1}\enspace.
\end{align*}
This density matrix represents the ensemble 
$\bigl[\,\ket{0}\mapsto
 \frac{1}{2},\, \ket{1}\mapsto \frac{1}{2}\,\bigr]$, or
$\bigl[\,\ket{0_{2}}\mapsto
 \frac{1}{2},\, \ket{1_{2}}\mapsto \frac{1}{2}\,\bigr]$ to be more
 explicit about which qubit in the original system we are referring to.


Two remarks are in order. Firstly, in the ensemble we have obtained, the
 first qubit in the original system has been discarded. This is a matter
 of choice: we could use a different QO that does retain the
 first qubit, resulting in the density matrix
\begin{displaymath}
   \frac{1}{2}\left(\footnotesize
 \begin{array}{cccc}
  1&0 &0 &0 \\
  0&0 &0 &0 \\
  0&0 &0 &0 \\
  0&0 &0 &1 
 \end{array}
\right)
\quad
\text{
that corresponds to the ensemble
$\bigl[\,\ket{0_{1}0_{2}}\mapsto
 \frac{1}{2},\, \ket{1_{1}1_{2}}\mapsto \frac{1}{2}\,\bigr]$.
}
\end{displaymath}
 Our choice
 above is because of the type $\qbit\limp\bit$ (rather than
 $\qbit\limp\bit\otimes \qbit$) of the measurement primitive in our
 calculus. 

The second remark is numbered for future reference.
\begin{myremark}\label{remark:splitMeasIntoProjs}
 For the purpose of denotational semantics introduced later, we
 find it useful to split up a measurement into two separate
 ``projections,'' each of which corresponds to a possible outcome of the
 measurement. For example,  the QO
 in~(\ref{eq:QOExampleForMeas}) would rather be thought of as a
 pair
 of projection QOs
 \begin{equation}\label{eq:QOExampleForMeasSeparated}
   \bra{0_{1}}\place\ket{0_{1}}
     \quad\text{and}\quad
  \bra{1_{1}}\place\ket{1_{1}}\enspace.
 \end{equation}
 The two projection QOs describe ``what happens to the quantum
 state when the measurement outcome is $\ket{0_{1}}$ (or $\ket{1_{1}}$,
 respectively).''  Having them separate allows us to model classical
 control structures that rely on the outcome of quantum measurements.
 This point will be more evident in~\S\ref{section:quantumBranchingMonad}.

 Given a density matrix $\rho$, the probability
 for observing $\ket{0_{1}}$ or $\ket{1_{1}}$ can then be calculated as
 \begin{equation}\label{eq:exampleTraceOut}
  \trace\bigl(\, \bra{0_{1}}\rho\ket{0_{1}}\,\bigr)
  \quad\text{or}\quad
  \trace\bigl(\, \bra{1_{1}}\rho\ket{1_{1}}\,\bigr)\enspace,
 \end{equation}
 respectively. For example, in the special case where
 $\rho=\ket{v}\bra{v}$, 
 \begin{displaymath}
  \trace\bigl(\, \bra{0_{1}}\rho\ket{0_{1}}\,\bigr)
  \;=\;
  \trace\bigl(\, 
 \bra{0_{1}}v\rangle 
 \bigl(\,
 \bra{0_{1}}v\rangle 
 \,\bigr)^{\dagger}
 \,\bigr)
  \;=\;
  \bigl\|\,\bra{0_{1}}v\rangle \,\bigr\|^{2}\enspace.
 \end{displaymath}
 See~\cite[\S{}8.2]{NielsenC00} for more details.
 This way we let density matrices implicitly carry probabilities
 (specifically by their trace values). This is why we allow density
 matrices with trace less than $1$. 
\end{myremark}

We extend the order  $\Lle$ in Definition~\ref{definition:LoewnerPartialOrder} 
in a pointwise manner
to obtain an order between QOs. This is done also 
in~\cite{Selinger04}.  
\begin{mydefinition}[Order $\Lle$ on $\QO_{m,n}$]
\label{definition:pointwiseorderOnQO}
Given $\Eop,\Fop\in\QO_{m,n}$, we define $  \Eop\Lle \Fop$
 if and only if $  \Eop(\rho)\Lle \Fop(\rho)$
for each $\rho\in \DM_{m}$.
  The latter $\Lle$ is the L\"{o}wner partial order
 (Definition~\ref{definition:LoewnerPartialOrder}).
\end{mydefinition}
\begin{myproposition}\label{proposition:QOisCPO}
 The order $\Lle$ on $\QO_{m,n}$ is an $\omega$-CPO.
\end{myproposition}
\begin{myproof}
 See \ref{appendix:CPOStrOfDensityMatrix};
 also~\cite[Lemma~6.4]{Selinger04}.  \myqed
\end{myproof}

For illustration, notice that  for any density matrix $\rho$,
\begin{displaymath}
 \bra{0_{1}}\rho\ket{0_{1}}
 \;\sqsubseteq\;
  \bra{0_{1}}\rho\ket{0_{1}} +
 \bra{1_{1}}\rho\ket{1_{1}}
 \quad\text{and}\quad
 \bra{1_{1}}\rho\ket{1_{1}}
 \;\sqsubseteq\;
  \bra{0_{1}}\rho\ket{0_{1}} +
 \bra{1_{1}}\rho\ket{1_{1}}\enspace,
\end{displaymath}
in the setting of~(\ref{eq:QOExampleForMeasSeparated}). This
establishes
\begin{displaymath}
 \bra{0_{1}}\place\ket{0_{1}}
 \;\sqsubseteq\;
  \bra{0_{1}}\place\ket{0_{1}} +
 \bra{1_{1}}\place\ket{1_{1}}
 \quad\text{and}\quad
 \bra{1_{1}}\place\ket{1_{1}}
 \;\sqsubseteq\;
  \bra{0_{1}}\place\ket{0_{1}} +
 \bra{1_{1}}\place\ket{1_{1}}
\end{displaymath}
where $\Lle$ is the order of Definition~\ref{definition:pointwiseorderOnQO}.
This example is prototypical of our use of the L\"{o}wner partial order
(Definition~\ref{definition:LoewnerPartialOrder}
\&~\ref{definition:pointwiseorderOnQO}): $\Eop\sqsubseteq\Fop$ means
that $\Eop$ is a projection (or ``partial measurement'') that is a
``component'' of $\Fop$.


\subsection{Monads for Branching}
\label{subsection:prelimMonadForBranching}
The notion of \emph{monad} is standard in category theory.  In
 computer science, after Moggi~\cite{Moggi91a}, the notion has been
 used for encapsulating \emph{computational effect} in functional
 programming. One such monad denoted by $T$ appears in this paper---at the
 last stage, as part of our categorical model.

 There is another monad $\Q$---called the \emph{quantum branching
 monad}---that marks the beginning of our development. It is introduced in~\S{}\ref{section:quantumBranchingMonad}.
 The idea is drawn
 from the coalgebraic study of state-based systems (see
 e.g.~\cite{Rutten00a,Jacobs12CoalgBook} for introduction); in
 particular from the use of a monad $B$ on $\Sets$ for modeling
 \emph{branching}, e.g.\ in~\cite{HasuoJS07b}.  
\begin{myexample}\label{example:branchingMonads}
 We list some examples of
 such ``branching monads'' $B$.
 \begin{itemize}
  \item The \emph{lift monad} 
\begin{displaymath}
 	\lift X=1+X
\end{displaymath}
 models potential
 	nontermination. Its unit 
	$\eta^{\lift}:X\to 1+X$ and multiplication
	$\mu^{\lift}:1+(1+X)\to 1+X$ are obvious.
  \item The \emph{powerset monad} 
\begin{displaymath}
 \pow X =\{X'\subseteq X\}
\end{displaymath} 
models
 	nondeterminism. Its unit 
	$\eta^{\pow}:X\to \pow X$ returns a singleton set and its multiplication
	$\mu^{\pow}:\pow(\pow X)\to \pow X$ takes the union.
  \item The \emph{subdistribution monad} 
\begin{displaymath}
 	\dist X =\bigl\{c:X\to[0,1]\,\bigl|\bigr.\,
 	\sum_{x}c(x)\le 1\bigr\}
\end{displaymath}	
models probabilistic branching.
	Its unit 
	$\eta^{\dist}:X\to \dist X$ carries $x\in X$ to the so-called 
	\emph{Dirac distribution} $[x\mapsto 1]$; and
	its multiplication
	$\mu^{\dist}:\dist(\dist X)\to \dist X$ ``suppresses'' a
 	distribution
	over distributions into a distribution (see also
 	(\ref{eq:multDist}) below):
       \begin{displaymath}
		\mu^{\dist}_{}(\xi) = \lambda x. \sum_{c\in \dist X}\xi(c)\cdot c(x)\enspace.
       \end{displaymath}
 \end{itemize}
\end{myexample}

 The monad structures (units and multiplications) of
the operations $\lift,\pow,\dist$ in the above list
 have a natural meaning in terms of branching. Among
others, a multiplication $\mu$ collapses ``branching twice'' into
``branching once,'' abstracting the internal branching structure.
 For example, $\pow$'s multiplication
\begin{displaymath}
    \mu^{\pow}_{X}:\pow\pow X\longrightarrow \pow X\enspace,
  \qquad\text{like}\quad
  \bigl\{\,\{x,y\},\{z\}\,\bigr\}\longmapsto \{x,y,z\}
\end{displaymath}
 can be understood as follows.
 \begin{displaymath}
   \vcenter{\xymatrix@R=-.5em@C=1.7em{
  &&
   {x}
  \\
  &
   {\bullet}
              \ar@{~>}[ru]\ar@{~>}[rd]
  \\
   {\bullet}
              \ar@{~>}[ru]\ar@{~>}[rdd]
  &&
   {y}
  \\
  \\
  &
   {\bullet}
              \ar@{~>}[r]
  &
   {z}
 }}
  \qquad
  \stackrel{\mu}{\longmapsto}
  \qquad
 \vcenter{\xymatrix@R=-.2em@C=4em{
  &
   {x}
  \\
   {\bullet}
              \ar@/^/@{~>}[ru]
	      \ar@/_/@{~>}[rd]
	      \ar@{~>}[r]
  &
   {y}
  \\
  &
   {z}
 }}
\end{displaymath}
For $\dist$, its multiplication
\begin{displaymath}
    \mu^{\dist}_{X}:\dist\dist X\longrightarrow \dist X\enspace,
  \qquad\text{like}\quad
	\left[
         \begin{array}{ccc}
	  \left[
           \begin{array}{l}
	    x\mapsto 1/2
           \\
            y\mapsto 1/2
	   \end{array}
          \right]
         &
          \mapsto 
         &
          1/3
         \\[+.3em]
	  [
	    z\mapsto 1
          ]
         &
          \mapsto 
         &
          2/3
	 \end{array}
	\right]
	   \;
	   \stackrel{\mu}{\longmapsto}
	   \;
        \left[
         \begin{array}{c}
	  x\mapsto 1/6
         \\
	  y\mapsto 1/6
         \\
	  z\mapsto 2/3
	 \end{array}
	\right]
\end{displaymath}
can be understood as follows.
       \begin{align}\label{eq:multDist}
	 &\vcenter{\xymatrix@R=-.3em@C=2em{
	   &&
	    {x}
	   \\
	   &
	    {\bullet}
		       \ar@{~>}[ru]^{1/2}\ar@{~>}[rd]_{1/2}
	   \\
	    {\bullet}
		       \ar@{~>}[ru]^{1/3}\ar@{~>}[rdd]_{2/3}
	   &&
	    {y}
	   \\
	   \\
	   &
	    {\bullet}
		       \ar@{~>}[r]_{1}
	   &
	    {z}
	 }}
	   \qquad
	   \stackrel{\mu}{\longmapsto}
	   \qquad
	  \vcenter{\xymatrix@R=.6em@C=4em{
	   &
	    {x}
	   \\
	    {\bullet}
		       \ar@{~>}[ru]^{1/6}\ar@{~>}[rd]_{2/3}\ar@{~>}[r]|{1/6}
	   &
	    {y}
	   \\
	   &
	    {z}
	 }}\enspace,
      \end{align}

Furthermore, these monads come with natural order structures which turn
out to be $\omega$-CPOs. This is exploited in~\cite{HasuoJS07b} to
prove---using a domain-theoretic technique from~\cite{SmythP82}---that
a final coalgebra in $\Kleisli{B}$ coincides with an initial algebra in
$\Sets$. The final coalgebra in $\Kleisli{B}$ thus identified provides a
fully abstract semantic domain for \emph{trace semantics}---execution trace-based
(i.e.\ ``linear-time'') semantics for state-based systems that is coarser than
(``branching-time'') \emph{bisimilarity}.  
See~\cite{HasuoJS07b}.   


\subsection{Geometry of Interaction}
\label{subsection:prelimGoI} Girard's \emph{Geometry of Interaction
(GoI)}~\cite{Girard89GoI} is an interpretation of proofs in linear logic
in terms of dynamic information flow.  It seems GoI's position as a tool
in denotational semantics is close to that of the game-based
interpretations of computation~\cite{AbramskyJM00,HylandO00}. Its
original formulation~\cite{Girard89GoI} utilizes a $C^{*}$-algebra;
later in~\cite{Mackie95} the same idea is given a more concrete
operational representation which is now commonly called \emph{token
machines}.  For an introduction to GoI, our favorite reference
is~\cite{Pinto01thesis}.

Besides these presentations of GoI by $C^{*}$-algebras and token
machines, particularly important for our developments is the
\emph{categorical} axiomatization of GoI by Abramsky, Haghverdi and
Scott~\cite{AbramskyHS02}.  They
 isolated some axiomatic properties of a category $\C$ on which one can
 build a GoI interpretation. Such a category $\C$ (together with some
 auxiliary data) is called a \emph{GoI situation}
 in~\cite{AbramskyHS02}: among other conditions, a crucial one is that
 $\C$ is a \emph{traced symmetric monoidal category
 (TSMC)}~\cite{JoyalSV96}. Then applying what they call the \emph{GoI
 construction} $\mathcal{G}$---it is isomorphic to the
 \emph{$\Int$-construction} in~\cite{JoyalSV96}---one is led to a
 compact closed category $\mathcal{G}(\C)$ of ``bidirectional
 computations'' or ``(stateless) games.''

The resulting category $\mathcal{G}(\C)$ comes close to a categorical
model of linear logic---a so-called \emph{linear
category}~\cite{Bierman95a,BentonW96}---but not quite, lacking an
appropriate operator that models the $\bang$ modality of linear logic. 
A step ahead is taken in~\cite{AbramskyHS02}: they extract a \emph{linear
combinatory algebra (LCA)} from $\mathcal{G}(\C)$. The notion of LCA is a variation 
of \emph{partial combinatory algebra (PCA)} and corresponds to a
Hilbert-style axiomatization of linear logic, including the
$\bang$ modality (see Definition~\ref{definition:LCA} later). 

A thorough introduction to the rich and deep theory of GoI is certainly 
out of the current paper's scope. We shall nevertheless provide further
intuitions---in a way tailored to categorical GoI and our use of
it---later in~\S{}\ref{subsection:LCAviaCategoricalGoI}. 

\begin{myremark}[Three ``traces'']
 \label{remark:threeTraces}
 In this paper we use three different notions of trace. One is the trace
 operator in linear algebra; in quantum mechanics a probability for a
 certain observation outcome is computed by ``tracing out'' a density
 matrix, like in~(\ref{eq:exampleTraceOut}). Another ``trace'' is
 in trace semantics in the context of process theory.
 See (the last paragraph of)~\S{}\ref{subsection:prelimMonadForBranching}. The other is in
 traced monoidal categories that play a central role in categorical
 GoI~\cite{AbramskyHS02}.

 These three notions are not unrelated. The first ``linear algebra
 trace'' is an example of the last ``monoidal trace'': namely in the
 category $\fdVect$ of finite-dimensional vector spaces and linear maps
 where the monoidal structure is given by the tensor product $\otimes$
 of vector spaces. The second trace---which we would like to call
 ``coalgebraic trace''---also yields an example of ``monoidal trace.''
 This result, shown in~\cite{Jacobs10trace}, will be exploited for
 construction of a traced monoidal category $\Kleisli{\Q}$ on which we
 run the machinery of categorical
 GoI. See~\S{}\ref{section:quantumBranchingMonad}.
\end{myremark}

\subsection{Realizability}
\label{subsection:prelimRealizability}
Roughly speaking, 
an LCA can be thought of as a collection of untyped closed 
linear $\lambda$-terms.  LCAs are, therefore, for interpreting
\emph{untyped} calculi.

What turns such a combinatory algebra into a model of a \emph{typed}
calculus is the technique of \emph{realizability}. It dates back to
Kleene; and its use in denotational semantics of programming languages
is advocated e.g.\ in~\cite{Longley94}.
We shall be based on its formulation found
in~\cite{AbramskyL05}. It goes as follows. Starting from 
an LCA $A$, we define the category $\PER_{A}$ of \emph{partial equivalence
relations (PERs)} on $A$; a PER on $A$ is roughly a subset of $A$ with some of
its elements mutually identified. An arrow of $\PER_{A}$ is represented 
by a \emph{code} $c\in A$.\footnote{Another standard technique is to use
\emph{$\omega$-sets} (also called \emph{assemblies}) in place of PERs.
This has been done for LCAs too; see~\cite{Hoshino07}.}

To turn $\PER_{A}$ into a model of a typed linear $\lambda$-calculus
(more specifically into a linear category) one needs type constructors
like $\otimes$, $\limp$ and $\bang$ on $\PER_{A}$. They can be introduced
by ``programming in untyped linear $\lambda$-calculus''---it is much
like encoding pairs, natural numbers, coproducts, etc.\ in the (untyped)
 $\lambda$-calculus ($\tuple{x,y}:=
\lambda z. zxy$, with a first projection $\lambda w. w(\lambda xy. x)$,
and so on).
More details can be found later in this paper; see also~\cite{AbramskyL05}.

This \emph{linear} version of realizability has been worked out e.g.\
in~\cite{AbramskyL05,Hoshino07}. The outcome of this construction is
a model of a typed linear $\lambda$-calculus---i.e.\ a model of linear
logic. There is a body of literature that seeks for what the latter
means exactly---including~\cite{CockettS99,Seely89,BentonW96,Bierman95a}---and there have
been a few different notions proposed.
It now seems that: the essence lies in what is called a
\emph{linear-non-linear adjunction} between a symmetric monoidal
closed category and a CCC; and that the different notions of model 
proposed earlier in
the literature are
different constructions of such an adjunction. See the extensive 
survey in~\cite{Mellies09}; also~\cite[\S{}9.6]{SelingerV09}.

Among those notions of ``model of linear logic,'' in this paper we use
the notion of \emph{linear category}~\cite{Bierman95a,BentonW96} since
its relationship to linear realizability has already been worked out
in~\cite{AbramskyL05}.

\section{The Language $\Hoq$}\label{section:syntaxAndOprSem}
Here we introduce our target calculus.
It is a variant of Selinger and Valiron's quantum
$\lambda$-calculus~\cite{SelingerV09}. The calculus shall be called
$\Hoq$---for \emph{higher-order quantum computation}.\footnote{$\Hoq$
is a minor modification of 
 the calculus $\qll$ that we used in the
conference version~\cite{HasuoH11}  of the current paper.}

The (only) major difference between $\Hoq$ and the calculus in~\cite{SelingerV09}
is separation of two tensors $\otimes$ and $\boxtimes$.  
\begin{itemize}
 \item The former
\emph{Hilbert space tensor} $\otimes$ denotes, as usual in quantum mechanics,
the tensor product $\mathcal{H}_{1}\otimes\mathcal{H}_{2}$ of Hilbert
spaces and designates compound quantum systems. 
 \item We use the latter
\emph{linear logic tensor} $\boxtimes$ for the ``multiplicative and''
connective in linear logic (hence in a linear $\lambda$-calculus). 
It is
       also denoted by $\otimes$ commonly in the literature; but we
choose to use the symbol $\boxtimes$.  
\end{itemize}
In fact, in $\Hoq$ the Hilbert space tensor $\otimes$ will not be visible since we
 let $\nqbit$ stand for $\qbit^{\otimes n}$. The difference between
 $\nqbit\boxtimes m\text{-}\qbit$ and $(n+m)\text{-}\qbit$ is: the
 former stands for two ($n$- and $m$-qubit) quantum states that are \emph{for
 sure not entangled}; the
 latter is for the composite system in which two states are
 \emph{possibly entangled}.

In contrast, in~\cite{SelingerV09} they use the same tensor operator
$\otimes$ for both---that is, the linear logic tensor is interpreted
using the Hilbert space tensor. The reason for this difference will be
explained in~\S\ref{subsection:HoqDesignChoice}, as well as the design
choices that we share with~\cite{SelingerV09}.

In this section we first introduce the syntax (including the type
system) of $\Hoq$ in~\S\ref{subsection:HoqSyntax}, followed by the 
operational semantics (\S\ref{subsection:HoqOprSemantics}). Then
in~\S{}\ref{subsection:HoqDesignChoice} we discuss our design choices,
especially the reason for the difference from the calculus
in~\cite{SelingerV09}. In~\S{}\ref{subsection:HoqSyntacticProperties} we
establish some properties on $\Hoq$, including some safety properties
such as substitution, subject reduction and progress.

\subsection{Syntax}\label{subsection:HoqSyntax}

\begin{mydefinition}[Types of {$\Hoq$}]\label{definition:TypesOfHoq}
The \emph{types} of $\Hoq$ are:
\begin{equation}\label{eq:HoqType}
  \begin{aligned}
  & A,B  
  \quad::=\quad
   \nqbit
    \mid
    \bang A
    \mid
    A\limp B
    \mid
    \top
    \mid
    A\boxtimes B
    \mid
    A+B\enspace, 
  \\
&\text{with conventions}\quad
  \qbit \;:\equiv\; 1\text{-}\qbit
 \quad\text{and}\quad
 \bit\; :\equiv\; \top + \top \enspace.
 \end{aligned}
\end{equation}
Here $n\in\nat$ is a natural number. 
\end{mydefinition}

\begin{mydefinition}[Terms of {$\Hoq$}]\label{definition:TermsOfHoq}
 The \emph{terms} of $\Hoq$ are:
\begin{equation}\label{eq:HoqTerm}
 \begin{aligned}
 & M,N,P\quad::= \quad
 \begin{array}[t]{l}
   x
  \mid \lambda x^{A}. M
  \mid MN \mid
 \\ 
  \tuple{M,N}
  \mid   \letcl{\tuple{x^{A},y^{B}}=M}{N}
  \mid
  \\ 
   *
  \mid
  \letcl{*=M}{N}
  \mid
 \\ 
 \injl^{B}\, M
  \mid \injr^{A}\, M
  \mid 
\matchcl{P}{(x^{A}\mapsto M\mid y^{B}\mapsto N)}
  \mid
  \\ 
  \letreccl{f^{A}x= M}{N}
   \mid 
  \\ 
\new_{\rho}
  \mid \meas^{n+1}_{i}
  \mid U
  \mid \cmp_{m,n}\enspace,
 \end{array}
\\
&\text{with conventions}
\quad
  \ttrue\;:\equiv\;\injl^{\top}(*)
 \quad\text{and}\quad
\ffalse\;:\equiv\;\injr^{\top}(*)\enspace.
\end{aligned}
\end{equation}
 Here $m,n\in\nat$ and $i\in [1,n+1]$ are natural numbers;
 $\rho\in \DM_{2^{k}}$ is a $2^{k}$-dimensional density matrix
 (corresponding to a $k$-qubit system); $U$ is a
 $2^{k}\times 2^{k}$ unitary matrix, for some $k\in\nat$; and $A$ and
 $B$ are \emph{type labels}.  The terms are almost the same as
 in~\cite{SelingerV09}; $\new_{\rho}$ designates \emph{preparation} of a
 new quantum state---more precisely deployment of some quantum
 apparatus that is capable of preparing the quantum state $\rho$. The additional
 \emph{composition} operator $\cmp$ will have the type
 $\mqbit\boxtimes\nqbit\limp(m+n)\text{-}\qbit$ and embed nonentangled
 states as possibly entangled states. For measurements we have operators
 $
 \meas^{1}_{1},
 \meas^{2}_{1},
 \meas^{2}_{2},
 \dotsc
$; $\meas^{n+1}_{i}$ takes an $(n+1)$-qubit system, measures its $i$-th
 qubit,
and returns the outcome (in the $\bit$ type) as well as the remaining
quantum state that consists of $n$ qubits.

 The set $\FV(M)$ of \emph{free variables} in $M$ is defined in the
 usual manner.
\end{mydefinition}

\begin{mydefinition}[Subtype relation $\subtp$ in $\Hoq$]\label{definition:SubtypeRelInHoq}
 For typing in $\Hoq$ we employ the same subtype relation $\subtp$ as
 in~\cite{SelingerV09} and implicitly track the $\bang $ modality (see~\S\ref{subsection:HoqDesignChoice}). The
 rules that derive $\subtp$ are as follows.
\begin{equation}\label{eq:HoqSubtypeRelation}
 { \renewcommand*{\arraystretch}{2}
 \begin{array}{l}
     \infer[(\kqbit)]{\bang ^{n}\kqbit\subtp \bang ^{m}\kqbit}{
  {n=0 \Rightarrow m=0}}
  \qquad
   \infer[(\top)]{\bang ^{n}\top\subtp \bang ^{m}\top}{{n=0 \Rightarrow m=0}}
 

  \\ 
   \infer[(\boxtimes)]{\bang ^{n}(A_{1}\boxtimes A_{2})\subtp
   \bang ^{m}(B_{1}\boxtimes B_{2})}{A_{1}\subtp B_{1} & A_{2}\subtp B_{2}
   &{n=0 \Rightarrow m=0}}
 
 \\
   \infer[(+) ]{\bang ^{n}(A_{1}+ A_{2})\subtp
   \bang ^{m}(B_{1}+ B_{2})}{A_{1}\subtp B_{1} & A_{2}\subtp B_{2}
   &{n=0 \Rightarrow m=0}}

  \\
   \infer[(\limp)]{\bang ^{n}(A_{1}\limp A_{2})\subtp
   \bang ^{m}(B_{1}\limp B_{2})}{B_{1}\subtp A_{1} & A_{2}\subtp B_{2}
   &{n=0 \Rightarrow m=0}} 
 \end{array}
 }
\end{equation}
 All the rules come with a condition $n=0 \Rightarrow m=0$,
  which is
 equivalent to $m=0\lor n\ge 1$.

\end{mydefinition}

We introduce the typing rules. They follow the ones
in~\cite{SelingerV09} and take subtyping into account. In the rules (\text{Ax.}1) and
(\text{Ax.}2)
the variables in the context can be thrown away, making the type system
\emph{affine} (weakening is allowed unconditionally while contraction is
regulated by $\bang$) rather than \emph{linear}. Due to these unconventional
features (subtyping and weakening) a derivation of a type judgment is
not necessarily unique in $\Hoq$---making it a delicate issue whether
the denotational semantics of a derivable type judgment is well-defined
(Lemma~\ref{lem:interpretationIsWellDfd}).
\begin{mydefinition}[Typing in $\Hoq$]\label{definition:TypingInHoq}
The typing rules of $\Hoq$ are as in Table~\ref{table:typingRules}. 
\begin{table}[tbp] \normalsize
\begin{displaymath}
\begin{array}{ll}
   \vcenter{\infer[(\text{Ax.}1)]{
   \Delta, x:A\vdash x:A'}{A\subtp A'}}
 &
   \vcenter{\infer[(\text{Ax.}2)]{
   \Delta\vdash c:A}{\bang
 \DType(c)\subtp A}}
 \\[+1em]
\multicolumn{2}{l}{  
\infer[(\mbox{$\limp$}.\text{I}_{1})]
  {\Delta\vdash\lambda x^{A}. M:A'\limp B}
  {
   x:A, \Delta\vdash M:B
  &\quad
   A'\subtp A
  }
}
 \\[+1em]
\multicolumn{2}{l}{  
  \infer[(\mbox{$\limp$}.\text{I}_{2})]
  {\bang\Delta,\Gamma\vdash\lambda x^{A}. M:\bang^{n}(A'\limp B)}
  {
   {x:A, \bang\Delta,\Gamma\vdash M:B}
   &\quad
   {\FV(M)\subseteq |\Delta|\cup\{x\}}
   &\quad
   {A'\subtp A}
  }
}
  \\[+1em]
  \multicolumn{2}{l}{\infer[(\mbox{$\limp$}.\text{E})] 
  {\bang\Delta,\Gamma_{1},\Gamma_{2}\vdash MN:B}
  {
  \bang\Delta,\Gamma_{1}\vdash M:A\limp B
  &\quad
  \bang\Delta,\Gamma_{2}\vdash N:C
  &\quad
  C\subtp A
  }}
  \\[+1em]
  \multicolumn{2}{l}{\infer[(\mbox{$\boxtimes$}.\text{I})] 
  {\bang\Delta,\Gamma_{1},\Gamma_{2}\vdash
  \tuple{M_{1},M_{2}}:\bang^{n}(A_{1}\boxtimes A_{2})}
  {
  \bang\Delta,\Gamma_{1}\vdash M_{1}:\bang^{n}A_{1}
  &\quad
  \bang\Delta,\Gamma_{2}\vdash M_{2}:\bang^{n}A_{2}
  }}
 \\[+1em]
  \multicolumn{2}{l}{\infer[(\mbox{$\boxtimes$}.\text{E})] 
  {\bang\Delta,\Gamma_{1},\Gamma_{2}\vdash
  \letcl{\tuple{x_{1}^{\bang^{n}A_{1}},x_{2}^{\bang^{n}A_{2}}}=M}{N}:A}
  {
    {\bang\Delta,\Gamma_{1}\vdash M:\bang^{n}(A_{1}\boxtimes
    A_{2})}
    &\quad
    {\bang\Delta,\Gamma_{2},x_{1}:\bang^{n}A_{1},x_{2}:\bang^{n}A_{2}\vdash N:A}
  }}
  \\[+1em]
  \infer[(\mbox{$\top$}.\text{I})]
  {\Delta\vdash *:\bang^{n}\top}
  {\phantom{\top}
  }
  &
  \infer[(\mbox{$\top$}.\text{E})] 
  {\bang\Delta,\Gamma_{1},\Gamma_{2}\vdash
  \letcl{*=M}{N}:A}
  {
  \bang\Delta,\Gamma_{1}\vdash M:\top
  &\quad
  \bang\Delta,\Gamma_{2}\vdash N:A
  }
 \\[+1em]
  \vcenter{\infer[(+.\text{I}_{1})]
  {\Delta\vdash\injl^{A_{2}} M:\bang^{n}(A_{1}+A'_{2})}
  {\Delta\vdash M:\bang^{n}A_{1}
   &\quad
   A_{2}\subtp A'_{2}}}
 \quad
 &
  \vcenter{\infer[(+.\text{I}_{2})]
  {\Delta\vdash\injr^{A_{1}} N:\bang^{n}(A'_{1}+A_{2})}
  {\Delta\vdash N:\bang^{n}A_{2}
  &\quad
   A_{1}\subtp A'_{1}}}
  \\[+1em]
 \multicolumn{2}{l}{  \infer[(+.\text{E})] 
  {
     \bang\Delta,\Gamma,\Gamma'
     \vdash
     \matchcl{P}{(x_{1}^{\bang^{n}A_{1}}\mapsto
      M_{1}\mid x_{2}^{\bang^{n}A_{2}}\mapsto M_{2})}:B
  }
  {
   \bang\Delta,\Gamma\vdash P:\bang^{n}(A_{1}+A_{2})
  &\quad
   \bang\Delta,\Gamma',x_{1}:\bang^{n}A_{1}\vdash M_{1}:B
  &\quad
   \bang\Delta,\Gamma',x_{2}:\bang^{n}A_{2}\vdash M_{2}:B
  }}
 \\[+1em]
 \multicolumn{2}{l}{
  \infer[(\text{rec})] 
  {\bang\Delta,\Gamma\vdash
  \letreccl{f^{A\limp B}x=M}{N}:C}
  {
  {\bang\Delta,f:\bang(A\limp B),x:A\vdash M:B}
  &\quad
         {\bang\Delta,\Gamma,f:\bang(A\limp B)\vdash N:C}
  }}
\end{array}
\end{displaymath}
 \caption{Typing rules for $\Hoq$}
 \label{table:typingRules}
\end{table}
Here $\Delta,\Gamma$, etc.\ denote
(unordered) \emph{contexts}. Given a context $\Delta=(x_{1}:A_{1},\dotsc,
 x_{m}:A_{m})$,
\begin{itemize}
 \item 
 $\bang \Delta$ denotes the context
 $(x_{1}:\bang A_{1},\dotsc,
 x_{m}:\bang A_{m})$; and
 \item $|\Delta|:=\{\seq{x}{m}\}$ is the \emph{domain} of $\Delta$.
\end{itemize}
 When we write $\Delta,\Gamma$ as the union of two contexts, we
 implicitly require that $|\Delta|\cap|\Gamma|=\emptyset$.
In the rule $(\text{Ax.}2)$, $c$ is a constant and its \emph{default
 type} $\DType(c)$ is defined as follows.
\begin{equation}\label{eq:HoqDefaultTypeForConst}
  \begin{array}{rcll}
   \DType(\new_{\rho})&:\equiv &k\text{-}\qbit \qquad\text{for a density
    matrix $\rho\in \DM_{2^{k}}$}
  \\
  \DType(\meas^{n+1}_{i}) &:\equiv& (n+1)\text{-}\qbit \limp(\bang\bit\boxtimes \nqbit)
  \qquad\text{for $n\ge 1$}
  \\
  \DType(\meas^{1}_{1}) &:\equiv& \qbit \limp \bang\bit
  \\
  \DType(U) &:\equiv & \kqbit\limp\kqbit
  \qquad\text{for a $2^{k}\times 2^{k}$ unitary matrix $U$}
  \\
  \DType(\cmp_{m,n}) &:\equiv&
  (\mqbit\boxtimes\nqbit)\limp (m+n)\text{-}\qbit
 \end{array} 
\end{equation}

 We shall write $\Pi\Vdash \Delta\vdash M:A$ if a derivation tree $\Pi$
 derives the type judgment. We write $\Vdash \Delta\vdash M:A$ if there
 exists such $\Pi$, that is, the type judgment is derivable.
\end{mydefinition}


\subsection{Operational Semantics}\label{subsection:HoqOprSemantics}
First we introduce small-step operational semantics, from which we
derive big-step one. The latter is given in the form of
probability distributions over the $\bit$ type and is to be compared
with the denotational semantics.

\auxproof{
\begin{mydefinition}[Extended $\Hoq$]\label{definition:extendedSyntax}
 For the purpose of  operational semantics, we extend $\Hoq$-terms by 
the following  additional 
set of constants:
 \begin{displaymath}
\begin{array}{ll}
   \qstate_{\rho} \quad&\text{for each $k\in\nat$ and $\rho\in\DM_{2^{k}}$.}
\end{array} 
\end{displaymath}
 Their default types are: 
\begin{displaymath}
 \label{eq:HoqDefaultTypeForAdditionalConst}
  \begin{array}{rcll}
   \DType(\qstate_{\rho})&:\equiv &\kqbit
   &\text{for $\rho\in
 \DM_{2^{k}}$.}
\end{array}
\end{displaymath}
\end{mydefinition}

\begin{myremark}\label{remark:qstateAndNew}
 The term $\qstate_{\rho}$ designates the (mixed) quantum state
 represented by the density matrix $\rho$; therefore
 $\qstate_{\ket{0}\bra{0}}$ and $\new\,\ttrue$ designate the same
 thing. They are distinguished in that the former is a value while the
 latter is not; this is important in consideration of $\Hoq$'s no-cloning
 property (see Remark~\ref{remark:nocloningForNewTT}). We note that the
 term $\qstate_{\rho}$ is not in the language $\Hoq$ itself but is an
 additional one used only in defining operational semantics. Therefore a
 programmer has no access to it.
 %
\end{myremark}
}

\begin{mydefinition}[Value, evaluation context]\label{definition:valueAndEvCtxt}
 The \emph{values} and \emph{evaluation contexts} of  $\Hoq$ are
 defined in the following (mostly standard) way.
 \begin{multline*}
  \begin{array}{l}
   \text{Values}\quad
   V,V_{1},V_{2}\;::= \;
\begin{array}[t]{l}
    x
   \mid \lambda x^{A}. M
   \mid \tuple{V_{1},V_{2}}
   \mid *\mid  
  \injl^{B}\, V
   \mid \injr^{A}\, V
   \mid
   \\ 
   \new_{\rho}
   \mid \meas^{n+1}_{i}
   \mid U
   \mid \cmp_{m,n}
  \enspace;
\end{array}  
\\
   \text{Evaluation contexts}
\\\quad
   E\;::= \;
\begin{array}[t]{l}
    [\place]
   \,\mid\, E[\,[\place]M\,]
   \,\mid\, E[\,V[\place]\,]
   \,\mid\, E[\,\tuple{[\place],M}\,]
   \,\mid\,
  E[\,\tuple{V,[\place]}\,]
   \,\mid\,
  \\
    E[\,\letcl{\tuple{x^{A},y^{B}}=[\place]}{M}\,]
   \,\mid\,
  E[\,\letcl{*=[\place]}{N}\,]
   \,\mid\,
  \\
 E[\,\injl^{B}[\place]\,]
   \,\mid\, E[\,\injr^{A}[\place]\,]
   \,\mid\, 
 E[\,\matchcl{[\place]}{(x^{A}\mapsto M\mid y^{B}\mapsto N)}\,]\enspace.
\end{array}  
\end{array}
 \end{multline*}
 Here $E[F]$ is the result of replacing $E$'s unique hole $[\place]$ 
   with the expression $F$. 
\end{mydefinition}
As usual, all the constants ($\new_{\rho}$, $\meas^{n+1}_{i}$, and so on) are
values. 

The definition of evaluation context is ``top-down.'' A ``bottom-up''
definition is also possible and will be used in later proofs.
\begin{mylemma}\label{lem:bottomUpDefOfEvalContext}
The following BNF notation defines the same notion of evaluation context 
as in Definition~\ref{definition:valueAndEvCtxt}. 
\begin{displaymath}
    D\;::= \;
\begin{array}[t]{l}
    [\place]
   \,\mid\, DM
   \,\mid\, VD
   \,\mid\, \tuple{D,M}
   \,\mid\, \tuple{V,D}
   \,\mid\,
  \\
    \letcl{\tuple{x^{A},y^{B}}=D}{M}
   \,\mid\,
  \letcl{*=D}{N}
   \,\mid\,
  \\
 \injl^{B}D
   \,\mid\, \injr^{A}D
   \,\mid\, 
 \matchcl{D}{(x^{A}\mapsto M\mid y^{B}\mapsto N)}\enspace.
\end{array}  
\end{displaymath}
Here $V$ is a value and $M,N$ are terms, as before. \myqed
\end{mylemma}

\begin{mydefinition}[Small-step semantics] \label{definition:operationalSemantics}
The \emph{reduction rules} of  $\Hoq$ are defined as follows. Each
 reduction $\longto$ is
 labeled with a real number from $[0,1]$. 
  \begin{align*}
\begin{array}{ll}
    E[\, (\lambda x^{A}.M)V\,] 
 \longto_{1}
   E[\,M[V/x]\,] 
  &(\limp)
 \\
   E[\,\letcl{\tuple{x^{A},y^{B}}=\tuple{V,W}}{M}\,] 
  \longto_{1}
   E[\,M[V/x,W/y]\,] 
  &(\boxtimes)
 \\
   E[\,\letcl{*=*}{M}\,] 
  \longto_{1}
   E[\,M\,] 
   &(\top)
 \\
   E[\,\matchcl{(\injl^{C}V)}{(x^{\bang^{n} A}\mapsto M\mid y^{\bang^{n} B}\mapsto N)}\,] 
  \longto_{1}
   E[\,M[V/x]\,] 
   &(+_{1})
 \\
   E[\,\matchcl{(\injr^{C}V)}{(x^{\bang^{n} A}\mapsto M\mid y^{\bang^{n}
    B}\mapsto N)}\,] 
  \longto_{1}
   E[\,N[V/y]\,] 
   &(+_{2})
 \\
   E[\,\letreccl{f^{A\limp B}x= M}{N}\,] 
  \longto_{1}
   E\bigl[\,N\bigl[\,(\lambda x^{A}.\,\letreccl{f^{A\limp B}x=
   M}{M})/f\,\bigr]\,\bigr] 
   &(\text{rec})
  \\
   E[\,U(\new_{\rho})\,] 
  \longto_{1}
   E[\,\new_{U\rho U^{\dagger}}\,] 
   &(U)
 \\
   E[\,\cmp_{m,n}\tuple{\new_{\rho},\new_{\sigma}}\,] 
  \longto_{1}
   E[\,\new_{\rho\otimes\sigma}\,] 
   &(\cmp)
 \\
   E[\,\meas^{n+1}_{i}(\new_{\rho})\,] 
  \longto_{1}
   E[\,\langle\,\ttrue,\,\new_{\bra{0_{i}}\rho\ket{0_{i}}}\,\rangle\,] 
   &(\meas_{1})
 \\
   E[\,\meas^{n+1}_{i}(\new_{\rho})\,] 
  \longto_{1}
   E[\,\langle\,\ffalse,\,\new_{\bra{1_{i}}\rho\ket{1_{i}}}\,\rangle\,] 
   &(\meas_{2})
 \\
   E[\,\meas^{1}_{1}(\new_{\rho})\,] 
  \longto_{\bra{0}\rho\ket{0}}
   E[\,\ttrue\,] 
   &(\meas_{3})
 \\
   E[\,\meas^{1}_{1}(\new_{\rho})\,] 
    \longto_{\bra{1}\rho\ket{1}}
   E[\,\ffalse\,] 
   &(\meas_{4})
\end{array}
\end{align*}
Here $M,N$ are terms, $V,W$ are values and $n\ge 1$ is a natural number.
 The reductions that involve $\new_{\rho}$---namely $(U), (\cmp), (\meas_{1}\text{--}\meas_{4})$---occur only when the
 dimensions match.  The last four rules are called \emph{measurement
 rules}. They always give rise to two reductions in a pair
 (corresponding to observing $\ttrue$ or $\ffalse$); the pair are said
 to be the \emph{buddy} to each other.

\end{mydefinition}
Observe that the label $p$ in reduction $\longto_{p}$ is like a probability but
not quite: from $\meas^{2}_{i}(\new_{\rho})$ there are two $\longto_{1}$
reductions, to $\new_{\bra{0_{i}}\rho\ket{0_{i}}}$ and to
$\new_{\bra{1_{i}}\rho\ket{1_{i}}}$. 
We understand that the  probabilities are implicitly
carried by the trace values of the matrices $\bra{0_{i}}\rho\ket{0_{i}}$
and $\bra{1_{i}}\rho\ket{1_{i}}$. See~(\ref{eq:exampleTraceOut})
and the  remarks that follow it.

As is standardly done,
we will prove adequacy of our denotational semantics focusing on
$\bit$-type closed terms. For that purpose we now introduce
big-step semantics for such terms.\footnote{We swapped the notations
$\oprred$ and $\denred$ from the previous version~\cite{HasuoH11}.}
\begin{mydefinition}[Big-step semantics]\label{definition:bigStepSemantics}
 For each $n\in\nat$ we define a relation $\oprred^{n}$ between closed $\bit$-terms $M$ and pairs
 $(p,q)$ of real numbers. This is by induction on $n$.

 For $n=0$, we define 
 \begin{displaymath}
\begin{array}{l}
   \ttrue\oprred^{0}(1,0)\,, \;
 \ffalse\oprred^{0}(0,1)\,, \; \text{and}\;
 M\oprred^{0}(0,0)\,\text{ for the other $M$. }
\end{array} 
\end{displaymath}

 For $n+1$,  if $M$ has a reduction $M\longto_{1}M'$ caused by a rule
 other than the measurement rules, we set:
 \begin{displaymath}
\begin{array}{l}
  M\oprred^{n+1} (p,q)\quad \defiff\quad M'\oprred^{n}(p,q)\enspace.
\end{array}
 \end{displaymath}
 If $M$ has a reduction $M\longto_{r}N$ caused by one of the
 measurement rules, there is always its buddy reduction
 $M\longto_{r'}N'$. 
 In this case we set
 \begin{displaymath}
\begin{array}{l}
  M\oprred^{n+1} (rp+r'p',rq+r'q')
  \quad\defiff\quad
  \text{$N\oprred^{n}(p,q)$ and $N'\oprred^{n}(p',q')$}\enspace.
\end{array}
 \end{displaymath}

Finally, we define a relation $\oprred$ as the supremum of
 $\oprred^{n}$.  That is, 
 \begin{displaymath}
\begin{array}{l}
   M\oprred (p,q)
     \quad\defiff\quad
   (p,q)\;=\;\sup\bigl\{\,(p',q')\mid M\oprred^{n}(p',q') \text{ for some
   $n$} \,\bigr\}\enspace,
\end{array}
 \end{displaymath}
 where $\sup$ is with respect to the pointwise order on
 $[0,1]\times[0,1]$. It is easy to see that for each $M$ and $n$, there is
 only one pair $(p,q)$ such that $M\oprred^{n}(p,q)$.  The same holds
 for $\oprred$, too.
\end{mydefinition}
The intuition of $M\oprred(p,q)$ is: the term $M$ (which is closed and
of  type $\bit$) reduces eventually to
$\ttrue$ with the probability $p$; to $\ffalse$ with the probability $q$.

\begin{myremark}\label{remark:simplerOprSem}
 The operational semantics of~\cite{SelingerV09} employs the notions of
 \emph{quantum array} and \emph{quantum closure}---it thus has the
 flavor of a language with \emph{quantum stores}
 (cf.~\cite{DelbecqueP08}). This is the very key in their setup that
 allows for using the Hilbert space tensor $\otimes$ as the linear
 logic tensor. We chose to separate the two tensors so that the $\bang$
 modality and recursion can be smoothly accommodated using known
 techniques (namely GoI and realizability). Accordingly, our operational
 semantics for $\Hoq$ is much more simplistic without quantum
 arrays.
\end{myremark}

\subsection{Design Choices}\label{subsection:HoqDesignChoice}
\subsubsection{What We Share with the Calculus of Selinger and Valiron}
Our calculus $\Hoq$ share
the following design choices with the original calculus
in~\cite{SelingerV09}:
\begin{itemize}
 \item building on linear $\lambda$-calculus---in particular the
       enforcement no-cloning by a linear type discipline;
 \item a \emph{call-by-value} reduction strategy;
 \item \emph{uniformity} of data, in the sense that classical and
       quantum data are dealt with in the same manner; 
 \item a formulation of the $\synt{letrec}$ operator,  as is
       usually done in a call-by-value setting (namely, recursion is
       only at function types, see the  (rec) rule in
       Table~\ref{table:typingRules}); and
 \item \emph{implicit linearity tracking}.
\end{itemize}
The last means the following (see also~\cite{SelingerV09}).  
  Linear $\lambda$-calculi,
  including the one in~\cite{Bierman95a,BentonW96}, typically have explicit syntax 
  for operating on
  the $\bang$ modality. An example is
  the $\synt{derelict}$ operator in 
\begin{equation}\label{eq:derelictOperator}
\vcenter{   \infer[]{\Gamma\,\vdash\, \synt{derelict}\, M\,:\, A}
{\Gamma\,\vdash\, M\,:\, \bang A}
}\quad.
\end{equation}
In~\cite{SelingerV09} a subtype relation $\subtp$ is introduced so that
such explicit operators can be dispensed with. For example, the subtype relation
$\bang A\subtp A$ replaces the $\synt{derelict}$ operator in the above.
This design choice is intended to aid programmers; and we
follow~\cite{SelingerV09} with regard to this
choice.

\subsubsection{What Are Different}
We now turn to the major difference from the original calculus, namely
the separation of $\boxtimes$ from $\otimes$ (mentioned already at the
beginning of the current section).  In~\cite{SelingerV09} they use the
same symbol $\otimes$ for both tensors; in other words, the linear logic
tensor is interpreted using the Hilbert space tensor. This leads to
their clean syntax: a 2-qubit system is naturally designated by the type
$\qbit\otimes\qbit$; and this is convenient when we translate quantum
circuits into programs.
Moreover, their ingenious operational
semantics---which carries the flavor of \emph{quantum store}---allows
such double usage of $\otimes$
(see Remark~\ref{remark:simplerOprSem}).

However, in developing interaction-based denotational semantics,
 we found this double usage of $\otimes$ inconvenient. We would like the
 linear logic tensor interpreted in the same way as it is interpreted in
 the conventional  interpretation of classical computation.
This seems to be a natural thing to do when working with a language with
``quantum data, classical control''---leaving the classical control
 part untouched. 
Moreover, there exists ample semantical machinery that provides natural
 interpretations of the operators like $\bang$ and $\limp$ and recursion that go
 along well with $\boxtimes$.

 The latter
 is not easily the case with $\otimes$.  While the duality
 $\mathcal{H}^{*}\cong \mathcal{H}$ gives a compact closed structure
 (hence the interpretation of $\limp$) to the category $\mathbf{fdHilb}$
 of \emph{finite-dimensional} Hilbert spaces, such is not available in
 the category $\mathbf{Hilb}$ of general Hilbert spaces, on the one
 hand. On the other hand, $\mathbf{Hilb}$ is a natural choice for a
 semantic domain: for interpreting the $\bang$ modality (``as many
 copies as requested''), we will need some kind of ``infinity,'' whose
 first candidate would be infinite-dimensional Hilbert
 spaces.\footnote{ In~\cite[\S{}1]{DelbecqueP08} it is argued---rather on
 a conceptual level---that the Hilbert space tensor $\otimes$ does not
 seem quite compatible with a closed structure (i.e.\ with respect to
 $\limp$).  }

\begin{myremark}[Other categorical models for higher-order quantum computation]\label{remark:relatedWorkCategoricallyOriented}
 A different approach is taken in~\cite{Malherbe10PhDThesis}.  The work
 keeps the original language of~\cite{SelingerV09}---where the monoidal and
 quantum tensors coincide---and starts from an axiomatic description of
 categorical models of the language. The latter is the notion of
 \emph{linear category for duplication}~\cite{SelingerV09} that combines
 a linear-non linear adjunction and monadic effects. 

 To construct a concrete instance of such models, the work~\cite{Malherbe10PhDThesis} employs a series of constructions known in
 category theory, notable among which is \emph{cocomplete completion}
 that embeds (via Yoneda) a monoidal category $\C$ in a monoidal closed
 category $[\C^{\op},\Sets]$~\cite{Day74}. The base category is
 $\C=\mathbf{Q}$ from~\cite{Selinger04} where arrows are
 essentially quantum operations and a monoidal structure is given by the
 quantum tensor $\otimes$.  

 The work~\cite{PaganiSV14} can be seen as a drastic simplification of
 the results in~\cite{Malherbe10PhDThesis} by: 1) simplifying the
 calculus (but still maintaining coincidence of the monoidal and quantum
 tensors); and 2) using \emph{quantitative semantics} for linear
 logic~\cite{Ehrhard05,DanosE11} instead of  completion by presheaves. The
 latter step is comparable to the simplification of Girard's normal functor
 semantics~\cite{Girard88} to quantitative semantics.

 In comparison to this \emph{categorical} and \emph{axiomatic} approach,
 our approach to a denotational model is \emph{operational}, 
 relying on intuitions from token abstract machines and transition systems.
 This approach is served well by
 the categorical formulation of 
 geometry of interaction, and the theory of coalgebras as
 a categorical theory of state-based dynamics. Moreover, it allows us to
 establish correspondences to operational semantics (soundness and
 adequacy),
 a feature that is lacking in~\cite{Malherbe10PhDThesis}.
\end{myremark}

 \begin{myremark}[Other GoI models for higher-order quantum computation]
  In contrast to the works discussed previously in
  Remark~\ref{remark:relatedWorkCategoricallyOriented},
  the line of
  work~\cite{DalLagoZ15,YoshimizuHFL14,DalLagoFHY14CSLLICS,DalLagoF11}
  aims at models for higher-order quantum
  computation with strong operational flavors---given by token-based
  presentations of GoI---rather than categorical models. Unlike the
  current work where we distinguish $\boxtimes$ and $\otimes$, they do
  use the same tensor $\otimes$ for both. The price to pay is that
  they
  in~\cite{DalLagoZ15,YoshimizuHFL14,DalLagoFHY14CSLLICS,DalLagoF11}
  need \emph{multiple} tokens---one for each qubit in entanglement---and they need
  to \emph{synchronize} from time to time, e.g.\ when they go through a
  multi-qubit unitary gate.

  Currently it is not clear how the multi-token GoI machines
  in~\cite{DalLagoZ15,YoshimizuHFL14,DalLagoFHY14CSLLICS,DalLagoF11} can
  be understood as instances of categorical GoI in~\cite{AbramskyHS02},
  or how those machines can be organized to form a categorical model.
  These research questions seem to be important ones, all the more since
  the significance of multi-token GoI machines seems to go beyond their
  roles in modeling quantum computation. For example, in~\cite{DalLagoFVY15}, it is
  shown that they successfully capture the difference between CBV and
  CBN evaluation strategies of PCF, via translations to a linear
  $\lambda$-calculus---much like CPS translation does in the classic
  work of Plotkin~\cite{Plotkin75}.
 \end{myremark}

  \begin{myremark}[``Quantum circuits'' in $\Hoq$]
   \label{remark:quantumCircuitsInHoq}
   Sequential and parallel compositions of unitary gates---much like
   those found in \emph{quantum circuits}---are pervasive in quantum
   computation.
   \begin{displaymath}
    \raisebox{-0.5\height}{\includegraphics[clip,trim=2.5in 0em 4.2in 0em,width=.3\textwidth]{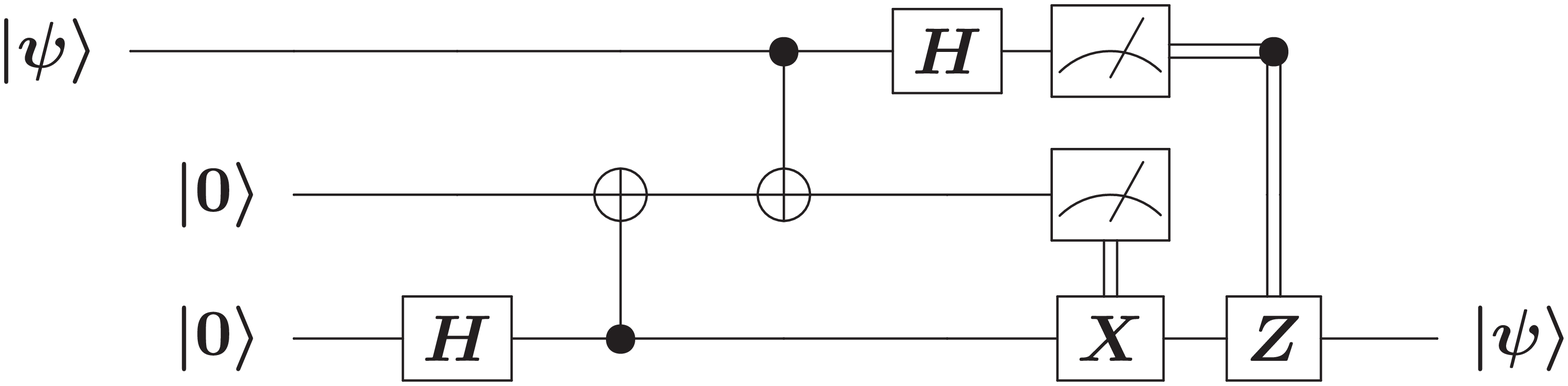}}
   \end{displaymath}
   Many higher-order functional languages for quantum computation (such
   as~\cite{SelingerV09,PaganiSV14,DalLagoZ15,YoshimizuHFL14,DalLagoFHY14CSLLICS})
   allow to express such compositions in a straightforward manner,
   exploiting their coincidence of $\otimes$ (for composite quantum
   systems) and $\boxtimes$ (as a type constructor). For example,
   in~\cite{YoshimizuHFL14} their proof nets (on which they define
   multi-token GoI machines) are claimed to be one realization of
   \emph{higher-order quantum circuits}. 

   This is not the
   case with $\Hoq$, unfortunately: due to the separation of $\boxtimes$ from
   $\otimes$, there is no type-theoretic infrastructure that supports
   the above  quantum circuit-like compositions of gates. See
   also an example in~\S{}\ref{subsection:HoqExamples}. 

   We can however foresee that the kind of (inherently first-order)
   typing disciplines that would be needed for the above quantum
   circuit-like compositions are fairly simple. It is not hard to add
   one of such typing principles to $\Hoq$---whose current type system's job is
   to take care of higher-order control structures that are much more
   complicated---as an additional layer. In other words, in such an
   extension of $\Hoq$ we will have a first-order fragment of the language that is
   devoted to composing quantum gates. We shall not describe such a
   straightforward extension in the current paper, since our focus here
   is on the integration of classical and quantum information using the
   GoI and realizability techniques. We note that, in such an extension,
   it will suffice to have a finite \emph{universal} set of quantum
   gates as primitives (any other gate can be approximately expressed in
   the language).
 \end{myremark}

\begin{myremark}
 In~\cite{PaganiSV14}, the previous
 version~\cite{HasuoH11} of the current paper is discussed, and the authors say:
``$\dotsc$ the model drops the possibility of
entangled states, and thereby fails to model one of the defining
features of quantum computation.'' We believe that this is not the case
 and the examples in~\S{}\ref{subsection:HoqExamples} will convince the
 reader. Entanglement may not be expressed by means of a type
 constructor, but is certainly there.
\end{myremark}

Besides the separation of $\boxtimes$ from $\otimes$,  $\Hoq$'s difference
from the calculus in~\cite{SelingerV09} is that bound variables and injections have explicit type
labels  (such as $A$ in $\lambda x^{A}. M$).
This choice is to ensure well-definedness of the interpretation
$\sem{\Delta\vdash M:A}$ of type judgments
(Lemma~\ref{lem:interpretationIsWellDfd})---a delicate issue with $\Hoq$
especially because of the subtype relation $\subtp$.

\begin{myremark}[Type labels and well-definedness of interpretations]
\label{remark:typeLabels}
 In general a derivable type judgment $\Delta\vdash M:A$ can have
 multiple derivations. Since denotational semantics is defined
 inductively on derivations, it is not always trivial if the
 interpretation $\sem{\Delta\vdash M:A}$ is well-defined or not.

 It is in fact nontrivial already for the simply typed
 $\lambda$-calculus in the Curry-style (i.e.\ variables' types are not
 predetermined but are specified in type contexts). An example is given
 by
 \begin{displaymath}
  x:A\vdash (\lambda y. x)(\lambda z.z): A\enspace,
 \end{displaymath}
 where the type of $z$ can be anything.  When we turn to classical
textbooks: in~\cite{LambekS86} a Church-style calculus is used
(variables come with their intrinsic types); in~\cite{Crole93} its
Curry-style calculus has explicit type labels
 (much like  in $\Hoq$).

 For the Curry-style simply typed $\lambda$-calculus, we can actually do
 without explicit type labels and still maintain well-definedness of
 $\sem{\Delta\vdash M:A}$. Its proof can be given exploiting strong
 normalization of the calculus.  The same proof strategy is used
 in~\cite[Chap.~9--11]{Valiron08PhD}---where the strategy is identified as
 \emph{normalization by evaluation}---to prove the well-definedness of
 $\sem{\Delta\vdash M:A}$ for the quantum lambda calculus
 of~\cite{SelingerV09}. (The proof is long and complicated,
 reflecting the complexity of the calculus.) We believe the same
 strategy can be employed and will get rid of type labels in $\Hoq$; however the
 proof will be very lengthy and it is therefore left as future work.

 We also note that all these troubles would be gone if we adopt
 explicit linearity tracking (explicit operators like
 $\synt{derelict}$ in~(\ref{eq:derelictOperator}), instead of the
 subtype relation $\subtp$), and move to the Church-style (variables
 have their own types). The reasons for not doing so are:
 \begin{itemize}
  \item we agree with~\cite{SelingerV09} in that 
	explicit tracking of the usage of the $\bang$ modality is a big burden to
	programmers; and
  \item our denotational model---based on GoI and realizability---has a
	merit of supporting implicit linearity tracking. This is not the
	case with every linear category, since we need
 certain type isomorphisms like $\bang\bang A\cong\bang A$
 (see Lemma~\ref{lem:bangVsProductTensorAndCoproduct}).
 \end{itemize}
\end{myremark}


\subsection{Syntactic and Operational Properties of $\Hoq$}\label{subsection:HoqSyntacticProperties}
Here we establish some syntactic and operational properties of $\Hoq$,
including some safety properties such as substitution, subject reduction
and progress. Although they are mostly parallel to
\cite[\S{}9.3]{SelingerV09}, syntax is
fragile and we have to redo all the proofs.
We shall defer most of the proofs to
\ref{section:appendixProofSyntProp}.

\begin{mylemma}[Properties of the subtype relation $\subtp$]\label{lem:propertiesOfTheSubtypeRelation}
 \begin{enumerate}
  \item\label{item:subtpIsPreorder} $\subtp$ is a preorder. 
	\auxproof{Notes 5/1/2011 2-1 p.7, Lemma 50.5}
  \item\label{item:bangIsMonotone}
    $\bang $ is monotone: $A\subtp B$ implies $\bang A \subtp \bang B$.
	\auxproof{Notes 5/1/2011 2-1 p.7.3, Lemma 50.6}
 \item\label{item:bangSubtp1}
    If $n=0 \Rightarrow m=0$ holds, we have $\bang ^{n}A\subtp \bang ^{m}A$.
	\auxproof{Notes 5/1/2011 2-1 p.7.3, Lemma 50.7}
  \item\label{item:bangSubtp2} 
       Assume that $\bang ^{n}A\subtp \bang ^{m}B$. If neither $A$ nor $B$ is of
	the form $\bang C$, we have $(n=0\Rightarrow m=0)$ and $A\subtp
       B$.
       	\auxproof{Notes 5/1/2011 2-1 p.7.47, Lemma 50.82}
 \item\label{item:subtpDirectedSupInf} The relation $\subtp$ has
       directed sups and
      infs. The former means the following (the latter is its dual). If 
      $A_{1}\subtp A$ and
      $A_{2}\subtp A$, 
      then there is a type $A_{0}$ such that: 1)
      $A_{1}\subtp A_{0}$ and
      $A_{2}\subtp A_{0}$; 
      2)
      $A_{1}\subtp A'$ and
      $A_{2}\subtp A'$ imply $A_{0}\subtp A'$. \myqed
       	\auxproof{Notes 5/1/2011 2-1 p.7.44, Lemma 50.81}
 \end{enumerate}
\end{mylemma}

\begin{mynotation}[Subtyping $\subtp$ between
 contexts]\label{notation:subtpBtwCtxt}
 We write
 $\Delta'\subtp \Delta$  when:
\begin{itemize}
 \item   $|\Delta'|=|\Delta|$, and
 \item   for each $(x:A')\in \Delta'$, there is
 $(x:A)\in\Delta$ with $A'\subtp A$.
\end{itemize}
\end{mynotation}

In Table~\ref{table:typingRules}, some rules including
$(\mbox{$\limp$}.\text{I}_{1})$ have \emph{type coercion}: while the
term $\lambda x^{A}.M$ has a type label $A$, its actual type is $A'\limp
B$ with $A'\subtp A$. This is so that the following holds.
\begin{mylemma}\label{lem:monotonicity}
The \emph{monotonicity} rule is admissible in $\Hoq$.
 \begin{displaymath}
   \vcenter{ \infer[(\text{Mon})]
  {\Delta'\vdash
  M:A'}
  {
  \Delta'\subtp \Delta
  &
  \Delta\vdash M:A
   &
   A\subtp A'
  }}
 \end{displaymath} 

\vspace*{-1.5em}
\myqed
\end{mylemma}

\begin{mycorollary}\label{corollary:derelictionAdmsble}
 The \emph{dereliction} and \emph{comultiplication} rules are admissible in $\Hoq$.
 \begin{displaymath}
   \vcenter{ \infer[(\text{Der})]
  {\Delta\vdash M: A}
  {\Delta\vdash M: \bang A}
  }
  \qquad
   \vcenter{ \infer[(\text{Comult})]
  {\Delta\vdash M: \bang \bang A}
  {\Delta\vdash M: \bang A}
  } \end{displaymath}

\vspace*{-1.5em}
\myqed
\end{mycorollary}

\begin{mylemma}\label{lem:lemmasInTyping} 
\begin{enumerate}
 \item\label{item:freeVarsAreInContext} If $\Vdash\Delta\vdash M:B$,
      then $\FV(M)\subseteq|\Delta|$.
 \item\label{item:removingUnusedContextInHoq} If $x\not\in\FV(M)$ and 
   $\Vdash\Delta,x:A\vdash M:B$, 
   then 
       $\Vdash\Delta\vdash M:B$.
 \item\label{item:weakeningInHoq} The following rule is admissible.
\begin{displaymath}
    \vcenter{ \infer[(\text{Weakening})]
  {\Gamma,\Delta\vdash M:  A}
  {\Delta\vdash M:  A}
  } 
\end{displaymath}
\end{enumerate}
\end{mylemma}
\begin{myproof}
Straightforward, by induction on  derivation.
\myqed
\end{myproof}

Many linear lambda calculi have the \emph{promotion} rule
\begin{displaymath}
 \vcenter{\infer[(\text{Prom})]{\bang\Delta\vdash \synt{promote}\, M:
 \bang A}{\bang\Delta\vdash M: A}}
\end{displaymath}
or its variant, like in~\cite{Bierman95a,BentonW96}. 
Much like the original calculus (see~\cite[Remark~9.3.27]{SelingerV09}), $\Hoq$ lacks the general
promotion rule but it has a restriction to values admissible.
\begin{mylemma}[Value promotion%
]\label{lem:promotionOnlyForValues}
Let $V$ be a value.
\begin{enumerate}
 \item\label{item:bangTypeThenFVAreBangType}
   If $\Vdash\Delta\vdash V:\bang A$, then for each
 $x\in\FV(V)$, we have $(x:\bang
 B)\in\Delta$ for some type $B$. 
 \item\label{item:promotionOnlyForValues} Conversely, the following rule
      is admissible in $\Hoq$. 
 \begin{displaymath}
 \vcenter{ \infer[(\text{ValProm})]{\bang\Delta,\Gamma\vdash V: \bang A}{
  {\bang\Delta,\Gamma\vdash V: A}
  &\quad
  {\FV(V)\subseteq|\Delta|}
}}\enspace,
 \quad\text{where $V$ is a value.}
 \end{displaymath}

\vspace*{-1.5em}
\myqed
\end{enumerate}
\end{mylemma}

 \begin{myremark}[No-cloning]\label{remark:nocloningForNewTT}
 We note that all constants are values
  (Definition~\ref{definition:valueAndEvCtxt}); therefore the last
  result (or ultimately  the rule $(\text{Ax.}2)$ in Table~\ref{table:typingRules})
  implies that any constant can have a type $\bang A$, where
   $A$ is the default type of the constant
  (Definition~\ref{definition:TypingInHoq}).
  This can raise a suspicion that our calculus $\Hoq$ does not respect
  \emph{no-cloning}, one of the most fundamental principles in quantum
  mechanics: it dictates that no quantum state should be duplicable
  (unless it is \emph{classical} in a suitable sense); still our primitive $\new_{\rho}$ can have a type $\bang k\text{-}\qbit$
  and hence is duplicable.

  We believe this is not problematic. As discussed briefly after
  Definition~\ref{definition:TermsOfHoq}, we understand the constant
  $\new_{\rho}$ to stand for ``deployment of some quantum
 apparatus that is capable of preparing the quantum state $\rho$.'' In
  this viewpoint it is no problem that the term
  \begin{displaymath}
   \bigl(\;\lambda x^{\bang \qbit}.\, \cmp_{1,1} \tuple{x,x}
   \;\bigr)
   (\new_{\ket{0}\bra{0}})
  \end{displaymath}
  is typable---it just denotes that we run an apparatus that prepares
  the state $\ket{0}$ twice to obtain the state $\ket{00}$.

  Still the no-cloning property is discerned in the calculus $\Hoq$.
  For example, the term
  \begin{displaymath}
   \lambda x^{\qbit}.\, \cmp_{1,1} \tuple{x,x}
   \enspace,
  \end{displaymath}
  without $\bang$ in its argument type, is not typable. This means
  we cannot duplicate a quantum state  whose preparation apparatus we do
  not have access to.
  
\end{myremark}

A general \emph{substitution} rule
 \begin{displaymath}
  \infer[(\text{Subst})]{\bang\Delta,\Gamma_{1},\Gamma_{2}\vdash N[M/x]: B}{
  {\bang\Delta,\Gamma_{1}\vdash M: A}
  &\quad
  {\bang\Delta,\Gamma_{2},x:A \vdash N:B}
}
 \end{displaymath}
is not admissible in $\Hoq$. A counter example can be given as follows.
The following two judgments are both derivable.
\begin{displaymath}
 x:\qbit\limp \bang\qbit, 
 y:\qbit
 \vdash
 xy:\bang \qbit
 \quad
\quad
 w:\bang\qbit\vdash \lambda z^{\bang\qbit}.\, w:
 \bang(\bang \qbit\limp\bang \qbit)
\end{displaymath}
In particular, the latter relies on the $(\mbox{$\limp$}.\text{I}_{2})$
rule. However, the result of substitution
\begin{displaymath}
 x:\qbit\limp \bang\qbit, 
 y:\qbit
 \vdash
 \lambda z^{\bang\qbit}.\, xy:
 \bang(\bang \qbit\limp\bang \qbit)
\end{displaymath}
is not derivable: since the types of the free variables $x,y$ are not
of the form $\bang\Delta$,  the $(\mbox{$\limp$}.\text{I}_{2})$
rule is not applicable. Therefore we impose some restrictions.

\begin{mylemma}[Substitution]\label{lem:substitutionLemma}
 The following rules are admissible in $\Hoq$. 
 \begin{displaymath}
\begin{array}{c}
   \infer[(\text{Subst}_{1})]{\bang\Delta,\Gamma_{1},\Gamma_{2}\vdash N[M/x]: B}{
  {\bang\Delta,\Gamma_{1}\vdash M: A}
  &\quad
  {\bang\Delta,\Gamma_{2},x:A \vdash N:B}
  &\quad
  {A\not\equiv\bang A'    \;\text{for any $A'$}}  
 }
\\[+.5em]
   \infer[(\text{Subst}_{2})]{\bang\Delta,
\Gamma_{2}\vdash N[M/x]: B}{
  {\bang\Delta 
  \vdash M: A}
  &\quad
  {\bang\Delta,\Gamma_{2},x:A \vdash N:B}
 }
\\[+.5em]
   \infer[(\text{Subst}_{3})]{\bang\Delta,\Gamma_{1},\Gamma_{2}\vdash N[V/x]: B}{
  {\bang\Delta,\Gamma_{1}\vdash V: A}
  &\quad
  {\bang\Delta,\Gamma_{2},x:A \vdash N:B}
   &\quad
   {\text{$V$ is a value}}  
 }
\\[+.5em]
   \infer[(\text{Subst}_{4})]{\bang\Delta,\Gamma_{1},\Gamma_{2}\vdash E[M]: B}{
  {\bang\Delta,\Gamma_{1}\vdash M: A}
  &\quad
  {\bang\Delta,\Gamma_{2},x:A \vdash E[x]:B}
   &\quad
   {\text{$x$ does not occur in $E$}}  
 }
\end{array} 
\end{displaymath}
Note that in the first assumption of the $(\text{Subst}_{2})$ rule, the
 whole context must
 be of the form $\bang\Delta$. In the $(\text{Subst}_{4})$ rule $E$ denotes an evaluation
 context (Definition~\ref{definition:valueAndEvCtxt}); the side condition
 means that $x$ occurs exactly once in the term $E[x]$.
\myqed
\end{mylemma}

\begin{mylemma}[Subject reduction]\label{lem:subjectReductionLemma}
Assume that $\Vdash \Delta\vdash M:A$, and that there is a reduction
 $M\longrightarrow_{p}N$ (we allow $p=0$). Then $\Vdash
 \Delta\vdash N:A$. \myqed
\end{mylemma}

\begin{mylemma}[Progress]\label{lem:progress}
 A typable closed term that is not a value has a reduction. More
 precisely: assume that $\Vdash\,\vdash M:A$ (therefore $M$ is closed by
 Lemma~\ref{lem:lemmasInTyping}.\ref{item:freeVarsAreInContext}), and that
 $M$ is not a value.
Then there exists a term $N$ and
 $p\in[0,1]$ such that $M\longrightarrow_{p}N$.
\myqed
\end{mylemma}
We note that, given $M$, the sum of the values $p$ for all possible reductions
$M\longrightarrow_{p}N$  is not necessarily equal to $1$. See
the remark right after Definition~\ref{definition:operationalSemantics}.

\auxproof{ \begin{mylemma}[No-cloning]\label{lem:nocloningInHoq}
  This is very complicated to formulate so is skipped (7/4/2013)
 \end{mylemma}
}

\subsection{$\Hoq$ Programs: an Example}\label{subsection:HoqExamples}
We give an example of a $\Hoq$ program that simulates quantum
teleportation: a procedure in which Alice sends a quantum state to Bob
using a classical communication channel.
The example, although first-order, will exemplify the expressivity as
well as inexpressivity of the calculus $\Hoq$: it fully supports
preparation of quantum states and unitary transformations; it also
features (classical) branching based on measurement outcomes; however,
due to the distinction between $\otimes$ and $\boxtimes$, composition of
unitary transformations---one that is much like in quantum
circuits---is less straightforward to express (cf.\ Remark~\ref{remark:quantumCircuitsInHoq}). See~(\ref{eq:EPR}) where
an explicit use of $\cmp$ is required, and some discussions that follow it.

Potential use of \emph{higher-order} quantum programs has already been
advocated by many authors; see
e.g.~\cite{SelingerV09,Valiron13NGC,DelbecqueP08}. They could also be
used in formalizing games in \emph{quantum game theory}, a
formalism that is attracting increasing  attention as a useful
presentation of quantum nonlocality (see e.g.~\cite{GisinMS07}). We here
use a first-order example of quantum teleportation, however, since it is
one of the most well-known quantum procedures.

In the quantum teleportation protocol, Alice and Bob start with
preparing an EPR-pair.  Alice keeps the first qubit of the EPR-pair; and
Bob keeps the second qubit. In $\Hoq$, we can prepare an EPR-pair by
applying a suitable unitary transformation to a qubit constructed by
$\new$:
\begin{equation}\label{eq:EPR}
  \mathtt{EPR}
  \;:\equiv\;
  U_{0}\, \bigl(\mathtt{cmp}_{1,1}\langle 
 \new_{\ket{0}\bra{0}}
 ,
 \new_{\ket{0}\bra{0}}
 \rangle\bigr)
  \qquad:\quad2\text{-}\mathtt{qbit}
\end{equation}
where $U_{0}$ is a unitary transformation given by
\begin{displaymath}
  U_{0} 
  \;:=\; \dfrac{1}{\sqrt{2}}
  \begin{pmatrix}
    1 & 0 & 1 & 0 \\
    0 & 1 & 0 & 1 \\
    0 & 1 & 0 & -1 \\
    1 & 0 & -1 & 0
  \end{pmatrix}\enspace.
\end{displaymath}

We note that $U_{0}$ is usually defined as the following composition of
the Hadamard gate $H$ and the conditional-not gate $N$:
\begin{displaymath}
  U_{0} = N (H \otimes \IM_{2})\enspace,
\end{displaymath}
where $\IM_{2}$ is the $2\times 2$ identity matrix.
However, since the tensor product $\boxtimes$ of $\Hoq$ is different from the tensor
product $\otimes$ of vector spaces, we cannot program $U_{0}$ as a
simple composition of $H$ and $N$
in $\Hoq$. We need to calculate $U_{0}$ 
outside $\Hoq$; see the discussions in
Remark~\ref{remark:quantumCircuitsInHoq}.


Then Alice applies a Bell measurement to the first two qubits of a quantum
state $\rho \boxtimes \mathtt{EPR}$ where $\rho$ is a quantum bit
that Alice wishes to send to Bob.
\begin{equation}\label{eq:bellMeasure}
 \begin{aligned}
  \mathtt{Bellmeasure} 
 \;:\equiv\qquad &\lambda w^{3\text{-}\mathtt{qbit}}. \
  \mathtt{let} \ \langle b_{0}^{\bit } ,
  p^{2\text{-}\mathtt{qbit}} \rangle = \mathtt{meas}^{3}_{1}\, (U_{1}
  w)
  \ \mathtt{in} \\
  & \hspace{60pt} \mathtt{let} \ \langle b_{1}^{\bit } ,
  q^{\mathtt{qbit}} \rangle = \mathtt{meas}^{2}_{1} \,p \ \mathtt{in}
  \ \langle b_{0},\langle b_{1},q \rangle \rangle \\
  & \hspace{150pt}:\quad 3\text{-}\mathtt{qbit} \limp
  \bit \boxtimes (\bit \boxtimes\mathtt{qbit})
\end{aligned}
\end{equation}
Here $U_{1}$ is the following unitary transformation:
\begin{align*}
  U_{1} \;:=\; \dfrac{1}{\sqrt{2}}
  \begin{pmatrix}
    1 & 0 & 0 & 0 & 0 & 0 & 1 & 0 \\
    0 & 1 & 0 & 0 & 0 & 0 & 0 & 1 \\
    0 & 0 & 1 & 0 & 1 & 0 & 0 & 0 \\
    0 & 0 & 0 & 1 & 0 & 1 & 0 & 0 \\
    1 & 0 & 0 & 0 & 0 & 0 & -1 & 0 \\
    0 & 1 & 0 & 0 & 0 & 0 & 0 & -1 \\
    0 & 0 & 1 & 0 & -1 & 0 & 0 & 0 \\
    0 & 0 & 0 & 1 & 0 & -1 & 0 & 0
  \end{pmatrix}\enspace,
\end{align*}
which is equal to $\bigl((H \otimes \mathcal{I}_{2}) N\bigr) \otimes \mathcal{I}_{2}$.
Although the third qubit has nothing to do with the Bell measurement,
we need to include the third qubit in the
program~(\ref{eq:bellMeasure}) because the type
$2\text{-}\mathtt{qbit}\boxtimes \qbit$ is different from 
$3\text{-}\mathtt{qbit}$ in $\Hoq$. This kind of awkwardness will also 
be gone, once we equip $\Hoq$ with constructs for composing unitary operations.

Alice tells the result $(i,j)$ of measurement to Bob, and Bob applies a
unitary transformation $U_{i,j}$ to his qubits:
\begin{align*}
  \mathtt{corr}
  \;: \equiv\quad & \lambda
  x^{\bit \boxtimes(\bit \boxtimes\mathtt{qbit})}. \,
  \mathtt{let} \ \langle b_{0}^{\bit },y^{\bit 
    \boxtimes \mathtt{qbit}} \rangle = x \ \mathtt{in} \ \mathtt{let} \
  \langle b_{1}^{\bit },q^\mathtt{qbit} \rangle = y \ \mathtt{in} \\
  & \hspace{90pt}\mathtt{match} \ b_{0} \
  \mathtt{with} \ (\\
  & \hspace{110pt} z_{0}^{\top} \mapsto \mathtt{match} \ b_{1}\
  \mathtt{with} \ ( w_{0}^{\top} \mapsto U_{00} q
  \mid w_{1}^{\top} \mapsto U_{01} q) \\
  & \hspace{110pt} \mid z_{1}^{\top} \mapsto \mathtt{match} \ b_{1}\
  \mathtt{with} \ (w_{0}^{\top} \mapsto U_{10} q \mid w_{1}^{\top} \mapsto
  U_{11} q)
  ) \\
  &\hspace{190pt}
  :\quad\bit \boxtimes(\bit \boxtimes\mathtt{qbit}) \limp
  \mathtt{qbit}
\end{align*}
where $U_{ij}$ are given as follows.
\begin{displaymath}
  U_{00} :=
  \begin{pmatrix}
    1 & 0 \\ 0 & 1
  \end{pmatrix} \qquad U_{01} :=
  \begin{pmatrix}
    0 & 1 \\ 1 & 0
  \end{pmatrix} \qquad U_{10} :=
  \begin{pmatrix}
    1 & 0 \\ 0 & -1
  \end{pmatrix} \qquad U_{11} :=
  \begin{pmatrix}
    0 & 1 \\ -1 & 0
  \end{pmatrix}
\end{displaymath}
The result is the qubit that Alice wishes to send to Bob.

We combine the above programs into one: we define a closed
value $\mathtt{qtel}:\mathtt{qbit} \limp \mathtt{qbit}$ to be
\begin{displaymath}
  \lambda x^{\mathtt{qbit}}. \
  \mathtt{corr}\,
  \bigl(\mathtt{Bellmeasure} \,\bigl(\,
  \mathtt{cmp}_{1,2}\,\langle x,\mathtt{EPR}\rangle\,\bigr)\bigr)\enspace.
\end{displaymath}

We can observe that Bob receives Alice's qubit.
\begin{myproposition}
  For any unitary transformation $U$, the
  reduction tree of $\mathtt{qtel}$ is
  of the following form.
  \begin{displaymath}
    \begin{xy}
      (0,13) *+{\mathtt{qtel} \, (U\, (\new_{\ket{0}\bra{0}}))}="source", %
      (-30,0) *+{\new_{\frac{1}{4}\rho}}="1", %
      (-10,0) *+{\new_{\frac{1}{4}\rho}}="2", %
      (10,0) *+{\new_{\frac{1}{4}\rho}}="3", %
      (30,0) *+{\new_{\frac{1}{4}\rho}}="4", %
      \ar_(.8){1}_{*} "source";"1" %
      \ar^(.55){1}_{*} "source";"2" %
      \ar_(.55){1}^{*} "source";"3" %
      \ar^(.8){1}^{*} "source";"4"
    \end{xy}
  \end{displaymath}
 Here $\rho \in \DM_{2}$ is $U \,| 0 \rangle \langle 0 |\,
  U^{\dagger}$ and
  $\xrightarrow{*}_{1}$ is the transitive closure of $\to_{1}$.
\myqed
\end{myproposition}

\subsubsection{Fair  Coin Toss}
This example is in~\cite{SelingerV09}: it simulates a fair coin toss
with
quantum primitives.
\begin{align*}
  \mathtt{fcoin} \;:\equiv\; H (\new\, \ttrue) && 
  \mathtt{toss} \;:\equiv\;  \lambda x^{\mathtt{qbit}}.\,\mathtt{meas}^{1}_{1}\, x
\end{align*}

\begin{myproposition}
The
  reduction tree of $\mathtt{toss} \ \mathtt{fcoin}$ is the following
 one. 
  \begin{displaymath}
    \begin{xy}
      (0,10) *+{\mathtt{toss} \ \mathtt{fcoin}}="fcoin",
      (-10,0) *+{\ttrue}="0",
      (10,0) *+{\ffalse}="1",
      \ar_(.8){\frac{1}{2}} "fcoin";"0"
      \ar^(.8){\frac{1}{2}} "fcoin";"1"
    \end{xy}
  \end{displaymath}
Here we omitted reductions $\to_{1}$. \myqed
\end{myproposition}

\auxproof{
\subsubsection{Grover's Algorithm}

We give a program that executes a quantum searching algorithm, called
Grover's algorithm. Since the algorithm has a process ``repeat ...
for $n$ times'', we first add natural numbers to $\Hoq$.


We extend $\Hoq$ by adding a type of natural numbers, as follows.
\begin{align*}
  \mathrm{Type} && A &\;::=\; \cdots \mid \mathtt{nat} \\ %
  \mathrm{Term} && M &\;::=\; \cdots \mid n \in \mathbb{N} \mid \mathtt{S}
  \mid \mathtt{isZero} \mid \mathtt{pred} \\%
  \mathrm{Value} && V &\;::=\; \cdots \mid n \in \mathbb{N} \mid \mathtt{S} \mid
  \mathtt{isZero} \mid \mathtt{pred} \\%
  \text{Evaluation \ context} && E &\;::=\; \cdots \mid E[\mathtt{S}[-]]
  \mid E[\mathtt{isZero}[-]] \mid E[\mathtt{pred}[-]]
\end{align*}
Default types of these constants are given as follows.
\begin{align*}
  \mathrm{DType}(n) &=  \mathtt{nat} &
  \mathrm{DType}(\mathtt{S}) &= \mathtt{nat} \limp \mathtt{nat} \\
  \mathrm{DType}(\mathtt{isZero})
  &= \mathtt{nat} \limp \bit  &
  \mathrm{DType}(\mathtt{pred}) &= \mathtt{nat} \limp \mathtt{nat}
\end{align*}
Operational semantics is extended in the standard way.
\begin{align*}
  E[\mathtt{S}\, n] &\to_{1} E[n+1] &
  E[\mathtt{isZero} \ 0] &\to_{1} E[\ttrue] \\
  E[\mathtt{isZero} \ (n+1)] &\to_{1} E[\ffalse] &
  E[\mathtt{pred} \ n] &\to_{1} E[n\mathbin{\dot{-}}1]
\end{align*}
Here $n\mathbin{\dot{-}}1$ denotes \emph{normalized subtraction}:
it is $n-1$ if $n>0$; and $0\mathbin{\dot{-}}1:=0$.

\begin{myremark}
  Since $\PER_{\Q}$ has a natural number object, we can extend the
  interpretation of $\Hoq$ in a canonical manner. Furthermore,
  Corollary~\ref{cor:adeq} holds for the extension.
\end{myremark}


In this extension of $\Hoq$ we express Grover's algorithm.
We suppose that there is an \emph{oracle} $U_{1}:\nqbit
\limp \nqbit$ that tells us which quantum state is an
answer $w \in \{0,1\}^{n}$:
\begin{align*}
  U_{1}|n\rangle &=
  \begin{cases}
    |n\rangle & (n \neq w), \\
    -|w\rangle & (n = w).
  \end{cases}
\end{align*}
The goal of Grover's algorithm is to find the answer $w$. 
The algorithm is described as follows.
\begin{enumerate}
\item Prepare a quantum state
  \begin{displaymath}
    |s\rangle = 2^{-\frac{n}{2}}\sum_{i \in 
      \{0,1\}^{n}}|i\rangle.
  \end{displaymath}
\item Do the following clauses sufficiently many times.
  \begin{enumerate}
  \item Apply $U_{1}$ to the quantum state.
  \item Apply $U_{2} = 2|s\rangle\langle s| - 
    \mathcal{I}_{2^{n}}$ to the
    quantum state.
  \end{enumerate}
\item Observe the quantum state.
\end{enumerate}
The result of observation will be $w$ with a high probability: given a
desired degree of precision, there is a formula that computes how many
iterations of Step 2 are needed.

We can express the above procedure as follows:
 \begin{multline*}
   \mathtt{Grover} \;:\equiv\; \lambda u^{\bang (\nqbit
     \limp \nqbit)}. \, \lambda
   q^{\nqbit}. \ \lambda m^{\bang \mathtt{nat}}. \,
   \mathtt{meas}_{n}(\mathtt{repeat} \, u \, m \, q) \\
   :\quad \bang (\nqbit \limp
   \nqbit) \limp \nqbit
   \limp \bang \mathtt{nat} \limp \bit ^{n}
  \end{multline*}
where
\begin{itemize}
\item $\bit ^{n} \equiv \overbrace{\bit  \boxtimes
    (\bit  \boxtimes (
    \cdots \boxtimes (\bit  \boxtimes \bit }^{n})\cdots ))$
\item $\mathtt{repeat}(u,m,q)$ computes $u^{m}(q)$. That is,
 \begin{align*}
  & \mathtt{repeat} \equiv \\
  &\lambda f^{\bang (\nqbit \limp
    \nqbit)}. \ \lambda m^{\bang \mathtt{nat}} . \
  \lambda x^{\nqbit}. \\
  & \hspace{5pt} 
  \mathtt{letrec} \ g^{\bang \mathtt{nat}\limp \nqbit} k =
  \mathtt{match} \ \mathtt{isZero} \ k \ \mathtt{with} \
  (y^{\top} \mapsto x \mid z^{\top} \mapsto
  f \, (g \, (\mathtt{pred} \, k))) \\
  &\hspace{320pt}\mathtt{in} \ g \ m \\
  & \hspace{130pt} :\quad\bang (\nqbit \limp \nqbit) 
  \limp \bang \mathtt{nat} \limp \nqbit
  \limp \nqbit
 \end{align*}
\item $\mathtt{meas}_{n}$ measures $\nqbit$:
  \begin{align*}
    \mathtt{meas}_{n}\;:\equiv\quad & \lambda q^{\nqbit}.\,
    \mathtt{let} \ \langle
    b_{1}^{\bit }, q_{1}^{(n-1)\text{-}\mathtt{qbit}}\rangle =
    \mathtt{meas}_{1}^{n}q \ \mathtt{in} \\ &\qquad \mathtt{let} \
    \langle b_{2}^{\bit },q_{2}^{(n-2)\text{-}\mathtt{qbit}}
    \rangle = \mathtt{meas}_{1}^{n-1}q_{1} \ \mathtt{in} \cdots \\
    &\qquad\qquad \mathtt{let} \ \langle
    b_{n-1}^{\bit },q_{n-1}^{\mathtt{qbit}} \rangle =
    \mathtt{meas}_{1}^{2} \ q_{n-2} \ \mathtt{in} \\
    &\qquad\qquad\qquad \langle b_{1}, \langle b_{2}, \cdots, \langle
    b_{n-1}, \mathtt{meas}_{1}^{1} \ q_{n-1}\rangle \cdots \rangle\rangle
   \quad :\quad\nqbit \limp \bit ^{n}\enspace.
  \end{align*}
\end{itemize}
The program $\mathtt{Grover}(u,q,m)$ applies $u^{m}$ to $q$, and
then measures the result. Therefore, we can run Grover's algorithm by
the following code:
\begin{displaymath}
  \mathtt{Grover}\, \bigl(\lambda q^{\nqbit}. \,
  U_{1}\,(U_{2}\,q)\bigr) \,
 \Bigl(U\, \bigl((
 \new_{\ket{0}\bra{0}}
 ) \boxtimes \cdots \boxtimes
 (
  \new_{\ket{0}\bra{0}}
 )\bigr) \,M_{0}
\end{displaymath}
where $U:\nqbit \limp \nqbit$ is a
unitary transformation that maps $|0\cdots 0\rangle$ to $|s\rangle$,
and $M_{0}$ is a sufficiently large number so that we can observe
the answer with a desired degree of precision. We suppressed 
occurrences of $\cmp$.
}

\section{The Quantum Branching Monad $\Q$ and The Category $\PER_{\Q}$}
\label{section:quantumBranchingMonad}
We now turn to denotational semantics of $\Hoq$. 

\subsection{Background}\label{subsection:quantumBranchingMonadBackground}
The starting point of our current work was Jacobs'
observation~\cite{Jacobs10trace} that relates: monads for branching
(\S{}\ref{subsection:prelimMonadForBranching}, used in coalgebraic
\emph{trace} semantics) and \emph{traced} monoidal categories that
appear in categorical GoI (\S{}\ref{subsection:prelimGoI}). See also
Remark~\ref{remark:threeTraces}. This relationship establishes the first among the
three steps in Figure~\ref{figure:constructionOfModel}.

 Examples of a traced monoidal category $\C$ used in categorical
GoI~\cite{AbramskyHS02} are divided into two groups: the so-called
\emph{wave-style} ones where $\C$'s monoidal structure is given by
products $\times$; and the \emph{particle-style} ones where it is given
by coproducts $+$. The former are of \emph{static} nature and includes
domain-theoretic examples like $\omega\text{-}\mathbf{CPO}$.  The latter
particle-style examples are often \emph{dynamic}, in the sense that we
can imagine a ``particle'' (or a ``token'') moving around (we will further
elaborate this point later). This is the class of examples we are more interested
in.  The examples include:
\begin{itemize}
 \item 
 the category $\mathbf{Pfn}$ of
 sets and partial functions; 
 \item  the category
$\mathbf{Rel}_{+}$ of sets and binary
 relations, where the subscript $+$ indicates that the relevant monoidal
	structure is the one given by disjoint unions of sets; and
 \item the category $\mathbf{SRel}$ of measurable spaces and stochastic
 relations.
\end{itemize} 
For us the crucial observation is that these examples are (close to) the
 Kleisli categories $\Kleisli{B}$ for the ``branching'' monads $B$
 in Example~\ref{example:branchingMonads},~\S{}\ref{subsection:prelimMonadForBranching}. Indeed, it is easy to
 see that $\Pfn$ and $\Rel$ are precisely $\Kleisli{\lift}$ and
 $\Kleisli{\pow}$, respectively; the category $\Kleisli{\dist}$ can be
 thought of as a discrete variant of $\mathbf{SRel}$.

Generalizing this observation, Jacobs~\cite{Jacobs10trace} proves that a
monad $B$ for branching---i.e.\ a monad on $\Sets$ with order
enrichment, subject to some additional conditions---has its Kleisli
category $\Kleisli{B}$ traced monoidal (see
Theorem~\ref{theorem:KleisliOfQIsTraced} later). Here the monoidal
structure is given by $+$, coproduct in $\Sets$ (and also in
$\Kleisli{B}$, since the Kleisli embedding $\Sets\to\Kleisli{B}$
preserves coproducts).

Let us elaborate on such a Kleisli category $\Kleisli{B}$.
We look at it as a category of \emph{piping}. An
arrow\footnote{We shall use $\relto$ to denote an arrow in a Kleisli
category.} $f:X\relto Y$ in $\Kleisli{B}$ is understood as a bunch of
pipes, with $|X|$-many entrances and $|Y|$-many exits.\footnote{ Our piping analogy is not completely faithful: in a
Kleisli arrow $f$ the two crossings
\includegraphics[width=1em]{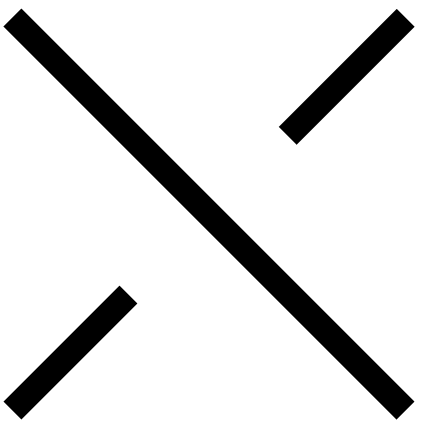} and
\includegraphics[width=1em]{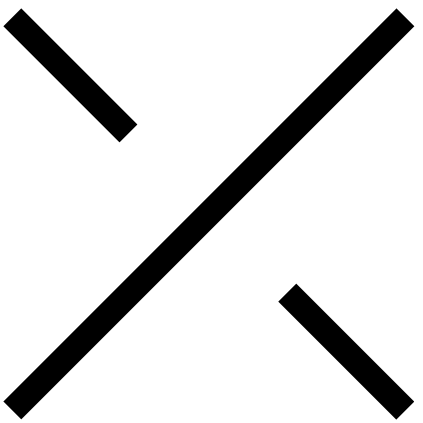}
are identified,  but  they are different as physical pipes.
}
The pipes are where a \emph{particle} (or \emph{token})
runs through.  See below; here the shaded box $f$ consists of a lot of pipes.
\begin{equation}\label{eq:pipingDrawing}
 \raisebox{-0.5\height}{\scalebox{.6} 
 {
 \begin{pspicture}(0,-3.4781835)(7.2781835,3.4781835)
 \psframe[linewidth=0.04,dimen=outer,fillstyle=solid,fillcolor=lightgray](7.126904,1.3575195)(0.32690418,-1.3218164)
 \usefont{T1}{ptm}{m}{n}
 \rput(3.617451,0.102519535){\LARGE $f$}
 \usefont{T1}{ptm}{m}{n}
 \rput(1.7074512,1.9825195){\LARGE $x$}
 \psline[linewidth=0.04cm](1.2869042,2.5175195)(1.2869042,1.3175194)
 \psline[linewidth=0.04cm](2.0869043,2.5175195)(2.0869043,1.3175194)
 \usefont{T1}{ptm}{m}{n}
 \rput(1.707451,-1.8174804){\LARGE $y$}
 \psline[linewidth=0.04cm](1.2869042,-1.2824805)(1.2869042,-2.4824805)
 \psline[linewidth=0.04cm](2.0869043,-1.2824805)(2.0869043,-2.4824805)
 \psellipse[linewidth=0.04,dimen=outer](1.6869042,2.5175195)(0.4,0.2)
 \usefont{T1}{ptm}{m}{n}
 \rput(0.2319432,2.0925195){\LARGE ...}
 \usefont{T1}{ptm}{m}{n}
 \rput(0.23194325,-1.9074805){\LARGE ...}
 \usefont{T1}{ptm}{m}{n}
 \rput(6.831943,2.0925195){\LARGE ...}
 \usefont{T1}{ptm}{m}{n}
 \rput(6.8319435,-1.9074805){\LARGE ...}
 \usefont{T1}{ptm}{m}{n}
 \rput(3.6319432,-1.9074805){\LARGE ...}
 \usefont{T1}{ptm}{m}{n}
 \rput(5.437451,-1.8174804){\LARGE $y'$}
 \psline[linewidth=0.04cm](5.086904,-1.2824805)(5.086904,-2.4824805)
 \psline[linewidth=0.04cm](5.8869042,-1.2824805)(5.8869042,-2.4824805)
 \usefont{T1}{ptm}{m}{n}
 \rput(5.487451,3.1825194){\LARGE $\downarrow$}
 \psellipse[linewidth=0.04,dimen=outer](1.6869042,-2.4824805)(0.4,0.2)
 \psellipse[linewidth=0.04,dimen=outer](5.486904,-2.4824805)(0.4,0.2)
 \usefont{T1}{ptm}{m}{n}
 \rput(1.6874511,-3.2174804){\LARGE $\downarrow$}
 \usefont{T1}{ptm}{m}{n}
 \rput(5.487451,-3.2174804){\LARGE $\downarrow$}
 \usefont{T1}{ptm}{m}{n}
 \rput(5.507451,1.9825195){\LARGE $x'$}
 \psline[linewidth=0.04cm](5.086904,2.5175195)(5.086904,1.3175194)
 \psline[linewidth=0.04cm](5.8869042,2.5175195)(5.8869042,1.3175194)
 \psellipse[linewidth=0.04,dimen=outer](5.486904,2.5175195)(0.4,0.2)
 \usefont{T1}{ptm}{m}{n}
 \rput(3.6319432,2.0925195){\LARGE ...}
 \usefont{T1}{ptm}{m}{n}
 \rput(1.6874511,3.1825194){\LARGE $\downarrow$}
 \end{pspicture} 
 }}
\end{equation}

According to the choice of a monad $B$ (see
Example~\ref{example:branchingMonads}), different ``branching'' of such
pipes is allowed.
\begin{itemize}
 \item 
 When
 $B=\lift$ a pipe can be ``stuck'' or ``looped.'' A pipe connects an
       entrance $x$ with the exit $f(x)$---hence a
token entering at
 $x$ comes out of $f(x)\in Y$---when $f(x)$ is defined. A token is caught in
   the piping and does not come out in case $f(x)$ is undefined, i.e.\ if $f(x)$
       belongs to $1$ in
       $\lift=1+(\place)$.
 \item 
 When $B=\pow$ a pipe can branch, with one
 entrance $x$ connected to possibly multiple (or zero) exits 
 (namely those in
       $f(x)\subseteq Y$).
 \item 
 When $B=\dist$ a pipe can branch too, but this time the branching is 
 probabilistic.
\end{itemize}
For all these monads $B$ it is shown in~\cite{Jacobs10trace} that the
Kleisli
category $\Kleisli{B}$ is symmetric traced monoidal, with respect to 
$+$ as a monoidal product and $0$ (the empty set) as a monoidal unit.
In view of Figure~\ref{figure:constructionOfModel}, all these Kleisli
categories can support  construction of a linear category via
categorical GoI and realizability. 

Moreover, it is plausible 
that the resulting linear category inherit some features from its
ingredients---ultimately from the branching monad $B$. For example, we
start with $B=\pow$ and the outcome would be a linear category with
some nondeterminism built-in, hence suited for interpreting a language
with
nondeterminism.

Therefore for our purpose of obtaining a linear category with a
quantum flavor---and interpreting a quantum lambda calculus in it---the
first question is to find a branching monad with a quantum flavor. 
Our answer is the \emph{quantum branching monad} $\Q$ that we introduce now.

\subsection{The Quantum Branching Monad $\Q$}
\label{subsection:quantumBranchingMonad}
%
%
The following formal definition of $\Q$ below is hardly illustrative. The
intuition will be explained shortly, using the piping analogy
(\ref{eq:pipingDrawing}) for arrows in the Kleisli
category $\Kleisli{\Q}$. 

\begin{mydefinition}[The quantum branching monad $\Q$]
\label{definition:quantumBranchingMonad}
The \emph{quantum branching monad} $\Q:\Sets\to\Sets$ is defined as
 follows. On objects,
\begin{displaymath}
 \Q
 X\;=\;\Bigl\{c:X\to\dirsum_{m,n\in\nat}
 \QO_{m,n}\;\Bigl|\Bigr.\;
 \text{the trace condition~(\ref{eq:traceCondForQuantumMonad})}\Bigr\}
\end{displaymath}
 where:
 $\QO_{m,n}$ is
the set of quantum operations of the type $\DM_{m}\to \DM_{n}$
 (Definition~\ref{definition:quantumOperation});
 $\dirsum_{m,n\in\nat}$ denotes a Cartesian product; and
 the \emph{trace condition} stands for the following.
\begin{equation}\label{eq:traceCondForQuantumMonad}
  \sum_{x\in X}\sum_{n\in\nat}\trace\Bigl(\bigl(c(x)\bigr)_{m,n}(\rho)\Bigr)\le
 1\enspace,\;
 \text{$\forall m\in\nat,\;\forall \rho\in \DM_{m}$.}
\end{equation}
 Here $(c(x))_{m,n}$ is the $(m,n)$-component of 
 $c(x)\in\dirsum_{m,n}\QO_{m,n}$.
On arrows, given  $f:X\to Y$ we define
$\Q f: \Q X \to \Q Y$ as follows. For $c\in\Q X$ and $y\in Y$:
\begin{equation}\label{eq:QuantumMonadOnArrows}
 \bigl(\,(\Q f)(c)(y)\,\bigr)_{m,n}\;:=\;
\sum_{x\in f^{-1}(\{y\})} \bigl(\,c(x)\,\bigr)_{m,n}\enspace.
\end{equation}
Note that the sum on the right-hand side is well-defined, because of the
 upper bound given by the trace condition~(\ref{eq:traceCondForQuantumMonad}).
As for the monad structure, its unit $\eta_{X}:X\to \Q X$ is:
\begin{equation}\label{eq:unitOfQuantumMonad}
 \bigl(\,\eta_{X}(x)(x')\,\bigr)_{m,n}
 \; :=\;
 \begin{cases}
  \id_{m} & \text{if $x=x'$ and $m=n$,}
 \\
  0 & \text{otherwise.}
 \end{cases}
\end{equation}
Here $\id_{m}$ is the identity map; $0$ is the constant QO that maps
 everything to $0$.
The multiplication $\mu_{X}:\Q\Q X \to \Q X$ is defined by:
\begin{equation}\label{eq:multOfQuantumMonad}
 \bigl(\,\mu_{X}(\gamma)(x)\,\bigr)_{m,n}
 \;:=\;
 \sum_{c\in\Q X}\sum_{k\in \nat}
\Bigl( \bigl(c(x)\bigr)_{k,n}  \co
 \bigl(\gamma(c)\bigr)_{m,k}
\Bigr)\enspace.
\end{equation}
 The QO  
\begin{math}
  \bigl(c(x)\bigr)_{k,n}  \co
 \bigl(\gamma(c)\bigr)_{m,k}
\end{math} 
on the RHS is the sequential composition of QOs:
 given a density matrix
 $\rho\in \DM_{m}$ it first applies
 $(\gamma(c))_{m,k}\in\QO_{m,k}$ and then applies
$(c(x))_{k,n} \in \QO_{k,n}$, transforming $\rho$ eventually into an
 $n$-dimensional density matrix.
\end{mydefinition}
\noindent In \ref{appendix:quantumBranchingMonad} we prove that 
the sums in~(\ref{eq:QuantumMonadOnArrows})
and~(\ref{eq:multOfQuantumMonad})
exist,  that $\Q$ is indeed a functor, and that $\Q$ is indeed a monad.

Let us first note 
a common pattern 
that is exhibited by $\Q$ and the previous examples of branching monads $B$,
namely:
\begin{displaymath}
BX\;=\; 
\{c:X\to W\mid \text{a \emph{normalizing condition}}\}\enspace,
\quad\text{where $W$ is a set of \emph{weights}.}
\end{displaymath}
Specifically:
\begin{itemize}
 \item For $B=\lift$ the set $W$ is $2=\{0,1\}$ (\emph{stuck} or \emph{through}); the normalizing condition is
       \begin{displaymath}
	c(x)=1 \quad\text{for at most one $x\in X$.}
       \end{displaymath}
 \item For $B=\pow$ the set $W$ is $2=\{0,1\}$ again, but there is no
       normalizing condition.
 \item For $B=\dist$ the set $W$ is the unit interval $[0,1]$ and 
       the normalizing condition is $\sum_{x\in X}c(x)\le 1$.
 \item For $B=\Q$ the weights are a tuple (or a block matrix) of quantum operations and the 
       normalizing condition is the trace
       condition~(\ref{eq:traceCondForQuantumMonad}). 
\end{itemize}

Let us continue~(\ref{eq:pipingDrawing}) and think of an arrow
$f:X\relto Y$ in $\Kleisli{\Q}$ as piping.  The piping analogy is still
valid for $\Q$; a crucial difference however is that, for $B=\Q$,
\begin{displaymath}
  \boxed{\txt{a token that runs
 through pipes is no longer a mere particle,\\ but it \emph{carries a
 quantum state}.}}  
\end{displaymath}
Each entrance $x\in X$ is ready for an incoming token that carries
$\rho\in \DM_{m}$ of any finite dimension $m$.  Such a token gives rise
to one outcoming token. However, its exit can be any $y\in Y$ and the
quantum state carried by the token can be of any finite dimension $n\in
\nat$. We think of the piping to be applying a certain QO to the quantum
state carried by the token; the QO to be applied is concretely given by
\begin{displaymath}
 \bigl(\,f(x)(y)\,\bigr)_{m,n}
 \;\in\;\QO_{m,n}\enspace,
\end{displaymath}
that is determined by: which exit $y\in Y$
the token takes; and what is the dimension $n$ of the resulting quantum state.
\begin{equation}\label{eq:quantumPiping}
\raisebox{-.5\height}{
 \scalebox{.6} 
 {
 \begin{pspicture}(0,-5.20078)(13.558184,4.520078)
 \psframe[linewidth=0.04,dimen=outer,fillstyle=solid,fillcolor=lightgray](11.726904,1.9994141)(2.1269042,-1.0799218)
 \usefont{T1}{ptm}{m}{n}
 \rput(6.817451,0.54441404){\LARGE $f$}
 \usefont{T1}{ptm}{m}{n}
 \rput(6.907451,2.624414){\LARGE $x$}
 \psline[linewidth=0.04cm](6.486904,3.159414)(6.486904,1.959414)
 \psline[linewidth=0.04cm](7.2869043,3.159414)(7.2869043,1.959414)
 \usefont{T1}{ptm}{m}{n}
 \rput(4.107451,-1.575586){\LARGE $y$}
 \psline[linewidth=0.04cm](3.6869042,-1.040586)(3.6869042,-2.240586)
 \psline[linewidth=0.04cm](4.486904,-1.040586)(4.486904,-2.240586)
 \psellipse[linewidth=0.04,dimen=outer](6.8869042,3.159414)(0.4,0.2)
 \usefont{T1}{ptm}{m}{n}
 \rput(4.831943,2.7344139){\LARGE ...}
 \usefont{T1}{ptm}{m}{n}
 \rput(3.0319433,-1.465586){\LARGE ...}
 \usefont{T1}{ptm}{m}{n}
 \rput(8.8319435,2.7344139){\LARGE ...}
 \usefont{T1}{ptm}{m}{n}
 \rput(10.431943,-1.465586){\LARGE ...}
 \usefont{T1}{ptm}{m}{n}
 \rput(6.831943,-1.465586){\LARGE ...}
 \usefont{T1}{ptm}{m}{n}
 \rput(9.437451,-1.575586){\LARGE $y'$}
 \psline[linewidth=0.04cm](9.086905,-1.040586)(9.086905,-2.240586)
 \psline[linewidth=0.04cm](9.886904,-1.040586)(9.886904,-2.240586)
 \usefont{T1}{ptm}{m}{n}
 \rput(6.887451,3.8){\LARGE $\downarrow$}
 \psellipse[linewidth=0.04,dimen=outer](4.086904,-2.240586)(0.4,0.2)
 \psellipse[linewidth=0.04,dimen=outer](9.486904,-2.240586)(0.4,0.2)
 \usefont{T1}{ptm}{m}{n}
 \rput(3.527451,-4.355858){\LARGE 
 $
  \begin{array}{r}
   \Bigl(\bigl(f(x)(y)\bigr)_{m,n}(\rho)\Bigr)_{n\in\nat}\quad
   \\
    \in\prod_{n}\DM_{n}
  \end{array}  
$
}
 \usefont{T1}{ptm}{m}{n}
 \rput(10.427451,-4.355858){\LARGE
 $
  \begin{array}{r}
   \Bigl(\bigl(f(x)(y')\bigr)_{m,n}(\rho)\Bigr)_{n\in\nat}\quad
   \\
    \in\prod_{n}\DM_{n}
  \end{array}  
$
}
 \usefont{T1}{ptm}{m}{n}
 \rput[l](6.8869042,4.524414){\LARGE $\rho\in\DM_{m}$}
 \usefont{T1}{ptm}{m}{n}
 \rput(4.087451,-2.975586){\LARGE $\downarrow$}
 \usefont{T1}{ptm}{m}{n}
 \rput(9.487452,-2.975586){\LARGE $\downarrow$}
 \end{pspicture} 
 }}
\end{equation}
The trace condition~(\ref{eq:traceCondForQuantumMonad}) now reads,
for an arrow $f:X\relto Y$ in $\Kleisli{\Q}$:
\begin{equation}\label{eq:InstanceOfTraceCondForQuantumMonad}
   \sum_{y\in
    Y}\sum_{n\in\nat}\trace\Bigl(\bigl(f(x)(y)\bigr)_{m,n}(\rho)\Bigr)
 \;\le\;
 1\enspace,\;
 \text{for each $ m\in\nat$, $\rho\in\DM_{m}$ and $x\in X$.}
\end{equation}
The trace value $\trace\Bigl(\bigl(f(x)(y)\bigr)_{m,n}(\rho)\Bigr)$ is
understood as the
probability with which a token  $\rho$ entering
at $x$ leads to an $n$-dimensional token at
$y$. These probabilities must add up to at most $1$ when 
the exit $y$ and the outcoming dimension $n$ vary. This
is precisely the condition~(\ref{eq:InstanceOfTraceCondForQuantumMonad}).

The composition $\Kco$ of Kleisli arrows can then be understood as 
 sequential connection of such piping,
one after another.
\begin{equation}\label{eq:kleisliComp}
 \raisebox{-.5\height}{
 \scalebox{.6} 
 {
 \begin{pspicture}(-3,-5.0194923)(8.716651,4.594923)
 \psframe[linewidth=0.04,dimen=outer,fillstyle=solid,fillcolor=lightgray](4.68,2.398828)(0.0,0.7588281)
 \rput(2.290547,1.5438281){\LARGE $f$}
 \rput(2.380547,3.023828){\LARGE $x$}
 \psline[linewidth=0.04cm](1.96,3.558828)(1.96,2.358828)
 \psline[linewidth=0.04cm](2.76,3.558828)(2.76,2.358828)
 \rput(1.3805468,0.2238281){\LARGE $y$}
 \psline[linewidth=0.04cm](0.96,0.7588281)(0.96,-0.44117188)
 \psline[linewidth=0.04cm](1.76,0.7588281)(1.76,-0.44117188)
 \psellipse[linewidth=0.04,dimen=outer](2.36,3.558828)(0.4,0.2)
 \rput(1.305039,3.133828){\LARGE ...}
 \rput(0.30503908,0.3338281){\LARGE ...}
 \rput(3.305039,3.133828){\LARGE ...}
 \rput(4.105039,0.3338281){\LARGE ...}
 \rput(2.1050391,0.3338281){\LARGE ...}
 \rput(3.1105468,0.2238281){\LARGE $y'$}
 \psline[linewidth=0.04cm](2.76,0.7588281)(2.76,-0.44117188)
 \psline[linewidth=0.04cm](3.56,0.7588281)(3.56,-0.44117188)
 \psframe[linewidth=0.04,dimen=outer,fillstyle=solid,fillcolor=lightgray](4.68,-0.40117186)(0.0,-2.0411718)
 \rput(2.3205469,-1.256172){\LARGE $g$}
 \rput(2.380547,-2.5761719){\LARGE $u$}
 \psline[linewidth=0.04cm](1.96,-2.0411718)(1.96,-3.241172)
 \psline[linewidth=0.04cm](2.76,-2.0411718)(2.76,-3.241172)
 \psellipse[linewidth=0.04,dimen=outer](2.36,-3.241172)(0.4,0.2)
 \rput(1.305039,-2.466172){\LARGE ...}
 \rput(3.305039,-2.466172){\LARGE ...}
 \rput[l](2.2,4.223828){\LARGE $\downarrow \; \rho\in\DM_{m}$}
 \rput[l](4.643232,0.22382812){\LARGE $\downarrow$ ($k$-dimensional)}
 \rput[l](2.20432324,-3.976172){\LARGE $\downarrow$ ($n$-dimensional)}
 \end{pspicture} 
 }}
\end{equation}
Here the numbers $m$, $k$ and $n$ stand for the dimension of the quantum states
carried by the token, at each stage of the piping.

Concretely, the Kleisli composition $\Kco$ is described as follows. 
\begin{mylemma}[Composition $\Kco$ in $\Kleisli{\Q}$]
\label{lem:compositionOfQKleisliArrowsInDetail}
 Given two successive arrows $f:X\relto Y$ and $g:Y\relto U$ in
 $\Kleisli{\Q}$, their composition $g\Kco f:X\relto U$ is concretely
 given as follows.
 \begin{displaymath}
\Bigl(\,( g \Kco f)(x)(u)\,\Bigr)_{m,n}
 \;=\;
  \sum_{y\in Y}\sum_{k\in\nat}
   \bigl(g(y)(u)\bigr)_{k,n}\co
  \bigl(f(x)(y)\bigr)_{m,k}\enspace.
\end{displaymath}
\end{mylemma}
\begin{myproof}
 See \ref{subsection:KleisliQ}.
\end{myproof}
This description of $\Kco$ in $\Kleisli{\Q}$ is ultimately
 due to the definition~(\ref{eq:multOfQuantumMonad}) of the
multiplication operation $\mu$.  We notice
its similarity
 to the multiplication operation of the
 subdistribution monad $\dist$.
 The latter 
is defined by
\begin{displaymath}
 \mu^{\dist}_{X}(\gamma)(x)\; =\; \sum_{c\in\dist X} \gamma(c)\cdot c(x)\enspace,
\end{displaymath}
where $\cdot$ denotes multiplication of real numbers.
This notably resembles~(\ref{eq:multOfQuantumMonad}).

\begin{myremark} \label{remark:QDCCbyQuantumBranchingMonad}
 The reason why $\Q$ is called a quantum branching monad is that a
 Kleisli arrow $f:X\relto Y$ in $\Kleisli{\Q}$---thought of as piping
 like~(\ref{eq:quantumPiping})---is a ``quantum branching
 function.'' This is in the same sense as an arrow $f:X\relto Y$ in
 $\Kleisli{\pow}$ is a ``nondeterministically branching function'' and 
 an arrow $f:X\relto Y$ in
 $\Kleisli{\dist}$ is a ``probabilistically branching function.'' 

 An example of such an arrow in $\Kleisli{\Q}$ is given by the
 following $f_{1}$. Here $k,l,m$ and $n$ are natural numbers, and $\rho\in\DM_{m}$ is
 an $m$-dimensional density matrix.
 \begin{align*}
  &f_{1}\;:\;\nat\longrelto\nat\enspace,
 \\
  &\bigl(f_{1}(k)(l)\bigr)_{m,n}(\rho)
 :=
  \begin{cases}
   \bra{0}\rho\ket{0}
  &
   \text{if $m=2$, $n=1$ and $l=2k$,}
  \\
   \bra{1}\rho\ket{1}
  &
   \text{if $m=2$, $n=1$ and $l=2k+1$,}
  \\
   0
 &
   \text{otherwise.}
  \end{cases}
 \end{align*}
 Imagine a token carrying a quantum state $\rho\in\DM_{m}$ entering this
 piping at the entrance $k\in\nat$. The token does not come out at all unless
 $\rho$ is 2-dimensional. If $\rho$ is 2-dimensional, 
 the token might come out of the exit $2k\in\nat$ or $2k+1\in\nat$. 
 To each of these exits the assigned value is $\bra{0}\rho\ket{0}$ and 
 $\bra{1}\rho\ket{1}$, respectively: these numbers in $\C^{1}$ (or
 rather $[0,1]$) are
 understood as the probabilities with which the token takes the exit.

 This way we are modeling \emph{branching structure that depends on
 quantum
 data}---or \emph{classical control and quantum data}. The principle is:
 \begin{itemize}
  \item a classical control structure is represented by the pipe the
	token is in; and
  \item quantum data is the one carried by the token.
 \end{itemize}
 Notice also that we are essentially relying on the separation of a
 measurement into projections (see Remark~\ref{remark:splitMeasIntoProjs}).

 A slightly more complicated example is  the
 following $f_{2}$. Here $N\in\nat$ is a natural number.
 \begin{align*}
  &f_{2}\;:\;\nat\longrelto\nat\enspace,
 \\
  &\bigl(f_{2}(k)(l)\bigr)_{m,n}(\rho)
 :=
  \begin{cases}
   \bra{0_{1}}\rho\ket{0_{1}}
  &
   \text{if $m=2^{N+1}$, $n=2^{N}$ and $l=2k$,}
  \\
   \bra{1_{1}}\rho\ket{1_{1}}
  &
   \text{if $m=2^{N+1}$, $n=2^{N}$ and $l=2k+1$,}
  \\
   0
 &
   \text{otherwise.}
  \end{cases}
 \end{align*}
Here an incoming token carries an $(N+1)$-qubit state
 $\rho\in\DM_{2^{N+1}}$, and the arrow $f_{2}$ measures its first qubit 
 (with respect to the basis consisting of
 $\ket{0_{1}}$ and $\ket{1_{1}}$), resulting in the token  sent to different
 exists according to the outcome. The outcoming token
 carries a state
\begin{displaymath}
    \bra{0_{1}}\rho\ket{0_{1}} 
    \quad\text{or}\quad
    \bra{1_{1}}\rho\ket{1_{1}}
    \quad\in\DM_{2^{N}}
\end{displaymath} 
that represents the qubits from the second to the $(N+1)$-th.
Here the trace value of each of the two density matrices implicitly represents
the probability with which the token is sent to the corresponding exit.
See Remark~\ref{remark:splitMeasIntoProjs}.

 It is not only measurements that we can model using arrows in
 $\Kleisli{\Q}$. Consider the following $f_{3}$.
 Here $K\in\nat$ is a natural number.
 \begin{align*}
  &f_{3}\;:\;\nat\longrelto\nat\enspace,
 \\
  &\bigl(f_{3}(k)(l)\bigr)_{m,n}(\rho)
 :=
  \begin{cases}
   \rho\otimes\ket{0}\bra{0}
  &
   \text{if $m=2^{N}$, $n=2^{N+1}$ and $k=l=2K$,}
  \\
   \rho
  &
   \text{if $m=n=2^{N}$ and $k=l=2K+1$,}
  \\
   0
 &
   \text{otherwise.}
  \end{cases}
 \end{align*}
This arrow models (conditional) \emph{state preparation}: it adds, to a token coming in at $k=2K$, a prepared state
$\ket{0}\bra{0}$ as an $(N+1)$-th qubit. 

Furthermore, the composition $f_{3}\Kco f_{2}:\nat\relto\nat$ represents
 the following operation: it measures the first qubit; and if the
 outcome is $\ket{0} $, it adjoins a new qubit.
\end{myremark}

The monad $\Q$ indeed satisfies the conditions
 in~\cite{Jacobs10trace}---much like $\lift,\pow$ and $\dist$, it is
 equipped with a suitable cpo structure---so that the Kleisli category
 $\Kleisli{\Q}$ is a \emph{traced symmetric monoidal category (TSMC)}.
\begin{mydefinition}[Order $\Lle$ on $\Q X$]
\label{definition:pointwiseorderOnQX}
We endow the set $\Q X$ with the pointwise extension of the
L\"{o}wner partial order in Definition~\ref{definition:pointwiseorderOnQO}. Namely:
given $c,d\in \Q X$, we have
$  c\Lle d$ if for each $x\in X,m,n\in\nat$,
$  (d(x))_{m,n}\Lle (c(x))_{m,n}$.

\end{mydefinition}

\begin{mytheorem}
\label{theorem:KleisliOfQIsTraced}
The category $\Kleisli{\Q}$ is partially
 additive  (a notion from \cite{ArbibM86}). Therefore
 by~\cite[Chap.~3]{Haghverdi00PhD}, $(\Kleisli{\Q},0,+)$ is a
 TSMC, with
 its trace operator given explicitly by Girard's execution formula.
\end{mytheorem}
\begin{myproof}
 By~\cite[Proposition~4.8]{Jacobs10trace}; see
 Theorem~\ref{theorem:KleisliOfQIsTracedInDetail}
for details. Notably, the Kleisli category $\Kleisli{\Q}$ is 
 $\omega\text{-}\mathbf{CPO}$ enriched: a homset $\Kleisli{\Q}(X,Y)$
 is equipped with the order that is the pointwise extension of that on $\Q
 Y$ (Definition~\ref{definition:pointwiseorderOnQX}). The trace operator
 will be described later, in the proof of Theorem~\ref{theorem:KleisliQGoISituation}.
\myqed
\end{myproof}

\begin{myremark}
 Continuing Remark~\ref{remark:QDCCbyQuantumBranchingMonad}, we note
 that in the field of quantum programming languages the study of
 \emph{quantum control} is emerging. With initial observations
 in~\cite{YingYF14,BadescuP15QPLtoAppear}, this line of work aims at
 extending the by-now accepted paradigm of \emph{classical control and
 quantum data}. Under quantum control, the current program counter of a
 program's execution can be a quantum superposition of multiple program
 locations. In the token analogue in the current section, this means
 that the token's position can be superposed---this is not possible with
 our current monad $\Q$, where a token's position can be a probabilistic
 ensemble of different positions but is never a quantum superposition.

 Some authors use the terminologies \emph{classical} and \emph{quantum
 branching} to distinguish the choices of classical/quantum control
 structures. In this view the name ``quantum branching monad'' of our
 monad $\Q$ is utterly inappropriate---it is more precisely a
 ``classical control and quantum data monad.'' We shall stick to this
 name, however, in view of other branching monads such as $\lift, \pow$
 and $\dist$.
\end{myremark}

\subsection{A Linear Combinatory Algebra via Categorical GoI
}
\label{subsection:LCAviaCategoricalGoI}

In the current section (\S{}\ref{subsection:LCAviaCategoricalGoI}) we
shall
elaborate on the high-level description
in~\S{}\ref{subsection:prelimGoI} and  review the general definitions and results
in~\cite{AbramskyHS02} on categorical GoI, using
the prototypical example
$\mathbf{Pfn}\cong\Kleisli{\lift}$ for providing some intuitions
(where $\lift$ is the lift monad given in Example~\ref{example:branchingMonads}).
 The constructions in~\cite{AbramskyHS02}, when
applied to $\mathbf{Pfn}\cong\Kleisli{\lift}$, lead to a model that 
is pretty close to token machines in~\cite{DanosR99,Mackie95}.
Later in~\S{}\ref{subsection:categoricaGoIQuantumInstance},  building on the categorical theory of branching
in~\S{}\ref{subsection:quantumBranchingMonadBackground}--\ref{subsection:quantumBranchingMonad},
we will observe that 
the Kleisli category 
$\Kleisli{\Q}$ for the quantum branching monad also constitutes 
an example of categorical GoI. This fact  suggests that the resulting model is a ``quantum
variant''---one among multiple possibilities, at least---of
token machine semantics.

\begin{mydefinition}[Retraction]\label{definition:retraction}
 Let $X$ and $Y$ be objects of a category $\C$. A \emph{retraction} from $X$
 to $Y$ is a pair of arrows $f:X\to Y$ and $g:Y\to X$ such that $g\co f
 = \id_{X}$, that is,
 \begin{displaymath}
  \vcenter{\xymatrix@1@C+0em{
   {X}
      \ar@(lu,ld)[]_{\id}
      \ar@/^/[r]^{f}
   &
   {Y}
   \shifted{2em}{0em}{.}
      \ar@/^/[l]^{g}
}}
 \end{displaymath}
 Such a retraction shall be denoted by $f:X\retr Y:g$, following~\cite{AbramskyHS02}.
\end{mydefinition}

Throughout the current paper all the examples of retractions are in fact
isomorphisms,  although retractions are sufficient for the axiomatic
developments. 

\begin{mydefinition}[GoI situation]\label{definition:GoIsituation}
A \emph{GoI situation} is a triple $(\C, F, U)$ where
\begin{itemize}
 \item $\C=(\C, I,\otimes)$ is a traced symmetric monoidal category
       (TSMC),
       see e.g.~\cite{JoyalSV96,Hasegawa09}. 
       This means that  $\C$ is a monoidal category  equipped with a \emph{trace} operator
       \begin{displaymath}
	\infer{X\xrightarrow{\trace^{Z}_{X,Y}(f)} Y\quad\text{in $\C$}}{X\otimes
	Z\xrightarrow{f}Y\otimes Z\quad\text{in $\C$}}
       \end{displaymath}
       that is subject to certain equational axioms.
 \item $F:\C\to\C$ is a traced symmetric monoidal functor, equipped with
       the following retractions (which are monoidal natural transformations).
       \begin{align*}
	e\;:\; FF &\retr F \;:\; e'&  
	&\text{Comultiplication}\\
	d\;:\; \id &\retr F \;:\; d'&  &\text{Dereliction}\\
	c\;:\; F\otimes F &\retr F \;:\; c'&  &\text{Contraction}\\
	w\;:\; K_{I} &\retr F \;:\; w'&&\text{Weakening}  
       \end{align*}
       Here $K_{I}$ is the constant functor into the monoidal unit
       $I$;
 \item $U\in\C$ is an object (called \emph{reflexive object}), equipped
       with the following retractions.
       \begin{align*}
        j\;:\; U\otimes U &\retr U \;:\; k
	\\
	I &\retr U 
	\\
	u\;:\; FU &\retr U \;:\; v
       \end{align*}
\end{itemize}
\end{mydefinition} 

Let us try to provide some intuitions on the notion of GoI
situation; in particular, on how this abstract categorical notion manages
to encapsulate the essence of GoI, that is, interaction-based semantics
of computation. We shall use a prototypical example of GoI situations from~\cite{AbramskyHS02}.
\begin{mylemma}[\cite{AbramskyHS02}]\label{lem:KleisliLGoISituation}
 The triple $\bigl(\,(\Kleisli{\lift},0,+),\, \nat\cdot\place\,,\, \nat\,\bigr)$ forms a GoI
 situation.
Here the functor
 $\nat\cdot\place:\Kleisli{\lift}\to\Kleisli{\lift}$ carries a set $X$ to the
 coproduct  $\nat\cdot X$ of countably many copies of $X$ (i.e.\ to
 the $\nat$-th copower of $X$).  
\myqed
\end{mylemma}
In the token-based presentation of GoI like in~\cite{DanosR99,Mackie95},
a term $M$ in a calculus is interpreted as a \emph{partial function} $\sem{M}$
from some countable set to another; let us say it is of type
$\sem{M}\colon \nat \relto \nat$. Note here that we used the general notation
$\relto$ for Kleisli arrows, since partial functions are nothing but
Kleisli arrows for the lift monad $\lift$.
 Recall the piping analogy
from~\S{}\ref{subsection:quantumBranchingMonadBackground}: in this
analogy, the interpretation $\sem{M}$ is a piping with countably many
entrances and exits. See the figure~(\ref{eq:pipingDrawing}).

What is intriguing about GoI is how  a function application
$MN$ is interpreted.
In this paper we aim at exploiting GoI in deriving a categorical and
denotational (hence algebraic and compositional) model---therefore we
should be able to derive $\sem{MN}\colon \nat \relto \nat$ 
from the interpretations
$\sem{M}\colon \nat \relto \nat$ and
$\sem{N}\colon \nat \relto \nat$ of the constituent
parts.\footnote{It is often emphasized (e.g.\ in~\cite{Girard89GoI}) that GoI semantics is different
from ``denotational semantics,'' in that the former explicitly uses 
the execution formula as a semantical counterpart of
$\beta$-reduction, while commonly in denotational semantics the
interpretation of terms is unchanged. In this paper we follow categorical
GoI~\cite{AbramskyHS02} and abstract away from this difference.}
This happens as follows. Here we use the \emph{piping} analogy
in~\S{}\ref{subsection:quantumBranchingMonadBackground}, identifying a
partial function $f\colon \nat\relto\nat$ with a piping with countably
many entrances and exits, with possible loops (a token enters and it
might not come out).
\begin{multline}\label{eq:parCompAndHiding}
 \raisebox{-.5\height}{
\includegraphics[width=13em]{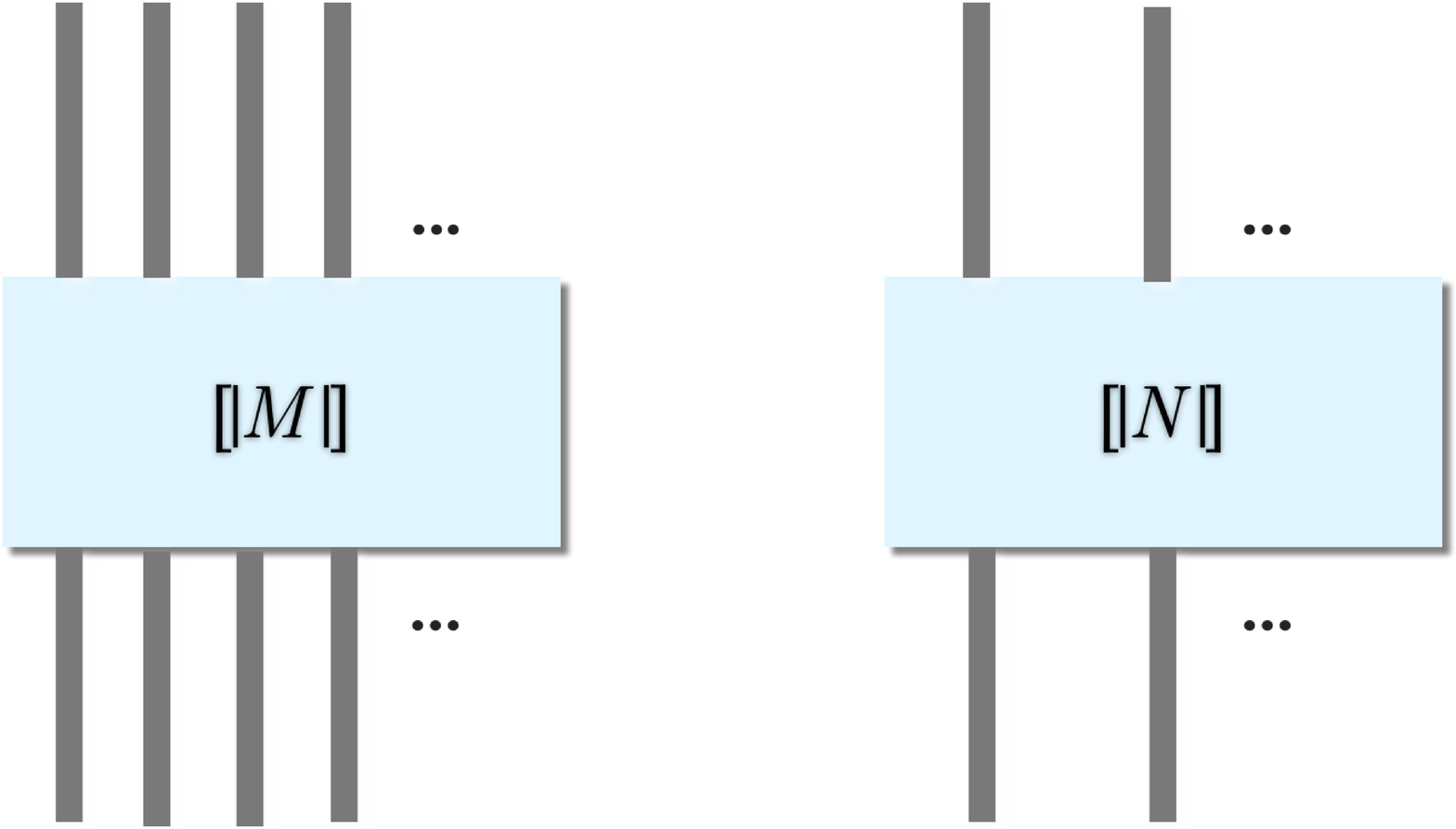}
}
\Longrightarrow
 \raisebox{-.5\height}{
\includegraphics[width=13em]{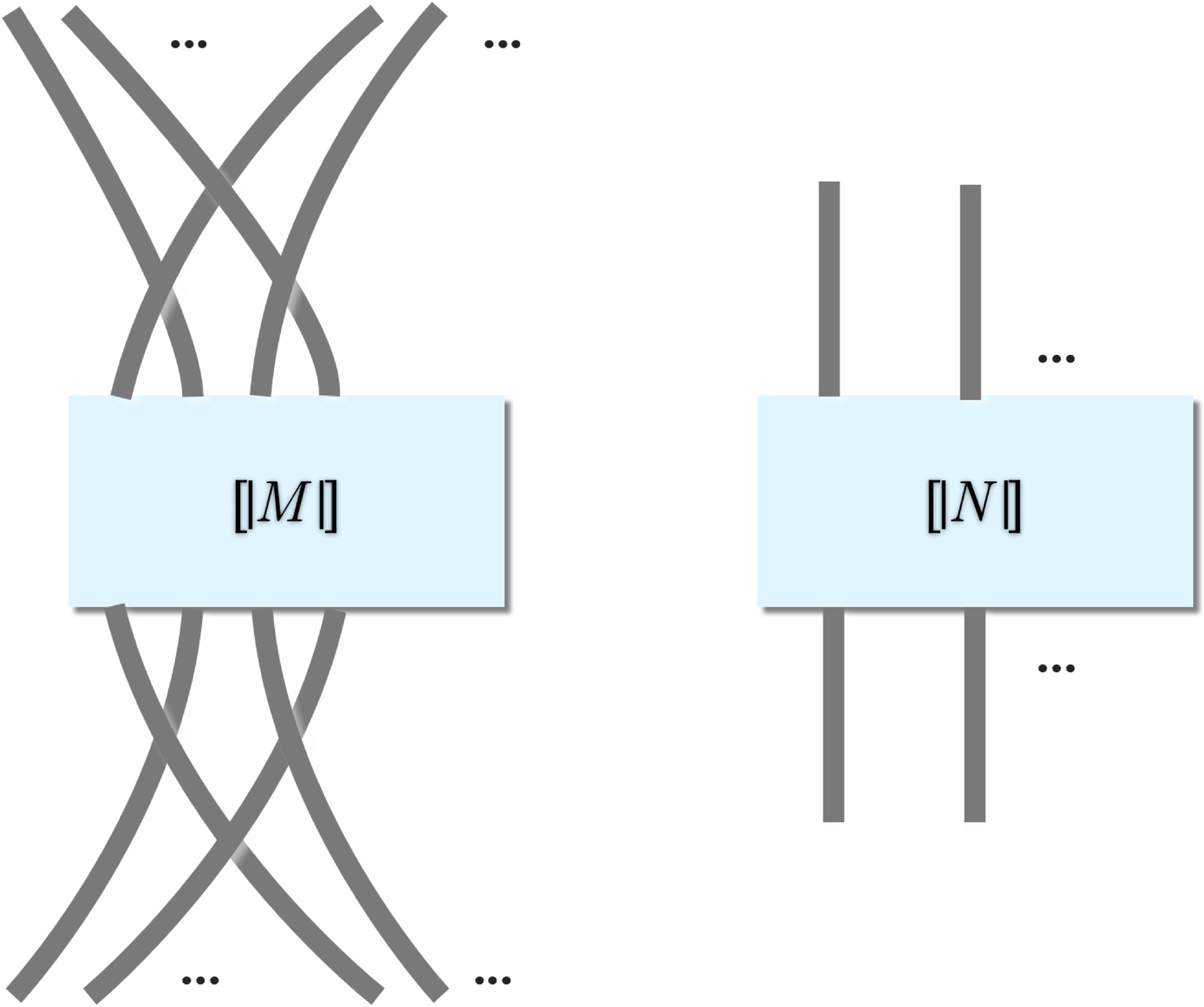}
}
 \Longrightarrow
 \\
 \raisebox{-.5\height}{
\includegraphics[width=13em]{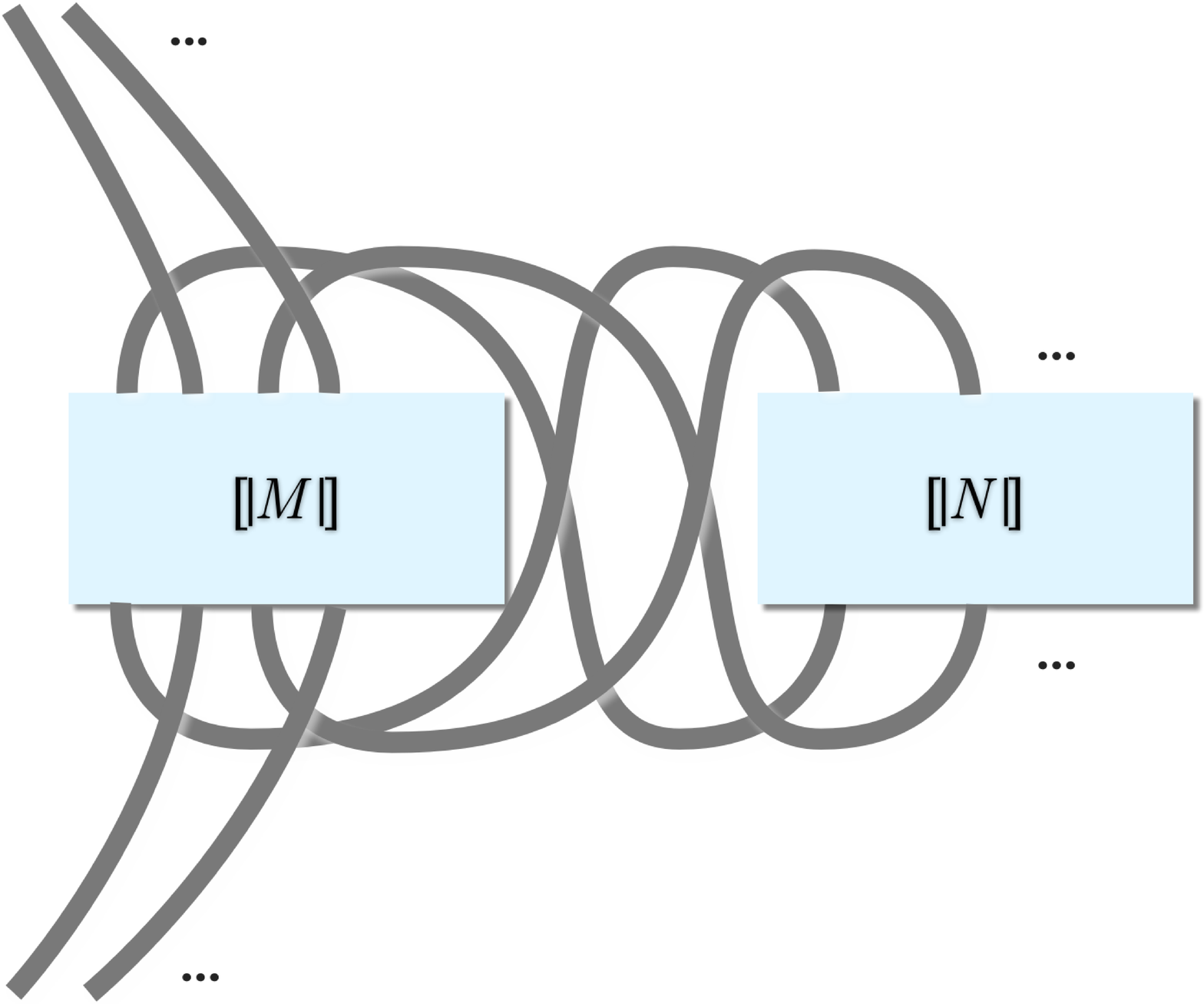}
}
\Longrightarrow
 \raisebox{-.5\height}{
\includegraphics[width=13em]{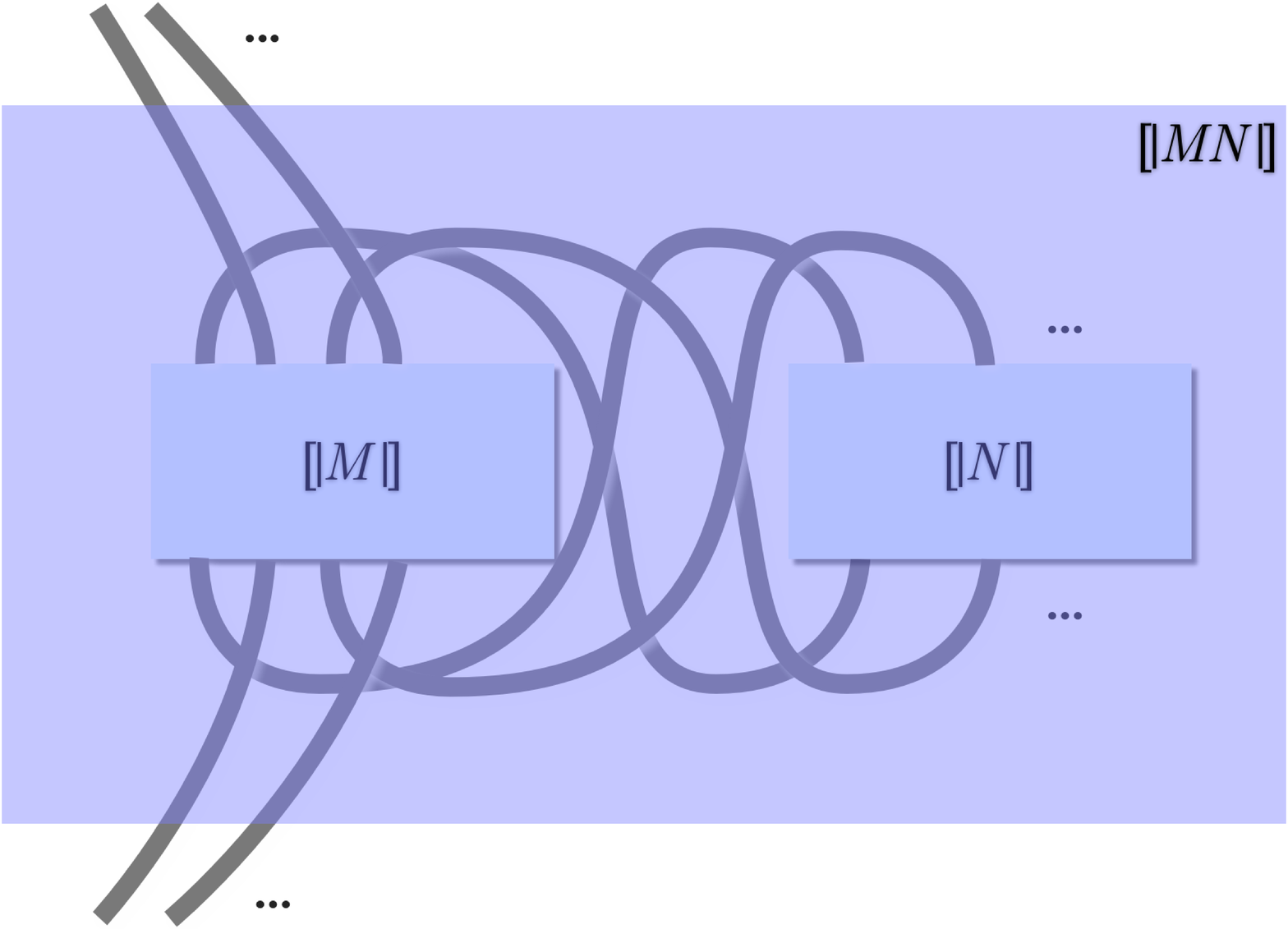}
}
\end{multline}
Let us go through the above process by one step after another.
In the first step ($\Longrightarrow$), a bundle of countably many pipes (for the entrances
of $\sem{M}$) is split into two bundles (left and right); and the same happens for the
exits of $\sem{M}$. This is possible because the set of natural numbers
is isomorphic to two copies of it ($\nat+\nat\cong\nat$); axiomatically
this
is what is required of the reflexive object $U$ in
Definition~\ref{definition:GoIsituation}. In the second step we
interconnect the bundles on the right into/from $\sem{M}$, and the
exits/entrances of $\sem{N}$, in the following way.
\begin{displaymath}
 \raisebox{-.5\height}{
\includegraphics[width=13em]{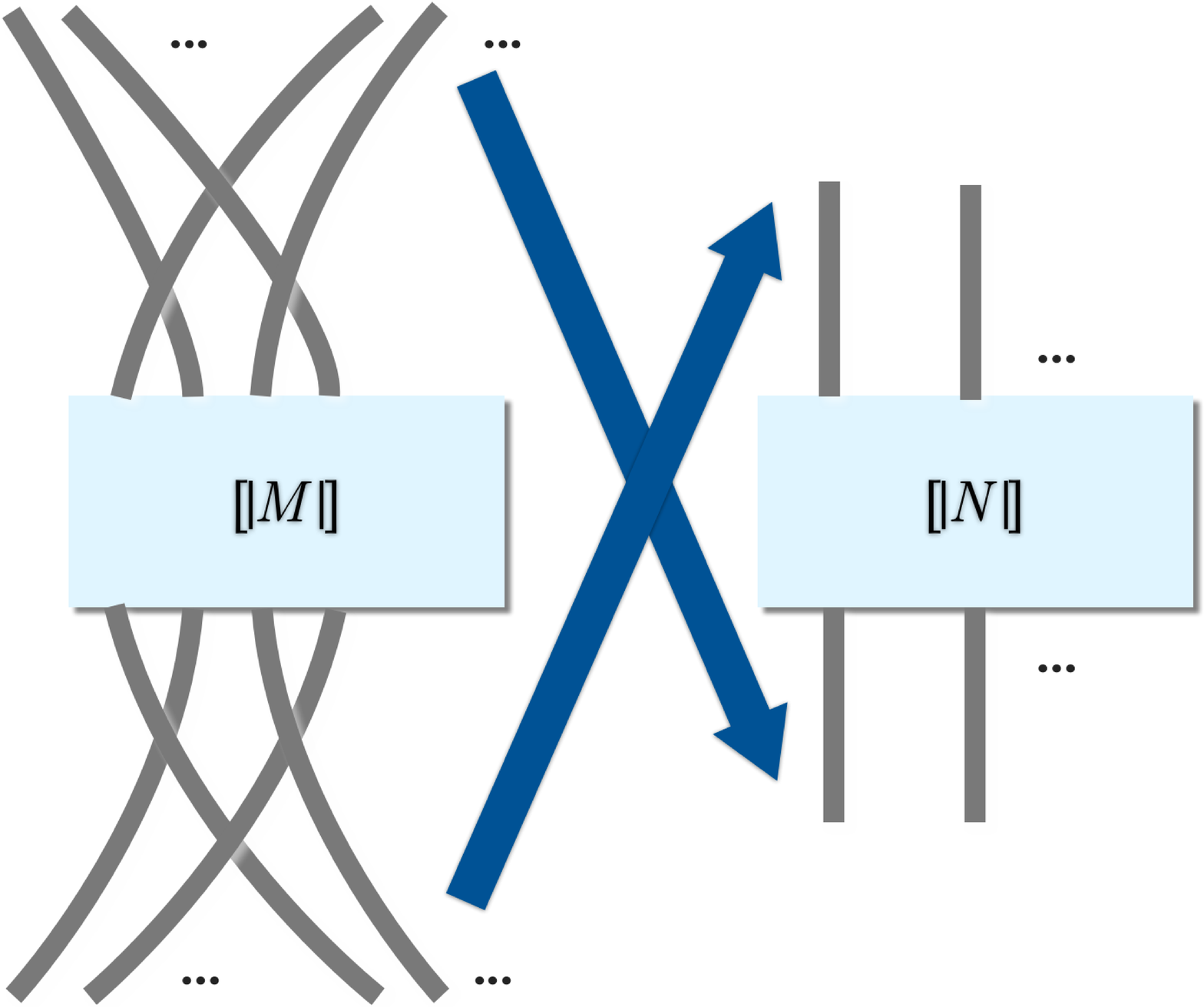}
}
\end{displaymath}
Notice that, after this second step ($\Longrightarrow$), we are now
looking at a piping with countably many entrances (top-left) and
countably many exits (bottom-left)---that is, a partial function
$\nat\relto\nat$. The third step ($\Longrightarrow$) just means that
we take the piping (i.e.\ a partial function) thus obtained as the
interpretation $\sem{MN}$ of the function application $MN$. 

One can also think of the process~(\ref{eq:parCompAndHiding}) as
``parallel composition and hiding,'' a design principle often heard in
game semantics~\cite{AbramskyJM00,HylandO00}. Specifically, the interpretations $\sem{M},\sem{N}$ of
the
 constituent terms operate in parallel, communicating with each other by passing a
token
(information is communicated by the choice of a pipe via which to pass the token, see
below);
and the interpretation $\sem{MN}$ is finally obtained by ``hiding''
the internal interactions between $\sem{M}$ and $\sem{N}$ (the third
$\Longrightarrow$ in~(\ref{eq:parCompAndHiding})).
 \begin{equation}\label{eq:tokensInParCompAndHiding}
   \raisebox{-.5\height}{
 \includegraphics[width=13em]{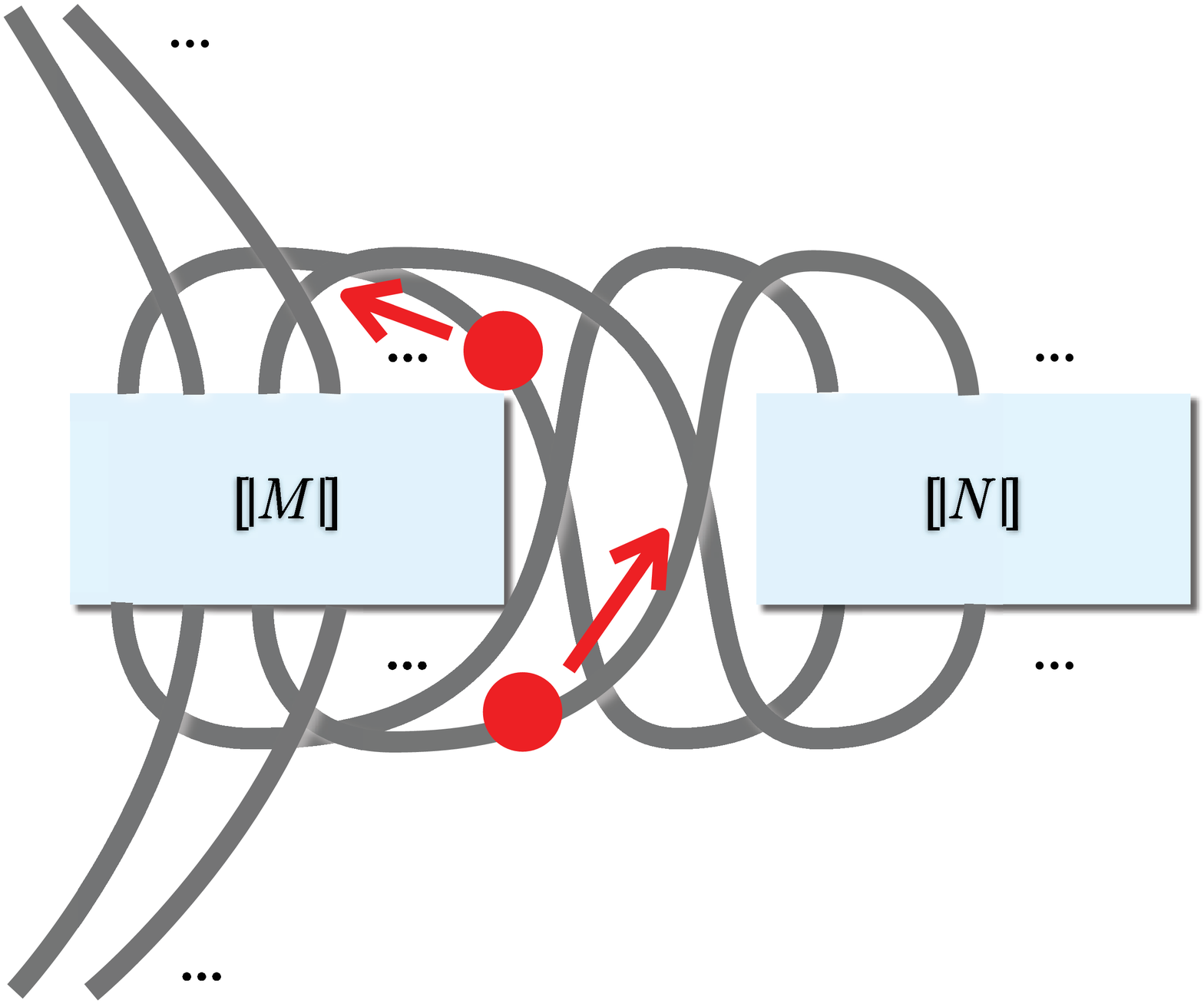}
}
 \end{equation}
 Let us also note that partiality is crucial in GoI. Even if $\sem{M}$
 and $\sem{N}$ are total functions from $\nat$ to $\nat$, the piping
 obtained in~(\ref{eq:tokensInParCompAndHiding}) does not necessarily
 correspond
 to a total function: a token that enters from one of the top-left pipes
 can be caught in an infinite loop between $\sem{M}$ and $\sem{N}$,
 never reaching one of the exits (the bottom-left pipes).

 As we already mentioned, a traced symmetric monoidal category
       (TSMC) is a main component of the notion of
 \emph{GoI situation} (Def.~\ref{definition:GoIsituation}), a
 categorical axiomatization of GoI given in~\cite{AbramskyHS02}.
 The reason is simple, 
in view of the above piping and token intuitions about
 GoI. Recall that a symmetric monoidal category $\C$ being \emph{traced}
 means that it is equipped with a trace operator; in our current setting
 where the tensor product of $\C=\Kleisli{\lift}$ is given by coproduct
 $+$, the type of the trace operator is
        \begin{displaymath}
	\vcenter{\infer{X\xrightarrow{\trace^{Z}_{X,Y}(f)} Y\quad\text{in $\Kleisli{\lift}$}}{X+
	Z\xrightarrow{f}Y+ Z\quad\text{in $\Kleisli{\lift}$}}}\enspace.
	\end{displaymath}
  Moreover the operator's action can be depicted as follows, using the
  string diagram formalism for (arrows in) monoidal
  categories~\cite{JoyalS91}.
  \begin{displaymath}
   \raisebox{-.5\height}{
\scalebox{.5} 
{
\begin{pspicture}(0,-1.53)(3.7381835,1.53)
\psframe[linewidth=0.04,dimen=outer](3.0991683,0.55)(0.67690426,-0.49)
\usefont{T1}{ptm}{m}{n}
\rput(1.7374512,-0.025){\LARGE $f$}
\psline[linewidth=0.04cm](1.0991684,1.51)(1.1169043,0.49)
\psline[linewidth=0.04cm](1.1169043,-0.51)(1.0991684,-1.49)
\psline[linewidth=0.04cm](2.6991684,1.51)(2.7169042,0.49)
\psline[linewidth=0.04cm](2.6991684,-0.49)(2.7169042,-1.51)
\usefont{T1}{ptm}{m}{n}
\rput(0.45745116,0.975){\LARGE $X$}
\usefont{T1}{ptm}{m}{n}
\rput(0.45745116,-1.025){\LARGE $Y$}
\usefont{T1}{ptm}{m}{n}
\rput(3.217451,0.975){\LARGE $Z$}
\usefont{T1}{ptm}{m}{n}
\rput(3.217451,-1.025){\LARGE $Z$}
\end{pspicture} 
}
   }   
   \quad\longmapsto\quad
   \raisebox{-.5\height}{
\scalebox{.5} 
{
\begin{pspicture}(0,-1.52)(3.5369043,1.52)
\psarc[linewidth=0.04](3.1169043,0.88){0.4}{0.0}{180.0}
\psarc[linewidth=0.04](3.1169043,-0.92){0.4}{-180.0}{6.262063}
\psframe[linewidth=0.04,dimen=outer](3.0991683,0.54)(0.67690426,-0.5)
\usefont{T1}{ptm}{m}{n}
\rput(1.9974512,-0.035){\LARGE $\trace^{Z}_{X,Y}(f)$}
\psline[linewidth=0.04cm](1.0991684,1.5)(1.1169043,0.48)
\psline[linewidth=0.04cm](1.1169043,-0.52)(1.0991684,-1.5)
\psline[linewidth=0.04cm](2.6991684,0.9)(2.6991684,0.5)
\psline[linewidth=0.04cm](2.6991684,-0.5)(2.6991684,-0.9)
\usefont{T1}{ptm}{m}{n}
\rput(0.45745116,0.965){\LARGE $X$}
\usefont{T1}{ptm}{m}{n}
\rput(0.45745116,-1.035){\LARGE $Y$}
\psline[linewidth=0.04cm](3.4991684,0.9)(3.4991684,-0.9)
\end{pspicture} 
}}
  \end{displaymath}
  A trace operator can be thought of as a \emph{feedback} operator in
  many examples. This is also the case in our current specific instance of
  $\Kleisli{\lift}\cong\Pfn$. Here a trace operator can be concretely given in the
  following straightforward manner, exploiting partiality. Given
  $f\colon X+Z\relto Y+Z$, let $f_{XY}\colon X\relto Y$ be its
  ``restriction'' to $X$ and $Y$, that is
  \begin{equation}\label{eq:traceInPfn}
   f_{XY}(u)\;:=\;
   \begin{cases}
    f(u) &\text{if $u\in X$ and $f(u)\in Y$,}
    \\
    \bot \text{ (undefined)} &\text{otherwise.}
   \end{cases}
  \end{equation}
  We define $f_{XZ}$, $f_{ZY}$ and $f_{ZZ}$ in a similar manner, and we let
  $\trace^{Z}_{X,Y}(f)\colon X\relto Y$ be
  \begin{equation}\label{eq:execFormulaPfn}
   \trace^{Z}_{X,Y}(f)(x)
   \;:=\;
   f_{XY}(x)\sqcup \bigsqcup_{n\in \nat}
   \bigl(\,f_{ZY}\co f_{ZZ}^{n}\co f_{XZ}
   \,\bigr)(x)\enspace,
  \end{equation}
  where $\sqcup$ denotes supremums in the flat cpo $\lift
  Y=\{\bot\}+Y$. The right-hand side informally reads: either the token
  immediately comes out of $Y$, or it comes out of $Z$, in which case it
  is fed back to the $Z$-entrance again. It bears notable similarity to
  Girard's \emph{execution formula}, too.

  This trace operator is what allows the series of transformations described
in~(\ref{eq:parCompAndHiding}). In~(\ref{eq:parCompAndHiding}) we used
pipes to informally depict partial functions; the same can be described
formally,
using string diagrams~\cite{JoyalS91} for the (traced) monoidal category
$(\Kleisli{\lift},0,+)$, as follows.
   \begin{multline}
       \raisebox{-.5\height}{
\scalebox{.5} 
{
\begin{pspicture}(0,-1.73)(5.5681834,1.73)
\psframe[linewidth=0.04,dimen=outer](2.3783593,0.49)(0.37835938,-0.51)
\usefont{T1}{ptm}{m}{n}
\rput(1.3474512,-0.005){\LARGE $\sem{M}$}
\psframe[linewidth=0.04,dimen=outer](4.998359,0.41)(3.36,-0.49)
\usefont{T1}{ptm}{m}{n}
\rput(4.1674514,-0.085){\LARGE $\sem{N}$}
\psline[linewidth=0.04cm](1.36,1.71)(1.3783593,0.49)
\usefont{T1}{ptm}{m}{n}
\rput(1.0179981,0.955){\LARGE $\nat$}
\psline[linewidth=0.04cm](4.16,-0.49)(4.16,-1.49)
\usefont{T1}{ptm}{m}{n}
\rput(4.517998,-1.045){\LARGE $\nat$}
\psline[linewidth=0.04cm](4.16,1.31)(4.16,0.39)
\usefont{T1}{ptm}{m}{n}
\rput(4.517998,0.755){\LARGE $\nat$}
\psline[linewidth=0.04cm](1.36,-0.49)(1.3783593,-1.71)
\usefont{T1}{ptm}{m}{n}
\rput(1.0179981,-1.245){\LARGE $\nat$}
\end{pspicture} 
}
    }
    \quad\longmapsto\quad
       \raisebox{-.5\height}{
\scalebox{.5} 
{
\begin{pspicture}(0,-2.93)(5.5681834,2.93)
\psframe[linewidth=0.04,dimen=outer](2.3783593,0.49)(0.37835938,-0.51)
\usefont{T1}{ptm}{m}{n}
\rput(1.3474512,-0.005){\LARGE $\sem{M}$}
\psframe[linewidth=0.04,dimen=outer](4.998359,0.41)(3.36,-0.49)
\usefont{T1}{ptm}{m}{n}
\rput(4.1674514,-0.085){\LARGE $\sem{N}$}
\pstriangle[linewidth=0.04,dimen=outer](1.3783593,-1.71)(1.6,1.0)
\usefont{T1}{ptm}{m}{n}
\rput(1.3679981,-1.405){\LARGE $k$}
\rput{-180.0}(2.7567186,2.38){\pstriangle[linewidth=0.04,dimen=outer](1.3783593,0.69)(1.6,1.0)}
\usefont{T1}{ptm}{m}{n}
\rput(1.327998,1.395){\LARGE $j$}
\psline[linewidth=0.04cm](1.3783593,0.69)(1.3783593,0.49)
\psline[linewidth=0.04cm](1.3783593,-0.51)(1.3783593,-0.71)
\psline[linewidth=0.04cm](0.96,2.91)(0.9783594,1.69)
\usefont{T1}{ptm}{m}{n}
\rput(0.5179981,2.155){\LARGE $\nat$}
\psline[linewidth=0.04cm](1.76,2.91)(1.7783594,1.69)
\psline[linewidth=0.04cm](4.16,-0.49)(4.16,-1.49)
\usefont{T1}{ptm}{m}{n}
\rput(2.1179981,2.155){\LARGE $\nat$}
\usefont{T1}{ptm}{m}{n}
\rput(4.517998,-1.045){\LARGE $\nat$}
\psline[linewidth=0.04cm](4.16,1.31)(4.16,0.39)
\usefont{T1}{ptm}{m}{n}
\rput(4.517998,0.755){\LARGE $\nat$}
\psline[linewidth=0.04cm](0.96,-1.69)(0.9783594,-2.91)
\usefont{T1}{ptm}{m}{n}
\rput(0.5179981,-2.445){\LARGE $\nat$}
\psline[linewidth=0.04cm](1.76,-1.69)(1.7783594,-2.91)
\usefont{T1}{ptm}{m}{n}
\rput(2.1179981,-2.445){\LARGE $\nat$}
\end{pspicture} 
}
    }
    \quad\longmapsto\quad
       \raisebox{-.5\height}{
\scalebox{.5} 
{
\begin{pspicture}(0,-3.72)(3.7681837,3.72)
\psframe[linewidth=0.04,dimen=outer](2.3783593,1.28)(0.37835938,0.28)
\usefont{T1}{ptm}{m}{n}
\rput(1.3474512,0.785){\LARGE $\sem{M}$}
\psframe[linewidth=0.04,dimen=outer](3.1983595,-1.8)(1.56,-2.7)
\usefont{T1}{ptm}{m}{n}
\rput(2.3674512,-2.295){\LARGE $\sem{N}$}
\pstriangle[linewidth=0.04,dimen=outer](1.3783593,-0.92)(1.6,1.0)
\usefont{T1}{ptm}{m}{n}
\rput(1.3679981,-0.615){\LARGE $k$}
\rput{-180.0}(2.7567186,3.96){\pstriangle[linewidth=0.04,dimen=outer](1.3783593,1.48)(1.6,1.0)}
\usefont{T1}{ptm}{m}{n}
\rput(1.327998,2.185){\LARGE $j$}
\psline[linewidth=0.04cm](1.3783593,1.48)(1.3783593,1.28)
\psline[linewidth=0.04cm](1.3783593,0.28)(1.3783593,0.08)
\psline[linewidth=0.04cm](0.96,3.7)(0.9783594,2.48)
\psline[linewidth=0.04cm](0.9783594,-0.92)(0.96,-3.7)
\usefont{T1}{ptm}{m}{n}
\rput(0.5179981,2.945){\LARGE $\nat$}
\usefont{T1}{ptm}{m}{n}
\rput(0.53799807,-2.635){\LARGE $\nat$}
\psline[linewidth=0.04cm](2.38,-1.84)(1.76,-0.9)
\psline[linewidth=0.04cm](1.76,3.7)(1.7783594,2.48)
\psline[linewidth=0.04cm](2.36,-2.7)(2.36,-3.7)
\usefont{T1}{ptm}{m}{n}
\rput(2.1179981,2.945){\LARGE $\nat$}
\usefont{T1}{ptm}{m}{n}
\rput(2.717998,-3.255){\LARGE $\nat$}
\end{pspicture} 
}
    }
    \\
    \quad\stackrel{\trace^{\nat}_{\nat,\nat}}{\longmapsto}\quad
       \raisebox{-.5\height}{
\scalebox{.5} 
{
\begin{pspicture}(0,-3.72)(3.98,3.72)
\psframe[linewidth=0.04,dimen=outer](2.3783593,1.28)(0.37835938,0.28)
\usefont{T1}{ptm}{m}{n}
\rput(1.3474512,0.785){\LARGE $\sem{M}$}
\psframe[linewidth=0.04,dimen=outer](3.1983595,-1.8)(1.56,-2.7)
\usefont{T1}{ptm}{m}{n}
\rput(2.3674512,-2.295){\LARGE $\sem{N}$}
\pstriangle[linewidth=0.04,dimen=outer](1.3783593,-0.92)(1.6,1.0)
\usefont{T1}{ptm}{m}{n}
\rput(1.3679981,-0.615){\LARGE $k$}
\rput{-180.0}(2.7567186,3.96){\pstriangle[linewidth=0.04,dimen=outer](1.3783593,1.48)(1.6,1.0)}
\usefont{T1}{ptm}{m}{n}
\rput(1.327998,2.185){\LARGE $j$}
\psline[linewidth=0.04cm](1.3783593,1.48)(1.3783593,1.28)
\psline[linewidth=0.04cm](1.3783593,0.28)(1.3783593,0.08)
\psline[linewidth=0.04cm](0.96,3.7)(0.9783594,2.48)
\psline[linewidth=0.04cm](0.9783594,-0.92)(0.96,-3.7)
\psarc[linewidth=0.04](3.1691797,-2.7108202){0.7908203}{180.0}{0.0}
\usefont{T1}{ptm}{m}{n}
\rput(0.5179981,2.945){\LARGE $\nat$}
\usefont{T1}{ptm}{m}{n}
\rput(0.53799807,-2.635){\LARGE $\nat$}
\psline[linewidth=0.04cm](2.38,-1.84)(1.76,-0.9)
\psarc[linewidth=0.04](2.7291796,2.4291797){0.7908203}{1.0492004}{178.9824}
\psline[linewidth=0.04cm](3.52,2.5)(3.96,-2.7)
\end{pspicture} 
}
    }
   \quad =\quad
       \raisebox{-.5\height}{
\scalebox{.5} 
{
\begin{pspicture}(0,-2.32)(5.7481837,2.32)
\psframe[linewidth=0.04,dimen=outer](2.3783593,0.5)(0.37835938,-0.5)
\usefont{T1}{ptm}{m}{n}
\rput(1.3474512,0.005){\LARGE $\sem{M}$}
\psframe[linewidth=0.04,dimen=outer](5.1783595,0.5)(3.5783594,-0.5)
\usefont{T1}{ptm}{m}{n}
\rput(4.347451,0.005){\LARGE $\sem{N}$}
\pstriangle[linewidth=0.04,dimen=outer](1.3783593,-1.7)(1.6,1.0)
\usefont{T1}{ptm}{m}{n}
\rput(1.3679981,-1.395){\LARGE $k$}
\rput{-180.0}(2.7567186,2.4){\pstriangle[linewidth=0.04,dimen=outer](1.3783593,0.7)(1.6,1.0)}
\usefont{T1}{ptm}{m}{n}
\rput(1.327998,1.405){\LARGE $j$}
\psline[linewidth=0.04cm](1.3783593,0.7)(1.3783593,0.5)
\psline[linewidth=0.04cm](1.3783593,-0.5)(1.3783593,-0.7)
\psline[linewidth=0.04cm](0.9783594,2.3)(0.9783594,1.7)
\psline[linewidth=0.04cm](0.9783594,-1.7)(0.9783594,-2.3)
\psarc[linewidth=0.04](2.1783593,1.7){0.4}{0.0}{180.0}
\psarc[linewidth=0.04](2.1783593,-1.7){0.4}{180.0}{0.0}
\psline[linewidth=0.04cm](3.5783594,1.3)(2.5783594,-1.7)
\psarc[linewidth=0.04](3.9783595,1.3){0.4}{0.0}{180.0}
\psline[linewidth=0.04cm](3.5783594,-1.3)(2.5783594,1.7)
\psarc[linewidth=0.04](3.9783595,-1.3){0.4}{180.0}{0.0}
\psline[linewidth=0.04cm](4.3783593,1.3)(4.3783593,0.5)
\psline[linewidth=0.04cm](4.3783593,-0.5)(4.3783593,-1.3)
\usefont{T1}{ptm}{m}{n}
\rput(0.5179981,1.965){\LARGE $\nat$}
\usefont{T1}{ptm}{m}{n}
\rput(0.53799807,-2.015){\LARGE $\nat$}
\end{pspicture} 
   }
    }
   \end{multline}
In the above, we start with two arrows $\sem{M},\sem{N}\colon
\nat\relto\nat$ in $\Kleisli{\lift}$; we compose $j$ and $k$, from the
retraction $\nat+\nat\retr\nat$ (Def.~\ref{definition:GoIsituation}),
with $\sem{M}$; to it we post-compose the arrow
\begin{displaymath}
 \nat +\sem{N}\;\colon \quad \nat+\nat\longrelto \nat+\nat
\end{displaymath}
(where the first $\nat$ stands for the identity arrow $\id_{\nat}\colon\nat\relto\nat$)
and obtain the arrow $(\nat+\sem{N})\co k\co\sem{M}\co j$ (the third
string diagram); here we crucially exploit the trace operator $\tr$ in
the category $\Kleisli{\lift}$ and obtain the arrow
\begin{displaymath}
 \trace^{\nat}_{\nat,\nat}\bigl(\,(\nat+\sem{N})\co k\co\sem{M}\co
 j\,\bigr)
 \;\colon\quad\nat\longrelto\nat\enspace;
\end{displaymath}
 finally some ``topological reasoning'' with string diagrams yields the
 last equality, and the last string diagram corresponds to the last
 ``piping'' diagram in~(\ref{eq:parCompAndHiding}). This way, a trace
 operator plays an essential role in axiomatizing ``two processes
 talking to each other (i.e.\ feeding one's answer back to the other).''

As we already discussed in~\S{}\ref{subsection:prelimGoI},
the workflow of categorical GoI~\cite{AbramskyHS02} turns a GoI
situation (Def.~\ref{definition:GoIsituation}) into an LCA, a model of
\emph{untyped} linear $\lambda$-calculus.

\begin{mydefinition}[Linear combinatory algebra, LCA]\label{definition:LCA}
 A \emph{linear combinatory algebra (LCA)} is a set $A$ equipped with
 \begin{itemize}
  \item a binary operator (called an \emph{applicative structure})
	$\cdot:A^{2}\to A$;
  \item a unary operator $\bang:A\to A$; and
  \item distinguished elements (called \emph{combinators}) 
$
\cmbt{B},
\cmbt{C},
\cmbt{I},
\cmbt{K},
\cmbt{W},
\cmbt{D},
\cmbtDelta$, and $\cmbt{F}
$, of $A$. These are required to satisfy the following equalities.
 \begin{align*}
    \cmbt{B}xyz&=x(yz)&&\text{Composition, Cut}
 \\
 \cmbt{C}xyz&=(xz)y &&\text{Exchange}
 \\
    \cmbt{I}x&=x&&\text{Identity}
 \\
 \cmbt{K}x\bang y&= x &&\text{Weakening}
 \\
    \cmbt{W}x\bang y&=x\bang y\bang y&&\text{Contraction}\\
 \cmbt{D}\bang x&= x  &&\text{Dereliction}\\
    \cmbtDelta\bang x&=\bang \bang x&&\text{Comultiplication}\\
 \cmbt{F}\bang x\bang y&= \bang (xy) &&\text{Monoidal functoriality}
\end{align*}
The notational convention is: $\cdot$ associates to the left; $\cdot$ is suppressed; and
	$\bang$ binds stronger than $\cdot$ does.
 \end{itemize}
\end{mydefinition}

\begin{mytheorem}[From GoI situations to LCAs~\cite{AbramskyHS02}]
 \label{theorem:fromGoISitToLCA}
 Let $(\C, F, U)$ be a GoI situation
 (Definition~\ref{definition:GoIsituation}). Then 
 the homset 
 \begin{displaymath}
  \C(U,U)
 \end{displaymath}
is a linear combinatory algebra
 (LCA).  \myqed
\end{mytheorem}

We shall review the proof of the previous result
(from~\cite{AbramskyHS02})
and describe the LCA structure of $\C(U,U)$ in some detail. 
Its application operator $\cdot$ is defined in the same way as what we already
described
for the special case of
$(\C,F,U)=(\Kleisli{\lift},\nat\cdot\place,\nat)$. That is, for each
$a,b\colon U\relto U$, 
\begin{equation}
 \label{eq:applicativeStrOfGoILCAInStringDiagrams}
 a\cdot b \;:=\;
 \trace^{U}_{U,U}\left(
\begin{array}{r}
   U\otimes U\stackrel{j}{\longrelto} 
  U\stackrel{a}{\longrelto}
  U
\stackrel{k}{\longrelto}
  U\otimes U
\stackrel{U\otimes b}{\longrelto}
  U\otimes U
\end{array}
\right)
  \;=\;
   \raisebox{-.5\height}{
\scalebox{.5} 
{
\begin{pspicture}(0,-2.32)(4.4,2.32)
\psframe[linewidth=0.04,dimen=outer](1.4,0.5)(0.2,-0.5)
\usefont{T1}{ptm}{m}{n}
\rput(0.8414551,0.0050){\LARGE $a$}
\psframe[linewidth=0.04,dimen=outer](4.4,0.5)(3.2,-0.5)
\usefont{T1}{ptm}{m}{n}
\rput(3.851455,0.0050){\LARGE $b$}
\pstriangle[linewidth=0.04,dimen=outer](0.8,-1.7)(1.6,1.0)
\usefont{T1}{ptm}{m}{n}
\rput(0.8514551,-1.395){\LARGE $k$}
\rput{-180.0}(1.6,2.4){\pstriangle[linewidth=0.04,dimen=outer](0.8,0.7)(1.6,1.0)}
\usefont{T1}{ptm}{m}{n}
\rput(0.8114551,1.405){\LARGE $j$}
\psline[linewidth=0.04cm](0.8,0.7)(0.8,0.5)
\psline[linewidth=0.04cm](0.8,-0.5)(0.8,-0.7)
\psline[linewidth=0.04cm](0.4,2.3)(0.4,1.7)
\psline[linewidth=0.04cm](0.4,-1.7)(0.4,-2.3)
\psarc[linewidth=0.04](1.6,1.7){0.4}{0.0}{180.0}
\psarc[linewidth=0.04](1.6,-1.7){0.4}{180.0}{0.0}
\psline[linewidth=0.04cm](3.0,1.3)(2.0,-1.7)
\psarc[linewidth=0.04](3.4,1.3){0.4}{0.0}{180.0}
\psline[linewidth=0.04cm](3.0,-1.3)(2.0,1.7)
\psarc[linewidth=0.04](3.4,-1.3){0.4}{180.0}{0.0}
\psline[linewidth=0.04cm](3.8,1.3)(3.8,0.5)
\psline[linewidth=0.04cm](3.8,-0.5)(3.8,-1.3)
\end{pspicture} 
}
}
 \enspace.
\end{equation}
Let us now describe the $\bang$ operator. Its string diagram
presentation,
using dashed boxes and double lines for denoting application of the
functor $F$ (like in~\cite{Mellies06,Mellies09}), is as follows.
\begin{equation}
 \label{eq:bangStrOfGoILCAInStringDiagrams}
 \begin{aligned}
& \bang a\; :=\;
\left(
\begin{array}{l}
 U  \stackrel{v}{\longrelto}  FU 
 \stackrel{ Fa}{\longrelto} FU  
 \stackrel{u}{\longrelto} U 
\end{array}
\right)
 \;=\;
      \raisebox{-.5\height}{
\scalebox{.5} 
{
\begin{pspicture}(0,-2.72)(4.5018945,2.72)
\psframe[linewidth=0.04,dimen=outer](1.6,0.5)(0.4,-0.5)
\usefont{T1}{ptm}{m}{n}
\rput(1.041455,0.0050){\LARGE $a$}
\usefont{T1}{ptm}{m}{n}
\rput(1.051455,1.605){\LARGE $v$}
\psline[linewidth=0.04cm,doubleline=true,doublesep=0.12](1.0,1.3)(1.0,0.7)
\psline[linewidth=0.04cm](1.0,-2.1)(1.0,-2.7)
\psline[linewidth=0.04cm](1.0,2.7)(1.0,2.1)
\pscircle[linewidth=0.04,dimen=outer](1.0,1.7){0.4}
\psframe[linewidth=0.04,linestyle=dashed,dash=0.16cm 0.16cm,dimen=outer](2.0,0.7)(0.0,-0.7)
\psline[linewidth=0.04cm,doubleline=true,doublesep=0.12](1.0,-0.7)(1.0,-1.3)
\usefont{T1}{ptm}{m}{n}
\rput(1.051455,-1.795){\LARGE $u$}
\pscircle[linewidth=0.04,dimen=outer](1.0,-1.7){0.4}
\usefont{T1}{ptm}{m}{n}
\rput(2,1.205){\Large$FU $}
\usefont{T1}{ptm}{m}{n}
\rput(3.,-0.795){\LARGE$F(\place)$}
\end{pspicture} 
}
}
 \enspace.
\end{aligned}\end{equation}
Now, for some intuition, let us instantiate the general definition
to the leading example of $(\C,F,U)=(\Kleisli{\lift},\nat\cdot\place,\nat)$. 
Since we have an isomorphism $\nat\cdot\nat\cong \nat$, a retraction
$u:\nat\cdot\nat\retr\nat:v$ is readily available; let us fix one such pair.
In the string diagram
in~(\ref{eq:bangStrOfGoILCAInStringDiagrams}) a plain string
represents $\nat$ and a double line represents $\nat\cdot \nat$; the
dashed box then represents making countably many copies of its content.
Using pipes in place of strings (i.e.\ thinking of a string of type
$\nat$ as a bunch of countably many pipes), the diagram
in~(\ref{eq:bangStrOfGoILCAInStringDiagrams}) comes to look like the following.
 \begin{displaymath}\label{eq:bangPicture}
  \includegraphics[height=10em]{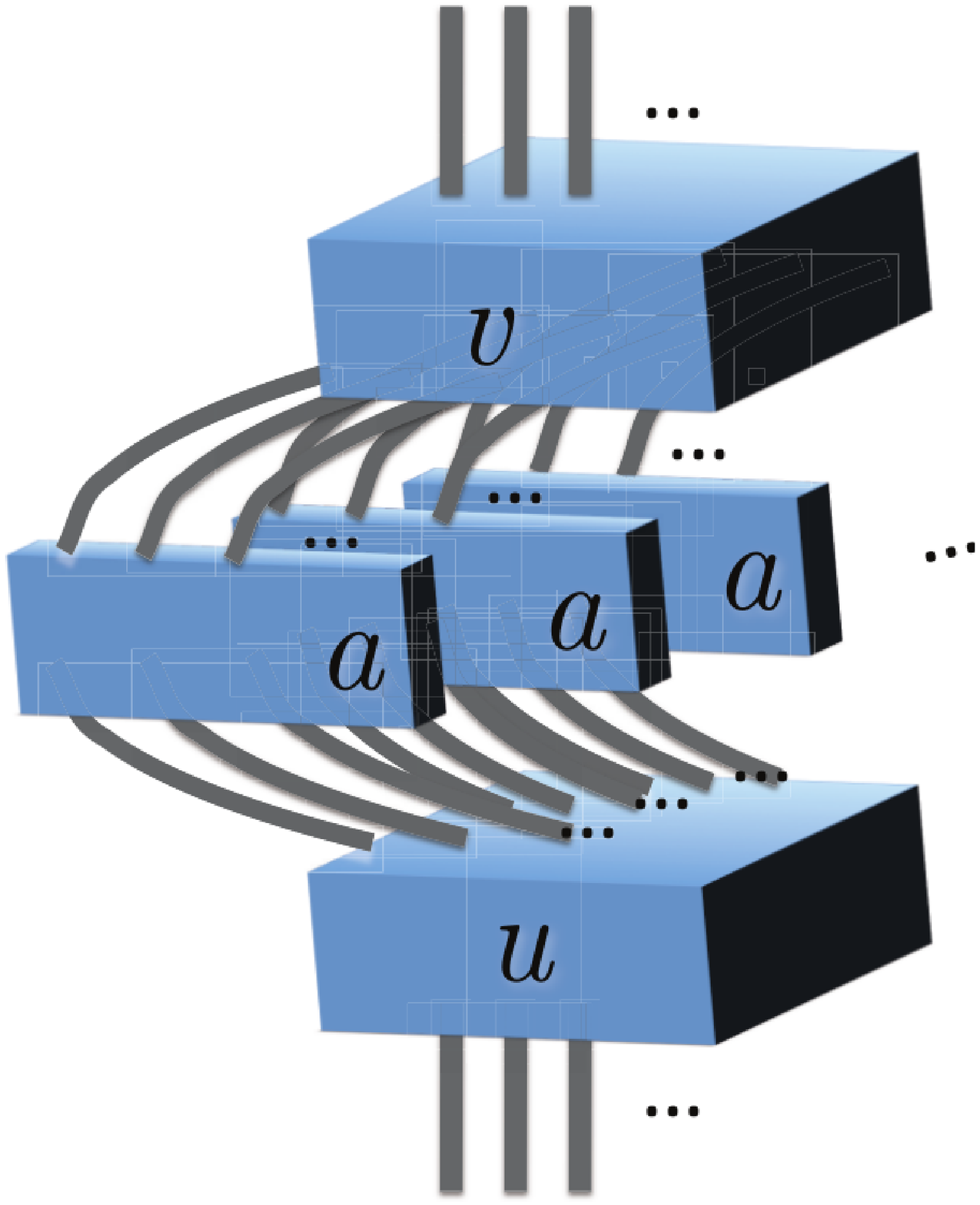}
 \end{displaymath}
Here a token moves from the top towards the bottom; the piping $v$
divides countably many pipes into countably many bunches (each of which
consists of countably many pipes), according to a fixed isomorphism
$\nat\cdot\nat\cong\nat$; in the middle we have a copy of $a$ for each
bunch;
and finally the bunches of pipes are unified by $u$ into a single bunch.

 Let us now go on to describing LCA combinators, 
 like $\cmbt{B},\cmbt{C}$ and $\cmbt{I}$, in the LCA $\C(U,U)$ in
 Theorem~\ref{theorem:fromGoISitToLCA}. 
Their definitions are given in~\cite[\S{}4]{AbramskyHS02}, uniformly for any GoI situation (whether it
is particle-style or wave-style),  by certain string diagrams. 
For example, the $\cmbt{B}$ combinator is given by the following element
of $\C(U,U)$.
\begin{equation}\label{eq:BCombAsStringDiagram}
 \raisebox{-.5\height}{
\scalebox{.5} 
{
\begin{pspicture}(0,-3.93)(6.137538,3.93)
\definecolor{color1b}{rgb}{0.8,0.8,0.8}
\pstriangle[linewidth=0.04,dimen=outer,fillstyle=solid,fillcolor=color1b](3.7,2.71)(4.6,0.6)
\pstriangle[linewidth=0.04,dimen=outer,fillstyle=solid,fillcolor=color1b](2.0,1.71)(3.6,0.6)
\pstriangle[linewidth=0.04,dimen=outer,fillstyle=solid,fillcolor=color1b](3.2,0.91)(1.6,0.4)
\pstriangle[linewidth=0.04,dimen=outer,fillstyle=solid,fillcolor=color1b](5.4,0.91)(1.6,0.4)
\pstriangle[linewidth=0.04,dimen=outer,fillstyle=solid,fillcolor=color1b](0.8,0.91)(1.6,0.4)
\rput{-180.0}(7.4,-5.98){\pstriangle[linewidth=0.04,dimen=outer,fillstyle=solid,fillcolor=color1b](3.7,-3.29)(4.6,0.6)}
\rput{-180.0}(4.0,-3.98){\pstriangle[linewidth=0.04,dimen=outer,fillstyle=solid,fillcolor=color1b](2.0,-2.29)(3.6,0.6)}
\rput{-180.0}(6.4,-2.18){\pstriangle[linewidth=0.04,dimen=outer,fillstyle=solid,fillcolor=color1b](3.2,-1.29)(1.6,0.4)}
\rput{-180.0}(10.8,-2.18){\pstriangle[linewidth=0.04,dimen=outer,fillstyle=solid,fillcolor=color1b](5.4,-1.29)(1.6,0.4)}
\rput{-180.0}(1.6,-2.18){\pstriangle[linewidth=0.04,dimen=outer,fillstyle=solid,fillcolor=color1b](0.8,-1.29)(1.6,0.4)}
\psline[linewidth=0.04](0.4,0.91)(0.4,0.51)(5.0,-0.49)(5.0,-0.89)
\psline[linewidth=0.04](1.2,0.91)(1.2,0.71)(3.6,-0.69)(3.6,-0.89)(3.6,-0.89)(3.6,-0.89)
\psline[linewidth=0.04](2.8,0.91)(2.8,0.71)(5.8,-0.69)(5.8,-0.89)
\psline[linewidth=0.04](3.6,0.91)(1.2,0.11)(1.2,-0.89)
\psline[linewidth=0.04](5.8,0.91)(5.8,0.71)(0.4,-0.69)(0.4,-0.89)(0.4,-0.89)
\psline[linewidth=0.04](2.8,-0.89)(2.8,-0.69)(5.0,0.11)(5.0,0.91)
\psline[linewidth=0.04cm](0.8,1.71)(0.8,1.31)
\psline[linewidth=0.04cm](3.2,1.71)(3.2,1.31)
\psline[linewidth=0.04cm](2.0,2.71)(2.0,2.31)
\psline[linewidth=0.04cm](5.4,2.71)(5.4,1.31)
\psline[linewidth=0.04cm](0.8,-1.29)(0.8,-1.69)
\psline[linewidth=0.04cm](3.2,-1.29)(3.2,-1.69)
\psline[linewidth=0.04cm](5.4,-1.29)(5.4,-2.69)
\psline[linewidth=0.04cm](2.0,-2.29)(2.0,-2.69)
\psline[linewidth=0.04cm](3.7,3.91)(3.7,3.31)
\psline[linewidth=0.04cm](3.72,-3.31)(3.72,-3.91)
\end{pspicture} 
}}
\end{equation}
This diagram is a string diagram in $\C$;
 the triangles denote the isomorphisms $j:U\otimes U\cong U:k$. 
 By expanding the application operation $\cdot$ 
according to~(\ref{eq:applicativeStrOfGoILCAInStringDiagrams}), it is not hard to see that the equation 
$\cmbt{B}xyz=x(yz)$ holds. See Figure~\ref{figure:BCmbtEq}.
In fact, it is a nice (and easy) puzzle to recover the string
 diagram~(\ref{eq:applicativeStrOfGoILCAInStringDiagrams}) from the
 specification $\cmbt{B}xyz=x(yz)$ of the combinator $\cmbt{B}$---the
 (seemingly complicated and arbitrary) wiring in the middle
 of~(\ref{eq:BCombAsStringDiagram}) can be deduced by working out which
 wire should be connected, in the end, to which wire.
An important point
 here is that two triangles pointing to each other cancel out, that is,
\begin{displaymath}
\raisebox{-.5\height}{ 
\scalebox{.3} 
{
\begin{pspicture}(0,-1.32)(3.4983447,1.32)
\definecolor{color1057b}{rgb}{0.8,0.8,0.8}
\pstriangle[linewidth=0.04,dimen=outer,fillstyle=solid,fillcolor=color1057b](1.8,-0.9)(3.6,0.6)
\psline[linewidth=0.04cm](0.6,-0.9)(0.6,-1.3)
\psline[linewidth=0.04cm](3.0,-0.9)(3.0,-1.3)
\psline[linewidth=0.04cm](1.8,0.1)(1.8,-0.3)
\rput{-180.0}(3.6,1.2){\pstriangle[linewidth=0.04,dimen=outer,fillstyle=solid,fillcolor=color1057b](1.8,0.3)(3.6,0.6)}
\psline[linewidth=0.04cm](1.8,0.3)(1.8,-0.1)
\psline[linewidth=0.04cm](0.6,1.3)(0.6,0.9)
\psline[linewidth=0.04cm](3.0,1.3)(3.0,0.9)
\end{pspicture} 
}
}
\;=\;
\raisebox{-.5\height}{
\scalebox{.3} 
{
\begin{pspicture}(0,-1.32)(2.42,1.32)
\psline[linewidth=0.04cm](0.0,-0.9)(0.0,-1.3)
\psline[linewidth=0.04cm](2.4,-0.9)(2.4,-1.3)
\psline[linewidth=0.04cm](0.0,1.3)(0.0,0.9)
\psline[linewidth=0.04cm](2.4,1.3)(2.4,0.9)
\psline[linewidth=0.04cm](0.0,1.3)(0.0,-1.3)
\psline[linewidth=0.04cm](2.4,1.3)(2.4,-1.3)
\end{pspicture} 
}}
\quad,
\qquad\text{since $k\co j=\id$.}
\end{displaymath}
\begin{figure}[tbp]
 \begin{align*}
 \cmbt{B}xyz \;=\;
 \scalebox{.3} 
 {\raisebox{-.5\height}{
 \begin{pspicture}(0,-7.42)(16.968359,7.42)
 \definecolor{color367b}{rgb}{0.8,0.8,0.8}
 \pstriangle[linewidth=0.04,dimen=outer,fillstyle=solid,fillcolor=color367b](4.8683596,2.8)(4.6,0.6)
 \pstriangle[linewidth=0.04,dimen=outer,fillstyle=solid,fillcolor=color367b](3.1683593,1.8)(3.6,0.6)
 \pstriangle[linewidth=0.04,dimen=outer,fillstyle=solid,fillcolor=color367b](4.3683596,1.0)(1.6,0.4)
 \pstriangle[linewidth=0.04,dimen=outer,fillstyle=solid,fillcolor=color367b](6.5683594,1.0)(1.6,0.4)
 \pstriangle[linewidth=0.04,dimen=outer,fillstyle=solid,fillcolor=color367b](1.9683594,1.0)(1.6,0.4)
 \rput{-180.0}(9.736719,-5.8){\pstriangle[linewidth=0.04,dimen=outer,fillstyle=solid,fillcolor=color367b](4.8683596,-3.2)(4.6,0.6)}
 \rput{-180.0}(6.3367186,-3.8){\pstriangle[linewidth=0.04,dimen=outer,fillstyle=solid,fillcolor=color367b](3.1683593,-2.2)(3.6,0.6)}
 \rput{-180.0}(8.736719,-2.0){\pstriangle[linewidth=0.04,dimen=outer,fillstyle=solid,fillcolor=color367b](4.3683596,-1.2)(1.6,0.4)}
 \rput{-180.0}(13.136719,-2.0){\pstriangle[linewidth=0.04,dimen=outer,fillstyle=solid,fillcolor=color367b](6.5683594,-1.2)(1.6,0.4)}
 \rput{-180.0}(3.9367187,-2.0){\pstriangle[linewidth=0.04,dimen=outer,fillstyle=solid,fillcolor=color367b](1.9683594,-1.2)(1.6,0.4)}
 \psline[linewidth=0.04](1.5683594,1.0)(1.5683594,0.6)(6.1683593,-0.4)(6.1683593,-0.8)
 \psline[linewidth=0.04](2.3683593,1.0)(2.3683593,0.8)(4.768359,-0.6)(4.768359,-0.8)(4.768359,-0.8)(4.768359,-0.8)
 \psline[linewidth=0.04](3.9683595,1.0)(3.9683595,0.8)(6.9683595,-0.6)(6.9683595,-0.8)
 \psline[linewidth=0.04](4.768359,1.0)(2.3683593,0.2)(2.3683593,-0.8)
\psline[linewidth=0.04](3.9683595,-0.8)(3.9683595,-0.6)(6.9683595,0.6)(6.9683595,1.0)
\psline[linewidth=0.04](6.1683593,1.0)(6.1683593,0.6)(1.5683594,-0.6)(1.5683594,-0.8)(1.5683594,-0.8)(1.5683594,-0.8)
 \psline[linewidth=0.04cm](1.9683594,1.8)(1.9683594,1.4)
 \psline[linewidth=0.04cm](4.3683596,1.8)(4.3683596,1.4)
 \psline[linewidth=0.04cm](3.1683593,2.8)(3.1683593,2.4)
 \psline[linewidth=0.04cm](6.5683594,2.8)(6.5683594,1.4)
 \psline[linewidth=0.04cm](1.9683594,-1.2)(1.9683594,-1.6)
 \psline[linewidth=0.04cm](4.3683596,-1.2)(4.3683596,-1.6)
 \psline[linewidth=0.04cm](6.5683594,-1.2)(6.5683594,-2.6)
 \psline[linewidth=0.04cm](3.1683593,-2.2)(3.1683593,-2.6)
 \psline[linewidth=0.04cm](4.8683596,4.0)(4.8683596,3.4)
 \psline[linewidth=0.04cm](4.8883595,-3.22)(4.8883595,-3.82)
 \psframe[linewidth=0.04,linestyle=dashed,dash=0.16cm 0.16cm,dimen=outer](7.768359,3.6)(0.76835936,-3.4)
 \usefont{T1}{ptm}{m}{n}
 \rput(1.3874512,2.975){\Huge $\cmbt{B}$}
 \rput{-180.0}(9.736719,8.2){\pstriangle[linewidth=0.04,dimen=outer,fillstyle=solid,fillcolor=color367b](4.8683596,3.8)(4.6,0.6)}
 \rput{-180.0}(6.3367186,10.2){\pstriangle[linewidth=0.04,dimen=outer,fillstyle=solid,fillcolor=color367b](3.1683593,4.8)(3.6,0.6)}
 \rput{-180.0}(3.9367187,12.0){\pstriangle[linewidth=0.04,dimen=outer,fillstyle=solid,fillcolor=color367b](1.9683594,5.8)(1.6,0.4)}
 \psline[linewidth=0.04cm](1.9683594,5.8)(1.9683594,5.4)
 \psline[linewidth=0.04cm](3.1683593,4.8)(3.1683593,4.4)
 \pstriangle[linewidth=0.04,dimen=outer,fillstyle=solid,fillcolor=color367b](4.8683596,-4.4)(4.6,0.6)
 \pstriangle[linewidth=0.04,dimen=outer,fillstyle=solid,fillcolor=color367b](1.9683594,-6.2)(1.6,0.4)
 \psline[linewidth=0.04cm](1.9683594,-5.4)(1.9683594,-5.8)
 \psline[linewidth=0.04cm](3.1683593,-4.4)(3.1683593,-4.8)
 \psframe[linewidth=0.04,dimen=outer](10.568359,0.8)(8.968359,-0.6)
 \usefont{T1}{ptm}{m}{n}
 \rput(9.787451,-0.025){\Huge $x$}
 \psframe[linewidth=0.04,dimen=outer](13.768359,0.8)(12.16836,-0.6)
 \usefont{T1}{ptm}{m}{n}
 \rput(12.987452,-0.025){\Huge $y$}
 \psframe[linewidth=0.04,dimen=outer](16.968359,0.8)(15.36836,-0.6)
 \usefont{T1}{ptm}{m}{n}
 \rput(16.167452,-0.025){\Huge $z$}
 \psline[linewidth=0.04](6.5683594,4.4)(6.5683594,4.8)(7.9683595,4.8)(8.768359,-1.6)(9.768359,-1.6)(9.768359,-0.6)(9.768359,-0.6)
 \psline[linewidth=0.04](9.768359,0.8)(9.768359,1.8)(8.768359,1.8)(7.9683595,-4.8)(6.5683594,-4.8)(6.5683594,-4.4)
 \psline[linewidth=0.04](12.968359,-0.6)(12.968359,-1.6)(11.968359,-1.6)(10.568359,5.8)(4.3683596,5.8)(4.3683596,5.4)(4.3683596,5.4)
 \psline[linewidth=0.04](12.968359,0.8)(12.968359,1.8)(11.968359,1.8)(10.568359,-5.6)(4.3683596,-5.6)(4.3683596,-5.2)(4.3683596,-5.2)
 \pstriangle[linewidth=0.04,dimen=outer,fillstyle=solid,fillcolor=color367b](3.1683593,-5.4)(3.6,0.6)
 \psline[linewidth=0.04](2.3683593,6.2)(2.3683593,6.6)(13.768359,6.6)(15.16836,-1.6)(16.16836,-1.6)(16.16836,-0.6)
 \psline[linewidth=0.04](16.16836,0.8)(16.16836,1.8)(15.16836,1.8)(15.16836,1.8)
 \psline[linewidth=0.04](2.3683593,-6.2)(2.3683593,-6.6)(13.768359,-6.6)(15.16836,1.8)(15.16836,1.8)
 \psline[linewidth=0.04cm](1.5683594,7.4)(1.5683594,6.2)
 \psline[linewidth=0.04cm](1.5683594,-6.2)(1.5683594,-7.4)
 \end{pspicture} 
 }}
 \quad=\quad
\raisebox{-.5\height}{
\scalebox{.3}{
\begin{pspicture}(0,-7.42)(15.8,7.42)
\definecolor{color1057b}{rgb}{0.8,0.8,0.8}
\pstriangle[linewidth=0.04,dimen=outer,fillstyle=solid,fillcolor=color1057b](2.0,1.8)(3.6,0.6)
\pstriangle[linewidth=0.04,dimen=outer,fillstyle=solid,fillcolor=color1057b](3.2,1.0)(1.6,0.4)
\pstriangle[linewidth=0.04,dimen=outer,fillstyle=solid,fillcolor=color1057b](5.4,1.0)(1.6,0.4)
\pstriangle[linewidth=0.04,dimen=outer,fillstyle=solid,fillcolor=color1057b](0.8,1.0)(1.6,0.4)
\rput{-180.0}(4.0,-3.8){\pstriangle[linewidth=0.04,dimen=outer,fillstyle=solid,fillcolor=color1057b](2.0,-2.2)(3.6,0.6)}
\rput{-180.0}(6.4,-2.0){\pstriangle[linewidth=0.04,dimen=outer,fillstyle=solid,fillcolor=color1057b](3.2,-1.2)(1.6,0.4)}
\rput{-180.0}(10.8,-2.0){\pstriangle[linewidth=0.04,dimen=outer,fillstyle=solid,fillcolor=color1057b](5.4,-1.2)(1.6,0.4)}
\rput{-180.0}(1.6,-2.0){\pstriangle[linewidth=0.04,dimen=outer,fillstyle=solid,fillcolor=color1057b](0.8,-1.2)(1.6,0.4)}
\psline[linewidth=0.04](0.4,1.0)(0.4,0.6)(5.0,-0.4)(5.0,-0.8)
\psline[linewidth=0.04](1.2,1.0)(1.2,0.8)(3.6,-0.6)(3.6,-0.8)(3.6,-0.8)(3.6,-0.8)
\psline[linewidth=0.04](2.8,1.0)(2.8,0.8)(5.8,-0.6)(5.8,-0.8)
\psline[linewidth=0.04](3.6,1.0)(1.2,0.2)(1.2,-0.8)
\psline[linewidth=0.04cm](0.8,1.8)(0.8,1.4)
\psline[linewidth=0.04cm](3.2,1.8)(3.2,1.4)
\psline[linewidth=0.04cm](2.0,2.8)(2.0,2.4)
\psline[linewidth=0.04cm](5.4,2.8)(5.4,1.4)
\psline[linewidth=0.04cm](0.8,-1.2)(0.8,-1.6)
\psline[linewidth=0.04cm](3.2,-1.2)(3.2,-1.6)
\psline[linewidth=0.04cm](5.4,-1.2)(5.4,-2.6)
\psline[linewidth=0.04cm](2.0,-2.2)(2.0,-2.6)
\rput{-180.0}(4.0,10.2){\pstriangle[linewidth=0.04,dimen=outer,fillstyle=solid,fillcolor=color1057b](2.0,4.8)(3.6,0.6)}
\rput{-180.0}(1.6,12.0){\pstriangle[linewidth=0.04,dimen=outer,fillstyle=solid,fillcolor=color1057b](0.8,5.8)(1.6,0.4)}
\psline[linewidth=0.04cm](0.8,5.8)(0.8,5.4)
\psline[linewidth=0.04cm](2.0,4.8)(2.0,4.4)
\pstriangle[linewidth=0.04,dimen=outer,fillstyle=solid,fillcolor=color1057b](0.8,-6.2)(1.6,0.4)
\psline[linewidth=0.04cm](0.8,-5.4)(0.8,-5.8)
\psline[linewidth=0.04cm](2.0,-4.4)(2.0,-4.8)
\psframe[linewidth=0.04,dimen=outer](9.4,0.8)(7.8,-0.6)
\usefont{T1}{ptm}{m}{n}
\rput(8.619092,-0.025){\Huge $x$}
\psframe[linewidth=0.04,dimen=outer](12.6,0.8)(11.0,-0.6)
\usefont{T1}{ptm}{m}{n}
\rput(11.819092,-0.025){\Huge $y$}
\psframe[linewidth=0.04,dimen=outer](15.8,0.8)(14.2,-0.6)
\usefont{T1}{ptm}{m}{n}
\rput(14.999092,-0.025){\Huge $z$}
\psline[linewidth=0.04](5.4,4.4)(5.4,4.8)(6.8,4.8)(7.6,-1.6)(8.6,-1.6)(8.6,-0.6)(8.6,-0.6)
\psline[linewidth=0.04](8.6,0.8)(8.6,1.8)(7.6,1.8)(6.8,-4.8)(5.4,-4.8)(5.4,-4.4)
\psline[linewidth=0.04](11.8,-0.6)(11.8,-1.6)(10.8,-1.6)(9.4,5.8)(3.2,5.8)(3.2,5.4)(3.2,5.4)
\psline[linewidth=0.04](11.8,0.8)(11.8,1.8)(10.8,1.8)(9.4,-5.6)(3.2,-5.6)(3.2,-5.2)(3.2,-5.2)
\pstriangle[linewidth=0.04,dimen=outer,fillstyle=solid,fillcolor=color1057b](2.0,-5.4)(3.6,0.6)
\psline[linewidth=0.04](1.2,6.2)(1.2,6.6)(12.6,6.6)(14.0,-1.6)(15.0,-1.6)(15.0,-0.6)
\psline[linewidth=0.04](15.0,0.8)(15.0,1.8)(14.0,1.8)(14.0,1.8)
\psline[linewidth=0.04](1.2,-6.2)(1.2,-6.6)(12.6,-6.6)(14.0,1.8)(14.0,1.8)
\psline[linewidth=0.04cm](0.4,7.4)(0.4,6.2)
\psline[linewidth=0.04cm](0.4,-6.2)(0.4,-7.4)
\psline[linewidth=0.04cm](2.0,4.4)(2.0,2.8)
\psline[linewidth=0.04cm](5.4,4.4)(5.4,2.4)
\psline[linewidth=0.04cm](5.4,-2.4)(5.4,-4.4)
\psline[linewidth=0.04cm](2.0,-2.6)(2.0,-4.6)
\psline[linewidth=0.04](5.0,1.0)(5.0,0.6)(0.4,-0.6)(0.4,-0.8)
\psline[linewidth=0.04](5.8,1.0)(5.8,0.6)(2.8,-0.6)(2.8,-0.8)
\end{pspicture} 
}
}
 \\
 =\quad
 \raisebox{-.5\height}{\scalebox{.3} 
 {
 \begin{pspicture}(0,-7.42)(15.8,7.42)
 \definecolor{color369b}{rgb}{0.8,0.8,0.8}
 \pstriangle[linewidth=0.04,dimen=outer,fillstyle=solid,fillcolor=color369b](3.2,1.0)(1.6,0.4)
 \pstriangle[linewidth=0.04,dimen=outer,fillstyle=solid,fillcolor=color369b](5.4,1.0)(1.6,0.4)
 \pstriangle[linewidth=0.04,dimen=outer,fillstyle=solid,fillcolor=color369b](0.8,1.0)(1.6,0.4)
 \rput{-180.0}(6.4,-2.0){\pstriangle[linewidth=0.04,dimen=outer,fillstyle=solid,fillcolor=color369b](3.2,-1.2)(1.6,0.4)}
 \rput{-180.0}(10.8,-2.0){\pstriangle[linewidth=0.04,dimen=outer,fillstyle=solid,fillcolor=color369b](5.4,-1.2)(1.6,0.4)}
 \rput{-180.0}(1.6,-2.0){\pstriangle[linewidth=0.04,dimen=outer,fillstyle=solid,fillcolor=color369b](0.8,-1.2)(1.6,0.4)}
 \psline[linewidth=0.04](0.4,1.0)(0.4,0.6)(5.0,-0.4)(5.0,-0.8)
 \psline[linewidth=0.04](1.2,1.0)(1.2,0.8)(3.6,-0.6)(3.6,-0.8)(3.6,-0.8)(3.6,-0.8)
 \psline[linewidth=0.04](2.8,1.0)(2.8,0.8)(5.8,-0.6)(5.8,-0.8)
 \psline[linewidth=0.04](3.6,1.0)(1.2,0.2)(1.2,-0.8)
 \psline[linewidth=0.04](5.0,1.0)(5.0,0.6)(0.4,-0.6)(0.4,-0.8)(0.4,-0.8)
 \psline[linewidth=0.04](2.8,-0.8)(2.8,-0.6)(5.8,0.6)(5.8,1.0)
 \psline[linewidth=0.04cm](0.8,1.8)(0.8,1.4)
 \psline[linewidth=0.04cm](3.2,1.8)(3.2,1.4)
 \psline[linewidth=0.04cm](5.4,2.8)(5.4,1.4)
 \psline[linewidth=0.04cm](0.8,-1.2)(0.8,-1.6)
 \psline[linewidth=0.04cm](3.2,-1.2)(3.2,-1.6)
 \psline[linewidth=0.04cm](5.4,-1.2)(5.4,-2.6)
 \rput{-180.0}(1.6,12.0){\pstriangle[linewidth=0.04,dimen=outer,fillstyle=solid,fillcolor=color369b](0.8,5.8)(1.6,0.4)}
 \psline[linewidth=0.04cm](0.8,5.8)(0.8,5.4)
 \pstriangle[linewidth=0.04,dimen=outer,fillstyle=solid,fillcolor=color369b](0.8,-6.2)(1.6,0.4)
 \psline[linewidth=0.04cm](0.8,-5.4)(0.8,-5.8)
 \psframe[linewidth=0.04,dimen=outer](9.4,0.8)(7.8,-0.6)
 \usefont{T1}{ptm}{m}{n}
 \rput(8.619092,-0.025){\Huge $x$}
 \psframe[linewidth=0.04,dimen=outer](12.6,0.8)(11.0,-0.6)
 \usefont{T1}{ptm}{m}{n}
 \rput(11.819092,-0.025){\Huge $y$}
 \psframe[linewidth=0.04,dimen=outer](15.8,0.8)(14.2,-0.6)
 \usefont{T1}{ptm}{m}{n}
 \rput(14.999092,-0.025){\Huge $z$}
 \psline[linewidth=0.04](5.4,4.4)(5.4,4.8)(6.8,4.8)(7.6,-1.6)(8.6,-1.6)(8.6,-0.6)(8.6,-0.6)
 \psline[linewidth=0.04](8.6,0.8)(8.6,1.8)(7.6,1.8)(6.8,-4.8)(5.4,-4.8)(5.4,-4.4)
 \psline[linewidth=0.04](11.8,-0.6)(11.8,-1.6)(10.8,-1.6)(9.4,5.8)(3.2,5.8)(3.2,5.4)(3.2,5.4)
 \psline[linewidth=0.04](11.8,0.8)(11.8,1.8)(10.8,1.8)(9.4,-5.6)(3.2,-5.6)(3.2,-5.2)(3.2,-5.2)
 \psline[linewidth=0.04](1.2,6.2)(1.2,6.6)(12.6,6.6)(14.0,-1.6)(15.0,-1.6)(15.0,-0.6)
 \psline[linewidth=0.04](15.0,0.8)(15.0,1.8)(14.0,1.8)(14.0,1.8)
 \psline[linewidth=0.04](1.2,-6.2)(1.2,-6.6)(12.6,-6.6)(14.0,1.8)(14.0,1.8)
 \psline[linewidth=0.04cm](0.4,7.4)(0.4,6.2)
 \psline[linewidth=0.04cm](0.4,-6.2)(0.4,-7.4)
 \psline[linewidth=0.04cm](5.4,4.4)(5.4,2.4)
 \psline[linewidth=0.04cm](5.4,-2.4)(5.4,-4.4)
 \psline[linewidth=0.04cm](0.8,5.4)(0.8,1.8)
 \psline[linewidth=0.04cm](3.2,5.4)(3.2,1.6)
 \psline[linewidth=0.04cm](3.2,-1.4)(3.2,-5.2)
 \psline[linewidth=0.04cm](0.8,-1.4)(0.8,-5.4)
 \end{pspicture} 
 }}
 \quad=\cdots=\quad
 \raisebox{-.5\height}{
 \scalebox{.3} 
 {
 \begin{pspicture}(0,-2.72)(8.0,2.72)
 \definecolor{color369b}{rgb}{0.8,0.8,0.8}
 \pstriangle[linewidth=0.04,dimen=outer,fillstyle=solid,fillcolor=color369b](4.0,-1.5)(1.6,0.4)
 \pstriangle[linewidth=0.04,dimen=outer,fillstyle=solid,fillcolor=color369b](0.8,-1.5)(1.6,0.4)
 \rput{-180.0}(8.0,2.6){\pstriangle[linewidth=0.04,dimen=outer,fillstyle=solid,fillcolor=color369b](4.0,1.1)(1.6,0.4)}
 \rput{-180.0}(1.6,2.6){\pstriangle[linewidth=0.04,dimen=outer,fillstyle=solid,fillcolor=color369b](0.8,1.1)(1.6,0.4)}
 \psframe[linewidth=0.04,dimen=outer](1.6,0.7)(0.0,-0.7)
 \usefont{T1}{ptm}{m}{n}
 \rput(0.8190918,-0.125){\Huge $x$}
 \psframe[linewidth=0.04,dimen=outer](4.8,0.7)(3.2,-0.7)
 \usefont{T1}{ptm}{m}{n}
 \rput(4.0190916,-0.125){\Huge $y$}
 \psframe[linewidth=0.04,dimen=outer](8.0,0.7)(6.4,-0.7)
 \usefont{T1}{ptm}{m}{n}
 \rput(7.199092,-0.125){\Huge $z$}
 \psline[linewidth=0.04](4.4,-1.5)(4.4,-1.9)(5.0,-1.9)(6.2,1.1)(7.2,1.1)(7.2,0.7)
 \psline[linewidth=0.04](7.2,-0.7)(7.2,-1.1)(6.2,-1.1)(5.0,1.9)(4.4,1.9)(4.4,1.5)
 \psline[linewidth=0.04](3.6,1.5)(3.6,1.9)(3.0,1.9)(1.8,-1.9)(1.2,-1.9)(1.2,-1.5)
 \psline[linewidth=0.04](1.2,1.5)(1.2,1.9)(1.8,1.9)(3.0,-1.9)(3.6,-1.9)(3.6,-1.5)
 \psline[linewidth=0.04cm](0.4,1.5)(0.4,2.7)
 \psline[linewidth=0.04cm](0.4,-1.5)(0.4,-2.5)
 \psline[linewidth=0.04cm](0.4,-2.5)(0.4,-2.7)
 \psline[linewidth=0.04cm](0.8,1.1)(0.8,0.7)
 \psline[linewidth=0.04cm](0.8,-0.7)(0.8,-1.1)
 \psline[linewidth=0.04cm](4.0,1.1)(4.0,0.7)
 \psline[linewidth=0.04cm](4.0,-0.7)(4.0,-1.1)
 \end{pspicture} 
 }
 }\enspace.
\end{align*}
\caption{Proof of $\cmbt{B}xyz=x(yz)$}
\label{figure:BCmbtEq}
\end{figure}

\begin{myremark}\label{rem:int}
 The definition~(\ref{eq:BCombAsStringDiagram}) can be derived by
 working backwards in the reasoning in Figure~\ref{figure:BCmbtEq}. A more
 ``logical'' derivation is possible, too; it works as follows. We first turn the
 TSMC $\C$ into a compact closed (hence symmetric monoidal closed) category 
 $\mathrm{Int}(\C)$, applying the
  $\Int$-construction~\cite{JoyalSV96} (or the GoI construction,
 see~\S{}\ref{subsection:prelimGoI}). The reflexive object $U$ will then
 give rise to an object $(U,U)$ in   $\mathrm{Int}(\C)$
 that is equipped with a retraction
 $(U,U)\limp(U,U)\retr (U,U)$. This retraction---of the function space
 $(U,U)\limp (U,U)$ in $(U,U)$ itself---will allow the interpretation of
 untyped linear $\lambda$-terms over it. We then interpret $\lambda xyz.
 \, x(yz)$, the $\lambda$-term for the combinator $\cmbt{B}$.
\end{myremark}
 
\begin{myremark}[wave-style GoI]\label{remark:wave-style}
 In~\S{}\ref{subsection:LCAviaCategoricalGoI} we have relied on  
 \emph{particle-style} examples of GoI situations for intuitions. This 
 leaves \emph{wave-style} examples---where tensor products are given by 
 products instead of coproducts,
 see~\S{}\ref{subsection:quantumBranchingMonadBackground}---untouched. 
 In fact we have preliminary observations that suggest the following
 correspondence:
 particle-style GoI situations naturally model \emph{forward}, \emph{state-based}
 description of tokens' dynamics, while wave-style ones model
 \emph{backward}, \emph{predicate-based}  description of it. Here the
 contrast is really the one between \emph{state-transformer} semantics
 and \emph{predicate-transformer} semantics---a classic topic in
 theoretical computer science~\cite{Dijkstra76,Hoare69} that has also proved
 to be  relevant in quantum dynamics (the Schr\"{o}dinger picture
 and the Heisenberg one; see e.g.~\cite{JacobsWW15}). Details are yet to
 be worked out.
 \end{myremark}

  \subsection{Categorical GoI Instantiated to a Quantum Setting}
\label{subsection:categoricaGoIQuantumInstance}

In~\S{}\ref{subsection:quantumBranchingMonadBackground} we observed that
``particle-style'' examples of GoI situations in~\cite{AbramskyHS02}
allow a uniform treatment as Kleisli categories. This generalizes to the
quantum branching monad $\Q$
in~\S{}\ref{subsection:quantumBranchingMonad}. 
\begin{mytheorem}\label{theorem:KleisliQGoISituation}
 The triple $\bigl(\,(\Kleisli{\Q},0,+),\, \nat\cdot\place\,,\, \nat\,\bigr)$ forms a GoI
 situation.
Here the functor
 $\nat\cdot\place:\Kleisli{\Q}\to\Kleisli{Q}$ carries $X$ to
 the coproduct  $\nat\cdot X$ of $\nat$-many copies of $X$ (i.e.\
 $\nat$-th copower of $X$).  
\end{mytheorem}
\begin{myproof}
 \auxproof{Notes 22/12/2010, p.3}
 The main challenge---namely, if $(\Kleisli{\Q},0,+)$ is traced or not---is
 already answered in Theorem~\ref{theorem:KleisliOfQIsTraced}.
 Its trace operator is given much like for the category
 $\Kleisli{\lift}\cong\Pfn$ of sets and partial functions. Namely,
 given $f\colon X+Z\relto Y+Z$ in $\Kleisli{\Q}$: its ``restrictions'' $f_{XY}, f_{XZ},
 f_{ZY}$ and $f_{ZZ}$ are defined much like in~(\ref{eq:traceInPfn});
 and we use Girard's execution formula
   \begin{displaymath}
   \trace^{Z}_{X,Y}(f)(x)
   \;:=\;
   f_{XY}(x) + \sum_{n\in \nat}
   \bigl(\,f_{ZY}\Kco f_{ZZ}^{n}\Kco f_{XZ}
   \,\bigr)(x)
  \end{displaymath}
 to define the trace operator. Notice  similarity to the formula~(\ref{eq:execFormulaPfn}).
 
The only nontrivial part that remains is to show that $\nat\cdot\place$ preserves
 traces. Since the trace operator in $\Kleisli{\Q}$ can be described
 using Girard's execution formula (much like in~(\ref{eq:traceInPfn})
 for $\Kleisli{\lift}$, due
 to the results in~\cite{Jacobs10trace,Haghverdi00PhD}), we can use a lemma that is
 similar to~\cite[Lemma~5.1]{AbramskyHS02}. \myqed
\end{myproof}

By Theorem~\ref{theorem:fromGoISitToLCA} (that is
from~\cite{AbramskyHS02}) we obtain the following LCA. It will be
denoted by $A_{\Q}$ and used in the rest of the paper.
\begin{mytheorem}[The quantum LCA $A_{\Q}$]
 \label{theorem:quantumLCA}
 The homset 
 \begin{displaymath}
 A_{\Q}\;:=\;\Kleisli{\Q}(\nat,\nat)
 \end{displaymath}
is a linear combinatory algebra
 (LCA).  \myqed
\end{mytheorem}
The LCA structure of $A_{\Q}$ (the operators $\cdot, \bang$ and
combinators like $\cmbt{B}$) can be described very much like for
$\Kleisli{\lift}$; we already described the latter
in~\S{}\ref{subsection:LCAviaCategoricalGoI}. In particular the piping
analogy is still valid, except for the difference in the notion of
``branching''
(possibly diverging vs.\ based on quantum operations) we explained
in~\S{}\ref{subsection:quantumBranchingMonad}.

The following special property is shared by the LCAs that arise from
particle-style GoI situations. For a proof see~\cite[\S{}2.2]{AbramskyL05}.
\begin{myproposition}\label{proposition:AQisAffine}
 The LCA $A_{\Q}$ in Theorem~\ref{theorem:quantumLCA}  is \emph{affine}:
it has the full $\cmbt{K}$ combinator such that
 $\cmbt{K}xy=x$. \myqed
\end{myproposition}

\subsection{A Linear Category via Realizability}
\label{subsection:LinCatViaRealizability}

According to our workflow in Figure~\ref{figure:constructionOfModel},
the next step is to
 employ the (linear) \emph{realizability} technique~\cite{AbramskyL05,Hoshino07} and turn an LCA
 (an \emph{untyped} model) into a linear category (a \emph{typed} model).
  Here in~\S{}\ref{subsection:LinCatViaRealizability} we describe
  how that happens, focusing on the constructions and observations
  already described in~\cite{AbramskyL05,Hoshino07}. Our specific linear
  category $\PER_{\Q}$ has some additional properties that result from
  the way we construct it and, at the same time, are exploited for
  interpreting
  some features of $\Hoq$ that go beyond standard linear
  $\lambda$-calculi. These additional features will be described
  separately, later in~\S{}\ref{subsection:additionalStrOfPERQ}.
  
Although a very brief introduction to realizability is
found in~\S{}\ref{subsection:prelimRealizability}, the current paper
would hardly be enough in  providing good intuitions behind the technical
constructions. Unfamiliar readers are referred
to~\cite{Longley94,vanOosten08} for realizability in general in
categorical settings, and to~\cite{AbramskyL05,Hoshino07} for linear
realizability in particular.
In what follows, for intuitions, it can be helpful to
imagine Kleene's first combinatory algebra---where elements are natural
numbers and application
$a\cdot b$ is defined by the outcome of the $a$-th recursive function applied to the
$b$-th tuple of natural numbers---in place of the LCA $A_{\Q}$. 

%
\begin{mydefinition}[PER]\label{definition:PER}
 A \emph{partial equivalence relation (PER)} over $A_{\Q}$ is a
 symmetric and transitive relation $X$ on the set $A_{\Q}$. The
 \emph{domain} of a PER $|X|$ is defined by 
\begin{displaymath}
 |X|\;:=\;\{x\mid (x,x)\in X\}\;=\;\{x\mid \exists y.\, (x,y)\in X\}\enspace,
\end{displaymath}
 where the last equality follows from symmetry and transitivity of $X$. 
 When restricted to its domain $|X|$,  $X$ is an equivalence relation;
 therefore  $X$ can be thought of as a subset $|X|\subseteq A_{\Q}$,
 suitably
 quotiented.
\end{mydefinition}
Intuitively: $X$ is a ``datatype,'' each element of which is represented
by some elements of $A_{\Q}$; not every element of $A_{\Q}$ represent an
entity of $X$; the set of those elements which do is the domain $|X|\subseteq
A_{\Q}$; and finally the equivalence relation $X$ (restricted to the
domain $|X|$) designates which elements of $A_{\Q}$ represent the same
entity of $X$. 

\begin{mydefinition}[The category $\PER_{\Q}$]\label{definition:PERQ}
 PERs over the LCA $A_{\Q}$ form a category; it is denoted by
 $\PER_{\Q}$.
 Its object $X$ is a PER over $A_{\Q}$.
 Its arrow $X\to Y$
 is defined to be an equivalence class of the  PER
  \begin{equation}\label{eq:homInPER}
  X\limp Y :=\bigl\{\,(c,c')\mid (x,x')\in X\Rightarrow (cx,c'x')\in Y\,\bigr\}\enspace,
  \end{equation}
 where $cx=c\cdot x$ denotes $c\in A_{\Q}$ applied to $x\in A_{\Q}$ via
 the applicative structure $\cdot$ of $A_{\Q}$.
 We denote by $[c]$ the equivalence class in $X\limp Y$ to which $c\in
 A_{\Q}$ belongs. That is, $[c]$ is an arrow that is ``realized by the
 code $c$.''

 Identity arrows and composition of arrows in $\PER_{\Q}$ are defined as
 usual (see e.g.~\cite{AbramskyL05,Hoshino07}). Explicitly,
 \begin{displaymath}
  \id_{X}\;:=\;[\cmbt{I}]\enspace;
  \quad\text{and}\quad
  [d]\co [c]\;:=\;[\cmbt{B}dc]
  \quad\text{for $X\stackrel{[c]}{\to} Y\stackrel{[d]}{\to}Z$ in $\PER_{\Q}$.}
 \end{displaymath}
 Observe that, for the latter, we indeed have
   \begin{displaymath}
    \bigl(\,\text{the code of $[d]\co [c]$}\,\bigr)\cdot
    x
    \;=\;
    \cmbt{B}dcx
    \;=\; d(cx)
    \quad\in |Z|\enspace.
   \end{displaymath}
 Note also that we use $\co$  for composition of arrows in
 $\PER_{\Q}$. This is to be distinguished from $\cdot$ (for application
 in the LCA $A_{\Q}$, that is often omitted); and from $\Kco$
 (for composition of arrows in $\Kleisli{\Q}$, like in~(\ref{eq:kleisliComp}).)
\end{mydefinition}
We elaborate further on the 
 definition~(\ref{eq:homInPER}).
 Its domain $|X\limp Y|$ is easily seen to be the set of $c\in A_{\Q}$
 such that $(x,x')\in X$ implies $(cx,cx')\in Y$.  This requirement is
 that the function $[c]:|X|/X \to |Y|/Y, [x]\mapsto [cx]$ is
 \emph{well-defined}:
 if $(x,x')\in X$, that is, if $x,x'\in A_{\Q}$ ``represent'' the same entity
 in the PER $X$, then applying the code $c$ to both elements must
 result in the elements $cx,cx'\in A_{\Q}$ that again represent the same
 entity in the PER $Y$.
 Furthermore, the PER $X\limp Y$ identifies $c$ and $c'$ such that
 $(cx,c'x)\in Y$ for each $|X|$. This is the \emph{extensionality} of
 the functions of the type $|X|/X\to|Y|/Y$.
 The situation
 is much like in recursion theory, where different natural numbers can
 be ``codes'' of the same recursive function.  

 In~(\ref{eq:homInPER}) $cx$ and $c'x'$ are short for 
$c\cdot x$ and $c'\cdot x'$, respectively. Recall that  $\cdot$ here is the
application operator in the LCA $A_{\Q}=\Kleisli{\Q}(\nat,\nat)$,
that is defined by
\begin{displaymath}
 a\cdot b
 \;:=\;\raisebox{-.5\height}{\includegraphics[height=10em]{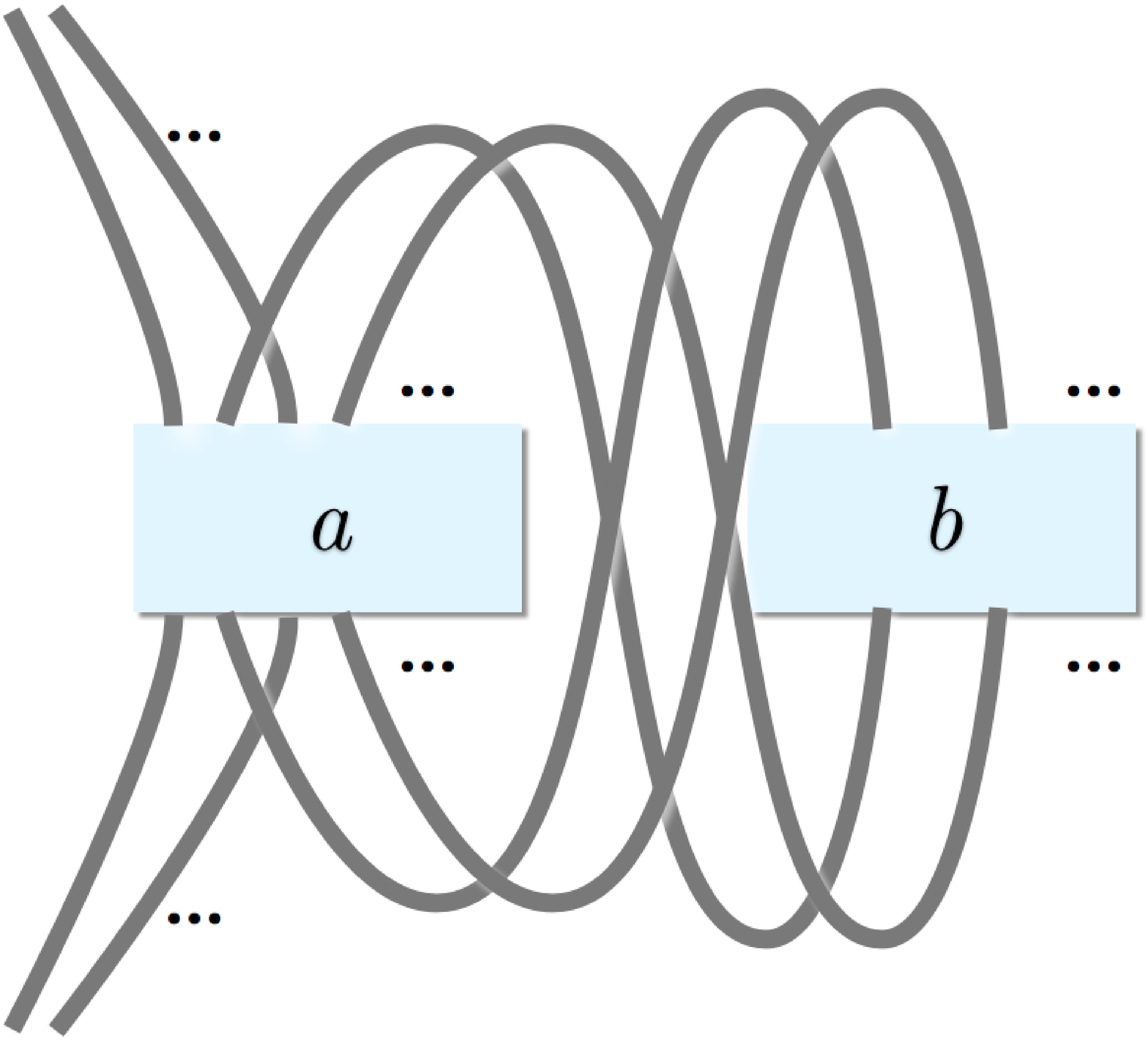}}
\end{displaymath}
in terms of pipes. 
See~\S{}\ref{subsection:LCAviaCategoricalGoI}; to repeat some of the
intuitions provided there, we imagine a token that takes one of the
entrances (top-left pipes) and comes out of one of the exits
(bottom-left pipes). The token exhibits certain branching: in the
example of $\Kleisli{\lift}\cong\Pfn$ it was simply possible
nontermination; in the current setting of quantum branching, a token
carries a quantum state and quantum measurements on the quantum state
give rise to probabilistic branching over different exits that the token
takes (\S{}\ref{subsection:quantumBranchingMonad}).

From now on we describe some properties and features of the category
$\PER_{\Q}$. It is a linear category, making it a categorical model for
usual typed linear $\lambda$-calculi. This fact, with some concrete
constructions of the linear category structure, is described in short.
The category $\PER_{\Q}$ also exhibits some additional
properties. This is for two principal reasons: 1) due to its construction via
realizability (Definition~\ref{definition:PERQ}); and 2) because the
underlying LCA $A_{\Q}$ is affine (rather than linear) and supports full
weakening (Proposition~\ref{proposition:AQisAffine}).
These additional features, separately presented later
in~\S{}\ref{subsection:additionalStrOfPERQ},  will be exploited for
interpreting various features of the language $\Hoq$.
%

Let us begin with some preparations.
\begin{mynotation}
 In what follows, an element of the LCA $A_{\Q}$ is often designated by 
 an untyped linear $\lambda$-term. This is justified by combinatory 
 completeness of LCAs. It is claimed e.g.\
 in~\cite{AbramskyL05,Simpson05} and not hard to establish; an explicit
 proof is found in~\cite{Yoshimizu12BSc}. 
\end{mynotation}
\begin{mydefinition}\label{def:combinatorsForParingAndSum}
 We introduce the following additional combinators in $A_{\Q}$; via
 combinatory completeness, they stand for certain terms composed of the basic
 combinators in Definition~\ref{definition:LCA}, together with the full
 $\cmbt{K}$ combinator in
 Proposition~\ref{proposition:AQisAffine}. 
\begin{displaymath}
  \begin{array}{llll}
  \cmbt{P}:= \lambda xyz. zxy
  \qquad &\text{Pairing}
 \\
  \cmbt{\bar{K}} := \cmbt{KI} &\text{Weakening, $\cmbt{\bar{K}}xy = y$}
 \\
  \cmbt{P_{l}}:= \lambda w. w\cmbt{K} &\text{Left Projection, $\cmbt{P_{l}}(\cmbt{P}xy)=x$}
 \\
  \cmbt{P_{r}}:= \lambda w. w\cmbt{\bar{K}} &\text{Right Projection,  $\cmbt{P_{r}}(\cmbt{P}xy)=y$}
  \end{array}
 \end{displaymath}
\end{mydefinition}
It is easy to see that 
the pairing in $\PER_{\Q}$ is \emph{extensional}:
 $\cmbt{P}xy=\cmbt{P}x'y'$ implies $x=x'$ and $y=y'$.

\begin{mytheorem}[$\PER_{\Q}$ as a linear category]
\label{theorem:PERQIsLinear}
The category  $\PER_{\Q}$ 
is a \emph{linear category}~\cite{Bierman95a,BentonW96}, equipped with a
 symmetric monoidal structure $(\punit,\ptensor)$ and a so-called
 \emph{linear exponential comonad} $\bang $. The latter means that
 $\bang$ is a symmetric monoidal comonad, with natural transformations
 \begin{equation}\label{eq:symMonComonadStrOfBang1}
  \begin{array}{rlrl}
   \der:&
   \bang X\to X\enspace,
   &
   \delta: &\bang X \to \bang\bang X \enspace,
  \\
\varphi:&\bang
  X\ptensor \bang Y\to \bang(X\ptensor Y)\enspace, 
 \qquad
  &
\varphi':&\punit\to\bang\punit\enspace,
  \end{array}
 \end{equation}
that is further equipped with
 monoidal natural transformations
 \begin{equation}\label{eq:symMonComonadStrOfBang2}
  \begin{array}{rlrl}
   \weak:&\bang X\to\punit \quad\text{and}\quad
   &
   \con: &\bang X \to \bang X\boxtimes \bang X\enspace,
  \end{array}
 \end{equation}
 subject to certain additional conditions (see~\cite{Bierman95a,BentonW96}).
\end{mytheorem}
\noindent We use the symbol $\ptensor$ for the monoidal product in $\PER_{\Q}$; it is
 distinguished from the tensor product of quantum states  denoted
 by $\otimes$. See~\S{}\ref{subsection:HoqDesignChoice} for the
 discussion on this issue.

 \begin{myproof}
  The result is due to~\cite[Theorem~2.1]{AbramskyL05}. For later use,
  we shall explicitly describe some of the structures of $\PER_{\Q}$.  

  The category $\PER_{\Q}$ is a symmetric monoidal closed category with
 respect to the following
 operations. 
 \begin{displaymath}
\begin{array}{l}
   X \ptensor Y :=\bigl\{\,(\cmbt{P}xy,
  \cmbt{P}x'y')\,\bigl|\bigr.\,(x,x')\in X\land (y,y')\in Y\,\bigr\}\enspace,
 \\
  \punit := \{(\cmbt{I},\cmbt{I})\}\enspace,
 \qquad
  X\limp Y := (\text{the same as (\ref{eq:homInPER})})\enspace.
\end{array} 
 \end{displaymath}
  Here $\cmbt{P}$ and $\cmbt{I}$ are combinators from
  Definitions~\ref{definition:LCA}
  and~\ref{def:combinatorsForParingAndSum}. 
The operations' action on arrows is defined in a straightforward
manner. For example, given $[c_{1}]:X_{1}\to Y_{1}$ and
$[c_{2}]:X_{2}\to Y_{2}$ in $\PER_{\Q}$,
\begin{displaymath}
[c_{1}]\boxtimes [c_{2}]
:=
\bigl[\lambda w. w\bigl(\lambda uv. \cmbt{P}(c_{1}u)(c_{2}v)\bigr)\bigr]
\;:\quad
X_{1}\boxtimes X_{2}
\longrightarrow
Y_{1}\boxtimes Y_{2}\enspace.
\end{displaymath}
With this definition we indeed have, for $\cmbt{P}x_{1}x_{2}\in
  |X_{1}\boxtimes X_{2}|$, 
  \begin{displaymath}
   \bigl(\,\text{the code of $[c_{1}]\boxtimes [c_{2}]$} \,\bigr) \cdot (\cmbt{P}x_{1}x_{2})
   \;=\; \bigl[\,\cmbt{P}(c_{1}x_{1})(c_{2}x_{2}) \,\bigr]
   \quad\in\; Y_{1}\boxtimes Y_{2}\enspace.
  \end{displaymath}

The linear exponential comonad $\bang$ (see~\cite{AbramskyL05}) is given
  as follows, via the $\bang$ operation on the LCA $A_{\Q}$:
\begin{equation}\label{eq:bangComonadConcretely}
 \bang X := \{(\bang x,\bang x')\mid (x,x')\in X\}\enspace,
 \quad
 \bang [c]:= [\cmbt{F}(\bang c)]\enspace.
\end{equation}
In particular, the use of the combinator $\cmbt{F}$ in the latter
  ensures the type $\bang[c]\colon \bang X\to \bang Y$ (in $\PER_{\Q}$)
  for $c\colon X\to Y$ (in $\PER_{\Q}$): indeed, for any $x\in |X|$, 
  \begin{displaymath}
   \bigl(\,\text{the code of $\bang[c]$} \,\bigr) \cdot (\bang x)
   \,=\,
   \cmbt{F}(\bang c)(\bang x)
   \,=\,\bang (cx) 
   \quad\in\; Y\enspace.
  \end{displaymath}

   The natural transformations $\der,\delta,\varphi, \varphi', \weak$ and $\con$ that
 accompany a linear exponential comonad
 (see~(\ref{eq:symMonComonadStrOfBang1}--\ref{eq:symMonComonadStrOfBang2})) are concretely given as
 follows.
\begin{equation}\label{eq:concreteDefOfLinearExpComonadStr}
 \begin{array}{rlll}
  \der
  &
  := [\cmbt{D}]
  \;
  &:\;\bang X\to X\enspace
\\
  \delta
  &
  :=[\cmbtDelta]
  \;
  &:\;\bang X\to \bang\bang X\enspace
\\
  \varphi
  &
  :=
  \bigl[\lambda w. w(\lambda uv. \cmbt{F}(\cmbt{F}(\bang \cmbt{P})u)v)\bigr]
  &:\;\bang X\boxtimes \bang Y\to \bang(X\boxtimes Y)\enspace
\\
  \varphi'
 &
  :=
  \bigl[\lambda w. w(\bang\cmbt{I})\bigr]
  &:\; \punit\to\bang\punit\enspace
 \\
   \weak
  &
   :=
   [\cmbt{K}\cmbt{I}]
  &:\; \bang X\to\punit\enspace
\\
  \con 
 &
  :=
  [\cmbt{W}\cmbt{P}]
 &:\; \bang X\to \bang X\boxtimes \bang X\enspace
 \end{array}
\end{equation}
 Recall $[c]$ denotes an arrow in $\PER_{\Q}$ that is realized by the
  code $c\in A_{\Q}$. 
 \myqed
 \end{myproof}
\noindent Note that the constructions in the previous proof are  the
  standard ones from~\cite{AbramskyL05} that work for any LCA.



\section{Interpretation of $\Hoq$}
\label{section:denotationalModel}
We now present our interpretation of $\Hoq$ in the category
$\PER_{\Q}$. 
 We have seen in Theorem~\ref{theorem:PERQIsLinear} that the category is a
 linear category,
 hence models (standard) linear $\lambda$-calculi. See
 e.g.~\cite{Mellies09}.

 However, the specific calculus $\Hoq$ calls for some extra features.
 Firstly, the linear exponential comonad $\bang$ on $\PER_{\Q}$ should
 be idempotent ($\bang\bang X\iso \bang X$). This is because in $\Hoq$
 we chose to implicitly track linearity by subtyping $\subtp$---in
 contrast to explicit tracking by constructs like $\synt{derelict}$ in
 standard linear $\lambda$-calculi.
 See~\S{}\ref{subsection:HoqDesignChoice}.  This issue is addressed
 in~\S{}\ref{subsection:additionalStrOfPERQ}.  Secondly, recursion in
 $\Hoq$ requires suitable cpo structures in the model. In the current
 style of realizability models the notion of \emph{admissibility} is a
 standard
 vehicle; this and some related issues are addressed in~\S{}\ref{subsection:furtherAdditionalStrOfPERQ}.
Thirdly we introduce
 some quantum mechanical constructs in $\PER_{\Q}$ for interpreting
 constants of $\Hoq$; see~\S{}\ref{subsection:QMconstructsInPERQ}.
 Finally, we need a strong monad $T$ on $\PER_{\Q}$ for the
 probabilistic effect that arises inevitably in quantum computation
 (more specifically through measurements). In fact we will use for
 $T$ a continuation monad $(\place\limp R)\limp R$ with the result type
 $R$ described as a final coalgebra;
 see~\S{}\ref{subsection:continuationMonad}. The actual interpretation
 of $\Hoq$ in $\PER_{\Q}$ is presented
 in~\S{}\ref{subsection:interpretationOfHoq}. 

The proofs for~\S{}\ref{section:denotationalModel} are deferred to \ref{section:proofsForTheFixedPointOperatorSection}.

\subsection{Additional Structures of $\PER_{\Q}$}
\label{subsection:additionalStrOfPERQ}
We go on to study some additional structures that are available in
$\PER_{\Q}$. These will be exploited later for interpreting various
features of $\Hoq$, such as subtyping and coproduct types.

Recall that in $A_{\Q}$ (that is an \emph{affine} LCA)
a full weakening combinator $\cmbt{K}$ is available
(Proposition~\ref{proposition:AQisAffine}).
  \begin{mylemma}[$\punit$ is terminal]
   \label{lem:PERQMonoidalUnitIsTerminal}
  The monoidal unit $\punit$ is terminal (i.e.\ final) in
 $\PER_{\Q}$, with a unique arrow $\weak:X\to \punit$ given by\footnote{We shall use the same notation,
 $\weak$, for both the unique arrow $X\to\punit$  and a structure
 morphism for a linear exponential comonad $\bang X\to \punit$
 (Theorem~\ref{theorem:PERQIsLinear}). They are indeed the same arrow $[\cmbt{K}\cmbt{I}]$.}
\begin{equation}
 \weak \;:=\; [\cmbt{K}\cmbt{I}]\enspace.
  \tag*{\myqed}
\end{equation}
\end{mylemma}

 \begin{mylemma}[Binary (co)products in $\PER_{\Q}$]
  \label{lem:productsAndCoproducts}
 The category $\PER_{\Q}$
 has binary products $\times$ and binary coproducts $+$.
 Products are realized by a CPS-like encoding.
 \begin{align*}
   X \times Y &:=\bigl\{\,\bigl(\cmbt{P}k_{1}(\cmbt{P}k_{2}u),\,
\cmbt{P}k'_{1}(\cmbt{P}k'_{2}u')\bigr)
   \;\bigl|\bigr.\;
  (k_{1}u,k'_{1}u')\in X\land (k_{2}u,k'_{2}u')\in Y\,\bigr\}\enspace,
 \\
   X + Y &:=
    \bigl\{\,(\cmbt{PK}x,\cmbt{PK}x')   \,\bigl|\bigr.\,
    (x,x')\in X \bigr\}
   \cup
    \bigl\{\,(\cmbt{P\bar{K}}y,\cmbt{P\bar{K}}y')   \,\bigl|\bigr.\,
    (y,y')\in Y \bigr\}\enspace.
 \end{align*}
Their accompanying structures are defined in a straightforward manner. 
For example, the projection maps are concretely as follows.
\begin{equation}\label{eq:realizerForProjectionMap}
\begin{array}{rclll}
  \pi_{1}
 \;&=&\;
 \bigl[\,
  \lambda w. w(\lambda kv. v(\lambda lu. ku))
 \,\bigr]
 &:\quad&
  X\times Y\longrightarrow X\enspace;
 \\
 \pi_{2}
 \;&=&\;
 \bigl[\,
  \lambda w. w(\lambda kv. v(\lambda lu. lu))
 \,\bigr]
 &:\quad&
  X\times Y\longrightarrow Y\enspace.
\end{array}
\end{equation}
\end{mylemma}
  \begin{myproof}
    Straightforward; see e.g.~\cite{AbramskyL05,Hoshino07}. 
  \end{myproof}
\noindent
Logically  $\ptensor$ is  ``multiplicative and'';  $\times$ is ``additive and.''

 \begin{mylemma}\label{lem:bangVsProductTensorAndCoproduct}
  The following canonical isomorphisms are available in
  any linear
  category with binary (co)products, hence in $\PER_{\Q}$. 
   \begin{equation}\label{eq:canonicalIsomorphismInLinCat}
\begin{array}{rcl}
    \bang(X\times Y)&\cong& \bang X\boxtimes \bang Y\enspace,
 \\
 (X+Y)\boxtimes Z &\cong& X\boxtimes Z + Y\boxtimes Z\enspace,
 \\
 \punit\limp X &\cong& X\enspace,
 \\
 (X+Y)\limp Z &\cong& (X\limp Z)\times (Y\limp Z)\enspace.
\end{array}
 \end{equation}
  
 Additionally, in $\PER_{\Q}$, we have the following  canonical isomorphisms.
 \begin{equation}\label{eq:additionalIsomorphismsInPERQ}
\begin{array}{rclrcl}
  \bang(X+Y)&\stackrel{\cong}{\longrightarrow}& \bang X+\bang Y  
 \\
 \der_{\bang X}:\; \bang \bang X&\stackrel{\cong}{\longrightarrow}&
  \bang X
 \; :\delta\enspace
 &
 \quad\bang\der_{ X}:\; \bang \bang X&\stackrel{\cong}{\longrightarrow}&
  \bang X
 \; :\delta\enspace
  \\
 \bang(X\boxtimes Y) &\stackrel{\cong}{\longrightarrow}& \bang
  X\boxtimes \bang Y
  \; :\varphi
  &
  \weak:\;
  \bang\punit
   &\stackrel{\cong}{\longrightarrow}& \punit
  \; :\varphi'
\end{array}
\end{equation}
Here the arrows $\der,\delta, \varphi,\varphi'$ are from
 Theorem~\ref{theorem:PERQIsLinear}.
Therefore  $\bang$ on $\PER_{\Q}$ is
 idempotent and strong monoidal; it also preserves coproducts.

\end{mylemma}
\begin{myproof}
 The ones in~(\ref{eq:canonicalIsomorphismInLinCat}) are standard;
 see~\cite[\S{}2.1.2]{Benton94}. The ones in~(\ref{eq:additionalIsomorphismsInPERQ}) also hold in any $\PER_{A}$ with an affine LCA $A$.  For the
 first ($\bang$ distributes over $+$), from right to left one takes
 $[\bang\kappa_{\ell},\bang\kappa_{r}]$ where $\kappa_{\ell},\kappa_{r}$
 are coprojections;\footnote{Note that we are using square brackets
 $[\place]$ to denote both: equivalence classes modulo PERs; and
 cotupling of arrows (like in $[\bang\kappa_{\ell},\bang\kappa_{r}]$).
 We hope this  does not lead to much confusion: the two usages have
 different arities. } from left to right one can take
 \begin{displaymath}
 [a]
  \;:\;
 \bang(X+Y)\longrightarrow \bang X +\bang Y
  \quad\text{with}\quad
  a:=\lambda
  w. \cmbt{WP}w\bigl(\lambda uv. 
  \cmbt{P}(\cmbt{D}u\cmbt{K})(\cmbt{F}(\bang
  \cmbt{P_{r}})v)\bigr)
\enspace;
 \end{displaymath}
 indeed it is straightforward to see that
 $a(\bang(\cmbt{P}\cmbt{K}x))=\cmbt{P}\cmbt{K}(\bang x)$
 and
 $a(\bang(\cmbt{P}\bar{\cmbt{K}}x))=\cmbt{P}\bar{\cmbt{K}}(\bang x)$
 for the above $a$.
 The second line of~(\ref{eq:additionalIsomorphismsInPERQ})
  ($\bang$ is idempotent) follows immediately from
 the definitions of $\der$ and $\delta$
 in~(\ref{eq:concreteDefOfLinearExpComonadStr}). For example,
 \begin{displaymath}
  \begin{array}{ll}
   (\bang\der_{X})(\bang\bang x)
  &
   = (\bang[\cmbt{D}])(\bang\bang x)
  \\
 &=
   \cmbt{F}(\bang\cmbt{D})(\bang\bang x)
   \quad\text{by~(\ref{eq:bangComonadConcretely}), def.\ of $\bang$'s action on
   arrows}
 \\
   &= 
   \bang(\cmbt{D}\bang x) = \bang x\enspace.
  \end{array}
   \end{displaymath}
For the third line ($\bang$ distributes over
 $\boxtimes$), an inverse of $\varphi$ can be given by the following
 composite, exploiting that $\punit$ is terminal
 (Lemma~\ref{lem:PERQMonoidalUnitIsTerminal}).
\begin{displaymath}
 \bang(X\boxtimes Y)
\stackrel{\con}{\longrightarrow}
 \bang(X\boxtimes Y)\boxtimes  \bang(X\boxtimes Y)
\stackrel{\weak}{\longrightarrow}
 \bang(X\boxtimes \punit)\boxtimes  \bang(\punit\boxtimes Y)
\stackrel{\cong}{\longrightarrow}
\bang X \boxtimes \bang Y\enspace
\end{displaymath}
This concludes the proof. \myqed
\end{myproof}
One consequence of the last result is that we have $\bang(X\boxtimes
Y)\cong \bang (X\times Y)$ in $\PER_{\Q}$. As stated in the proof, this
is true in $\PER_{A}$ for any \emph{affine} LCA $A$. 

We go ahead and show that the category $\PER_{\Q}$ has countable limits
and colimits.
\begin{mydefinition}[Combinators $(x_{i})_{i\in\nat}$, $\cmbt{D}_{i}$]
 \label{definition:sequenceAndNThDereliction}
 Let $x_{0},x_{1},\dotsc\in A_{\Q}$. We define
 the element $(x_{i})_{i\in\nat}\in
 A_{\Q}$ by:
 \begin{displaymath}
(x_{i})_{i\in\nat}\;:=\;
\Bigl(\;  \nat\stackrel{v}{\longrelto}\nat\cdot\nat
  \stackrel{\coprod_{i}x_{i}}{\longrelto}
  \nat\cdot\nat
  \stackrel{u}{\longrelto}
  \nat
\;\Bigr)
\quad
=
\quad
     \raisebox{-.5\height}{
\scalebox{.4} 
{
\begin{pspicture}(0,-2.72)(5.879824,2.72)
\psframe[linewidth=0.04,dimen=outer](1.2,0.6)(0.0,-0.4)
\usefont{T1}{ptm}{m}{n}
\rput(0.6305468,0.105){\LARGE $x_0$}
\usefont{T1}{ptm}{m}{n}
\rput(2.8605468,1.605){\LARGE $v$}
\psline[linewidth=0.04cm](2.84,-2.1)(2.84,-2.7)
\psline[linewidth=0.04cm](2.84,2.7)(2.84,2.1)
\pscircle[linewidth=0.04,dimen=outer](2.84,1.7){0.4}
\usefont{T1}{ptm}{m}{n}
\rput(2.8605468,-1.795){\LARGE $u$}
\pscircle[linewidth=0.04,dimen=outer](2.84,-1.7){0.4}
\usefont{T1}{ptm}{m}{n}
\rput(4.829092,0.065){\LARGE $\cdots$}
\psframe[linewidth=0.04,dimen=outer](3.2,0.6)(2.0,-0.4)
\usefont{T1}{ptm}{m}{n}
\rput(2.6305463,0.105){\LARGE $x_1$}
\psline[linewidth=0.04cm](2.48,1.4)(0.68,0.6)
\psline[linewidth=0.04cm](0.68,-0.38)(2.54,-1.48)
\psline[linewidth=0.04cm](2.68,1.4)(2.48,0.6)
\psline[linewidth=0.04cm](2.48,-0.4)(2.74,-1.32)
\end{pspicture} 
}
}
 \end{displaymath}
 where $u:\nat\cdot\nat\cong\nat:v$ are (fixed) isomorphisms in
 Theorem~\ref{theorem:quantumLCA}.
 
 For each $i\in\nat$, we define an element $ \cmbt{D}_{i}\in A_{\Q}$ by
\begin{multline*}
\cmbt{D}_{i}
\;:=\;
\left(\begin{array}{l}
  \nat 
 \stackrel{k}{\longrelto}
 \nat + \nat
 \stackrel{\kappa_{i}\cdot \nat + v}{\longrelto}
 \nat\cdot \nat +  \nat\cdot \nat
   \\\qquad
 \stackrel{u+p_{i}\cdot \nat}{\longrelto}
 \nat+\nat
 \stackrel{[\kappa_{r},\kappa_{\ell}]}{\longrelto}
 \nat +\nat
 \stackrel{j}{\longrelto}
 \nat
\end{array}
\right)
 \quad=\quad
\raisebox{-.5\height}{
\scalebox{.4} 
{
\begin{pspicture}(0,-4.3)(10.955742,4.3)
\definecolor{color1865b}{rgb}{0.8,0.8,0.8}
\pstriangle[linewidth=0.04,dimen=outer,fillstyle=solid,fillcolor=color1865b](1.9870257,3.28)(3.6,0.6)
\psline[linewidth=0.04cm](1.9870257,3.88)(1.9870257,4.28)
\psarc[linewidth=0.04](0.88744986,1.9074498){0.88744986}{0.0}{180.0}
\psline[linewidth=0.04cm](0.0,1.9)(0.0,1.1)
\psline[linewidth=0.04cm](0.0,1.1)(1.7748997,1.1)
\psline[linewidth=0.04cm](1.7748997,1.9)(1.7748997,1.1)
\usefont{T1}{ptm}{m}{n}
\rput(.9,1.76){\LARGE $\kappa_i \cdot \mathbb{N}$}
\pscircle[linewidth=0.04,dimen=outer](0.9,-0.2){0.7}
\usefont{T1}{ptm}{m}{n}
\rput(.914668,-0.24){\Huge $u$}
\psarc[linewidth=0.04](3.0874498,-0.21255016){0.88744986}{170.70195}{12.3849325}
\psline[linewidth=0.04cm](2.2,-0.20510033)(2.2,0.59489965)
\psline[linewidth=0.04cm](2.2,0.59489965)(3.9748998,0.59489965)
\psline[linewidth=0.04cm](3.9748998,-0.20510033)(3.9748998,0.59489965)
\usefont{T1}{ptm}{m}{n}
\rput(3.034668,-0.24){\LARGE $p_i \cdot \mathbb{N}$}
\pscircle[linewidth=0.04,dimen=outer](3.1,1.8948997){0.7}
\usefont{T1}{ptm}{m}{n}
\rput(3.014668,1.76){\Huge $v$}
\psline[linewidth=0.04cm](0.8,3.3)(0.8,2.9)
\psline[linewidth=0.04cm](0.8,1.1)(0.8,0.5)
\psline[linewidth=0.04cm](1.0,1.1)(1.0,0.5)
\psline[linewidth=0.04cm](0.8,-0.9)(0.8,-1.7)
\psline[linewidth=0.04cm](3.0,-1.1)(3.0,-1.7)
\psline[linewidth=0.04cm](2.98,1.26)(2.98,0.6)
\psline[linewidth=0.04cm](3.18,1.18)(3.18,0.62)
\psline[linewidth=0.04cm](3.12,3.26)(3.12,2.58)
\rput{-180.0}(3.9340515,-7.16){\pstriangle[linewidth=0.04,dimen=outer,fillstyle=solid,fillcolor=color1865b](1.9670258,-3.88)(3.6,0.6)}
\psline[linewidth=0.04cm](1.9670258,-3.88)(1.9670258,-4.28)
\psline[linewidth=0.04cm](0.8,-1.7)(3.0,-2.9)
\psline[linewidth=0.04cm](3.0,-1.7)(0.8,-2.9)
\psline[linewidth=0.04cm](0.8,-2.9)(0.8,-3.3)
\psline[linewidth=0.04cm](3.0,-2.9)(3.0,-3.3)
\end{pspicture} 
}
}\enspace.
\end{multline*}
Here $\kappa_{i}$ (for $i\in\nat$), $\kappa_{\ell},\kappa_{r}$ (for
 ``left'' and ``right'') are coprojections; and 
 $p_{i}:\nat\relto 1$ is a ``projection'' map in~\cite{Jacobs10trace}
 (see also Theorem~\ref{theorem:KleisliOfQIsTracedInDetail}) that satisfies
 \begin{displaymath}
  p_{i}\Kco \kappa_{j} =
 \begin{cases}
  \id_{1} &\text{if $i=j$,}
 \\
  \bot &\text{otherwise,}
 \end{cases}
 \end{displaymath}
 where $\Kco$ denotes composition of arrows in $\Kleisli{\Q}$ and $\bot$
 is the least element (the ``zero map'') in the homset $\Kleisli{\Q}(1,1)$.
\end{mydefinition}

\begin{mylemma}\label{lem:sequenceVsIThDereliction}
 We have $\cmbt{D}_{j} \cdot (x_{i})_{i\in\nat}=x_{j}$. Here $\cdot$ denotes
 the applicative structure of $A_{\Q}$ (Theorem~\ref{theorem:quantumLCA}).
\end{mylemma}
\begin{myproof}
 By easy manipulation of string diagrams, the claim boils down to
 the equality
 \begin{displaymath}
  \left(\begin{array}{l}
  \nat 
 \stackrel{\kappa_{j}\cdot \nat}{\longrelto}
 \nat\cdot \nat
   \stackrel{\coprod_{i}x_{i}}{\longrelto}
 \nat\cdot \nat
 \stackrel{p_{j}\cdot \nat}{\longrelto}
 \nat
\end{array}
\right)
 \;=\; x_{j}\enspace.
 \end{displaymath}
This follows from the fact that
\begin{displaymath}
 (\coprod_{i}x_{i})\Kco (\kappa_{j}\cdot\nat)
=
 \bigl[\,(\kappa_{i}\cdot\nat)\Kco x_{i}\,\bigr]_{i}\Kco (\kappa_{j}\cdot\nat)
=
\kappa_{j}\cdot\nat\Kco x_{j}
\end{displaymath}
and that $p_{j}\Kco \kappa_{j}=\id_{1}$. \myqed
\end{myproof}

\begin{myproposition}\label{proposition:PERQIsCountablyComplAndCoCompl}
 The category $\PER_{\Q}$ has countable limits and colimits.
\end{myproposition}
\begin{myproof}
 Constructions of equalizers and coequalizers in realizability
 categories are described in~\cite{Longley94}; they work in the current
 setting of PERs over an LCA, too. Concretely: given $[c],[d]:X\rightrightarrows Y$
 in $\PER_{\Q}$, let
 \begin{align*}
 E\;&:=\;
 \bigl\{\,
 (x,x')
\,\bigl|\bigr.\,
  (x,x')\in X
  \land
  (cx,dx')\in Y 
\,\bigr\}
 \enspace;\;\text{and}
\\
 C\;&:=\;
 \bigl(\,\text{the symmetric and transitive closure of}\;
 \bigl\{\,
 (y,y')
\,\bigl|\bigr.\,
 (y,y')\in Y
\,\bigr\}
\cup
 \bigl\{\,
 (cx,dx')
\,\bigl|\bigr.\,
  (x,x')\in X
\,\bigr\}
\,\bigr)
 \enspace.
 \end{align*}
Then it is straightforward to see that 
\begin{math}
 [\cmbt{I}]:E\to X
\end{math}
and 
\begin{math}
  [\cmbt{I}]:Y\to C
\end{math}---where $\cmbt{I}$ is the identity combinator (Definition~\ref{definition:LCA})---are an equalizer and a coequalizer of $[c]$ and $[d]$, respectively.
\auxproof{
It is easy. Let us just check  that if $(cx,dx')\in Y$, then $(cx,dx)\in Y$. This is proved as follows.
Since $d$ tracks a function $X\to Y$, $(dx',dx)\in Y$, hence by
 transitivity of $Y$, $(cx,dx)\in Y$. 
}

 It suffices to show that  $\PER_{\Q}$ has countable
 products and coproducts (since (co)products and (co)equalizers give all
 (co)limits). Given a countable family $(X_{i})_{i\in\nat}$
 of  objects of $\PER_{\Q}$, we use the constructs in
 Definition~\ref{definition:sequenceAndNThDereliction} and define
 \begin{align*}
  \prod_{i\in\nat} X_{i}
 \;&:=\;
 \bigl\{\,\bigl(
  \cmbt{P}(k_{i})_{i\in\nat}u,\,
  \cmbt{P}(k'_{i})_{i\in\nat}u'
 \bigr)
\,\bigl|\bigr.\,
  (k_{i}u,k'_{i}u')\in X_{i} \text{ for each $i\in\nat$}\,\bigr\}
 \enspace;
 \\
 \pi_{i}
 \;&:=\;
 \bigl[\,
  \lambda w. w(\lambda vu.(\cmbt{D}_{i}v)u)
\,\bigr]
 \quad:\quad
  \prod_{i\in\nat} X_{i}
  \longrightarrow 
  X_{i} \enspace.
 \end{align*}
 Then, given a family $\bigl(\,[c_{i}]:Y\to X_{i}\,\bigr)_{i\in\nat}$ of
 arrows, its tupling
can be given by
 \begin{displaymath}
  \bigl\langle[c_{i}]\bigr\rangle_{i\in\nat}
  \;:=\;
 \bigl[\lambda y.\cmbt{P}(c_{i})_{i\in\nat} \,y\bigr]
  \quad:\quad Y\longrightarrow
 \prod_{i}X_{i}\enspace.
 \end{displaymath}
 The uniqueness of such a tupling directly follows from the definition
 of the PER $\prod_{i\in\nat} X_{i}$. 
 
On coproducts, we define
 \begin{align*}
  \coprod_{i\in\nat} X_{i}
 \;&:=\;
  \bigl\{\,(\cmbt{P}\cmbt{D}_{i}x_{i},\cmbt{P}\cmbt{D}_{i}x'_{i})\,\bigl|\bigr. \,i\in\nat,
  (x_{i},x'_{i})\in
  X_{i}\,\bigr\}\enspace;
 \\
 \kappa_{i}
 \;&:=\;
 \bigl[\,
   \cmbt{P}\cmbt{D}_{i}
\,\bigr]
 \quad:\quad
  X_{i}
  \longrightarrow 
  \coprod_{i\in\nat} X_{i}
 \enspace.
 \end{align*}
Given a family $\bigl(\,[c_{i}]:X_{i}\to Y\,\bigr)_{i\in\nat}$ of
 arrows, its cotupling
 can be given by
\begin{displaymath}
 \bigl[\,[c_{i}]\,\bigr]_{i\in\nat}
 \;:=\;
 \bigl[\lambda w. w(\lambda d x. d(c_{i})_{i\in\nat}\, x)\bigr]
  \quad:\quad
\coprod_{i}X_{i}\longrightarrow
 Y\enspace.
\end{displaymath}
 This concludes the proof. \myqed
\end{myproof}


\begin{myremark} 
 Although we have focused on the specific linear category $\PER_{\Q}$,
 what are said in the current~\S{}\ref{subsection:additionalStrOfPERQ}
 are true in more general settings. 

 One point (that is already mentioned) is that the extra canonical
 isomorphisms in~(\ref{eq:additionalIsomorphismsInPERQ}) hold in
 any $\PER_{A}$ with an affine LCA $A$. This makes such categories
 $\PER_{A}$ suitable for modeling linear $\lambda$-calculi with implicit 
 linearity tracking. 

 Another point is about the constructions in
 Definition~\ref{definition:sequenceAndNThDereliction},
 Lemma~\ref{lem:sequenceVsIThDereliction} and
 Proposition~\ref{proposition:PERQIsCountablyComplAndCoCompl}. It is not hard
 to see that these are all possible in the category $\PER_{A_{B}}$,
 where the LCA $A_{B}$ is obtained via  categorical
 GoI~\cite[Proposition~4.2]{AbramskyHS02} from the Kleisli category
 $\Kleisli{B}$ for any ``branching monad'' like $B=\lift,\pow,\dist$ and
 $\Q$ (see~\S{}\ref{subsection:prelimMonadForBranching}).


\end{myremark}

\subsection{Order Enrichment and Further Additional Constructs in $\PER_{\Q}$}
\label{subsection:furtherAdditionalStrOfPERQ}
Here in~\S{}\ref{subsection:furtherAdditionalStrOfPERQ} we describe
another additional  construct---namely an alternative pairing combinator
$\cmbt{\dot{P}}$---in our categorical model
$\PER_{\Q}$. What we describe here are mostly concerned about order/cpo
structures, which we exploit for the purpose of interpreting recursion
in $\Hoq$.

Let us first note that (the underlying set of) the LCA $A_{\Q}$ is
equipped
with an $\omega$-CPO structure $\sqsubseteq$. This is because:
$A_{\Q}=\Kleisli{\Q}(\nat,\nat)$ (Theorem~\ref{theorem:quantumLCA}); and
the category $\Kleisli{\Q}$ is $\omega\text{-}\mathbf{CPO}$ enriched
(Theorem~\ref{theorem:KleisliOfQIsTraced}). Recalling from
Definition~\ref{definition:pointwiseorderOnQX}, the order on $A_{\Q}$ is
(a suitable pointwise extension of) L\"{o}wner partial order
(Definition~\ref{definition:LoewnerPartialOrder}). For the record:
\begin{mylemma}[$A_{\Q}$ is an $\omega$-CPO]\label{lem:LCAIsCPO}
The set $A_{\Q}$ is an $\omega$-CPO with the smallest element $\bot$.
Furthermore: 
\begin{enumerate}
 \item \label{item:applBangConti} 
the application operator
 $\cdot:A_{\Q}^{2}\to A_{\Q}$  and 
 the $\bang$ operator are continuous; and
 \item \label{item:applLeftStrict}
   application      $\cdot$ is left strict, that is, $\bot\cdot a=\bot$
       for each $a\in A_{\Q}$. \myqed
\end{enumerate}
\end{mylemma}
This order structure on $A_{\Q}$, unfortunately, does \emph{not} give
rise to orders on $\PER_{\Q}$ in a categorically structured
manner. Certainly not so nicely as for the Kleisli category
$\Kleisli{\Q}$---we do not see any straightforward way to make the
category $\PER_{\Q}$ $\omega\text{-}\mathbf{CPO}$ enriched. 
To see it, consider the following notion of \emph{admissibility} of
PERs---this is a standard vehicle for interpreting recursion in
realizability models (see e.g.~\cite{AbadiP90,Abadi00}).
\begin{mydefinition}[Admissible PER]\label{definition:admissiblePER}
 A PER $U\in\PER_{\Q}$ is said to be \emph{admissible} if: 
\begin{itemize}
 \item{} (strictness)
 $(\bot,\bot)\in U$ for
 the least element $\bot\in A_{\Q}$; and

 \item{} (inductiveness)
 $x_{0}\Lle x_{1}\Lle\cdots$, $y_{0}\Lle
 y_{1}\Lle\cdots$ and
 $(x_{i},y_{i})\in U$ for each $i\in\nat$ imply $(\sup_{i}x_{i},\sup_{i}y_{i})\in
 U$.
\end{itemize} 
\end{mydefinition}
The trickiness of order structures on $\PER_{\Q}$ is exemplified by the
fact that admissibility is not preserved by isomorphisms in
$\PER_{\Q}$.
\begin{myexample}[Admissibility not preserved by isomorphisms]\label{example:admissibilityAndIso}
 It is easy to see that we have an isomorphism
 $\punit = \{(\cmbt{I},\cmbt{I})\}\cong
 \{(\bot,\bot)\}=:\Bt$ in $\PER_{\Q}$; here $\Bt$ is admissible while
 $\punit$ is not. 
\end{myexample}



It is for this trickiness of the order structures on $\PER_{\Q}$ that we
are introducing an alternative $\cmbt{\dot{P}}$  to the pairing
combinator $\cmbt{P}$
(Definition~\ref{def:combinatorsForParingAndSum})---the former leads to 
 a different ``implementation'' $\dtimes$ of products.  The merit of
 $\dtimes$ is that it exhibits a better order-theoretic property
 (namely it preserves admissibility, Lemma~\ref{lem:closurePropertyOfAdmissibility}); we will
need them for recursion. In contrast, $\cmbt{P}$ enjoys a useful
combinatorial property: $(\cmbt{P}xy)z = zxy$.

\begin{mydefinition}[Combinator $\cmbt{\dot{P}}$, binary product
 $X\dtimes Y$]\label{definition:dotProduct} 
  We define an element $\cmbt{\dot{P}}\in A_{\Q}$ by the string diagram
  in $\Kleisli{\Q}$ shown below on the left. The triangles denote
  $j:\nat +\nat\cong \nat:k$ in Theorem~\ref{theorem:quantumLCA}.  Then
  $\cmbt{\dot{P}}x y$ becomes as shown bottom on the right.  
\begin{equation}\label{eq:dotPinStringDiagrams}
\cmbt{\dot{P}}\;:=\;
 \raisebox{-.5\height}{
\scalebox{.5} 
{
\begin{pspicture}(0,-2.82)(2.937538,2.82)
\definecolor{color244b}{rgb}{0.8,0.8,0.8}
\pstriangle[linewidth=0.04,dimen=outer,fillstyle=solid,fillcolor=color244b](1.2,1.2)(1.6,0.4)
\pstriangle[linewidth=0.04,dimen=outer,fillstyle=solid,fillcolor=color244b](2.2,2.0)(1.6,0.4)
\psline[linewidth=0.04cm](2.2,2.8)(2.2,2.4)
\psline[linewidth=0.04cm](1.8,2.0)(1.2,1.6)
\pstriangle[linewidth=0.04,dimen=outer,fillstyle=solid,fillcolor=color244b](0.8,0.4)(1.6,0.4)
\psline[linewidth=0.04cm](0.8,1.2)(0.8,0.8)
\rput{-180.0}(2.4,-2.8){\pstriangle[linewidth=0.04,dimen=outer,fillstyle=solid,fillcolor=color244b](1.2,-1.6)(1.6,0.4)}
\rput{-180.0}(4.4,-4.4){\pstriangle[linewidth=0.04,dimen=outer,fillstyle=solid,fillcolor=color244b](2.2,-2.4)(1.6,0.4)}
\psline[linewidth=0.04cm](2.2,-2.8)(2.2,-2.4)
\psline[linewidth=0.04cm](1.8,-2.0)(1.2,-1.6)
\rput{-180.0}(1.6,-1.2){\pstriangle[linewidth=0.04,dimen=outer,fillstyle=solid,fillcolor=color244b](0.8,-0.8)(1.6,0.4)}
\psline[linewidth=0.04cm](0.8,-1.2)(0.8,-0.8)
\psline[linewidth=0.04](2.6,2.0)(2.6,0.8)(0.4,-0.2)(0.4,-0.4)
\psline[linewidth=0.04](1.6,1.2)(1.8,0.8)(1.8,-0.2)(1.2,-0.4)
\psline[linewidth=0.04](0.4,0.4)(0.4,0.2)(2.6,-0.8)(2.6,-2.0)
\psline[linewidth=0.04](1.2,0.4)(2.0,0.2)(2.0,-0.8)(1.6,-1.2)
\end{pspicture} 
}
}
\qquad
\cmbt{\dot{P}}xy\;=\;
\raisebox{-.5\height}{
\scalebox{.5} 
{
\begin{pspicture}(0,-2.12)(4.0,2.12)
\definecolor{color244b}{rgb}{0.8,0.8,0.8}
\pstriangle[linewidth=0.04,dimen=outer,fillstyle=solid,fillcolor=color244b](2.0,1.1)(3.6,0.6)
\rput{-180.0}(4.0,-2.8){\pstriangle[linewidth=0.04,dimen=outer,fillstyle=solid,fillcolor=color244b](2.0,-1.7)(3.6,0.6)}
\psline[linewidth=0.04cm](0.8,1.1)(0.8,0.7)
\psline[linewidth=0.04cm](3.2,1.1)(3.2,0.7)
\psline[linewidth=0.04cm](2.0,2.1)(2.0,1.7)
\psline[linewidth=0.04cm](0.8,-0.7)(0.8,-1.1)
\psline[linewidth=0.04cm](3.2,-0.7)(3.2,-1.1)
\psframe[linewidth=0.04,dimen=outer](1.6,0.7)(0.0,-0.7)
\usefont{T1}{ptm}{m}{n}
\rput(0.81466794,-0.04){\Huge $x$}
\psframe[linewidth=0.04,dimen=outer](4.0,0.7)(2.4,-0.7)
\usefont{T1}{ptm}{m}{n}
\rput(3.214668,-0.04){\Huge $y$}
\psline[linewidth=0.04cm](2.0,-1.7)(2.0,-2.1)
\end{pspicture} 
}
}
\end{equation}
 Let 
 $\cmbt{\dot{P}_{l}}$ 
 and
$\cmbt{\dot{P}_{r}}$ 
 be the following elements of $A_{\Q}$.
 \begin{displaymath}
 \cmbt{\dot{P}_{l}}
 :=
    \raisebox{-.5\height}{
\scalebox{.4} 
{
\begin{pspicture}(0,-2.42)(3.7375379,2.42)
\definecolor{color244b}{rgb}{0.8,0.8,0.8}
\pstriangle[linewidth=0.04,dimen=outer,fillstyle=solid,fillcolor=color244b](1.8,1.4)(3.6,0.6)
\pstriangle[linewidth=0.04,dimen=outer,fillstyle=solid,fillcolor=color244b](3.0,0.6)(1.6,0.4)
\rput{-180.0}(3.6,-3.4){\pstriangle[linewidth=0.04,dimen=outer,fillstyle=solid,fillcolor=color244b](1.8,-2.0)(3.6,0.6)}
\rput{-180.0}(6.0,-1.6){\pstriangle[linewidth=0.04,dimen=outer,fillstyle=solid,fillcolor=color244b](3.0,-1.0)(1.6,0.4)}
\psline[linewidth=0.04cm](0.6,1.4)(0.6,1.0)
\psline[linewidth=0.04cm](3.0,1.4)(3.0,1.0)
\psline[linewidth=0.04cm](0.6,-1.0)(0.6,-1.4)
\psline[linewidth=0.04cm](3.0,-1.0)(3.0,-1.4)
\psline[linewidth=0.04cm](1.8,-2.0)(1.8,-2.4)
\psline[linewidth=0.04cm](1.8,2.4)(1.8,2.0)
\psline[linewidth=0.04](2.6,0.6)(2.6,0.4)(0.6,-0.4)(0.6,-1.0)(0.6,-0.8)
\psline[linewidth=0.04](0.6,1.0)(0.6,0.4)(2.6,-0.4)(2.6,-0.6)
\psline[linewidth=0.04](3.4,0.6)(3.4,0.2)(3.4,0.2)
\psline[linewidth=0.04](3.4,-0.2)(3.4,-0.6)(3.4,-0.6)
\pscircle[linewidth=0.04,dimen=outer,fillstyle=solid,fillcolor=black](3.4,0.2){0.1}
\pscircle[linewidth=0.04,dimen=outer,fillstyle=solid,fillcolor=black](3.4,-0.2){0.1}
\end{pspicture} 
}
}
\qquad
 \cmbt{\dot{P}_{r}}
 :=
  \raisebox{-.5\height}{
\scalebox{.4} 
{
\begin{pspicture}(0,-2.62)(3.7375379,2.62)
\definecolor{color244b}{rgb}{0.8,0.8,0.8}
\pstriangle[linewidth=0.04,dimen=outer,fillstyle=solid,fillcolor=color244b](1.8,1.6)(3.6,0.6)
\pstriangle[linewidth=0.04,dimen=outer,fillstyle=solid,fillcolor=color244b](3.0,0.8)(1.6,0.4)
\rput{-180.0}(3.6,-3.8){\pstriangle[linewidth=0.04,dimen=outer,fillstyle=solid,fillcolor=color244b](1.8,-2.2)(3.6,0.6)}
\rput{-180.0}(6.0,-2.0){\pstriangle[linewidth=0.04,dimen=outer,fillstyle=solid,fillcolor=color244b](3.0,-1.2)(1.6,0.4)}
\psline[linewidth=0.04cm](0.6,1.6)(0.6,1.2)
\psline[linewidth=0.04cm](3.0,1.6)(3.0,1.2)
\psline[linewidth=0.04cm](0.6,-1.2)(0.6,-1.6)
\psline[linewidth=0.04cm](3.0,-1.2)(3.0,-1.6)
\psline[linewidth=0.04cm](1.8,-2.2)(1.8,-2.6)
\psline[linewidth=0.04cm](1.8,2.6)(1.8,2.2)
\psline[linewidth=0.04](2.6,0.8)(2.6,0.4)(2.6,0.4)
\psline[linewidth=0.04](2.6,-0.4)(2.6,-0.8)(2.6,-0.8)
\pscircle[linewidth=0.04,dimen=outer,fillstyle=solid,fillcolor=black](2.6,0.4){0.1}
\pscircle[linewidth=0.04,dimen=outer,fillstyle=solid,fillcolor=black](2.6,-0.4){0.1}
\psline[linewidth=0.04](3.4,0.8)(3.4,0.4)(0.6,-0.8)(0.6,-1.2)
\psline[linewidth=0.04](3.4,-0.8)(3.4,-0.4)(0.6,0.8)(0.6,1.6)
\end{pspicture} 
}
}
 \end{displaymath}
Here the nodes
\begin{math}
\scalebox{.5} 
{
\begin{pspicture}(0,-0.32)(0.4,0.3)
\psline[linewidth=0.04](0.2,0.1)(0.2,-0.3)(0.2,-0.3)
\pscircle[linewidth=0.04,dimen=outer,fillstyle=solid,fillcolor=black](0.2,0.1){0.1}
\end{pspicture} 
}
\end{math}
and
\begin{math}
\scalebox{.5} 
{
\begin{pspicture}(0,-0.32)(0.4,0.3)
\psline[linewidth=0.04](0.2,0.1)(0.2,-0.3)(0.2,-0.3)
\pscircle[linewidth=0.04,dimen=outer,fillstyle=solid,fillcolor=black](0.2,-0.3){0.1}
\end{pspicture} 
}
\end{math}
denote the unique arrows $\nat\relto 0$ and $0\relto \nat$ in
$\Kleisli{\Q}$, respectively. Furthermore, let us introduce the
following conversion combinators.
\begin{displaymath}
  \cmbt{C_{P\mapsto\dot{P}}}
  := \lambda w. w \cmbt{\dot{P}}
  \qquad
  \cmbt{C_{\dot{P}\mapsto P}}
  :=
  \raisebox{-.5\height}{
    \scalebox{.4} 
    {
      \begin{pspicture}(0,-5.02)(4.4498243,5.02)
        \definecolor{color244b}{rgb}{0.8,0.8,0.8}
        \pstriangle[linewidth=0.04,dimen=outer,fillstyle=solid,fillcolor=color244b](2.0,4.0)(3.6,0.6)
        \pstriangle[linewidth=0.04,dimen=outer,fillstyle=solid,fillcolor=color244b](3.2,3.2)(1.6,0.4)
        \rput{-180.0}(4.0,2.6){\pstriangle[linewidth=0.04,dimen=outer,fillstyle=solid,fillcolor=color244b](2.0,1.0)(3.6,0.6)}
        \rput{-180.0}(1.6,4.4){\pstriangle[linewidth=0.04,dimen=outer,fillstyle=solid,fillcolor=color244b](0.8,2.0)(1.6,0.4)}
        \psline[linewidth=0.04cm](0.8,4.0)(0.8,3.6)
        \psline[linewidth=0.04cm](3.2,4.0)(3.2,3.6)
        \psline[linewidth=0.04cm](2.0,5.0)(2.0,4.6)
        \psline[linewidth=0.04cm](0.8,2.0)(0.8,1.6)
        \psline[linewidth=0.04cm](3.2,2.0)(3.2,1.6)
        \psline[linewidth=0.04cm](2.0,1.0)(2.0,0.6)
        \psframe[linewidth=0.04,dimen=outer](2.8,0.6)(1.2,-0.6)
        \usefont{T1}{ptm}{m}{n}
        \rput(2.0919,-0){\Huge $\cmbt{P}$}
        \psline[linewidth=0.04](0.8,3.6)(0.4,2.8)(0.4,2.4)(0.4,2.4)
        \psline[linewidth=0.04](1.2,2.4)(1.2,2.6)(3.6,3.0)(3.6,3.2)
        \psline[linewidth=0.04](2.8,3.2)(2.8,2.8)(3.2,2.0)
        \rput{-180.0}(4.0,-8.6){\pstriangle[linewidth=0.04,dimen=outer,fillstyle=solid,fillcolor=color244b](2.0,-4.6)(3.6,0.6)}
        \rput{-180.0}(6.4,-6.8){\pstriangle[linewidth=0.04,dimen=outer,fillstyle=solid,fillcolor=color244b](3.2,-3.6)(1.6,0.4)}
        \pstriangle[linewidth=0.04,dimen=outer,fillstyle=solid,fillcolor=color244b](2.0,-1.6)(3.6,0.6)
        \pstriangle[linewidth=0.04,dimen=outer,fillstyle=solid,fillcolor=color244b](0.8,-2.4)(1.6,0.4)
        \psline[linewidth=0.04cm](0.8,-4.0)(0.8,-3.6)
        \psline[linewidth=0.04cm](3.2,-4.0)(3.2,-3.6)
        \psline[linewidth=0.04cm](2.0,-5.0)(2.0,-4.6)
        \psline[linewidth=0.04cm](0.8,-2.0)(0.8,-1.6)
        \psline[linewidth=0.04cm](3.2,-2.0)(3.2,-1.6)
        \psline[linewidth=0.04cm](2.0,-1.0)(2.0,-0.6)
        \psline[linewidth=0.04](0.8,-3.6)(0.4,-2.8)(0.4,-2.4)(0.4,-2.4)
        \psline[linewidth=0.04](1.2,-2.4)(1.2,-2.6)(3.6,-3.0)(3.6,-3.2)
        \psline[linewidth=0.04](2.8,-3.2)(2.8,-2.8)(3.2,-2.0)
      \end{pspicture}
    }
  }
\end{displaymath}

We define $X\dtimes Y$ by replacing $\cmbt{P}$ with $\cmbt{\dot{P}}$
in $X\times Y$ (Lemma~\ref{lem:productsAndCoproducts}).
\end{mydefinition}

\begin{mylemma}\label{lem:timesVsDtimes}
  We have, for each $x,y\in A_{\Q}$,
  \begin{displaymath}
    \cmbt{\dot{P}_{l}}
    (
    \cmbt{\dot{P}} xy
    )
    = x\enspace,
    \qquad 
    \cmbt{\dot{P}_{r}}
    (
    \cmbt{\dot{P}} xy
    )
    = y\enspace;
    \qquad
    \cmbt{C_{P\mapsto\dot{P}}}(\cmbt{P}xy)=\cmbt{\dot{P}}xy\enspace,
    \qquad
    \cmbt{C_{\dot{P}\mapsto P}}(\cmbt{\dot{P}}xy)=\cmbt{P}xy\enspace.
  \end{displaymath}
  The latter two result in a canonical natural isomorphism
  $X\times Y\iso X\dtimes Y$ in $\PER_{\Q}$.  Therefore in what
  follows we shall use $\times$ and $\dtimes$ interchangeably. That
  is, we suppress use of the conversion combinators
  $\cmbt{C_{P\mapsto\dot{P}}}$ and $\cmbt{C_{\dot{P}\mapsto P}}$.
\end{mylemma}
\auxproof{see scannedNotes/productConv.eps}
 \begin{myproof}
  In the proof we shall rely heavily on the reasoning in string
  diagrams in the traced monoidal category $\Kleisli{\Q}$.
  
  For the first equality, the proof goes as follows.
  \begin{align*}
    \cmbt{\dot{P}_{l}}
    (
    \cmbt{\dot{P}} xy
    )
    & =
      \raisebox{-.5\height}{
      \scalebox{.3} 
      {
      \begin{pspicture}(0,-0.1)(8.698345,7.35)
        \definecolor{color1279b}{rgb}{0.8,0.8,0.8}
        \pstriangle[linewidth=0.04,dimen=outer,fillstyle=solid,fillcolor=color1279b](1.8,5.07)(3.6,0.6)
        \pstriangle[linewidth=0.04,dimen=outer,fillstyle=solid,fillcolor=color1279b](3.0,4.27)(1.6,0.4)
        \rput{-180.0}(3.6,3.94){\pstriangle[linewidth=0.04,dimen=outer,fillstyle=solid,fillcolor=color1279b](1.8,1.67)(3.6,0.6)}
        \rput{-180.0}(6.0,5.74){\pstriangle[linewidth=0.04,dimen=outer,fillstyle=solid,fillcolor=color1279b](3.0,2.67)(1.6,0.4)}
        \psline[linewidth=0.04cm](0.6,5.07)(0.6,4.67)
        \psline[linewidth=0.04cm](3.0,5.07)(3.0,4.67)
        \psline[linewidth=0.04cm](0.6,2.67)(0.6,2.27)
        \psline[linewidth=0.04cm](3.0,2.67)(3.0,2.27)
        \psline[linewidth=0.04cm](1.8,1.67)(1.8,1.27)
        \psline[linewidth=0.04cm](1.8,6.07)(1.8,5.67)
        \psline[linewidth=0.04](2.6,4.27)(2.6,4.07)(0.6,3.27)(0.6,2.67)(0.6,2.87)
        \psline[linewidth=0.04](0.6,4.67)(0.6,4.07)(2.6,3.27)(2.6,3.07)
        \psline[linewidth=0.04](3.4,4.27)(3.4,3.87)
        \psline[linewidth=0.04](3.4,3.47)(3.4,3.07)
        \pscircle[linewidth=0.04,dimen=outer,fillstyle=solid,fillcolor=black](3.4,3.87){0.1}
        \pscircle[linewidth=0.04,dimen=outer,fillstyle=solid,fillcolor=black](3.4,3.47){0.1}
        \pstriangle[linewidth=0.04,dimen=outer,fillstyle=solid,fillcolor=color1279b](6.6983447,4.71)(3.6,0.6)
        \rput{-180.0}(13.396689,4.42){\pstriangle[linewidth=0.04,dimen=outer,fillstyle=solid,fillcolor=color1279b](6.6983447,1.91)(3.6,0.6)}
        \psline[linewidth=0.04cm](5.498345,4.71)(5.498345,4.31)
        \psline[linewidth=0.04cm](7.8983445,4.71)(7.8983445,4.31)
        \psline[linewidth=0.04cm](6.6983447,5.71)(6.6983447,5.31)
        \psline[linewidth=0.04cm](5.498345,2.91)(5.498345,2.51)
        \psline[linewidth=0.04cm](7.8983445,2.91)(7.8983445,2.51)
        \psframe[linewidth=0.04,dimen=outer](6.2983446,4.31)(4.6983447,2.91)
        \usefont{T1}{ptm}{m}{n}
        \rput(5.4676805,3.57){\Huge $x$}
        \psframe[linewidth=0.04,dimen=outer](8.698345,4.31)(7.098345,2.91)
        \usefont{T1}{ptm}{m}{n}
        \rput(7.8676805,3.57){\Huge $y$}
        \psline[linewidth=0.04cm](6.6983447,1.91)(6.6983447,1.51)
        \pstriangle[linewidth=0.04,dimen=outer,fillstyle=solid,fillcolor=color1279b](1.8,0.53)(3.6,0.74)
        \psline[linewidth=0.04cm](1.8,1.67)(1.8,1.27)
        \rput{-180.0}(3.6,12.8){\pstriangle[linewidth=0.04,dimen=outer,fillstyle=solid,fillcolor=color1279b](1.8,6.07)(3.6,0.66)}
        \psline[linewidth=0.04cm](1.8,5.67)(1.8,6.07)
        \psline[linewidth=0.04](2.6983447,6.73)(2.6983447,7.13)(3.6983447,7.13)(4.8983445,1.53)(6.6983447,1.53)
        \psline[linewidth=0.04](6.6983447,5.73)(4.8983445,5.73)(3.6983447,0.13)(2.6983447,0.13)(2.6983447,0.53)(2.6983447,0.53)
        \psline[linewidth=0.04cm](0.89834476,7.33)(0.89834476,6.73)
        \psline[linewidth=0.04cm](0.89834476,0.53)(0.89834476,-0.07)
      \end{pspicture}
      }
      }
      =
      \raisebox{-.5\height}{
\scalebox{.3} 
{
\begin{pspicture}(0,-3.42)(4.62,3.42)
\definecolor{color1514b}{rgb}{0.8,0.8,0.8}
\pstriangle[linewidth=0.04,dimen=outer,fillstyle=solid,fillcolor=color1514b](2.2793317,2.2)(1.6,0.52)
\rput{-180.0}(4.5586634,1.92){\pstriangle[linewidth=0.04,dimen=outer,fillstyle=solid,fillcolor=color1514b](2.2793317,0.72)(1.6,0.48)}
\psline[linewidth=0.04cm](2.2793317,0.72)(2.2793317,0.32)
\psline[linewidth=0.04cm](2.68,2.2)(2.68,1.84)
\pscircle[linewidth=0.04,dimen=outer,fillstyle=solid,fillcolor=black](2.689231,1.8498992){0.10989922}
\pstriangle[linewidth=0.04,dimen=outer,fillstyle=solid,fillcolor=color1514b](2.2776763,-0.1)(3.6,0.6)
\rput{-180.0}(4.5553517,-5.2){\pstriangle[linewidth=0.04,dimen=outer,fillstyle=solid,fillcolor=color1514b](2.2776759,-2.9)(3.6,0.6)}
\psline[linewidth=0.04cm](1.0776765,-0.1)(1.0776765,-0.5)
\psline[linewidth=0.04cm](3.4776762,-0.1)(3.4776762,-0.5)
\psline[linewidth=0.04cm](1.0776765,-1.9)(1.0776765,-2.3)
\psline[linewidth=0.04cm](3.4776762,-1.9)(3.4776762,-2.3)
\psframe[linewidth=0.04,dimen=outer](1.8776761,-0.5)(0.27767628,-1.9)
\usefont{T1}{ptm}{m}{n}
\rput(1.00168,-1.24){\Huge $x$}
\psframe[linewidth=0.04,dimen=outer](4.2776766,-0.5)(2.6776767,-1.9)
\usefont{T1}{ptm}{m}{n}
\rput(3.40168,-1.24){\Huge $y$}
\psline[linewidth=0.04](2.2,-2.8)(2.2,-3.2)(4.6,-3.2)(4.6,3.2)(2.2,3.2)(2.2,2.6)
\psline[linewidth=0.04](1.8,2.2)(1.8,2.2)(1.8,2.2)(1.8,2.2)(1.8,2.0)(0.0,0.0)(0.0,-3.4)(0.0,-3.4)
\psline[linewidth=0.04](1.8,1.2)(1.8,1.4)(0.0,2.4)(0.0,3.4)
\psline[linewidth=0.04cm](2.68,1.16)(2.68,1.52)
\pscircle[linewidth=0.04,dimen=outer,fillstyle=solid,fillcolor=black](2.689231,1.5101007){0.10989922}
\end{pspicture} 
}
}
=
\raisebox{-.5\height}{
\scalebox{.3} 
{
\begin{pspicture}(0,-3.42)(5.04,3.42)
\definecolor{color1514b}{rgb}{0.8,0.8,0.8}
\pstriangle[linewidth=0.04,dimen=outer,fillstyle=solid,fillcolor=color1514b](2.6993318,2.2)(1.6,0.52)
\psline[linewidth=0.04cm](3.1,2.2)(3.1,1.84)
\pscircle[linewidth=0.04,dimen=outer,fillstyle=solid,fillcolor=black](3.1092308,1.8498992){0.10989922}
\rput{-180.0}(5.395352,-5.2){\pstriangle[linewidth=0.04,dimen=outer,fillstyle=solid,fillcolor=color1514b](2.6976757,-2.9)(3.6,0.6)}
\psline[linewidth=0.04cm](3.897676,-0.1)(3.897676,-0.5)
\psline[linewidth=0.04cm](1.4976766,-1.9)(1.4976766,-2.3)
\psline[linewidth=0.04cm](3.897676,-1.9)(3.897676,-2.3)
\psframe[linewidth=0.04,dimen=outer](2.297676,-0.5)(0.6976763,-1.9)
\usefont{T1}{ptm}{m}{n}
\rput(1.4216801,-1.24){\Huge $x$}
\psframe[linewidth=0.04,dimen=outer](4.6976767,-0.5)(3.0976765,-1.9)
\usefont{T1}{ptm}{m}{n}
\rput(3.82168,-1.24){\Huge $y$}
\psline[linewidth=0.04](2.62,-2.8)(2.62,-3.2)(5.02,-3.2)(5.02,3.2)(2.62,3.2)(2.62,2.6)
\psline[linewidth=0.04](2.22,2.2)(2.22,2.2)(2.22,2.2)(2.22,2.2)(2.22,2.0)(0.42,0.0)(0.42,-3.4)(0.42,-3.4)
\psline[linewidth=0.04](1.5,-0.52)(1.5,0.18)(0.42,2.4)(0.42,3.4)
\psline[linewidth=0.04cm](3.9,-0.14)(3.9,0.22)
\pscircle[linewidth=0.04,dimen=outer,fillstyle=solid,fillcolor=black](3.9092307,0.21010077){0.10989922}
\psframe[linewidth=0.04,linestyle=dashed,dash=0.16cm 0.16cm,dimen=outer](4.86,1.4)(0.0,-3.02)
\end{pspicture} 
}
}
\stackrel{(*)}{=}
\raisebox{-.5\height}{
\scalebox{.3} 
{
\begin{pspicture}(0,-3.94)(5.26,3.94)
\definecolor{color1514b}{rgb}{0.8,0.8,0.8}
\pstriangle[linewidth=0.04,dimen=outer,fillstyle=solid,fillcolor=color1514b](2.7193317,-3.18)(1.6,0.52)
\psline[linewidth=0.04cm](3.12,-3.18)(3.12,-3.54)
\pscircle[linewidth=0.04,dimen=outer,fillstyle=solid,fillcolor=black](3.1292307,-3.5301008){0.10989922}
\rput{-180.0}(5.395352,-4.16){\pstriangle[linewidth=0.04,dimen=outer,fillstyle=solid,fillcolor=color1514b](2.6976757,-2.38)(3.6,0.6)}
\psline[linewidth=0.04cm](3.897676,0.42)(3.897676,0.02)
\psline[linewidth=0.04cm](1.4976766,-1.38)(1.4976766,-1.78)
\psline[linewidth=0.04cm](3.897676,-1.38)(3.897676,-1.78)
\psframe[linewidth=0.04,dimen=outer](2.297676,0.02)(0.6976763,-1.38)
\usefont{T1}{ptm}{m}{n}
\rput(1.4216801,-0.72){\Huge $x$}
\psframe[linewidth=0.04,dimen=outer](4.6976767,0.02)(3.0976765,-1.38)
\usefont{T1}{ptm}{m}{n}
\rput(3.82168,-0.72){\Huge $y$}
\psline[linewidth=0.04](2.32,-3.12)(2.32,-3.8)(5.24,-3.8)(5.24,3.0)(2.22,3.0)(2.22,2.6)
\psline[linewidth=0.04](2.22,2.72)(2.22,2.72)(2.22,2.72)(2.22,2.72)(2.22,2.52)(0.42,0.52)(0.44,-2.64)(0.42,-3.92)
\psline[linewidth=0.04](1.5,0.0)(1.5,0.7)(0.42,2.92)(0.42,3.92)
\psline[linewidth=0.04cm](3.9,0.38)(3.9,0.74)
\pscircle[linewidth=0.04,dimen=outer,fillstyle=solid,fillcolor=black](3.9092307,0.73010075){0.10989922}
\psframe[linewidth=0.04,linestyle=dashed,dash=0.16cm 0.16cm,dimen=outer](4.86,1.92)(0.0,-2.5)
\psline[linewidth=0.04cm](2.7,-2.38)(2.7,-2.68)
\end{pspicture} 
}
}
\\
&=
\raisebox{-.5\height}{
\scalebox{.3} 
{
\begin{pspicture}(0,-3.94)(4.84,3.94)
\psline[linewidth=0.04cm](3.44,-1.36)(3.44,-1.72)
\pscircle[linewidth=0.04,dimen=outer,fillstyle=solid,fillcolor=black](3.449231,-1.7101008){0.10989922}
\psline[linewidth=0.04cm](3.4776762,0.42)(3.4776762,0.02)
\psframe[linewidth=0.04,dimen=outer](1.8776761,0.02)(0.27767628,-1.38)
\usefont{T1}{ptm}{m}{n}
\rput(1.00168,-0.72){\Huge $x$}
\psframe[linewidth=0.04,dimen=outer](4.2776766,0.02)(2.6776767,-1.38)
\usefont{T1}{ptm}{m}{n}
\rput(3.40168,-0.72){\Huge $y$}
\psline[linewidth=0.04](1.08,-1.38)(1.08,-2.2)(4.82,-2.2)(4.82,3.0)(1.8,3.0)(1.8,2.6941175)
\psline[linewidth=0.04](1.8,2.72)(1.8,2.72)(1.8,2.72)(1.8,2.72)(1.8,2.52)(0.0,0.52)(0.02,-2.64)(0.0,-3.92)
\psline[linewidth=0.04](1.08,0.0)(1.08,0.7)(0.0,2.92)(0.0,3.92)
\psline[linewidth=0.04cm](3.48,0.38)(3.48,0.74)
\pscircle[linewidth=0.04,dimen=outer,fillstyle=solid,fillcolor=black](3.4892309,0.73010075){0.10989922}
\end{pspicture} 
}
}
\stackrel{(\dagger)}{=}
\raisebox{-.5\height}{
\scalebox{.3} 
{
\begin{pspicture}(0,-2.22)(4.0000005,2.22)
\psline[linewidth=0.04cm](3.1623237,-0.68)(3.1623237,-1.04)
\pscircle[linewidth=0.04,dimen=outer,fillstyle=solid,fillcolor=black](3.1715546,-1.0301008){0.10989922}
\psline[linewidth=0.04cm](3.1999998,1.1)(3.1999998,0.7)
\psframe[linewidth=0.04,dimen=outer](1.5999999,0.7)(0.0,-0.7)
\usefont{T1}{ptm}{m}{n}
\rput(0.7240038,-0.04){\Huge $x$}
\psframe[linewidth=0.04,dimen=outer](4.0000005,0.7)(2.4000003,-0.7)
\usefont{T1}{ptm}{m}{n}
\rput(3.124004,-0.04){\Huge $y$}
\psline[linewidth=0.04cm](3.2023237,1.06)(3.2023237,1.42)
\pscircle[linewidth=0.04,dimen=outer,fillstyle=solid,fillcolor=black](3.2115545,1.4101008){0.10989922}
\psline[linewidth=0.04cm](0.76232374,0.68)(0.7823237,2.2)
\psline[linewidth=0.04cm](0.76232374,-2.2)(0.7823237,-0.68)
\end{pspicture} 
}
}
\stackrel{(\ddagger)}{=}
\raisebox{-.5\height}{
\scalebox{.3} 
{
\begin{pspicture}(0,-2.22)(2.0000005,2.22)
\psframe[linewidth=0.04,dimen=outer](1.5999999,0.7)(0.0,-0.7)
\usefont{T1}{ptm}{m}{n}
\rput(0.7240038,-0.04){\Huge $x$}
\usefont{T1}{ptm}{m}{n}
\psline[linewidth=0.04cm](0.76232374,0.68)(0.7823237,2.2)
\psline[linewidth=0.04cm](0.76232374,-2.2)(0.7823237,-0.68)
\end{pspicture} 
}
}
=x\enspace,
\end{align*}
where $(*)$ and $(\dagger)$ hold because of the \emph{dinaturality} (also called
 \emph{sliding}) and
 \emph{yanking} axioms of traced monoidal categories, respectively
 (see~\cite{JoyalSV96,AbramskyHS02}); and $(\ddagger)$ holds by a direct
 calculation in $\Kleisli{\Q}$. The second equality
\begin{math}
   \cmbt{\dot{P}_{r}}
  (
\cmbt{\dot{P}} xy
  )
  = y
\end{math}
 is similar.

The third equality is easy exploiting the combinatorial property
$(\cmbt{P}xy)z = zxy$
 of
 $\cmbt{P}$:
\begin{displaymath}
 \cmbt{C_{P\mapsto\dot{P}}}(\cmbt{P}xy)=
 (\cmbt{P}xy)
\cmbt{\dot{P}}
 = \cmbt{\dot{P}} xy\enspace.
\end{displaymath}
The last equality is shown as follows, where 
$(*)$ holds due to the dinaturality axiom.
\begin{align*}
  \cmbt{C_{\dot{P}\mapsto P}}(\cmbt{\dot{P}}xy)&=
\raisebox{-.5\height}{
\scalebox{.3} 
{
\begin{pspicture}(0,-4.44)(8.58,4.44)
\definecolor{color1865b}{rgb}{0.8,0.8,0.8}
\pstriangle[linewidth=0.04,dimen=outer,fillstyle=solid,fillcolor=color1865b](6.8,1.14)(3.6,0.6)
\pstriangle[linewidth=0.04,dimen=outer,fillstyle=solid,fillcolor=color1865b](3.2,3.2)(1.6,0.4)
\rput{-180.0}(4.0,2.6){\pstriangle[linewidth=0.04,dimen=outer,fillstyle=solid,fillcolor=color1865b](2.0,1.0)(3.6,0.6)}
\rput{-180.0}(1.6,4.4){\pstriangle[linewidth=0.04,dimen=outer,fillstyle=solid,fillcolor=color1865b](0.8,2.0)(1.6,0.4)}
\psline[linewidth=0.04cm](0.7983062,4.42)(0.8,3.6)
\psline[linewidth=0.04cm](3.2,4.0)(3.2,3.6)
\psline[linewidth=0.04cm](6.8,2.14)(6.8,1.74)
\psline[linewidth=0.04cm](0.8,2.0)(0.8,1.6)
\psline[linewidth=0.04cm](3.2,2.0)(3.2,1.6)
\psline[linewidth=0.04cm](2.0,1.0)(2.0,0.6)
\psframe[linewidth=0.04,dimen=outer](2.8,0.6)(1.2,-0.6)
\usefont{T1}{ptm}{m}{n}
\rput(2.046568,0.0){\Huge $\cmbt{P}$}
\psline[linewidth=0.04](0.8,3.6)(0.4,2.8)(0.4,2.4)
\psline[linewidth=0.04](1.2,2.4)(1.2,2.6)(3.6,3.0)(3.6,3.2)
\psline[linewidth=0.04](2.8,3.2)(2.8,2.8)(3.2,2.0)
\rput{-180.0}(13.6,-3.0){\pstriangle[linewidth=0.04,dimen=outer,fillstyle=solid,fillcolor=color1865b](6.8,-1.8)(3.6,0.6)}
\rput{-180.0}(6.4,-6.8){\pstriangle[linewidth=0.04,dimen=outer,fillstyle=solid,fillcolor=color1865b](3.2,-3.6)(1.6,0.4)}
\pstriangle[linewidth=0.04,dimen=outer,fillstyle=solid,fillcolor=color1865b](2.0,-1.6)(3.6,0.6)
\pstriangle[linewidth=0.04,dimen=outer,fillstyle=solid,fillcolor=color1865b](0.8,-2.4)(1.6,0.4)
\psline[linewidth=0.04cm](0.7983062,-4.42)(0.8,-3.6)
\psline[linewidth=0.04cm](3.2,-4.0)(3.2,-3.6)
\psline[linewidth=0.04cm](6.8,-2.2)(6.8,-1.8)
\psline[linewidth=0.04cm](0.8,-2.0)(0.8,-1.6)
\psline[linewidth=0.04cm](3.2,-2.0)(3.2,-1.6)
\psline[linewidth=0.04cm](2.0,-1.0)(2.0,-0.6)
\psline[linewidth=0.04](0.8,-3.6)(0.4,-2.8)(0.4,-2.4)
\psline[linewidth=0.04](1.2,-2.4)(1.2,-2.6)(3.6,-3.0)(3.6,-3.2)
\psline[linewidth=0.04](2.8,-3.2)(2.8,-2.8)(3.2,-2.0)
\psline[linewidth=0.04cm](5.7983065,1.16)(5.8,0.6)
\psframe[linewidth=0.04,dimen=outer](6.6,0.6)(5.0,-0.6)
\usefont{T1}{ptm}{m}{n}
\rput(5.7265677,-0.02){\Huge $x$}
\psline[linewidth=0.04cm](5.7983065,-1.22)(5.8,-0.6)
\psline[linewidth=0.04cm](7.7783065,1.16)(7.78,0.6)
\psframe[linewidth=0.04,dimen=outer](8.58,0.6)(6.98,-0.6)
\usefont{T1}{ptm}{m}{n}
\rput(7.746568,0.04){\Huge $y$}
\psline[linewidth=0.04cm](7.7783065,-1.22)(7.78,-0.6)
\psline[linewidth=0.04](3.1983063,4.02)(3.9783063,4.02)(4.9783063,-2.16)(6.7983065,-2.16)
\psline[linewidth=0.04](3.1983063,-3.98)(3.9783063,-3.98)(4.9783063,2.12)(6.7983065,2.12)
\end{pspicture} 
}
}
=
\raisebox{-.5\height}{
\scalebox{.3} 
{
\begin{pspicture}(0,-6.28)(5.218306,6.28)
\definecolor{color1865b}{rgb}{0.8,0.8,0.8}
\pstriangle[linewidth=0.04,dimen=outer,fillstyle=solid,fillcolor=color1865b](3.2,-2.7)(3.6,0.6)
\pstriangle[linewidth=0.04,dimen=outer,fillstyle=solid,fillcolor=color1865b](3.2,5.04)(1.6,0.4)
\rput{-180.0}(4.0,6.28){\pstriangle[linewidth=0.04,dimen=outer,fillstyle=solid,fillcolor=color1865b](2.0,2.84)(3.6,0.6)}
\rput{-180.0}(1.6,8.08){\pstriangle[linewidth=0.04,dimen=outer,fillstyle=solid,fillcolor=color1865b](0.8,3.84)(1.6,0.4)}
\psline[linewidth=0.04cm](0.7983062,6.26)(0.8,5.44)
\psline[linewidth=0.04cm](3.2,5.84)(3.2,5.44)
\psline[linewidth=0.04cm](3.2,-1.7)(3.2,-2.1)
\psline[linewidth=0.04cm](0.8,3.84)(0.8,3.44)
\psline[linewidth=0.04cm](3.2,3.84)(3.2,3.44)
\psline[linewidth=0.04cm](2.0,2.84)(2.0,2.44)
\psframe[linewidth=0.04,dimen=outer](2.8,2.44)(1.2,1.24)
\usefont{T1}{ptm}{m}{n}
\rput(2.046568,1.84){\Huge $\cmbt{P}$}
\psline[linewidth=0.04](0.8,5.44)(0.4,4.64)(0.4,4.24)
\psline[linewidth=0.04](1.2,4.24)(1.2,4.44)(3.6,4.84)(3.6,5.04)
\psline[linewidth=0.04](2.8,5.04)(2.8,4.64)(3.2,3.84)
\rput{-180.0}(6.4,-10.68){\pstriangle[linewidth=0.04,dimen=outer,fillstyle=solid,fillcolor=color1865b](3.2,-5.64)(3.6,0.6)}
\rput{-180.0}(6.4,-3.12){\pstriangle[linewidth=0.04,dimen=outer,fillstyle=solid,fillcolor=color1865b](3.2,-1.76)(1.6,0.4)}
\pstriangle[linewidth=0.04,dimen=outer,fillstyle=solid,fillcolor=color1865b](2.0,0.24)(3.6,0.6)
\pstriangle[linewidth=0.04,dimen=outer,fillstyle=solid,fillcolor=color1865b](0.8,-0.56)(1.6,0.4)
\psline[linewidth=0.04cm](0.75830626,-6.26)(0.8,-1.76)
\psline[linewidth=0.04cm](3.2,-5.98)(3.2,-5.58)
\psline[linewidth=0.04cm](0.8,-0.16)(0.8,0.24)
\psline[linewidth=0.04cm](3.2,-0.16)(3.2,0.24)
\psline[linewidth=0.04cm](2.0,0.84)(2.0,1.24)
\psline[linewidth=0.04](0.8,-1.76)(0.4,-0.96)(0.4,-0.56)
\psline[linewidth=0.04](1.2,-0.56)(1.2,-0.76)(3.6,-1.16)(3.6,-1.36)
\psline[linewidth=0.04](2.8,-1.36)(2.8,-0.96)(3.2,-0.16)
\psline[linewidth=0.04cm](2.1983063,-2.68)(2.2,-3.24)
\psframe[linewidth=0.04,dimen=outer](3.0,-3.24)(1.4,-4.44)
\usefont{T1}{ptm}{m}{n}
\rput(2.126568,-3.86){\Huge $x$}
\psline[linewidth=0.04cm](2.1983063,-5.06)(2.2,-4.44)
\psline[linewidth=0.04cm](4.178306,-2.68)(4.18,-3.24)
\psframe[linewidth=0.04,dimen=outer](4.98,-3.24)(3.38,-4.44)
\usefont{T1}{ptm}{m}{n}
\rput(4.146568,-3.8){\Huge $y$}
\psline[linewidth=0.04cm](4.178306,-5.06)(4.18,-4.44)
\psline[linewidth=0.04](3.1983063,-5.98)(5.198306,-5.98)(5.198306,5.88)(3.1983063,5.84)(3.1983063,5.84)
\end{pspicture} 
}
}
\stackrel{(*)}{=}
\raisebox{-.5\height}{
\scalebox{.3} 
{
\begin{pspicture}(0,-3.83)(5.218306,3.83)
\definecolor{color1865b}{rgb}{0.8,0.8,0.8}
\rput{-180.0}(4.0,3.7){\pstriangle[linewidth=0.04,dimen=outer,fillstyle=solid,fillcolor=color1865b](2.0,1.55)(3.6,0.6)}
\rput{-180.0}(1.6,5.5){\pstriangle[linewidth=0.04,dimen=outer,fillstyle=solid,fillcolor=color1865b](0.8,2.55)(1.6,0.4)}
\psline[linewidth=0.04cm](0.8,2.55)(0.8,2.15)
\psline[linewidth=0.04cm](3.2,2.55)(3.2,2.15)
\psline[linewidth=0.04cm](2.0,1.55)(2.0,1.15)
\psframe[linewidth=0.04,dimen=outer](2.8,1.15)(1.2,-0.05)
\usefont{T1}{ptm}{m}{n}
\rput(2.046568,0.55){\Huge $\cmbt{P}$}
\pstriangle[linewidth=0.04,dimen=outer,fillstyle=solid,fillcolor=color1865b](2.0,-1.05)(3.6,0.6)
\pstriangle[linewidth=0.04,dimen=outer,fillstyle=solid,fillcolor=color1865b](0.8,-1.85)(1.6,0.4)
\psline[linewidth=0.04cm](0.8,-1.45)(0.8,-1.05)
\psline[linewidth=0.04cm](2.0,-0.45)(2.0,-0.05)
\psframe[linewidth=0.04,dimen=outer](4.46,-1.63)(2.86,-2.83)
\usefont{T1}{ptm}{m}{n}
\rput(3.5865679,-2.25){\Huge $x$}
\psframe[linewidth=0.04,dimen=outer](2.68,-2.21)(1.08,-3.41)
\usefont{T1}{ptm}{m}{n}
\rput(1.846568,-2.77){\Huge $y$}
\psline[linewidth=0.04](3.6383061,-1.63)(2.9583063,-1.05)(2.9583063,-1.05)
\psline[linewidth=0.04](3.6383061,-2.79)(3.6383061,-3.09)(4.8583064,-3.09)(4.8383064,2.57)(3.1783063,2.55)(3.1783063,2.55)
\psline[linewidth=0.04](1.1983062,-1.85)(1.8383063,-2.21)(1.8383063,-2.21)
\psline[linewidth=0.04](1.8583063,-3.39)(1.8583063,-3.65)(5.198306,-3.63)(5.198306,3.43)(1.1583062,3.41)(1.1783062,2.93)
\psline[linewidth=0.04cm](0.39830625,2.95)(0.39830625,3.81)
\psline[linewidth=0.04cm](0.37830624,-1.87)(0.37830624,-3.81)
\end{pspicture} 
}
}
=
\raisebox{-.5\height}{
\scalebox{.3} 
{
\begin{pspicture}(0,-3.25)(8.2,3.25)
\definecolor{color1865b}{rgb}{0.8,0.8,0.8}
\rput{-180.0}(4.0,2.54){\pstriangle[linewidth=0.04,dimen=outer,fillstyle=solid,fillcolor=color1865b](2.0,0.97)(3.6,0.6)}
\rput{-180.0}(1.6,4.34){\pstriangle[linewidth=0.04,dimen=outer,fillstyle=solid,fillcolor=color1865b](0.8,1.97)(1.6,0.4)}
\psline[linewidth=0.04cm](0.8,1.97)(0.8,1.57)
\psline[linewidth=0.04cm](3.2,1.97)(3.2,1.57)
\psline[linewidth=0.04cm](2.0,0.97)(2.0,0.57)
\psframe[linewidth=0.04,dimen=outer](2.8,0.57)(1.2,-0.63)
\usefont{T1}{ptm}{m}{n}
\rput(2.046568,-0.03){\Huge $\cmbt{P}$}
\pstriangle[linewidth=0.04,dimen=outer,fillstyle=solid,fillcolor=color1865b](2.0,-1.63)(3.6,0.6)
\pstriangle[linewidth=0.04,dimen=outer,fillstyle=solid,fillcolor=color1865b](0.8,-2.43)(1.6,0.4)
\psline[linewidth=0.04cm](0.8,-2.03)(0.8,-1.63)
\psline[linewidth=0.04cm](2.0,-1.03)(2.0,-0.63)
\psframe[linewidth=0.04,dimen=outer](6.08,0.57)(4.48,-0.63)
\usefont{T1}{ptm}{m}{n}
\rput(5.206568,-0.05){\Huge $x$}
\psframe[linewidth=0.04,dimen=outer](8.2,0.57)(6.6,-0.63)
\usefont{T1}{ptm}{m}{n}
\rput(7.366568,0.01){\Huge $y$}
\psline[linewidth=0.04cm](0.39830625,2.37)(0.39830625,3.23)
\psline[linewidth=0.04cm](0.37830624,-2.43)(0.37830624,-3.23)
\psline[linewidth=0.04](3.1983063,1.95)(3.9383063,1.95)(4.3983064,-1.01)(5.238306,-1.01)
\psline[linewidth=0.04](5.238306,-0.61)(5.238306,-0.99)(5.238306,-0.99)(5.238306,-1.01)
\psline[linewidth=0.04](5.238306,0.57)(5.238306,1.03)(4.3583064,1.01)(3.9183064,-1.99)(3.1583064,-1.99)(3.1583064,-1.61)
\psline[linewidth=0.04](1.1783062,2.37)(1.1783062,2.65)(6.0183063,2.67)(6.578306,-0.99)(7.3983064,-0.99)(7.3983064,-0.61)
\psline[linewidth=0.04](7.3983064,0.55)(7.3983064,1.03)(6.558306,1.03)(5.9783063,-2.81)(1.1583062,-2.81)(1.1583062,-2.41)(1.1583062,-2.41)
\end{pspicture} 
}
}
\\
&=
\cmbt{P}xy\enspace.
\tag*{\myqed}
\end{align*}
 \end{myproof}

\begin{mylemma}\label{lem:closurePropertyOfAdmissibility}
 For admissible  $U,V$ and any $X$, we have
 $U\dtimes V$ and $X\limp U$  admissible. 
\myqed
\end{mylemma}
\noindent
As we observed
earlier in Example~\ref{example:admissibilityAndIso}, admissibility is not
necessarily preserved by isomorphisms in $\PER_{\Q}$. Therefore
replacing $\dtimes$ with $\times$ (that are isomorphic, see
Lemma~\ref{lem:timesVsDtimes}) would make the last result
(Lemma~\ref{lem:closurePropertyOfAdmissibility}) fail.


Admissibility of a PER (Definition~\ref{definition:admissiblePER}) gives
rise to a fixed-point construction, in the following sense. 
\begin{mydefinition}[Fixed point operator]\label{definition:fixedPointOperator}
 Let $U,X\in \PER_{\Q}$; assume that $U$ is admissible.
 We introduce a \emph{fixed point operator} (denoted by $\fix$) that carries
 \begin{displaymath}
  f\;:\;\bang U\boxtimes\bang X\longrightarrow U
  \qquad\text{to}\qquad
  \fix(f)\;:\;\bang X\longrightarrow U
 \end{displaymath}
 in the following way.
 Let $c$ be a code of $f$. We define $c_{0},c_{1},\dotsc\in |\bang
 X\limp U|$ by
 \begin{align*}
   c_{0}&:=\bot\enspace;\qquad   
  \\
  c_{n+1}&:=\;\text{the canonical code of}
  \;
\bigl(\,
   \bang X\stackrel{\con}{\longto}
   \bang X\boxtimes\bang X
   \stackrel{\delta\boxtimes\id}{\longto}
   \bang   \bang X\boxtimes\bang X
   \stackrel{\bang[c_{n}]\boxtimes\id}{\longto}
   \bang   U\boxtimes\bang X
   \stackrel{[c]}{\to}
   U
\,\bigr)\enspace.
 \end{align*}
A concrete description of $c_{n+1}$ in terms of $c_{n}$ can be easily 
given using in
 particular~(\ref{eq:concreteDefOfLinearExpComonadStr}). 
 Since $U$ is admissible and $\bot\cdot x =\bot$, $c_{0}=\bot$ is a
 valid code.  It is not hard to show that
$c_{0}\Lle c_{1}\Lle\cdots$ by induction;
 since $\bang X\limp U$ is admissible its supremum $\sup_{i}c_{i}$
 belongs to the domain $|\bang X\limp U|$. 
 Finally, 
we define
\begin{displaymath}
  \fix(f)\;:=\;[\,\sup_{i}c_{i}\,]\enspace.
\end{displaymath}
It is easily seen too that the above definition of $\fix(f)$
does not depend on the choice of a code $c$ of $f$. Here admissibility
 of $U$ is crucial.
\end{mydefinition}

\subsection{Quantum Mechanical Constructs in $\PER_{\Q}$}
\label{subsection:QMconstructsInPERQ}
Here we introduce some constructs in $\PER_{\Q}$ that we use for
interpreting quantum primitives in $\Hoq$. 

We start with the following combinator $\cmbt{A}$ that allows to
``juxtapose''
piping; see~(\ref{eq:kleisliComp}).
\begin{mydefinition}[Combinator $\cmbt{A}$]\label{definition:theApplCombinator}
  We define $\cmbt{A}\in A_{\Q}$ by the string diagram in $\Kleisli{\Q}$
  shown below. 
 The triangles are
   $j:\nat +\nat\cong \nat:k$ in Theorem~\ref{theorem:quantumLCA}. 
It is easily seen to satisfy the equation
 \begin{equation}\label{eq:AComb}
 \cmbt{A}xy=x\Kco y\enspace,
\end{equation}
 where $\Kco$ denotes composition of arrows in
 $\Kleisli{Q}$ (see \S{}\ref{subsection:quantumBranchingMonad}).
 \hfill
\begin{displaymath}
 \cmbt{A}\;:=\;
 \raisebox{-.5\height}{
\scalebox{.4} 
{
\begin{pspicture}(0,-2.22)(2.5375378,2.22)
\definecolor{color1844b}{rgb}{0.8,0.8,0.8}
\pstriangle[linewidth=0.04,dimen=outer,fillstyle=solid,fillcolor=color1844b](0.8,0.6)(1.6,0.4)
\pstriangle[linewidth=0.04,dimen=outer,fillstyle=solid,fillcolor=color1844b](1.8,1.4)(1.6,0.4)
\psline[linewidth=0.04cm](1.8,2.2)(1.8,1.8)
\psline[linewidth=0.04cm](1.4,1.4)(0.8,1.0)
\rput{-180.0}(1.6,-1.6){\pstriangle[linewidth=0.04,dimen=outer,fillstyle=solid,fillcolor=color1844b](0.8,-1.0)(1.6,0.4)}
\rput{-180.0}(3.6,-3.2){\pstriangle[linewidth=0.04,dimen=outer,fillstyle=solid,fillcolor=color1844b](1.8,-1.8)(1.6,0.4)}
\psline[linewidth=0.04cm](1.8,-2.2)(1.8,-1.8)
\psline[linewidth=0.04cm](1.4,-1.4)(0.8,-1.0)
\psline[linewidth=0.04cm](0.4,0.6)(1.2,-0.6)
\psline[linewidth=0.04](2.2,1.4)(2.2,0.4)(0.4,-0.4)(0.4,-0.6)(0.4,-0.6)
\psline[linewidth=0.04](1.2,0.6)(2.2,-0.4)(2.2,-1.4)
\end{pspicture} 
}
}
\qquad
\cmbt{A}xy
\;=\;
\raisebox{-.5\height}{\scalebox{.4} 
{
\begin{pspicture}(0,-2.02)(1.6,2.02)
\psframe[linewidth=0.04,dimen=outer](1.6,1.6)(0.0,0.2)
\usefont{T1}{ptm}{m}{n}
\rput(0.8190918,0.775){\Huge $y$}
\psframe[linewidth=0.04,dimen=outer](1.6,-0.2)(0.0,-1.6)
\usefont{T1}{ptm}{m}{n}
\rput(0.8190918,-1.025){\Huge $x$}
\psline[linewidth=0.04cm](0.8,2.0)(0.8,1.6)
\psline[linewidth=0.04cm](0.8,0.2)(0.8,-0.2)
\psline[linewidth=0.04cm](0.8,0.2)(0.8,-0.2)
\psline[linewidth=0.04cm](0.8,-1.6)(0.8,-2.0)
\end{pspicture} 
}}
\end{displaymath}
\end{mydefinition}

\begin{mydefinition}[Combinators $\cmbt{Q}_{\rho}$, $\cmbt{Q}_{U}$,
 $\cmbt{Q}^{N+1}_{\ket{0_{i}}}$, $\cmbt{Q}^{N+1}_{\ket{1_{i}}}$]
\label{definition:quantumCombinators}
We define elements
$\cmbt{Q}_{\rho},\cmbt{Q}_{U},\cmbt{Q}^{N+1}_{\ket{0_{i}}}
,\cmbt{Q}^{N+1}_{\ket{1_{i}}}
\in
 A_{\Q}$ as follows. Here $N\in \nat$, $\rho\in \DM_{2^{N}}$,  $U$ is an $2^{N}\times 2^{N}$
 unitary matrix, and $i\in [1,N+1]$.
  \begin{align*}
   \cmbt{Q}_{\rho},\cmbt{Q}_{U},\cmbt{Q}^{N+1}_{\ket{0_{i}}},
\cmbt{Q}^{N+1}_{\ket{1_{i}}}
:\;
  \nat
  &\longrelto\nat\;
   \text{ in $\Kleisli{\Q}$;}
  \quad
  \text{given $\sigma\in \DM_{m}$,}
 \\
  \bigl(\cmbt{Q}_{\rho}(k)(l)\bigr)_{m,n}(\sigma)
  &:=
  \begin{cases}
   \rho\otimes \sigma 
  &
   \text{if $k=l\land n=2^{N}\cdot m$}
  \\
   0
 &
   \text{otherwise,}
  \end{cases}
 \\
 \bigl(\cmbt{Q}_{U}(k)(l)\bigr)_{m,n}(\sigma)
 &:=
\begin{cases}
  (U(\place)U^{\dagger}\otimes \id_{j})\sigma
 &
  \text{if $k=l$ and $\exists j.\, (n=m=2^{N}\cdot j)$}
 \\
  0 &\text{otherwise,}
 \end{cases}
 \\
 \bigl(\cmbt{Q}^{N+1}_{\ket{0_{i}}}(k)(l)\bigr)_{m,n}(\sigma)
 &:=
 \begin{cases}
  \bigl(\bra{0_{i}}\place\ket{0_{i}}\otimes \id_{j}\bigr)\sigma
  \quad
  \text{if $k=l$ and  $\exists j. \,(m=2^{N+1}\cdot j \land n = 2^{N}\cdot j)$}
 \\
  0 \qquad\text{otherwise.}
 \end{cases}
 \\
 \bigl(\cmbt{Q}^{N+1}_{\ket{1_{i}}}(k)(l)\bigr)_{m,n}(\sigma)
 &:=
 \begin{cases}
  \bigl(\bra{1_{i}}\place\ket{1_{i}}\otimes \id_{j}\bigr)\sigma
  \quad
  \text{if $k=l$ and  $\exists j. \,(m=2^{N+1}\cdot j \land n = 2^{N}\cdot j)$}
 \\
  0 \qquad\text{otherwise.}
 \end{cases}
\end{align*}
In the definition of $\cmbt{Q}_{\rho}$, $\otimes$ denotes tensor product
 of matrices. 
The combinator $\cmbt{Q}_{\rho}$ ``adjoins an auxiliary state $\rho$'':
 an incoming token carrying $\sigma$ comes out of the same pipe, with
 its state composed with $\rho$. In particular the state $1\in \DM_{1}$
 comes out as $\rho$.  Similarly, the combinator $\cmbt{Q}_{U}$ applies
 the unitary transformation $U$ to (the first $N$ qubits of) the
 incoming quantum state; and $\cmbt{Q}^{N+1}_{\ket{0_{i}}}$ and
 $\cmbt{Q}^{N+1}_{\ket{1_{i}}}$ apply suitable projections (cf.\
 Remark~\ref{remark:splitMeasIntoProjs}),
 focusing on the first
 $N+1$ qubits of the incoming quantum state $\sigma$ and projecting
 its $i$-th qubit.
 Indeed:
\begin{mylemma}\label{lem:KleisliCompAndQuantumCombinators}
We have the following equalities. 
  \begin{equation}
 \cmbt{Q}_{\rho}\Kco\cmbt{Q}_{\sigma}
  =\cmbt{Q}_{\rho\otimes\sigma}
 \quad
    \cmbt{Q}_{U}\Kco\cmbt{Q}_{\rho}=\cmbt{Q}_{U\rho U^{\dagger}}
 \quad
  \cmbt{Q}^{N+1}_{\ket{0_{i}}}\Kco\cmbt{Q}_{\sigma}=\cmbt{Q}_{\bra{0_{i}}\sigma\ket{0_{i}}}
 \quad
  \cmbt{Q}^{N+1}_{\ket{1_{i}}}\Kco\cmbt{Q}_{\sigma}=\cmbt{Q}_{\bra{1_{i}}\sigma\ket{1_{i}}}
 \tag*{\myqed}
 \end{equation}
\end{mylemma}

\begin{mydefinition}[$\sem{\Nqbit}$,$\sem{\bit}$]\label{definition:semQbitAndSemBit}
For each $N\in\nat$ we define a PER $\sem{\Nqbit}$ over $A_{\Q}$ by:
\begin{displaymath}
  \sem{\Nqbit}\;:=\;\bigl\{(\cmbt{Q}_{\rho},\cmbt{Q}_{\rho})\mid \rho\in
 \DM_{2^{N}}\bigr\}\enspace.
\end{displaymath}
 In particular,
  $\sem{0\text{-}\qbit}=\{\,(\cmbt{Q}_{p},\cmbt{Q}_{p})\mid p\in
  [0,1]\,\}$  (cf.\ Definition~\ref{definition:densityMatrix}). 
 This type can be thought of as the unit interval $[0,1]$.

 A PER $\sem{\bit}$ is defined to be $\punit + \punit$ (see
 Lemma~\ref{lem:productsAndCoproducts}).
\end{mydefinition}

%
%
\end{mydefinition}

 The following fact supports the idea that $\bang$ stands for duplicable, hence
 classical, data.
\begin{mylemma}\label{lem:bitAndBangBit}
 There is a canonical isomorphism $\sem{\bit}\iso\bang\sem{\bit}$
 in $\PER_{\Q}$.
\end{mylemma}
\begin{myproof}
 Use the isomorphisms~(\ref{eq:additionalIsomorphismsInPERQ}) in
 Lemma~\ref{lem:bangVsProductTensorAndCoproduct}. \myqed
\end{myproof}

The quantum combinators in Definition~\ref{definition:quantumCombinators} 
are
combined with the $\cmbt{A}$ combinator in
Definition~\ref{definition:theApplCombinator} and yield the following
combinators for quantum operations.

\begin{mydefinition}[Combinators $\cmbt{U}_{U}, \cmbt{Pr}^{N+1}_{\ket{0_{i}}},\cmbt{Pr}^{N+1}_{\ket{1_{i}}}$]\label{definition:quantumOprCombinator}
 We define
\begin{displaymath}
 \cmbt{U}_{U}:= \cmbt{A}\cmbt{Q}_{U}\enspace,
 \quad
\cmbt{Pr}^{N+1}_{\ket{0_{i}}}:=\cmbt{A}\cmbt{Q}^{N+1}_{\ket{0_{i}}}\enspace,
 \quad
\cmbt{Pr}^{N+1}_{\ket{1_{i}}}:=\cmbt{A}\cmbt{Q}^{N+1}_{\ket{1_{i}}}\enspace.
\end{displaymath}
\end{mydefinition}

\begin{mylemma}\label{lem:quantumOprCombinatorEq}
  For combinators in Definition~\ref{definition:quantumCombinators}
 and~\ref{definition:quantumOprCombinator}, and $\rho,\sigma,U$ of
 suitable dimensions,
 we have
 \begin{align*}
 \cmbt{A}\cmbt{Q}_{\rho}\cmbt{Q}_{\sigma}
 & =\cmbt{Q}_{\rho\otimes\sigma}\enspace,
 &
   \cmbt{U}_{U}\cmbt{Q}_{\rho}&=\cmbt{Q}_{U\rho U^{\dagger}}\enspace,
 \\
  \cmbt{Pr}^{N+1}_{\ket{0_{i}}}\cmbt{Q}_{\sigma}&=\cmbt{Q}_{\bra{0_{i}}\sigma\ket{0_{i}}}
 \enspace,
 &
  \cmbt{Pr}^{N+1}_{\ket{1_{i}}}\cmbt{Q}_{\sigma}&=\cmbt{Q}_{\bra{1_{i}}\sigma\ket{1_{i}}}
 \enspace.
 \end{align*} 
\end{mylemma}
\begin{myproof}
Obvious from Lemma~\ref{lem:KleisliCompAndQuantumCombinators}.
\myqed
\end{myproof}

 \begin{myremark}[No-cloning in the category $\PER_{\Q}$]
  \label{remark:nocloningInPERQ}
  We noted in Remark~\ref{remark:nocloningForNewTT} that: the
  ``preparation'' primitive $\new_{\rho}$ in $\Hoq$ can be typable with
  the type $\bang k\text{-}\qbit$; hence the constant $\new_{\rho}$  is
  duplicable; nevertheless ``no-cloning'' is enforced by the linear
  typing discipline in $\Hoq$, in the sense that a given quantum
  state---whose preparation apparatus we do not have access to---cannot
  be duplicated. Here we shall discuss how these design choices are
  reflected in our model $\PER_{\Q}$. 

  It is important to note that, in $\Hoq$ and in its model $\PER_{\Q}$
  alike, the quantum tensor $\otimes$ (for composed and entangled
  systems) and the linear-logic tensor $\boxtimes$ are
  distinguished. Let us speak in the piping analogy
  in~(\ref{eq:quantumPiping}): an arrow in $\PER_{\Q}$ is ``realized''
  by a code $c\in A_{\Q}$ (Definition~\ref{definition:PERQ}); and each
  element $c$ of $A_{\Q}$ is a Kleisli arrow $c\colon \nat\relto\nat$ in
  $\Kleisli{\Q}$ (Theorem~\ref{theorem:quantumLCA}), that is, piping
  like in~(\ref{eq:quantumPiping}) with countably infinite numbers of
  entrances and exits. Then the quantum tensor $\otimes$ resides solely
  in the quantum states carried by tokens; importantly it has nothing to
  do with \emph{how tokens move around} (except for the indirect
  relationship in which measurements on quantum states result in
  branching of tokens). In contrast, the type constructors derived from
  linear logic---namely $\boxtimes$, $\limp$ and $\bang$---are all
  concerned about how pipes are connected.

  Concretely, the denotation of $\new_{\ket{0}\bra{0}}\colon \bang\qbit$
  will look like the piping in~(\ref{eq:bangPicture}), with
  $a=\cmbt{Q}_{\ket{0}\bra{0}}$ where the latter is from
  Definition~\ref{definition:quantumCombinators}. Therefore $\bang$ in
  the type $\bang\qbit$ here merely makes infinitely many copies
  $a=\cmbt{Q}_{\ket{0}\bra{0}}$; it does nothing like duplicating
  quantum states (that are carried by tokens that go through the
  pipes). It is easy to see that $\bang \cmbt{Q}_{\ket{0}\bra{0}} =
  \cmbt{Q}_{\ket{0}\bra{0}}$ as arrows $\nat\relto\nat$ in
  $\Kleisli{\Q}$,   as a matter of fact, since
  $\cmbt{Q}_{\ket{0}\bra{0}}$ does not alter the path taken by a token but
  only adjoins the quantum state $\ket{0}\bra{0}$ to the one carried by
  the token.

\end{myremark}

\subsection{Continuation Monad $T$}
\label{subsection:continuationMonad} 
In order to capture probabilistic
branching in Hoq, we use a strong monad $T$ on $\PER_{\Q}$
following Moggi's idea~\cite{Moggi91a}: 
the interpretation of a type judgment
$\Delta\vdash M:A$ will be an arrow $\sem{\Delta}\to T\sem{A}$ in the
category $\PER_{\Q}$. The monad $T$ is in fact a continuation monad
$T=(\place\limp R)\limp R$ with a suitable result type $R$; hence our
semantics is in the \emph{continuation-passing style (CPS)}.  The
resulting CPS model is fairly complex as a matter of fact, but our
efforts for its simplification have so far been barred by technical
problems, leading us to believe that CPS is a right way to go.
Informally, the reason is as follows.
\auxproof{See trialJune2012.eps}

Think of the construct $\meas^{1}_{1}$ that measures one qubit; 
for the purpose of case-distinction based on the outcome, it is
desired  that  $\meas^{1}_{1}$ is of the type  $\qbit\limp\bit$.
Therefore it is natural to use \emph{monadic semantics}:
 we use a monad $T$---with a probabilistic flavor---so that
we have $\sem{\meas^{1}_{1}}:\sem{\qbit}\to T\sem{\bit}$.

 For our GoI semantics based on local interaction, however, a
simple ``probability distribution'' monad (something like $\dist$
in~\S{}\ref{subsection:prelimMonadForBranching}) would not do.  One
explanation is as follows. Think of the construct
$\meas^{2}_{1}: 2\text{-}\qbit\limp\bit\boxtimes \qbit$: it takes a
state $\rho$ of a 2-qubit system; measures the first qubit; and returns
its outcome ($\ttrue$ or $\ffalse$) together with the remaining qubit. 
The probability of observing $\ttrue$ is calculated by
\begin{math}
 \trace\Bigl(\,\bigl(\,\bra{0_{1}}\place\ket{0_{1}}\otimes \id_{2}\,\bigr)\rho\,\Bigr)
\end{math}, and use of a naive  ``probability distribution monad''
requires calculation of the explicit value of this probability. However,
the
calculation traces out the second qubit, destroys and leaves it
inept for further quantum operations. 

(To put it differently: since we let a quantum state $\rho$ implicitly
carry a probability in the form of its trace value $\trace(\rho)$, a
naive interpretation of $\meas^{2}_{1}$ would have the codomain
$\bit\otimes \qbit$---with entanglement---rather than the desired
codomain $\bit\boxtimes \qbit$ that goes along well with $\bang$ and recursion.)

Hence we need to postpone such calculation of probabilities
until the very end of computation. Use of \emph{continuations} is a
standard way to do so.
As a result type $R$, we take that of infinite complete binary trees with
each edge labeled with a real number $p\in [0,1]$---obtained as a final
coalgebra.

\begin{mydefinition}[The functor $\Fpbt$]\label{definition:functorForProbBinaryTree}
We define an endofunctor $\Fpbt:\PER_{\Q}\to\PER_{\Q}$ 
by
\begin{displaymath}
 \Fpbt \quad:=\quad\sem{\bit}\limp(\sem{\zeroqbit}\dtimes\place)
\end{displaymath} 
where the objects $\sem{\bit}$ and $\sem{\zeroqbit}$ are as in
 Definition~\ref{definition:semQbitAndSemBit}. 
\end{mydefinition}
In the functor $\Fpbt$ we use  $\dtimes$  instead of
$\times$ for Cartesian products. This ensures a good order-theoretic
property (namely admissibility) of the carrier $R$ of the final
coalgebra.

The functor $\Fpbt$ represents
the
branching type of \emph{probabilistic binary trees} like the one shown below.
In the functor, the $\sem{\bit}$
part designates which of the left and right successors; and
the $\sem{\zeroqbit}$ part designates the value $p_{i}\in[0,1]$
that
is assigned to the edge (here $i\in\{0,1\}$). 
\begin{equation}\label{eq:probabilisticBinaryTree}
\includegraphics{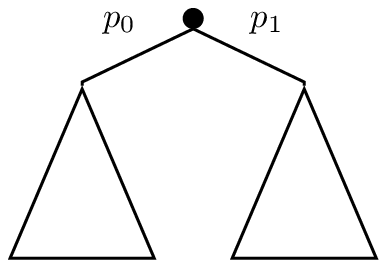}
\end{equation}
The values $p_{0}$ and $p_{1}$ can be thought of as probabilities,
although they might not add up to $1$.

As is usual in the theory of coalgebras (see e.g.~\cite{Rutten00a,Jacobs12CoalgBook}), the collection of such trees
is identified with (the carrier of) a final coalgebra.
\begin{mylemma}[The result type
 $R$]\label{lem:resultTypeAsFinalCoalg}
The functor $\Fpbt$
has a final coalgebra
\begin{displaymath}
r\;:\; R\stackrel{\cong}{\longrightarrow} \Fpbt R\enspace.
\end{displaymath}
\end{mylemma}
\begin{myproof} 
 We use the standard construction by a \emph{final sequence} (see
 e.g.~\cite{Pattinson03,Worrell05}). Let 
\begin{displaymath}
  \Bt\;:=\;\{(\bot,\bot)\}
\end{displaymath} 
be
a
 final object; here $\bot\in A_{\Q}$ is the least element with respect
 to the order $\sqsubseteq$ (Lemma~\ref{lem:LCAIsCPO}).
 Although $\Bt\iso \punit$, we use $\Bt$ due to its
 order-theoretic property (namely: $\Bt$ is admissible while $\punit$ is
 not, see Example~\ref{example:admissibilityAndIso}).
 Consider the final sequence
 \begin{equation}\label{eq:finalSeqInPERQ}
  \vcenter{\xymatrix@1@C+2em{
    {\Bt}
  &
    {\Fpbt \Bt}
      \ar[l]_-{\weak}
  &
    {\Fpbt ^{2}\Bt}
      \ar[l]_-{\Fpbt \weak}
  &
    {\cdots}
      \ar[l]_-{\Fpbt ^{2}\weak}
}}
 \end{equation}
 in $\PER_{\Q}$. Here $\weak$ denotes the unique arrow to final $\Bt$.
 For $j\le i$, let $c_{i,j}$ be (the canonical choice of) a realizer of the 
 arrow $\Fpbt^{i}\Bt\to \Fpbt ^{j}\Bt$ in the final sequence. 

 By Proposition~\ref{proposition:PERQIsCountablyComplAndCoCompl} there is a
 limit $R$ of the sequence~(\ref{eq:finalSeqInPERQ}). Moreover the
 PER $R$  can be concretely described as:
 the symmetric closure of
 \begin{equation}\label{eq:finalCoalgConcretelyAsPER}
\begin{aligned}
 &  \bigl\{\,\bigl(\cmbt{\dot{P}}(k_{i})_{i}u,\cmbt{\dot{P}}(k'_{i})_{i}u'\bigr)\,\bigl|\bigr.\,
   j \le i\;\text{implies}\;
 \\ 
&\qquad\qquad\bigl(c_{i,j}(k_{i}u), k'_{j}u'\bigr)\in \Fpbt ^{j}\Bt
\;\text{and}\;
 \bigl(k_{j}u,c_{i,j}(k'_{i}u') \bigr)\in \Fpbt ^{j}\Bt
 \,\bigr\}\enspace.
\end{aligned} 
\end{equation}
Now the functor $\Fpbt $ preserves  limits:
 $\sem{\zeroqbit}\dtimes\place$ does since $\dtimes$ is for products;
 and
 $\sem{\bit}\limp\place$ does since it has a left adjoint
 $\sem{\bit}\boxtimes\place$
 (Theorem~\ref{theorem:PERQIsLinear}). Therefore the well-known
 argument (see
 e.g.~\cite{Pattinson03,Worrell05}) proves that $R$ carries a final 
$\Fpbt$-coalgebra, with the coalgebraic structure $r:R\iso \Fpbt R$
 obtained as a suitable mediating arrow.
 \myqed
\end{myproof}

\begin{mylemma}\label{lem:continuationMonad}
 Let 
\begin{displaymath}
 T\;:=\; (\place\limp R)\limp R\enspace. 
\end{displaymath}
Then $T$ is a strong monad on
 $\PER_{\Q}$.
\end{mylemma}
\begin{myproof}
 This is a standard fact that is true---as one would readily prove---for any symmetric
 monoidal closed category
 $(\C,\punit,\ptensor,\limp)$ and for any $R\in\C$.
 \myqed
\end{myproof}

We introduce a map
\begin{displaymath}
   \mult\;:\;\sem{\zeroqbit}\boxtimes R\longrightarrow R
\end{displaymath}
that will be needed later. Intuitively, what it does is to
receive $p\in[0,1]$ and a binary tree $t$
like~(\ref{eq:probabilisticBinaryTree}) and returns the tree
in which the probabilities assigned to all the edges are multiplied
by $p$. For example,
\begin{align*}
\includegraphics{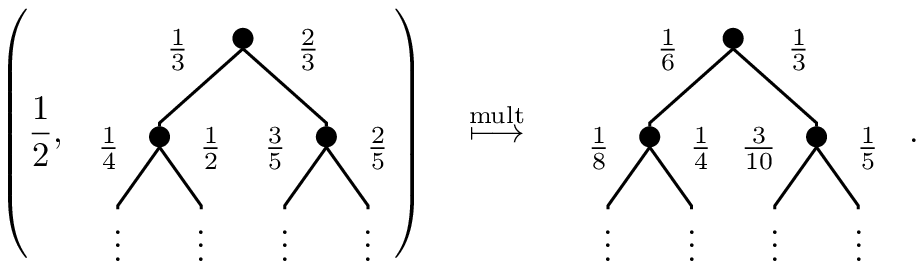}
\end{align*}
The precise definition of $\mult$ is by  coinduction  as a definition principle.
\begin{mydefinition}[$\mult$]\label{definition:mult}
 Let a coalgebra 
\begin{displaymath}
 c_{\mult}\;:\; \sem{\zeroqbit}\boxtimes R
  \longto \Fpbt \bigl(\,\sem{\zeroqbit}\boxtimes R\,\bigr) \quad\text{in $\PER_{\Q}$}
\end{displaymath}
 be defined as the adjoint transpose of the following composite.
 \begin{displaymath}
  \begin{array}{l}
   \sem{\zeroqbit}\boxtimes R\boxtimes\sem{\bit}
   \stackrel{\id\boxtimes r\boxtimes \id}{\longto}
   \sem{\zeroqbit}\boxtimes (\sem{\bit}\limp(\sem{\zeroqbit}\dtimes R))\boxtimes\sem{\bit}
    \\\qquad
    \stackrel{\id\boxtimes\ev}{\longto}
    \sem{\zeroqbit}\boxtimes (\sem{\zeroqbit}\dtimes R)
    \stackrel{\tuple{\id\boxtimes \pi_{\ell},\id\boxtimes\pi_{r}}}{\longto}
    (\sem{\zeroqbit}\boxtimes \sem{\zeroqbit})\dtimes
    (\sem{\zeroqbit}\boxtimes R)
    \\\qquad
    \stackrel{[\lambda w. w\cmbt{A}]\dtimes\id}{\longto}    
    \sem{\zeroqbit}\dtimes
    (\sem{\zeroqbit}\boxtimes R)  
  \end{array}
 \end{displaymath}
Here the map $[\lambda w. w\cmbt{A}]:
 \sem{\zeroqbit}\boxtimes  \sem{\zeroqbit}
\to  \sem{\zeroqbit}
$---that carries $\cmbt{P}xy$ to $x\Kco y$---plays the role of multiplication over $[0,1]$.
 By finality, this coalgebra $c_{\mult}$ induces a unique
 coalgebra homomorphism
 from $c_{\mult}$ to $r$. This is denoted by $\mult$.
\begin{displaymath}
 \vcenter{\xymatrix@R=1em@C+3em{
  {\Fpbt \bigl(\,\sem{\zeroqbit}\boxtimes R\,\bigr)}
    \ar@{-->}[r]^-{\Fpbt (\mult)}
 &
  {\Fpbt R}
 \\
  {\sem{\zeroqbit}\boxtimes R}
    \ar[u]^{c_{\mult}}
    \ar@{-->}[r]_-{\mult}
 &
  {R}
    \ar[u]^{\cong}_{r}
}}
\end{displaymath} 
\end{mydefinition}

%


For the purpose of interpreting recursion we need the following
property.
The notion of admissibility is from
Definition~\ref{definition:admissiblePER}. 
\begin{mylemma}\label{lem:resultTypeIsAdmissible}
 The PER $R$ in~(\ref{eq:finalCoalgConcretelyAsPER})
is admissible. Therefore by Lemma~\ref{lem:closurePropertyOfAdmissibility},
 $TX=(X\limp R)\limp R$ is admissible for each $X$; so is
$X\limp TY$.
\myqed
\end{mylemma}

The last result (Lemma~\ref{lem:resultTypeIsAdmissible}) implies that the following ``fixed-point'' construction
is available, by Definition~\ref{definition:fixedPointOperator} (where
$U$ was required to be admissible).
\begin{displaymath}
\vcenter{ \infer{\fix(f)\;:\;\bang X\longrightarrow R}{f\;:\;\bang
 R\boxtimes\bang X\longrightarrow R}}\enspace,\quad
 \text{for each $X\in\PER_{\Q}$.}
\end{displaymath}

\begin{myremark}
  The need for CPS-style semantics, as is the case here, does not seem to
 be a phenomenon unique to (``classical control and quantum data'')
 quantum computation. The same kind of difficulties are already observed
 with nondeterministic branching (see the leading example
 in~\cite{HoshinoMH14CSLLICS}), and they seem to occur with \emph{any}
 computational effect---at least the \emph{algebraic} ones in the sense
 of~\cite{PlotkinP01}. 

 The work~\cite{HoshinoMH14CSLLICS} presents an alternative approach to
 the one used here (namely the CPS-style semantics by a continuation
 monad): it identifies the underlying problem to be the ``memoryless''
 nature of processes (i.e.\ links in proof nets, boxes in string
 diagrams, etc.), and solves the problem by systematically equipping the
 processes with internal states (called \emph{memories}) exploiting
 \emph{coalgebraic component calculus}.

 The framework in~\cite{HoshinoMH14CSLLICS} is categorical and general,
 parametrized by a monad and algebraic operations interpreted over it.
 We expect its application to the current question (of higher-order quantum
 computation) will simplify the current CPS model by a great deal. This
 is however left as future work.
\end{myremark}

\subsection{Interpretation}\label{subsection:interpretationOfHoq}
Standing on all the constructs and properties exhibited
in~\S{}\ref{subsection:additionalStrOfPERQ}--\ref{subsection:continuationMonad},
we shall now interpret $\Hoq$ in the category $\PER_{\Q}$. 
\begin{mydefinition}[Interpretation of types]\label{definition:interpretationOfTypes}
 For each $\Hoq$-type $A$, we assign $\sem{A}\in\PER_{\Q}$ as follows,
 using
  the structures of $\PER_{\Q}$ we described in previous sections.
 For base types, $\sem{\Nqbit}$ is as in Definition~\ref{definition:semQbitAndSemBit}.
\begin{displaymath}
 \begin{array}{rclrcl}
 \sem{\bang A}&:= &\bang\sem{A} \quad
&
 \sem{A\limp B} & := & \sem{A}\limp T\sem{B}
\\
 \sem{\top} &:= &\punit
&
 \sem{A\boxtimes B} &:=& \sem{A}\boxtimes \sem{B}
\\
 \sem{A+B} &:=& \sem{A} + \sem{B}\qquad
 \end{array}
\end{displaymath}
\end{mydefinition}

\begin{mydefinition}[Interpretation of the subtype relation]\label{definition:interpretationOfSubtypeRel}
 We shall assign, to each derivable subtype relation $A\subtp B$, an arrow 
 \begin{displaymath}
  \sem{A\subtp B}\;:\;
  \sem{A}\longrightarrow \sem{B}
  \qquad\text{in $\PER_{\Q}$.}
 \end{displaymath}
 For that purpose we first introduce a natural transformation
 \begin{displaymath}
  \delta_{n,m}\;:\; \bang^{n}X\longrightarrow\bang^{m}X\enspace,
  \quad\text{natural in $X$,}
 \end{displaymath}
  for each $n,m\in\nat$ that satisfy $n=0\Rightarrow m=0$. This is as follows.
 \begin{displaymath}
  \delta_{n,m}
  :=
  \begin{cases}
   \id
   & \text{if $n=m$}
  \\
   \delta\co\cdots \co \delta
   &\text{if $n<m$ (note that in this case $n>0$)}
   \\
   \der \co\cdots\co\der
   &\text{if $n>m$}
  \end{cases}
 \end{displaymath}
 Using $\delta_{n,m}$, an arrow $\sem{A\subtp B}$ is defined by
 induction on the derivation (that is according to the rules in~(\ref{eq:HoqSubtypeRelation})).
 \begin{displaymath}
  \begin{array}{rlll}
   \sem{\bang^{n}\kqbit\subtp \bang^{m}\kqbit}:=
  &
   \bigl(\;
   \bang^{n}\sem{\kqbit}
   \stackrel{\delta_{n,m}}{\longrightarrow}
   \bang^{m}\sem{\kqbit}
   \;\bigr)\enspace,
   \\
   \sem{\bang^{n}\top\subtp \bang^{m}\top}:=
  &
   \bigl(\;
   \bang^{n}\sem{\top}
   \stackrel{\delta_{n,m}}{\longrightarrow}
   \bang^{m}\sem{\top}
   \;\bigr)\enspace,
   \\
   \sem{\bang^{n}(A_{1}\boxtimes A_{2})\subtp
       \bang^{m}(B_{1}\boxtimes B_{2})}
    :=
  &
   \\
   \multicolumn{2}{l}{
    \qquad
   \Bigl(\;
   \bang^{n}\bigl(\sem{A_{1}}\boxtimes \sem{A_{2}}\bigr)
   \stackrel{
   \bang^{n}\bigl(\,\sem{A_{1}\subtp B_{1}}\boxtimes\sem{A_{2}\subtp
   B_{2}}\,\bigr)
   }{\longrightarrow}
   \bang^{n}\bigl(\sem{B_{1}}\boxtimes \sem{B_{2}}\bigr)
   \stackrel{\delta_{n,m}}{\longrightarrow}
   \bang^{m}\bigl(\sem{B_{1}}\boxtimes \sem{B_{2}}\bigr)
   \;\Bigr)\enspace.
   }
  \end{array}
 \end{displaymath}
\begin{math}
     \sem{\bang^{n}(A_{1}+ A_{2})\subtp
       \bang^{m}(B_{1}+ B_{2})}
\end{math}    
and
\begin{math}
     \sem{\bang^{n}(A_{1}\limp A_{2})\subtp
       \bang^{m}(B_{1}\limp B_{2})}
\end{math}    
 are defined in a similar manner.  

 It is obvious from the rules in~(\ref{eq:HoqSubtypeRelation})
that a derivable judgment $A\subtp B$ has
 only one derivation. Therefore $\sem{A\subtp B}$ is well-defined.
\end{mydefinition}

\begin{mylemma}\label{lem:interpSubtpRelTransitive}
 Let $A\subtp B$ and $B\subtp C$. Then $A\subtp C$ by
 Lemma~\ref{lem:propertiesOfTheSubtypeRelation}.\ref{item:subtpIsPreorder};
 moreover
 \begin{displaymath}
  \sem{A\subtp C} = \sem{B\subtp C}\co \sem{A\subtp B}\enspace.
 \end{displaymath}
\end{mylemma}
\begin{myproof}
 Much like the proof of
 Lemma~\ref{lem:propertiesOfTheSubtypeRelation}.\ref{item:subtpIsPreorder}, 
 in whose course we exploit
 naturality of $\delta_{n,m}$, and that
 $\delta_{n,k}=\delta_{m,k}\co \delta_{n,m}$. The latter follows from Lemma~\ref{lem:bangVsProductTensorAndCoproduct}.
 \myqed
 \auxproof{See scannedNotes/welldfd.eps, p. 20.}
\end{myproof}

We now interpret constants.  In the general definition
(Definition~\ref{definition:interpretationOfTypeJudgments})
 a typed term $\Delta\vdash M:A$ will be interpreted as an arrow
$\sem{M}:\sem{\Delta}\to T\sem{A}$---the monad
$T$ is there because of our CPS semantics. For constants however we do
not need $T$: intuitively this is because a
 constant $c$ can always have the type $\bang \DType(c)$ (see (\text{Ax.}2)
 in Table~\ref{table:typingRules}). Therefore we first define
 $\semConst{c}:\punit\to \sem{\DType(c)}$ whose descriptions are simpler,
 and then $\sem{c}:\punit\to T\sem{\DType(c)}$ will be defined to be the
 embedding via the unit $\eta^{T}:\id\Rightarrow T$ of the monad $T$.

The technical core is in the interpretation of measurements. We explain its
idea after its formal definition. 
\begin{mydefinition}[Interpretation of constants]\label{definition:interpretationOfConstants}
To each constant $c$ in $\Hoq$
we assign 
 an arrow
\begin{displaymath}
 \semConst{c}\;:\; \punit\longrightarrow \sem{\DType(c)}
\end{displaymath}
 as follows. For $c\equiv\new_{\rho}$,
 $\semConst{\new_{\rho}}$ is given by
  \begin{displaymath}
    \punit
 \stackrel{[\lambda x. \cmbt{Q}_{\rho}]}{\longto}
 \sem{\nqbit}\enspace.
  \end{displaymath} 

 For $c\equiv\meas^{n+1}_{i}$ with $n\ge 1$, 
by transpose we need an arrow
 \begin{equation}\label{eq:measurementDefHelper}
   \sem{(n+1)\text{-}\qbit}\boxtimes
    \bigl((\sem{\bit}\boxtimes\sem{\nqbit})\limp R\bigr)\stackrel{m}{\longto }
    R
    \enspace,
 \end{equation}
where we also used $\bang\sem{\bit}\cong\sem{\bit}$ (Lemma~\ref{lem:bitAndBangBit}).
 By   $R$'s fixed point property (namely $r:R\iso
 \sem{\bit}\limp(\sem{\zeroqbit}\times R)$), 
 this is further reduced to an arrow
 \begin{displaymath} 
   \sem{(n+1)\text{-}\qbit}\boxtimes
    \bigl((\sem{\bit}\boxtimes\sem{\nqbit})\limp R\bigr)
    \boxtimes \sem{\bit}
    \;\stackrel{}{\longto }\;
    \sem{\zeroqbit}\times R
    \enspace.
 \end{displaymath}
  This can be obtained as follows. 
 \begin{displaymath}
 \begin{array}{l}
   \sem{(n+1)\text{-}\qbit}\boxtimes
    \bigl((\sem{\bit}\boxtimes\sem{\nqbit})\limp R\bigr)
      \boxtimes \sem{\bit}
    \\
 \cong\;
   \sem{(n+1)\text{-}\qbit}\boxtimes
    \bigl(\,(\sem{\nqbit}+\sem{\nqbit})\limp R\,\bigr)
    \boxtimes \sem{\bit}
 \qquad\text{by~(\ref{eq:canonicalIsomorphismInLinCat}) and
 $\punit\boxtimes X\cong X$}
 \\
 \cong\;
   \sem{(n+1)\text{-}\qbit}\boxtimes
    \bigl(\,(\sem{\nqbit}\limp R)\times (\sem{\nqbit}\limp R)\,\bigr)
     \boxtimes \sem{\bit}
 \qquad\text{by~(\ref{eq:canonicalIsomorphismInLinCat}) }
 \\
 \cong\;
  \sem{(n+1)\text{-}\qbit}\boxtimes (\sem{\nqbit}\limp R)^{\times
  2}
 +
 \sem{(n+1)\text{-}\qbit}\boxtimes (\sem{\nqbit}\limp R)^{\times
  2}
 \qquad\text{by~(\ref{eq:canonicalIsomorphismInLinCat})}
\\
 \stackrel{
    \bigl[
          \cmbt{Pr}^{n+1}_{\ket{0_{i}}}
    \bigr]
     \boxtimes \pi_{\ell}+
     \bigl[    
      \cmbt{Pr}^{n+1}_{\ket{1_{i}}}
     \bigr]\boxtimes\pi_{r}}{\longto}
 \sem{n\text{-}\qbit}\boxtimes (\sem{\nqbit}\limp R)
 +
 \sem{n\text{-}\qbit}\boxtimes (\sem{\nqbit}\limp R)
 \\
 \stackrel{\ev+\ev}{\longto}
  R+R
 \stackrel{[\tuple{[\lambda x.\cmbt{Q}_{0}],\id},
 \tuple{[\lambda x.\cmbt{Q}_{0}],\id}]}{\longto}
 \sem{\zeroqbit}\times R\enspace.
\end{array} 
\end{displaymath}
Here 
\begin{math}
           \cmbt{Pr}^{n+1}_{\ket{0_{i}}}
\end{math}
and
\begin{math}
           \cmbt{Pr}^{n+1}_{\ket{1_{i}}}
\end{math}
are from Definition~\ref{definition:quantumOprCombinator}, and
$\cmbt{Q}_{0}$ is from Definition~\ref{definition:quantumCombinators} (see
 also Definition~\ref{definition:semQbitAndSemBit}).

 For $c\equiv\meas^{1}_{1}$, similarly, by transpose we need an arrow
 \begin{equation}\label{eq:measurementDefHelperTwo}
  \sem{\qbit}\boxtimes \bigl(\sem{\bit}\limp R\bigr) 
 \;\stackrel{m'}{\longto} \;R\enspace,
 \end{equation}
which is  equivalent to  (by $R$ being a final coalgebra)
 \begin{displaymath}
  \sem{\qbit}\boxtimes \bigl(\sem{\bit}\limp R\bigr) \boxtimes\sem{\bit}
 \;\stackrel{}{\longto} \;
  \sem{\zeroqbit}\times R\enspace.
 \end{displaymath}
This is obtained as follows. 
\begin{displaymath}
 \begin{array}{l}
\sem{\qbit}\boxtimes \bigl(\sem{\bit}\limp R\bigr) \boxtimes\sem{\bit}
  \\
\cong\;
\sem{\qbit}\boxtimes \bigl(\sem{\bit}\limp R\bigr) +
\sem{\qbit}\boxtimes \bigl(\sem{\bit}\limp R\bigr)
 \qquad\text{by~(\ref{eq:canonicalIsomorphismInLinCat})}
  \\
\cong\;
\sem{\qbit}\boxtimes R^{\times 2} +
\sem{\qbit}\boxtimes R^{\times 2} 
 \qquad\text{by~(\ref{eq:canonicalIsomorphismInLinCat})}
 \\
 \stackrel{
\bigl[\cmbt{Pr}^{1}_{\ket{0}}
\bigr]
\boxtimes\pi_{\ell}+
\bigl[ \cmbt{Pr}^{1}_{\ket{1}}
\bigr]
\boxtimes\pi_{r}}{\longto}
\;
   \sem{\zeroqbit}\boxtimes R
 +    \sem{\zeroqbit}\boxtimes R
 \\
 \stackrel{[\kappa_{\ell},\kappa_{r}]}{\longrightarrow}\;
   \sem{\zeroqbit}\boxtimes R
 \stackrel{\tuple{[\lambda x.\cmbt{Q}_{0}],\mult}}{\longrightarrow}
  \sem{\zeroqbit}\times R\enspace;
 \end{array}
\end{displaymath}
here $\mult$
 is from Definition~\ref{definition:mult}.

For the other constants
 we use Theorem~\ref{theorem:PERQIsLinear},
 Lemma~\ref{lem:PERQMonoidalUnitIsTerminal},
 Definition~\ref{definition:quantumCombinators} and Definition~\ref{definition:quantumOprCombinator}.
The arrow $\semConst{U}$ is the transpose of
\begin{displaymath}
 \sem{\nqbit}
 \stackrel{[\cmbt{U}_{U}]}{\longto}
 \sem{\nqbit}
 \stackrel{\eta^{T}}{\longto}
 T\sem{\nqbit}
 \enspace;
\end{displaymath}
$\semConst{\cmp_{m,n}}$ is the transpose of
\begin{displaymath}
 \sem{\mqbit}\boxtimes\sem{\nqbit}
 \stackrel{[\lambda w. w\cmbt{A}]}{\longto}
 \sem{(m+n)\text{-}\qbit}
 \stackrel{\eta^{T}}{\longto}
 T \sem{(m+n)\text{-}\qbit}\enspace;
\end{displaymath}
\end{mydefinition}

The use of $\cmbt{Q}_{0}$ (that stands for the value $0\in [0,1]$) in
 the last line of the definition of $\semConst{\meas^{n+1}_{i}}$ indicates that a tree $m(t)\in |R|$ that can arise as an outcome of
 the map 
$m$ in~(\ref{eq:measurementDefHelper}) looks as follows.
\begin{equation}\label{eq:treeForMeasNPlusOne}
 \includegraphics{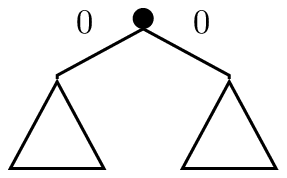}
\end{equation}
This is strange if we think of the values attached to edges as
 probabilities. In fact they are not probabilities: as we discussed in
 the beginning of~\S{}\ref{subsection:continuationMonad}, the actual
 probabilities are carried implicitly by the remaining quantum states
 (consisting of $n$ qubits) as their trace values.  The labels $0$
 in~(\ref{eq:treeForMeasNPlusOne}) mean calculation of
 probabilities is postponed; they are done later and probabilities occur
 on some lower level in the tree~(\ref{eq:treeForMeasNPlusOne}).

More specifically, 
the idea for $\semConst{\meas^{n+1}_{i}}$ is as follows. Let $n\ge 1$ and
consider the map $m$ in~(\ref{eq:measurementDefHelper}), which can 
be identified (via Lemma~\ref{lem:bangVsProductTensorAndCoproduct})
with an arrow
\begin{displaymath}
    \sem{(n+1)\text{-}\qbit}\boxtimes
    (\sem{\nqbit}\limp R)^{\times 2}
\stackrel{\overline{m}}{\longto }
    R\enspace.
\end{displaymath}
Roughly speaking its
input is a triple $(\rho, f_{\ttrue},f_{\ffalse})$ of $\rho\in
\DM_{2^{n+1}}$ and $f_{\ttrue},f_{\ffalse}: \DM_{2^{n}}\to R$. Then $\overline{m}$'s
output is the following tree. 
\begin{displaymath}
\includegraphics{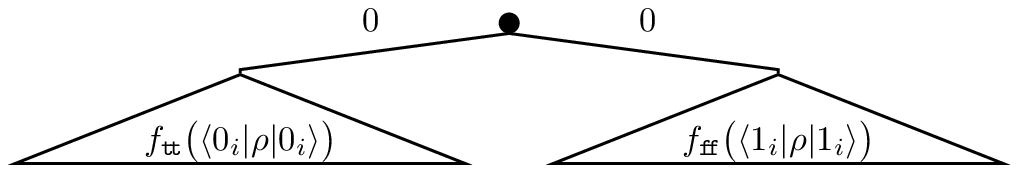}
\end{displaymath}
Here we put $0$ as the
labels on the edges of depth one; the probabilities for observing
$\ket{0_{i}}$ or $\ket{1_{i}}$ are implicitly passed down in the form of
the trace of the projected matrices.

When there is only one qubit left, we finally compute actual probabilities.
This is what $\semConst{\meas^{1}_{1}}$ does. Consider $m'$
in~(\ref{eq:measurementDefHelperTwo}),
which can be identified with
\begin{displaymath}
 \sem{\qbit}\boxtimes (R\times R)
 \;\stackrel{\overline{m'}}{\longrightarrow}\;
R\enspace.
\end{displaymath}
Its input is
roughly a triple $(\rho,t_{\ttrue},t_{\ffalse})$ of $\rho\in \DM_{2}$
and trees $t_{\ttrue},t_{\ffalse}\in R$. Let $p=\bra{0}\rho\ket{0}$ and
$q=\bra{1}\rho\ket{1}$; these are the probabilities for each outcome of
the measurement. Then the output of $\overline{m'}$  is the following
tree.
\begin{displaymath}
\includegraphics{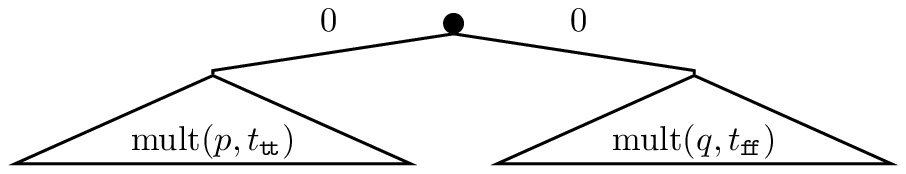}
\end{displaymath}
 Recall that $\mult(p,t)$ multiplies all
the labels of the input tree $t$ by $p$.

This way we only generate edges with its label $0$. This is no problem: once
we use trees with nonzero labels as $t_{\ttrue}$ and $t_{\ffalse}$ 
in the above,
we observe nonzero probabilities. 

The following definition of interpretation of type judgments looks
rather complicated. It is essentially the usual definition, as with
other typed (linear) calculi.   The subtype relation needs
careful handling, however---especially so that well-definedness
(Lemma~\ref{lem:interpretationIsWellDfd}) holds---and this adds all the
details.
\begin{mydefinition}[Interpretation of contexts]\label{definition:interpOfContexts}
We fix an enumeration of variables, i.e.\ 
a predetermined linear order $\prec$ between variables.
Given an (unordered) context $\Delta =
(x_{i}:A_{i})_{i\in[1,n]}$, we define $\sem{\Delta}\in
 \PER_{\Q}$ by
 $\sem{A_{\sigma(1)}}\boxtimes\cdots\boxtimes \sem{A_{\sigma(n)}}$,
where $\sigma$ is a bijection s.t.\  $x_{\sigma(1)}\prec\cdots\prec x_{\sigma(n)} $.
\end{mydefinition}

\begin{mydefinition}[Interpretation of type judgments]
\label{definition:interpretationOfTypeJudgments}
 For each derivation $\Pi\Vdash \Delta\vdash M:A$ of a type judgment in
 $\Hoq$, we assign an arrow
 \begin{displaymath}
  \sem{\Pi}\;:\;\sem{\Delta}\longrightarrow T\sem{A} 
 \end{displaymath}
 in the following way.
 Let $\Delta|_{\FV(M)}$ denote the obvious restriction of a context
 $\Delta$ to the set $\FV(M)$ of free variables in $M$.
 First we define
 \begin{displaymath}
  \fvsem{\Pi}\;:\;\sem{\Delta|_{\FV(M)}}\longrightarrow T\sem{A}
 \end{displaymath} 
 by induction on the derivation, which is used in
\begin{displaymath}
   \bigl(\;
  \sem{\Delta} 
   \stackrel{\sem{\Pi}}{\longrightarrow}
  T\sem{A}
  \;\bigr)
  \quad:=\quad
  \bigl(\;
   \sem{\Delta}
   \stackrel{\weak}{\longrightarrow}
   \sem{\Delta|_{\FV(M)}}
   \stackrel{\fvsem{\Pi}}{\longrightarrow} T\sem{A}
  \;\bigr)\enspace.
\end{displaymath}
 The definition of $\fvsem{\Pi}$ is as shown below.
 It is by
 induction---therefore the definition relies on the interpretations
 $\fvsem{\Pi'},\fvsem{\Pi''},\dotsc$ of the sub-derivation(s)
of $\Pi$
 that derive the second last typing judgment(s).
 \begin{displaymath}
  \infer{\Delta\vdash M\colon A}
  {
  \cdots
  &&
  \infer*[\Pi']{\Delta'\vdash M'\colon A'}{}
  &&
  \cdots
  &&
    \infer*[\Pi'']{\Delta''\vdash M''\colon A''}{}
  &&
  \cdots
  }
 \end{displaymath}
 In such cases, for simplicity of presentation, we shall refer to
 $\fvsem{\Pi'}$ as $\fvsem{M'}$, letting
 a term stand for the derivation tree that assigns a type to it.
 This will not cause confusion.
 \begin{align*}
 \boxed{\text{Ax.}1}    
 \quad&
 \sem{A}
 \stackrel{\sem{A\subtp A'}}{\longrightarrow}
 \sem{A'}
 \stackrel{\eta^{T}}{\longrightarrow}
 T\sem{A'}
 \\   
 \boxed{\text{Ax.}2}    
 \quad&
 \punit
 \stackrel{\varphi'}{\longrightarrow}
 \bang\punit
 \stackrel{\bang\semConst{c}\; \text{(cf.\ Definition~\ref{definition:interpretationOfConstants})}}{\longrightarrow}
 \bang\sem{\DType(c)}
 \stackrel{\eta^{T}}{\longrightarrow}
 T \bang\sem{\DType(c)}
 \stackrel{T\sem{\bang \DType(c)\subtp A}}{\longrightarrow}
 T\sem{A}
 \\
 \boxed{\mbox{$\limp$}.\text{I}_{1}}\quad&
  \sem{\Delta|_{\FV(\lambda x^{A}.M)}}
  \stackrel{g}{\longto}
  \sem{A}\limp T\sem{B}
  \stackrel{\sem{A'\subtp A}}{\longrightarrow}
 \sem{A'}\limp T\sem{B}
  =
  \sem{A'\limp B}
  \stackrel{\eta^{T}}{\longrightarrow}
  T \sem{A'\limp B}\enspace, \quad
 \\&
  \text{
  where $g=\fvsem{M}^{\wedge}$  if $x\in\FV(M)$; otherwise $g$ is the
  adjoint transpose of
  }
 \\
 &\qquad
   \sem{A}\boxtimes\sem{\Delta|_{\FV(M)}}
   \stackrel{\weak}{\longrightarrow} 
   \sem{\Delta|_{\FV(M)}}
   \stackrel{\fvsem{M}}{\longrightarrow}
   \sem{B}
 \\
 \boxed{\mbox{$\limp$}.\text{I}_{2}}\quad& 
 \sem{(\bang\Delta,\Gamma)|_{\FV(\lambda x^{A}.M)}}
 =
 \bang\sem{\Delta|_{\FV(\lambda x^{A}.M)}}
 \stackrel{\delta}{\longrightarrow}
 \bang^{n+1}\sem{\Delta|_{\FV(\lambda x^{A}.M)}}
 \stackrel{\bang^{n}g}{\longto}
 \bang^{n}(\sem{A}\limp T\sem{B})
 \\&
 \stackrel{\sem{A'\subtp A}}{\longrightarrow}
  \bang^{n}(\sem{A'}\limp T\sem{B})
 =
 \sem{\bang^{n}(A'\limp B)}
 \stackrel{\eta^{T}}{\longrightarrow}
 T \sem{\bang^{n} (A'\limp B)}\enspace, \quad
 \\&
 \text{where $g$ is defined as in the case $\mbox{$\limp$}.\text{I}_{1}$}
 \\
 \boxed{\mbox{$\limp$}.\text{E}}\quad& 
  \sem{(\bang\Delta,\Gamma_{1},\Gamma_{2})|_{\FV(MN)}}
 \stackrel{\con}{\longto}
  \sem{(\bang\Delta,\Gamma_{1})|_{\FV(M)}}
  \boxtimes
  \sem{(\bang\Delta,\Gamma_{2})|_{\FV(N)}}
 \\\quad&
 \stackrel{\fvsem{M}\boxtimes\fvsem{N}}{\longto}
 T(\sem{A}\limp T\sem{B})\boxtimes T\sem{C}
 \stackrel{\sem{C\subtp A}}{\longto}
 T(\sem{A}\limp T\sem{B})\boxtimes T\sem{A}
 \stackrel{\str'}{\longto}
 \\\quad&
 T\bigl((\sem{A}\limp T\sem{B})\boxtimes T\sem{A}\bigr)
 \stackrel{T\str}{\longto}
 TT\bigl((\sem{A}\limp T\sem{B})\boxtimes \sem{A}\bigr)
 \stackrel{\ev,\mu,\mu}{\longto}
 T\sem{B}
 \\
 \boxed{\mbox{$\boxtimes$}.\text{I}}\quad&
  \sem{(\bang\Delta,\Gamma_{1},\Gamma_{2})|_{\FV(\tuple{M_{1},M_{2}})}}
 \stackrel{\con}{\longto}
  \sem{(\bang\Delta,\Gamma_{1})|_{\FV(M_{1})}}
  \boxtimes
  \sem{(\bang\Delta,\Gamma_{2})|_{\FV(M_{2})}}
  \stackrel{\fvsem{M_{1}}\boxtimes\fvsem{M_{2}}}{\longto}
 \\\quad&
 T\bang^{n}\sem{A_{1}}
 \boxtimes
 T\bang^{n}\sem{A_{2}}
 \stackrel{\str',\text{ and then }\str,\mu}{\longto}
 T(\bang^{n}\sem{A_{1}}\boxtimes\bang^{n}\sem{A_{2}})
 \stackrel{\text{Lemma~\ref{lem:bangVsProductTensorAndCoproduct}}}{\longto}
 T\bang^{n}(\sem{A_{1}}\boxtimes\sem{A_{2}})
 \\ 
 \boxed{\mbox{$\boxtimes$}.\text{E}}\quad&
  \sem{(\bang\Delta,\Gamma_{1},\Gamma_{2})|_{\FV(
\letcl{\tuple{x_{1}^{\bang^{n}A_{1}},x_{2}^{\bang^{n}A_{2}}}=M}{N}
)}}
 \stackrel{\con}{\longto}
  \sem{(\bang\Delta,\Gamma_{1})|_{\FV(M)}}
  \boxtimes
  \sem{(\bang\Delta,\Gamma_{2})|_{\FV(N)}}
 \\\quad& 
 \stackrel{\fvsem{M}\boxtimes\id}{\longto}
   T\bang^{n}(\sem{A_{1}}\boxtimes\sem{A_{2}})
 \boxtimes
    \sem{(\bang\Delta,\Gamma_{2})|_{\FV(N)}}
 \\\quad& 
  \stackrel{\str',\text{ Lemma~\ref{lem:bangVsProductTensorAndCoproduct}},
  }{\longto}
 T\bigl(\bang^{n}\sem{A_{1}}\boxtimes\bang^{n}\sem{A_{2}}\boxtimes
 \sem{(\bang\Delta,\Gamma_{2})|_{\FV(N)}} \bigr)
 \stackrel{\fvsem{N}, (*)}{\longrightarrow}
 T^{2}\sem{A}
 \stackrel{\mu}{\longrightarrow}
 T\sem{A}
\enspace,
 \\\quad&
\text{where, in $(*)$, $\weak$ is suitably applied in case $x_{1}$ or $x_{2}$ is not in $\FV(N)$}
 \\
\boxed{\mbox{$\top$}.\text{I}}\quad&
  \punit
  \stackrel{\varphi'}{\longto}
  \bang\punit
  \stackrel{\delta,\der}{\longto}
  \bang^{n}\punit
 \stackrel{\eta^{T}}{\longrightarrow}
   T    \bang^{n}\punit
 \\
 \boxed{\mbox{$\top$}.\text{E}}\quad&\text{Similar}
 \\
 \boxed{+.\text{I}_{1}} 
 \quad&
  \sem{\Delta|_{\FV(\injl^{A_{2}}M)}}
 =
  \sem{\Delta|_{\FV(M)}}
 \stackrel{\fvsem{M}}{\longto} T(\bang^{n}\sem{A_{1}})
    \stackrel{T\bang^{n}\kappa_{\ell}}{\longto}
       T\bang^{n}(\sem{A_{1}}+\sem{A_{2}})
 \\\quad&
  \stackrel{\sem{A_{2}\subtp A'_{2}}}{\longto}
       T\bang^{n}(\sem{A_{1}}+\sem{A'_{2}})
 \\
 \boxed{+.\text{I}_{2}}\quad&\text{Similar}
 \\
 \boxed{+.\text{E}}\quad&
  \sem{(\bang\Delta,\Gamma,\Gamma')|_{\FV(    
    \matchcl{P}{(x_{1}^{\bang^{n}A_{1}}\mapsto
      M_{1}\mid x_{2}^{\bang^{n}A_{2}}\mapsto M_{2})})}}
 \\\quad&
  \stackrel{\con}{\longto}
  \sem{(\bang\Delta,\Gamma)|_{\FV(P)}}
  \boxtimes
  \sem{(\bang\Delta,\Gamma')|_{\FV(M_{1})\cup\FV(M_{2})}}
 \\\quad&
  \stackrel{\fvsem{P}}{\longrightarrow}
  T\bang^{n}(\sem{A_{1}}+\sem{A_{2}})
  \boxtimes
  \sem{(\bang\Delta,\Gamma')|_{\FV(M_{1})\cup\FV(M_{2})}}
 \\\quad&
  \stackrel{\str'}{\longrightarrow}
  T\Bigl(\bang^{n}(\sem{A_{1}}+\sem{A_{2}})
  \boxtimes
  \sem{(\bang\Delta,\Gamma')|_{\FV(M_{1})\cup\FV(M_{2})}}
  \Bigr)
 \\\quad&
  \stackrel{
  \text{Lemma~\ref{lem:bangVsProductTensorAndCoproduct}}}{\longto}
  T\Bigl((\bang^{n}\sem{A_{1}}+\bang^{n}\sem{A_{2}})
  \boxtimes
  \sem{(\bang\Delta,\Gamma')|_{\FV(M_{1})\cup\FV(M_{2})}}
  \Bigr)
 \\\quad&
  \stackrel{
  \text{Lemma~\ref{lem:bangVsProductTensorAndCoproduct}}}{\longto}
  T\Bigl(
  \bang^{n}\sem{A_{1}}
  \boxtimes
  \sem{(\bang\Delta,\Gamma')|_{\FV(M_{1})\cup\FV(M_{2})}}
+\bang^{n}\sem{A_{2}}
  \boxtimes
  \sem{(\bang\Delta,\Gamma')|_{\FV(M_{1})\cup\FV(M_{2})}}
  \Bigr)
 \\\quad&
  \stackrel{T\bigl[\fvsem{M_{1}},\fvsem{M_{2}}\bigr], (*)}{\longrightarrow}
  T^{2}\sem{B}
  \stackrel{\mu}{\rightarrow} TB\enspace,
 \quad
 \text{where, in $(*)$, $\weak$ is  applied if needed}
 \\
 \boxed{\text{rec}}\quad& 
  \sem{(\bang\Delta,\Gamma)|_{\FV(\letreccl{f^{A\limp B}x=M}{N})}}
  \stackrel{\con,\delta}{\longto}
  \bang\bang\sem{\Delta|_{\FV(M)}}\boxtimes\sem{(\bang\Delta,\Gamma)|_{\FV(N)}}
  \stackrel{\bang g}{\longto}
  \\ \quad&
  \bang\sem{A\limp B}\boxtimes\sem{(\bang\Delta,\Gamma)|_{\FV(N)}}
  \stackrel{\fvsem{N}, \weak}{\longrightarrow}
  T\sem{C}\enspace,
 \\\quad&
 \text{where $g:\bang\sem{\Delta|_{\FV(M)}}\to \sem{A\limp B}$ is
  obtained as follows.}
 \\\quad&
 \begin{array}{l}
  \bang\sem{\Delta|_{\FV(M)}}\boxtimes \bang(\sem{A}\limp T\sem{B})
  \boxtimes \sem{A} \stackrel{}{\longrightarrow}
  T\sem{B}
  \quad\text{is obtained as $\fvsem{M}$}
  \\
\multicolumn{1}{r}{  \text{ 
 (possibly with $\weak$ applied too);}
}  \\
  \bang\sem{\Delta|_{\FV(M)}}\boxtimes \bang(\sem{A}\limp T\sem{B})
 \stackrel{}{\longrightarrow}
  \sem{A}\limp T\sem{B}
  \quad\text{as its adjoint transpose; and then}
  \\
  \bang\sem{\Delta|_{\FV(M)}}
 \stackrel{}{\longrightarrow}
  \sem{A}\limp T\sem{B} = \sem{A\limp B}
    \quad\text{via the fixed point operator $\fix$ in Definition~\ref{definition:fixedPointOperator}.}
 \end{array}
\end{align*}
%
  Recall that $\weak$ denotes a unique map $X\to \punit$ to the tensor
 unit $\punit$ that is terminal (Lemma~\ref{lem:PERQMonoidalUnitIsTerminal}).  
The arrows
 $\der,\delta,\varphi,\varphi'$ and $\con$ are from
 Theorem~\ref{theorem:PERQIsLinear} (see also
 Lemma~\ref{lem:bangVsProductTensorAndCoproduct}).  In the above some
 obvious elements are omitted: we write $\weak$ in place of
 $\weak\boxtimes\id$, $\sem{M}$ in place of $\sem{\Delta\vdash M:A}$, etc.
 We denote $f$'s transpose by $f^{\wedge}$. The strength $X\boxtimes
 TY\to T(X\boxtimes Y)$ is denoted by $\str$; $\str'$ stands for
 $TX\boxtimes Y\to T(X\boxtimes Y)$.  
  For the rule (rec) we use the
  fixed point operator from Definition~\ref{definition:fixedPointOperator};
  note that the PER $\sem{A}\limp T\sem{B}$ is admissible
 (Lemma~\ref{lem:resultTypeIsAdmissible}).

\end{mydefinition}

The proof of the following important lemma is rather complicated due to 
implicit linearity tracking. It is deferred to \ref{section:wellDefinednessOfInterpretation}.
\begin{mylemma}[Interpretation of well-typed terms is well-defined]
\label{lem:interpretationIsWellDfd} If $\Pi,\Pi'$ are derivations of
the same type judgment $\Delta\vdash M:A$, their interpretations are the
same: $\sem{\Pi}=\sem{\Pi'}$.  Therefore the interpretation
$\sem{\Delta\vdash M:A}$ of a derivable type judgment is
well-defined. \myqed
\end{mylemma}

To compare with operational semantics (introduced
in~\S{}\ref{subsection:HoqOprSemantics}),  the interpretation
$\sem{\Delta\vdash M:A}:\sem{\Delta}\to T\sem{A}$
thus
obtained
is too fine. Hence we go further and extract $M$'s \emph{denotation} which is
given by a probability distribution.  We do so only for closed terms $M$
of type $\bit$. This is standard: for non-$\bit$ terms one will find
distinguishing contexts of type $\bit$.

In the following definition, the intuitions are:
\begin{align*}
 &  t_{0} \;=\; (\text{the infinite binary tree whose labels are all $0$})\enspace,
\\
 &  t_{\ttrue}\;= \;
\raisebox{-1em}{
\includegraphics{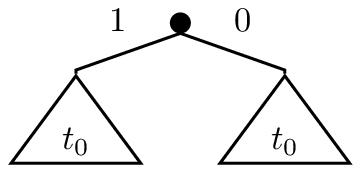}
}
\quad\text{and}\quad
 t_{\ffalse}
\;= \;
\raisebox{-1em}{
\includegraphics{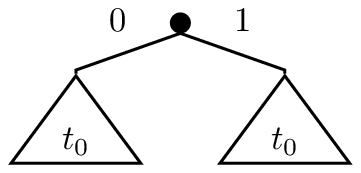}
}
\enspace.
\end{align*}
\begin{mydefinition}[Trees $t_{0}, t_{\ttrue},t_{\ffalse}$, and $\test$]
 Let
 $c_{\test}$ be the coalgebra
\begin{displaymath}
  \punit+\punit+\punit
  \stackrel{c_{\test}}{\longrightarrow}
  \Fpbt(\punit+\punit+\punit)
\end{displaymath}
 whose transpose
\begin{displaymath}
 \sem{\bit}\boxtimes (\punit+\punit+\punit)
 \longrightarrow
  \sem{\zeroqbit}\times (\punit+\punit+\punit)
\end{displaymath}
 is described as follows, using  informal notations. 
 \begin{align*}
  \tuple{\ttrue,\kappa_{1}(*)}
  &\longmapsto \tuple{1,\kappa_{2}(*)}\enspace,
  &
  \tuple{\ffalse,\kappa_{1}(*)}
  &\longmapsto \tuple{0,\kappa_{2}(*)}\enspace,
  \\
  \tuple{\ttrue,\kappa_{2}(*)}
  &\longmapsto \tuple{0,\kappa_{2}(*)}\enspace,
  &
  \tuple{\ffalse,\kappa_{2}(*)}
  &\longmapsto \tuple{0,\kappa_{2}(*)}\enspace,
  \\
  \tuple{\ttrue,\kappa_{3}(*)}
  &\longmapsto \tuple{0,\kappa_{2}(*)}\enspace,
  &
  \tuple{\ffalse,\kappa_{3}(*)}
  &\longmapsto \tuple{1,\kappa_{2}(*)}\enspace.
 \end{align*}
By coinduction we obtain the following arrow $\overline{c_{\test}}$.
\begin{displaymath}
 \vcenter{\xymatrix@R=1em@C+3em{
  {\Fpbt (\punit+\punit+\punit)}
    \ar@{-->}[r]^-{\Fpbt(\overline{c_{\test}})}
 &
  {\Fpbt R}
 \\
  {(\punit+\punit+\punit)}
    \ar[u]^{c_{\test}}
    \ar@{-->}[r]_{\overline{c_{\test}}}
 &
  {R}
    \ar[u]^{\cong}_{r}
}}
\end{displaymath} 
Now the trees 
$t_{0}, t_{\ttrue},t_{\ffalse}:\punit\to R$
are defined by
\begin{displaymath}
 t_{0}:=\overline{c_{\test}}\co \kappa_{2}\enspace,\quad
 t_{\ttrue}:=\overline{c_{\test}}\co \kappa_{1}\enspace,\quad
 t_{\ffalse}:=\overline{c_{\test}}\co \kappa_{3}\enspace.
\end{displaymath}


 The arrow
 \begin{displaymath}
  \test\;:\; \punit\longrightarrow (\sem{\bit}\limp R)
 \end{displaymath}
 in $\PER_{\Q}$ is defined to be the adjoint transpose of
 $[t_{\ttrue},t_{\ffalse}]:\punit+\punit\to R$.
\end{mydefinition}

\begin{mydefinition}[Operation $\prb$ on trees]
\label{definition:probOperationOnTrees}
 For each arrow $t:\punit\to R$ thought of as a tree, we define $\prb(t)\in
 (\mathbb{R}\cup\{\infty\})^{2}$ by:
  \begin{multline*}
   \prb(t):=
   \bigl(\,
  \textstyle\sum\left\{
       \text{labels on edges going down-left}
   \right\}
   \,,\,
   \\
  \textstyle\sum\left\{\text{labels on edges going down-right}\right\}  
  \,\bigr)
  \end{multline*}
For example,
 $\prb(t_{\ttrue})=(1,0)$ and 
 $\prb(t_{\ffalse})=(0,1)$.
 See also Example~\ref{eg:two_measurement} later.
\end{mydefinition}
\auxproof{\noindent 
Implicit in the last definition is that: if $(x,x')\in R$
 then $x$ and $x'$ represents the same binary tree. This is easy.}

The operation $\prb$ quotients the interpretation $\sem{\Delta\vdash
M:A}:\sem{\Delta}\to T\sem{A}$ and yields a denotation relation
$\denred$, which is to be compared with the big-step operational
semantics $\oprred$ (Definition~\ref{definition:bigStepSemantics}).  We note
again that we swapped the notations $\oprred$ and $\denred$ from the
previous version~\cite{HasuoH11}.
\begin{mydefinition}[Denotation relation $\denred$]\label{definition:denotationRelation}
 We define a relation $\denred$ between closed $\bit$-terms $M$---i.e.\
 those terms for which
 $\vdash M:\bit$ is derivable---and pairs $(p,q)$ of real numbers, as follows.
 Such a term $M$ gives rise to an arrow $\tree(M):\punit\to R$ in
 $\PER_{\Q}$  by:
 \begin{equation}\label{eq:treeOfM}
\tree(M)
\;:=\;
\left(\begin{array}{l}
   \punit
  \stackrel{\cong}{\longto}
  \punit\boxtimes\punit
  \stackrel{\test\boxtimes\sem{\vdash M:\bit}}{\longto}
  (\sem{\bit}\limp R)\boxtimes T\sem{\bit}
 \\\qquad
  \;=\;
  (\sem{\bit}\limp R)\boxtimes ((\sem{\bit}\limp R)\limp R)
  \stackrel{\ev}{\longto}
  R
\end{array} 
\right)
\enspace.\end{equation}
 We say $M\denred (p,q)$ if $\prb(\tree(M))=(p,q)$. Obviously such $(p,q)$
 is uniquely determined by $M$. 
\end{mydefinition}
The infinite tree $\tree(M)$ always satisfies
the following conditions:
\begin{itemize}
\item every branching is either
  \begin{center}
\includegraphics{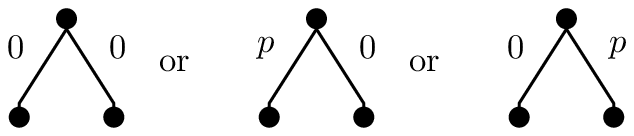}
  \end{center}
  for some $p \in (0,1]$, and
\item every non-zero branching is followed by the
  tree whose labels are all $0$:
  \begin{align*}
\includegraphics{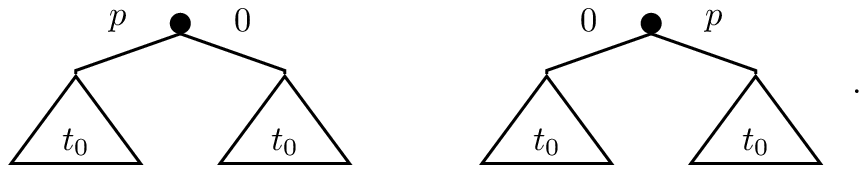}
\end{align*}
\end{itemize}
Intuitively, each zero-zero branch preceding a non-zero branch
corresponds to measurement in the evaluation of the term $M$. We will
see that the summation of labels on edges going down-left (going
down-right) is the probability to get $\ttrue$ (to get $\ffalse$) by
evaluating $M$. We note that when there are infinitely many measurements
in the evaluation sequence of $M$, the tree associated to $M$
may have infinitely many non-zero branching.

\begin{myexample}\label{eg:tree}
 Let $H$ be the Hadamard matrix.
 The term $M:=\meas^{1}_{1}(H(\new_{\ket{0}\bra{0}}))$ is a closed $\bit$-term; it
 measures the qubit $(\ket{0}+\ket{1})/\sqrt{2}$. The associated tree
 $\tree(M)$ is
 \begin{equation*}
\includegraphics{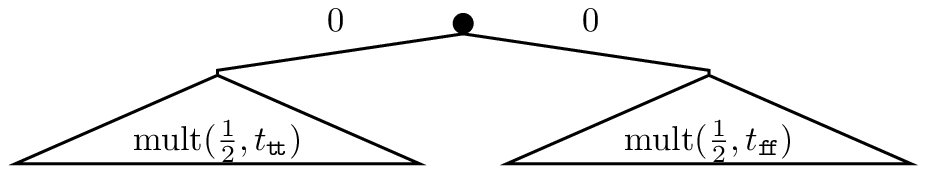}
\end{equation*}
Indeed we have $M\denred(1/2,1/2)$.
Similarly, $\tree(\meas^{1}_{1}(
 \new_{\ket{0}\bra{0}}
 ))$ is
 \begin{equation*}
\includegraphics{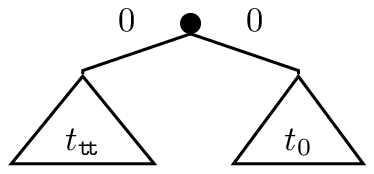}
\end{equation*}
 and we have $\meas^{1}_{1}(
 \new_{\ket{0}\bra{0}}
 ) \Downarrow (1,0)$.  The
zero-zero branches at the top node of $\tree(M)$ and
 $\tree(\meas^{1}_{1}(
 \new_{\ket{0}\bra{0}}
 ))$ correspond to measurement in
 the evaluation sequences of $M$ and $\meas^{1}_{1}(
\new_{\ket{0}\bra{0}}
 )$.
\end{myexample}
\begin{myexample}\label{eg:two_measurement}
  Let $M$ be 
  the closed term of type $\bit$ given as follows.
  \begin{enumerate}
  \item The term $M$ prepares two qubits $v = \frac{1}{\sqrt{2}}\ket{0}+\frac{1}{\sqrt{2}}\ket{1}$
    and $u = \frac{1}{\sqrt{3}}\ket{0}+\sqrt{\frac{2}{3}}\ket{1}$
  \item The term $M$ measures $v$.
  \item If the result of the measurement of $v$ is $\ffalse$,
    then $M$ outputs $\ttrue$.
  \item If the result of the measurement of $v$ is $\ttrue$, then $M$ measures
    the other qubit $u$ and outputs the result.
  \end{enumerate}
  The tree associated to $M$ is
 \begin{equation*}
\includegraphics{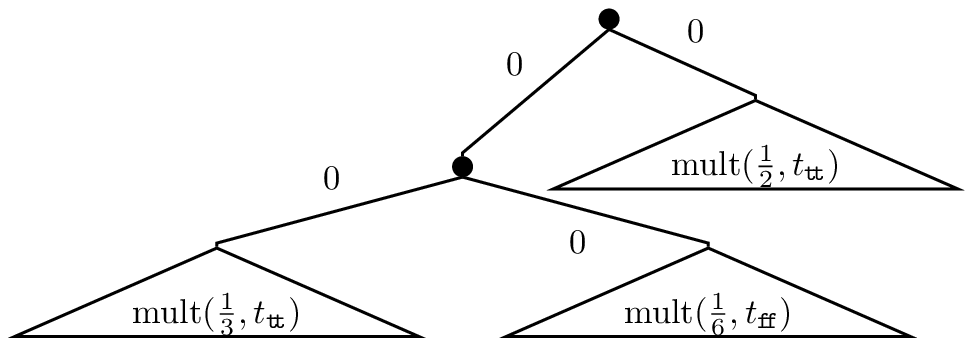}
 \end{equation*}
 and we have $M \Downarrow (\frac{5}{6},\frac{1}{6})$.
 The branch at the top node corresponds to measurement of $v$,
 and the branch of the node at the lower-left 
 corresponds to the measurement of $u$.
\end{myexample}

As we observed in Example~\ref{eg:tree} and
Example~\ref{eg:two_measurement}, $\tree(M)$ associated to closed
$\bit$-term $M$ has intentional information: we can see how measurement
is done in the evaluation of $M$. For instance, 
$\tree(\ttrue)=t_{\ttrue}$ is different from
$\tree(\meas^{1}_{1}(
\new_{\ket{0}\bra{0}}
))$.


\section{Adequacy}\label{section:adequacy}

As we use a continuation monad to capture probabilistic branching raised
by measurements, our interpretation of $\Hoq$-terms contains intentional
data. For example, the interpretation of a term $\vdash M:\mathtt{bit}$
is a tree in the result type $R$
(Lemma~\ref{lem:resultTypeAsFinalCoalg}) that reflects the evaluation
tree of $M$. In this section, we show that the operation $\prb$ in
Definition~\ref{definition:probOperationOnTrees}---that reduces a tree in $R$
to a pair $(p,q)$ of probabilities---correctly extracts the evaluation
result of $M$, that is, we have $M \Downarrow (p,q)$ if and only if $M
\oprred (p,q)$.

We list several basic properties of our denotational
semantics. Their proofs are found in
\ref{section:proofsForAdequacySection}. Many of them 
follow common patterns found in the study of call-by-value languages,
although  we
need to be careful about the fact that a term can have multiple types 
(due to  subtyping $\subtp$)

\begin{mylemma}\label{lem:07121946}
  Let $E$ be  an evaluation context, and $x$ be a variable that does
 not occur in $E$. Assume that $x:A \vdash E[x]:B$ is derivable.
  Then for any term $M$ such that $\Vdash\Gamma \vdash M:A$, the interpretation
  $\llbracket \Gamma \vdash E[M]:B\rrbracket:\sem{\Gamma}\to T\sem{B}$ is calculated by
 \begin{equation}
  \llbracket \Gamma \vdash E[M]:B\rrbracket
  \quad=\quad
  \mu^{T}_{\llbracket B \rrbracket} \co T\llbracket x:A \vdash
  E[x]:B\rrbracket \co \llbracket \Gamma \vdash M:A
  \rrbracket\enspace.
  \tag*{\myqed}
 \end{equation}
\end{mylemma}

\begin{mylemma}\label{lem:07122017}
  For a closed term $\vdash M:A$, if there is a reduction $M \to_{1}
  N$ that is not due to a measurement rule (($\meas_{1}$--$\meas_{4}$)
 in Definition~\ref{definition:operationalSemantics}), then 
\begin{displaymath}
 \llbracket \vdash M
  :A\rrbracket
 \quad=\quad
 \llbracket \vdash N
  :A\rrbracket\enspace.
\end{displaymath}
Note
  that $\vdash N:A$ is derivable by
  Lemma~\ref{lem:subjectReductionLemma}.
\myqed
\end{mylemma}
If we allow reduction rules $\meas_{1}$--$\meas_{4}$,
Lemma~\ref{lem:07122017} is no longer correct. This is because our
semantics contains some intentional information. In fact,
$\tree(\ttrue)$ is different from $\tree(\meas_{1}(
\new_{\ket{0}\bra{0}}
))$ 
since the latter tree contains information about measurement in the
evaluation of $\meas_{1}(
\new_{\ket{0}\bra{0}}
)$
we observed in Example~\ref{eg:tree}. Therefore,
$\llbracket \vdash \ttrue : \bit \rrbracket$ is different from
$\llbracket \vdash \meas_{1}(
\new_{\ket{0}\bra{0}}
) : \bit\rrbracket$.
However, at the base type $\bit$, we can kill such intentionality by
forgetting branching information in $\tree(M)$ by means of $\prb(\place)$.

\begin{mylemma}\label{lem:E[meas^n+1]}
  Let $E$ be an evaluation context.
  If $\vdash E[\meas^{n+1}_{i}\, \new_{\rho}] : \bit$
  is derivable and
  \begin{align*}
    E[\langle \ttrue, \new_{\langle 0_{i}\mid \rho \mid
      0_{i}\rangle}\rangle] \Downarrow (p_{0},q_{0}) && E[\langle
    \ffalse, \new_{\langle 1_{i}\mid \rho \mid
      1_{i}\rangle}\rangle] \Downarrow (p_{1},q_{1}),
  \end{align*}
  then $E[\mathtt{meas}_{i}^{n+1}\,\new_{\rho}] \Downarrow
  (p_{0}+p_{1},q_{0}+q_{1})$.
\end{mylemma}
\begin{myproof}
  By
  Lemma~\ref{lem:07121946} we have
  \begin{equation}\label{eq:07122032}
\begin{aligned}
&   \sem{\,\vdash E[\mathtt{meas}_{i}^{n+1}\,\new_{\rho}]:\mathtt{bit}\,}
  \quad=
\\
&\qquad
    \mu^{T}_{\llbracket \mathtt{bit}\rrbracket}
    \co T\llbracket x:\bang \mathtt{bit}\boxtimes n\text{-}\mathtt{qbit}
    \vdash E[x]:\mathtt{bit}\rrbracket
    \co
    \llbracket \vdash \mathtt{meas}_{i}^{n+1}\,
    \new_{\rho} : \bang \mathtt{bit} \boxtimes 
    n\text{-}\mathtt{qbit}\rrbracket
\\
&\qquad\qquad:\quad \punit\longrightarrow T\sem{\bit}\enspace.
\end{aligned}  
\end{equation}
  By the definition of the interpretation of  $\mathtt{meas}_{i}^{n+1}$,
the transpose of the interpretation
  \begin{displaymath}
    \llbracket\, \vdash \mathtt{meas}_{i}^{n+1}\,
    \new_{\rho} : \bang \mathtt{bit} \boxtimes 
    n\text{-}\mathtt{qbit}\,\rrbracket \quad:\quad
    \punit \longto 
    T\llbracket \bang \mathtt{bit}\boxtimes n\text{-}\mathtt{qbit} \rrbracket
  \end{displaymath}
  is equal to
\begin{equation}
 \label{eq:meas-qstate}
  \begin{aligned}
 &(\,\llbracket \bang \bit \rrbracket 
    \boxtimes \llbracket \nqbit \rrbracket\,) \limp R
 \\
 & \stackrel{\cong}{\longto}\quad
 \punit
 \boxtimes
 \bigl(\,
 (\llbracket \bang \bit \rrbracket 
    \boxtimes \llbracket \nqbit \rrbracket) \limp R
 \,\bigr)
 \\ &
  \stackrel{\tuple{\llbracket \new_{\rho}
   \rrbracket_{\mathrm{const}}\boxtimes\id}}{\longto}
 \quad
 \sem{(n+1)\text{-}\qbit}
 \boxtimes
 \bigl(\,
 (\llbracket \bang \bit \rrbracket 
    \boxtimes \llbracket \nqbit \rrbracket) \limp R
 \,\bigr)
 \quad \stackrel{m}{\longto}\quad
 R\enspace,
\end{aligned}\end{equation}
  where 
$m$ is from
  (\ref{eq:measurementDefHelper}). Recall that
 $\sem{\bit}\iso\bang\sem{\bit}$; see Lemma~\ref{lem:bitAndBangBit}.
 Under the following
  identifications
  \begin{align*}
    (\llbracket \bang \bit \rrbracket \boxtimes \llbracket \nqbit
    \rrbracket) \limp R 
  \;\stackrel{\cong}{\longto}\;
   (\llbracket \nqbit \rrbracket
    \limp R)^{\times 2} 
    \quad\text{and}\quad
    R
   \;
   \stackrel{\cong}{\longto}
   \;
   (\llbracket \zeroqbit \rrbracket \times
    R)^{\times 2}
  \end{align*}
  that are derived from
 Lemma~\ref{lem:bangVsProductTensorAndCoproduct},~\ref{lem:bitAndBangBit}
 and~\ref{lem:resultTypeAsFinalCoalg},
  the value of (\ref{eq:meas-qstate}) at $\langle f,g\rangle
  :\punit \to (\llbracket \nqbit \rrbracket \limp R)^{\times 2}$ is
\begin{equation}\label{eq:fourTuple}
   \begin{aligned}
&    \Bigl\langle\;
   \punit\stackrel{[\lambda x.\cmbt{Q}_{0}]}{\longrightarrow}\llbracket
   \zeroqbit\rrbracket ,\quad
   \punit
   \stackrel{f \boxtimes
    [\lambda x.\cmbt{Pr}_{|0_{i}\rangle}^{n+1}\,
   \cmbt{Q}_{\rho}]}{\longrightarrow}    
   (\llbracket \nqbit \rrbracket \limp R)\boxtimes\llbracket \nqbit
   \rrbracket
   \stackrel{\ev}{\longrightarrow}
   R,\;
    \\ %
&\qquad   \punit\stackrel{[\lambda x.\cmbt{Q}_{0}]}{\longrightarrow}\llbracket
   \zeroqbit\rrbracket ,\quad
   \punit
   \stackrel{g \boxtimes
    [\lambda x.\cmbt{Pr}_{|1_{i}\rangle}^{n+1}\,
   \cmbt{Q}_{\rho}]}{\longrightarrow}    
   (\llbracket \nqbit \rrbracket \limp R)\boxtimes\llbracket \nqbit
   \rrbracket
   \stackrel{\ev}{\longrightarrow}
   R
\;\Bigl\rangle
  \\
   &\qquad\qquad
   : \qquad \punit \longrightarrow (\llbracket \zeroqbit 
    \rrbracket \times R)^{\times  2}\enspace.
\end{aligned}\end{equation}
Here combinators like $\cmbt{Q}_{0}$, $\cmbt{Q}_{\rho}$ and 
$\cmbt{Pr}_{|0_{i}\rangle}$ are from
 Definition~\ref{definition:quantumCombinators} and~\ref{definition:quantumOprCombinator}; affine $\lambda$-terms like
$\lambda x.\cmbt{Q}_{0}$ denote suitable elements of $A_{\Q}$ by
 combinatory completeness; and
the arrow $[\lambda x.\cmbt{Q}_{0}]$ is the one in $\PER_{\Q}$ that is
realized by $(\lambda x.\cmbt{Q}_{0})\in A_{\Q}$.
From Lemma~\ref{lem:quantumOprCombinatorEq},
it is easy to see that the last arrow~(\ref{eq:fourTuple}) is equal to
  \begin{multline}\label{eq:E-meas-qstate}
    \bigl\langle\; [\lambda x.\cmbt{Q}_{0}],\; %
    \mathsf{ev}
   \co ( f \boxtimes
    \llbracket \new_{\langle 0_{i} | \rho | 0_{i} \rangle}
    \rrbracket_{\mathrm{const}}), \\
    [\lambda x.\cmbt{Q}_{0}],\; %
    \mathsf{ev}
   \co ( g \boxtimes
    \llbracket \new_{\langle 1_{i} | \rho | 1_{i} \rangle}
    \rrbracket_{\mathrm{const}}
    ) %
    \;\bigr\rangle\enspace,
  \end{multline}
 where $\semConst{\place}$ is from Definition~\ref{definition:interpretationOfConstants}.
 Let us now define $f_{0},f_{1}:\llbracket \nqbit
  \rrbracket\to T\llbracket \mathtt{bit}\rrbracket$ to be the
  following arrows:
  \begin{align*}
    f_{0} \;&:=\; 
  \left(\;
\begin{array}{l}
   \sem{\nqbit}
  \stackrel{\cong}{\longrightarrow}
  \punit\boxtimes\sem{\nqbit}
  \stackrel{\varphi'\boxtimes\id}{\longrightarrow}
  \bang\punit\boxtimes\sem{\nqbit}
  \stackrel{\bang\kappa_{\ell}\boxtimes\id}{\longrightarrow}
  \bang\sem{\bit}\boxtimes\sem{\nqbit}
  \\
  \quad
  \stackrel{\llbracket x:\bang \mathtt{bit}\boxtimes
    \nqbit \vdash E[x]:\mathtt{bit}\rrbracket}{\longrightarrow}
  T\sem{\bit}
\end{array}
\;\right)\enspace,
%
\\
    f_{1} \;&:=\; 
  \left(\;
\begin{array}{l}
   \sem{\nqbit}
  \stackrel{\cong}{\longrightarrow}
  \punit\boxtimes\sem{\nqbit}
  \stackrel{\varphi'\boxtimes\id}{\longrightarrow}
  \bang\punit\boxtimes\sem{\nqbit}
  \stackrel{\bang\kappa_{r}\boxtimes\id}{\longrightarrow}
  \bang\sem{\bit}\boxtimes\sem{\nqbit}
  \\
  \quad
  \stackrel{\llbracket x:\bang \mathtt{bit}\boxtimes
    \nqbit \vdash E[x]:\mathtt{bit}\rrbracket}{\longrightarrow}
  T\sem{\bit}
\end{array}
\;\right)\enspace,
%
  \end{align*}
where $\varphi':\punit\iso\bang\punit$ is from Theorem~\ref{theorem:PERQIsLinear}.
  By (\ref{eq:E-meas-qstate}), the transpose of (\ref{eq:07122032})
  is equal to
  \begin{displaymath}
    \bigl\langle\; [\lambda x.\cmbt{Q}_{0}],\, %
    g_{0},\,
    [\lambda x.\cmbt{Q}_{0}], \,%
    g_{1}  %
    \;\bigr\rangle\quad:\quad\llbracket \bit  \rrbracket \limp R
   \; \longrightarrow \; R
  \end{displaymath}
  where we  identified $R$ with  $(\llbracket \zeroqbit 
  \rrbracket \times R)^{\times 2}$, and $g_{k}:\llbracket \bit 
  \rrbracket  \limp R  \longrightarrow R$  is  the transpose  of
 \begin{displaymath}
  \punit
  \stackrel{[\lambda  x.\cmbt{Q}_{\langle k_{i}  | \rho  |
  k_{i}\rangle}]}{\longrightarrow}
  \sem{\nqbit}
  \stackrel{f_{k}}{\longrightarrow}
  T\sem{\bit}\enspace.
 \end{displaymath}
  Hence,
  \begin{align*}
&\Bigl(\;
\punit
\stackrel{\tree(E[\mathtt{meas}_{i}^{n+1}\,\new_{\rho}])}{\longrightarrow}
R
\stackrel{\cong}{\longrightarrow}
(\llbracket \zeroqbit 
  \rrbracket \times R)^{\times 2}
\;
\Bigr)
  \\
&=
    \Bigl\langle\; 
   [\lambda x.\cmbt{Q}_{0}],\, %
    \tree\bigl(f_{0} \co \llbracket \new_{
      \langle 0_{i}\mid \rho \mid 0_{i}\rangle} \rrbracket_{\mathrm{const}}\bigr) ,\,
    [\lambda x.\cmbt{Q}_{0}], \,%
    \tree\bigl(f_{1} \co \llbracket \new_{
      \langle 1_{i}\mid \rho \mid 1_{i}\rangle} \rrbracket_{\mathrm{const}}\bigr) %
    \;\Bigr\rangle\enspace,
  \end{align*}
where we abused the notation  $\tree$  from~(\ref{eq:treeOfM}).
This means that the tree
 $\tree(E[\mathtt{meas}_{i}^{n+1}\,\new_{\rho}])$
can be illustrated as follows.
\begin{displaymath}
\includegraphics{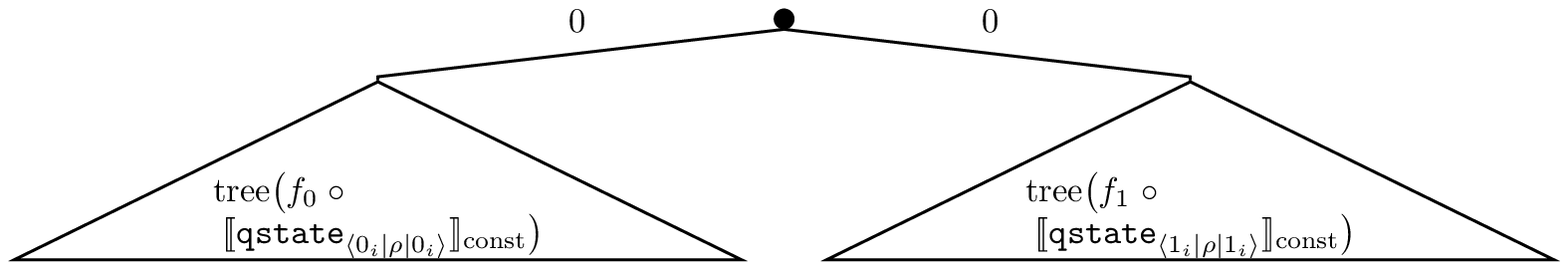}
\end{displaymath}
Therefore
\begin{displaymath}
   \prb\Bigl(\tree\bigl(\llbracket\, \vdash
  E[\mathtt{meas}_{i}^{n+1}\,\new_{\rho}]:\bit 
  \rrbracket\bigr)\Bigr)
\end{displaymath}
is equal to
  \begin{displaymath}
    (0,0) +
    \prb(\tree(f_{0} \co \llbracket \new_{
      \langle 0_{i}\mid \rho \mid 0_{i}\rangle} \rrbracket_{\mathrm{const}})) +
    \prb(\tree(f_{1} \co \llbracket \new_{
      \langle 1_{i}\mid \rho \mid 1_{i}\rangle} \rrbracket_{\mathrm{const}}))
  \end{displaymath}
  where the summation is pointwise. By Lemma~\ref{lem:07121946}, we
  have the following equalities:
  \begin{align*}
    f_{0} \co \llbracket \new_{ \langle 0_{i}\mid \rho
      \mid 0_{i}\rangle} \rrbracket_{\mathrm{const}} &\;=\; \llbracket\,
    \vdash E[\langle \ttrue, \new_{\langle 0_{i}\mid \rho
      \mid
      0_{i}\rangle}\rangle] \rrbracket\enspace, \\
    f_{1} \co \llbracket \new_{ \langle 1_{i}\mid \rho
      \mid 1_{i}\rangle} \rrbracket_{\mathrm{const}} &\;=\; \llbracket\,
    \vdash E[\langle \ffalse, \new_{\langle 1_{i}\mid \rho
      \mid 1_{i}\rangle}\rangle] \rrbracket\enspace.
  \end{align*}
  Therefore, if
  \begin{align*}
    E[\langle \ttrue, \new_{\langle 0_{i}\mid \rho \mid
      0_{i}\rangle}\rangle] \Downarrow (p_{0},q_{0}) && E[\langle
    \ffalse, \new_{\langle 1_{i}\mid \rho \mid
      1_{i}\rangle}\rangle] \Downarrow (p_{1},q_{1}),
  \end{align*}
  then $E[\mathtt{meas}_{i}^{n+1}\,\new_{\rho}] \Downarrow
  (p_{0}+p_{1},q_{0}+q_{1})$.
  \myqed
\end{myproof}
We can similarly prove the following lemma.
\begin{mylemma}\label{lem:E[meas^1]}
  Let $E$ be an evaluation context.
  If $\vdash E[\meas^{1}_{1} \new_{\rho}] : \bit$
  is derivable and
  \begin{align*}
    E[\ttrue] \Downarrow (p_{0},q_{0}) && E[\ffalse] \Downarrow (p_{1},q_{1}),
  \end{align*}
  then $E[\mathtt{meas}_{1}^{1}\,\new_{\rho}] \Downarrow
  (\langle 0 \mid \rho \mid 0 \rangle p_{0}+
  \langle 1 \mid \rho \mid 1 \rangle p_{1},\langle 0 \mid \rho \mid 0 \rangle q_{0}+
  \langle 1 \mid \rho \mid 1 \rangle q_{1})$.
\end{mylemma}

Our soundness result
is restricted to closed $\bit$-terms because 
we do not know how to relate evaluation results of
closed terms $\vdash M : A$ with its interpretation
$\llbracket \vdash M : A \rrbracket \colon I \to T\llbracket A
\rrbracket$ when $A$ is not $\bit$.
\begin{mytheorem}[Soundness]\label{thm:sound}
  For any closed $\bit$-term $M$ (meaning that $\vdash M: \mathtt{bit}$
 is derivable), and for any $k\in\nat$,
 \begin{displaymath}
  M \oprred^{k} (p,q)
  \quad\text{and}\quad
M \Downarrow (p',q')
  \quad\text{imply}\quad
(p,q) \leq (p',q')\enspace.
 \end{displaymath}
Here the last inequality is the pointwise one and means
$p \leq p'$ and $q \leq
  q'$.
\end{mytheorem}
\begin{myproof}
  By induction on $k$. When $k = 0$, if $M$ is neither $\ttrue$ nor
  $\ffalse$, then $p,q=0$ and the statement is true. If $M \equiv
  \ttrue$, then $p = p '= 1$ and $q = q' = 0$. If $M \equiv \ffalse$,
  then $p = p '= 0$ and $q = q' = 1$.  

 When $k > 0$, if there is a reduction $M \to_{1} N$ that is not due to
  a measurement rule, then $\llbracket \vdash M : \mathtt{bit}
  \rrbracket$ is equal to $\llbracket \vdash N : \mathtt{bit}
  \rrbracket$ by Lemma~\ref{lem:07122017}.  Therefore, $M \Downarrow
  (p',q')$ if and only if $N \Downarrow (p',q')$. Since $M \oprred^{k}
  (p,q)$ if and only if $N \oprred^{k-1} (p,q)$, we obtain $p \leq p'$
  and $q \leq q'$ from the induction hypothesis. If $M$ is of the form
  $E[\mathtt{meas}_{i}^{n+1}\,\new_{\rho}]$, then we have $M
  \oprred^{k} (p_{0}+p_{1},q_{0}+q_{1})$ where $E[\langle
  \ttrue,\new_{\langle 0_{i}\mid \rho \mid
    0_{i}\rangle}\rangle] \oprred^{k-1} (p_{0},q_{0})$ and $E[\langle
  \ffalse,\new_{\langle 1_{i}\mid \rho \mid
    1_{i}\rangle}\rangle] \oprred^{k-1} (p_{1},q_{1})$.
  By Lemma~\ref{lem:E[meas^n+1]}, if
  \begin{align*}
    E[\langle \ttrue, \new_{\langle 0_{i}\mid \rho \mid
      0_{i}\rangle}\rangle] \Downarrow (p'_{0},q'_{0}) && E[\langle
    \ffalse, \new_{\langle 1_{i}\mid \rho \mid
      1_{i}\rangle}\rangle] \Downarrow (p'_{1},q'_{1}),
  \end{align*}
  then $E[\mathtt{meas}_{i}^{n+1}\,\new_{\rho}] \Downarrow
  (p'_{0}+p'_{1},q'_{0}+q'_{1})  \geq (p_{0}+p_{1},q_{0}+q_{1})$.
  We can similarly show the statement when the reductions are due to
 the $(\mathtt{meas}_{3})$ and $(\mathtt{meas}_{4})$ rules in Definition~\ref{definition:operationalSemantics}
 by Lemma~\ref{lem:E[meas^1]}.
 \myqed
\end{myproof}

We shall now show the other direction: if $M \oprred (p,q)$ and $M
\Downarrow (p',q')$, then $(p',q') \leq (p,q)$. Our proof employs
the techniques of
logical relations (see e.g.~\cite{Mitchell96a,PlotkinPST00}) and
$\top\top$-lifting 
(see e.g.~\cite{Abadi00,Katsumata13}). 
We write
$\Val(A)$ for the set of closed values of type $A$ and
$\ClTerm(A)$ for the set of closed terms of type $A$. We write
$\mathrm{EC}(A)$ for the set of evaluation contexts $E$ such that $x:A
\vdash E[x]:\bit $ is derivable. 

Firstly, we introduce a relation $\lessdot$
between $\PER_{\Q}(\punit,T\llbracket \bit  \rrbracket)$
and $\ClTerm(\bit )$. It is defined by by
\begin{displaymath}
  t \lessdot M \quad\defiff\quad
  \text{if }M \oprred (p,q) \text{ then }
  \prb(\tree(t))\leq (p,q)\enspace.
\end{displaymath}

Secondly we introduce the operation of $\top\top$-lifting.
Given a relation
\begin{displaymath}
  S \;\subseteq\; \PER_{\Q}(\punit,\llbracket A \rrbracket) 
  \times \Val(A)\enspace,  
\end{displaymath}
we define a relation $S^{\top} \subseteq \PER_{\Q}(\llbracket A
\rrbracket , T\llbracket \bit  \rrbracket) \times
\mathrm{EC}(A)$ by
\begin{displaymath}
  S^{\top} 
  \;:=\; \bigl\{\,(k,E) \,\bigl|\bigr.\, \forall (t,V) \in S.\;
  k \co t \lessdot E[V]\,\bigr\}\enspace;
\end{displaymath}
and we define a relation $S^{\top\top} \subseteq \PER_{\Q}(\punit,
T\llbracket A \rrbracket) \times \ClTerm(A)$ by
\begin{displaymath}
  S^{\top\top} 
  \;:=\; \bigl\{\,(t,M) \,\bigl|\bigr.\, \forall (k,E) \in S^{\top}.\;
  \mu^{T}_{\llbracket \bit  \rrbracket} \co 
  Tk \co t \lessdot E[M]\,\bigr\}\enspace.
\end{displaymath}

This operation of $\top\top$-lifting is applied to the following
relation $R_{A}\subseteq
\PER_{\Q}(\punit,\llbracket A \rrbracket) \times
\Val(A)$. It is inductively defined for each type $A$.
\begin{equation}
 \label{eq:defOfLogicalRelationR}
\begin{aligned}
  R_{\punit} &\;:=\; \bigl\{(\mathrm{id}_{\punit},*)\bigr\} \\
  R_{\nqbit} &\;:=\;
  \bigl\{\,\bigl(\llbracket\new_{\rho}\rrbracket_{\mathrm{const}},
  \new_{\rho}\bigr) \,\bigl|\bigr.\, \rho \in \DM_{2^{n}}\,\bigr\} \\
  R_{A \boxtimes B} &\;:=\; \bigl\{\,\bigl(t \boxtimes s , \langle V,W\rangle\bigr) \,\bigl|\bigr.\,
  (t,V) \in R_{A} \text{ and } (s,W) \in R_{B}\,\bigr\} \\
  R_{A \limp B} &\;:=\; \bigl\{\,(t,V) \,\bigl|\bigr.\, \forall (s,W) \in R_{A}.\,
  \bigl(\mathsf{ev}_{\llbracket A\rrbracket,\llbracket B \rrbracket} \co
  (t \boxtimes s) ,\,
  V \, W\bigr)\in R_{B}^{\top\top}\,\bigr\} \\
  R_{\bang A} &\;:=\; \bigl\{\,(\bang t \co \varphi', V) \,\bigl|\bigr.\,  (t,V) \in R_{A}\,\bigr\} \\
  R_{A+B} &\;:=\; \bigl\{\,\bigl(\kappa_{\ell} \co t, \mathtt{inj}_{\ell}^{B'}(V)\bigr)
  \,\bigl|\bigr.\,
  (t,V) \in R_{A} \text{ and }B' \subtp B\,\bigr\} \\
  & \qquad \cup \bigl\{\,\bigl(\kappa_{r} \co t, \mathtt{inj}_{r}^{A'}(V)\bigr) \,\bigl|\bigr.\,
  (t,V) \in R_{B} \text{ and }A' \subtp A\,\bigr\}
\end{aligned}
\end{equation}
Here $\varphi':\punit\iso\bang\punit$ is from Theorem~\ref{theorem:PERQIsLinear}.
In order to prove the \emph{basic lemma} for the logical relation
$\{R_{A}\}_{A : \mathrm{type}}$, we show some properties of $R_{A}$.
\begin{mylemma}\label{lem:07141800}
  \begin{enumerate}
  \item \label{enu:07130423} %
    If $(t,V)$ is in $R_{A}$, then $(\eta_{\llbracket A
      \rrbracket}^{T} \co t,V)$ is in $R_{A}^{\top\top}$.
  \item \label{enu:07130422} %
    If $(t,V) \in R_{A}$ and $A \subtp A'$, then
    $(\llbracket A \subtp A' \rrbracket \co t,V) \in R_{A'}$.
   \myqed
  \end{enumerate}
\end{mylemma}

The following property is much like the \emph{admissibility}
requirement. See e.g.~\cite{Abadi00}.
\begin{mylemma}\label{lem:bot}
  Let $M$ be a closed term of type $A$.
  \begin{enumerate}
  \item\label{enu:bot} $([\bot],M) \in R_{A}^{\top\top}$.
  \item\label{enu:lab} If there exists a sequence of realizers $a_{1}
       \sqsubseteq a_{2} \sqsubseteq \cdots$ of arrows
      $[a_{1}],[a_{2}],\dotsc$ 
       in
    $\PER_{\Q}(\punit,T\llbracket A \rrbracket)$, such that
    $([a_{n}],M) \in R_{A}^{\top\top}$ for each $n$, then we have $\bigl(\,[\bigvee_{n
      \geq 1}a_{n}],\,M\,\bigr) \in R_{A}^{\top\top}$. 
       \myqed
  \end{enumerate}
\end{mylemma}

\begin{mytheorem}[Basic Lemma]\label{thm:basiclemma}
  Let $M$ be a term such that  $x_{1}:A_{1},\cdots,x_{n}:A_{n} \vdash
 M:A$ is derivable. If
  $(t_{i},V_{i})$ is in $R_{A_{i}}$ for each $i\in[1,n]$, then the pair
  \begin{displaymath}
    \bigl(\;\llbracket
    x_{1}:A_{1},\cdots,x_{n}:A_{n} \vdash M:A \rrbracket \co (t_{1}
    \boxtimes \cdots \boxtimes t_{n}),\quad M[
    V_{1}/x_{1},\cdots,V_{n}/x_{n}]\;\bigr)    
  \end{displaymath}
  is in $R_{A}^{\top\top}$.
\end{mytheorem}
\begin{myproof}
  By induction on $M$. When $M\equiv x_{i}$, we have
  \begin{equation}\label{eq:07130424}
    \llbracket x_{1}:A_{1},\cdots,x_{n}:A_{n} \vdash M:A \rrbracket
    \co (t_{1} \boxtimes \cdots \boxtimes t_{n})  
    = \eta_{\llbracket A \rrbracket}^{T} \co \llbracket
    A_{i}\subtp A \rrbracket \co t_{i}.
  \end{equation}  
  By (\ref{enu:07130422}) in Lemma~\ref{lem:07141800}, the pair
  $(\llbracket A_{i}\subtp A \rrbracket \co t_{i},V_{i})$ is in
  $R_{A}$. Therefore, by (\ref{enu:07130423}) in
  Lemma~\ref{lem:07141800} and (\ref{eq:07130424}), we see that
  \begin{equation*}
    \bigl(\;\llbracket x_{1}:A_{1},\cdots,x_{n}:A_{n} \vdash M:A \rrbracket
    \co (t_{1} \boxtimes \cdots \boxtimes t_{n}),\;
    V_{i}\;\bigr)
  \end{equation*}
  is in $R_{A}^{\top\top}$. When $M$ is a constant, see
  Lemma~\ref{lem:new}--\ref{lem:meas2}. 
  When $M$ is an application
    $\bang \Delta,\Gamma_{1},\Gamma_{2} \vdash N_{0} N_{1} : B$ for
    $\bang \Delta,\Gamma_{1} \vdash N_{0} : A \limp B$ and
    $\bang \Delta,\Gamma_{2} \vdash N_{1} : A \limp B$, we suppose
    that $\Delta$, $\Gamma_{1}$, $\Gamma_{2}$ are empty lists for
    simplicity. Generalization is straightforward. Since
    \begin{equation*}
      \bigl(
      \mu_{\llbracket \bit \rrbracket}^{T} \circ 
      Tk \circ \mathsf{ev}_{\llbracket A\rrbracket,\llbracket B \rrbracket} \co
      (t \boxtimes \llbracket A \rrbracket) ,\,
      E[V[\place]] \bigr)\in R_{A}^{\top}
    \end{equation*}
    for any $(t,V) \in R_{A \limp B}$ and $(k,E) \in R_{B}^{\top}$, we
    have
    \begin{multline*}
      \bigl(\,
      \mu_{\llbracket \bit \rrbracket}^{T} \co 
      T\mu_{\llbracket \bit \rrbracket}^{T} \co TTk \co
      T\mathsf{ev}_{\llbracket A\rrbracket,\llbracket B \rrbracket} \co
      \str \co (\llbracket A  \limp B \rrbracket \boxtimes (\llbracket \vdash N_{1} 
      : A \rrbracket )) ,\, \\
      E[[\place]N_{1}] \,\bigr) \quad\in\; R_{A \limp B}^{\top}
    \end{multline*}
    for any $(k,E) \in R_{B}^{\top}$. Therefore,
    \begin{multline*}
      \mu_{\llbracket \bit \rrbracket}^{T} \co 
      T\mu_{\llbracket \bit \rrbracket}^{T} \co 
      TT\mu_{\llbracket \bit
        \rrbracket}^{T} \co TTTk \co TT\mathsf{ev}_{\llbracket
        A\rrbracket,\llbracket B \rrbracket} \co T\str \co \str' \co
      \\ (\llbracket \vdash N_{0} :A \limp B \rrbracket \boxtimes
      \llbracket \vdash N_{1} : A \rrbracket) \lessdot E[N_{0} N_{1}]
    \end{multline*}
    holds for any $(k,E) \in R_{B}^{\top}$, and we obtain
    \begin{equation*}
      (\llbracket \vdash N_{0}N_{1} : B \rrbracket , N_{0}N_{1}) \in R_{B}^{\top\top}.
    \end{equation*}
    When $M$ is a lambda abstraction
    $x_{1}:A_{1},\cdots,x_{n}:A_{n} \vdash \lambda x^{A}.N: A' \limp
    B$,
    we suppose that $x$ is the only free variable of $N$ and $n=0$ for
    simplicity. Also in this case, generalization is straightforward. We have
    \begin{equation*}
      \bigl(\,\llbracket x:A \vdash M : B \rrbracket \co \llbracket A' \subtp A \rrbracket \co t,\,
      M[V/x]\,\bigr) \in R_{B}^{\top\top}
    \end{equation*}  
    for any $(t,V) \in R_{A'}$. By the definition of $R_{A \limp B}$,
    \begin{equation*}
      \bigl(g,\, \lambda x^{A}.M\bigr) \in R_{A \limp B}
    \end{equation*}  
    where $g$ is the adjoint transpose of
    $\llbracket x:A \vdash M : B \rrbracket \co \llbracket A' \subtp A
    \rrbracket$.
    Hence, by Lemma~\ref{lem:07141800},
    $\bigl(\llbracket M \rrbracket,\, \lambda x^{A}.M\bigr)$ is in
    $R_{A \limp B}$.
    When $M$ is
  $\mathtt{letrec} \ f^{A}\,x = N \ \mathtt{in} \ L$, the statement
  follows from Lemma~\ref{lem:bot}. Note that the interpretation of
  $\mathtt{letrec} \ f^{A}\,x = N \ \mathtt{in} \ L$ is given by the
  least upper bound of a sequence of realizers. The other cases are
  easy.
 \myqed
\end{myproof}

\begin{mycorollary}[Adequacy]\label{cor:adeq}
  For a closed term $\vdash M:\bit $, we have
  \begin{displaymath}
    M \oprred (p,q) \iff
    M \Downarrow (p,q).
  \end{displaymath}
\end{mycorollary}
\begin{myproof}
  We suppose that $M \oprred (p,q)$ and $M \Downarrow (p',q')$.  By
  Theorem~\ref{thm:sound}, we have $(p,q) \leq (p',q')$ on the one hand.  On the other
  hand, by Theorem~\ref{thm:basiclemma} (consider its special case where
 $M$ is closed), we have $\bigl(\llbracket\, \vdash
  M:\bit  \rrbracket, M\bigr) \in R_{\bit }^{\top\top}$.
  Since $(\eta^{T}_{\llbracket \bit  \rrbracket}, [\place])$ is
  easily shown to be
  in $R_{\bit }^{\top}$, we obtain $\llbracket \vdash
  M:\bit  \rrbracket \lessdot M$. Hence $(p',q') \leq
  (p,q)$. \myqed
\end{myproof}

\section{Conclusions and Future Work}\label{section:concl}
We presented a concrete denotational model of a quantum
$\lambda$-calculus that supports the calculus' full features including
the $\bang$ modality and recursion. The model's construction is via
known semantical techniques like GoI and realizability.  The
current work is a demonstration of the generality of these techniques
in the sense that, with a suitable choice of a parameter (namely $B=\Q$ in
Figure~\ref{figure:constructionOfModel}), the known techniques for
classical computation apply also to quantum computation (or more
precisely ``quantum data, classical control''). Our model is also one answer
to the question ``Quantum GoI?'' raised in~\cite{Scott04Tutorial}.

Our semantics is based on so-called \emph{particle-style GoI} and hence
on local interaction of agents, passing a token to each other.  This is
much like in \emph{game semantics}~\cite{AbramskyJM00,HylandO00}; our
denotational model, therefore, has a strong operational flavor. We are
currently working on extracting abstract machines for quantum
computation, much like the classical cases
in~\cite{Mackie95,GhicaS10,GhicaS11,GhicaSS11,MuroyaKHH14LOLA}.\footnote{Such
``compilation'' from a quantum program to an abstract machine is
presented in~\cite{Grattage11}. The functional language considered there
is a first-order one, drastically easing the challenge of dealing with
classical control.} In doing so, our current use of the continuation
monad $T$ (see~\S{}\ref{section:denotationalModel}) is a technical
burden; it seems we need such continuation monads not only for quantum effects
(in the current paper) but also for various computational effects (in
general). 
Endowing realizers with an explicit
notion of \emph{memory} (or
\emph{state})~\cite{HoshinoMH14CSLLICS,MuroyaKHH14LOLA}---in a
systematic manner using \emph{coalgebraic component calculus}~\cite{Barbosa01,HasuoJ11TracesForCoalgCompo}---seems to
be a potent alternative to use of continuation monads.

\section*{Acknowledgments} 
Thanks are due to Peter Selinger for a number of insightful comments; to
 Kazuyuki Asada, Ugo Dal Lago, Claudia Faggian, Bart Jacobs, Prakash Panangaden, Peter Selinger, Phil Scott,
 Beno\^{\i}t Valiron,
 Mingsheng Ying and the anonymous reviewers (for the earlier
 version~\cite{HasuoH11} as well as for the current version) for
 useful discussions and comments; and to Andy Pitts for his ETAPS'07
 talk that inspired the current work. 

\newpage

\appendix


\section{CPO Structures of Density Matrices and Quantum Operations}
\label{appendix:CPOStrOfDensityMatrix} \auxproof{ There is also the
\emph{operator norm} that is standard: see the Wikipedia entry on
``matrix norms'' } For the limit properties of density matrices and
quantum operations (such as
Lemma~\ref{lem:LoewnerPartialOrderIsOrderAndCPO}) we employ some basic
facts from matrix analysis. 

There are various notions of norms
for matrices but they are known to coincide in finite-dimensional
settings. We will be using the following two.
\begin{mydefinition}[Norms $\trNorm{\place}$ and $\FrNorm{\place}$]
 \label{definition:normsForMatrices}
 Given a matrix $A\in M_{m}$, its \emph{trace norm} $\trNorm{\place}$ is  defined by
 \begin{displaymath}
  \trNorm{A}:=\tr(\sqrt{A^{\dagger}A})\enspace.
 \end{displaymath}
 Here the matrix $A^{\dagger}A$ is positive
\auxproof{ (Lemma~\ref{lem:charOfPositiveMatrix})}
 hence its square root is well defined
 (see e.g.~\cite[\S{}2.1.8]{NielsenC00}).  In particular, we have
 \begin{equation}\label{eq:traceNormForPositiveMatrix}
  \trNorm{A}=\trace(A) \quad\text{when $A$ is positive.}
 \end{equation}

 The \emph{Frobenius norm} $\FrNorm{A}$ of a matrix $A$
 is defined by
 \begin{displaymath}
  \FrNorm{A}:=\sqrt{\textstyle\sum_{i,j}|A_{i,j}|^{2}}
  =\sqrt{\trace(A^{\dagger}A)}\enspace.
 \end{displaymath}
Here $A_{i,j}$ is the $(i,j)$-entry of the matrix $A$,
 hence the Frobenius norm coincides with the standard norm on $M_{m}\cong\C^{m\times m}$.
 The latter
 equality  is immediate by a direct calculation. 
\end{mydefinition}
\noindent The metric induced by $\trNorm{\place}$ is called the \emph{trace
distance} and heavily used
 in~\cite[\S{}9.2]{NielsenC00}.
\auxproof{
 Among the axioms of a norm,  we prove that
 $\trNorm{A}=0$ implies $A=0$ (the other norm axioms are straightforward). 
Assume $\trNorm{A}=0$. Then the eigenvalues of the positive matrix 
 $A^{\dagger}A$ must be all $0$; thus $A^{\dagger}A=0$. 
 Now, for each vector $\ket{v}$,
\begin{displaymath}
  \abs{A\ket{v}}^{2}
 =
  \bra{v}A^{\dagger}A\ket{v}
 =
  \bra{v}0\ket{v} = 0\enspace.
\end{displaymath}
 Thus $A\ket{v}$ is the zero vector for each $\ket{v}$; hence  $A =0$.
}

\begin{mylemma}\label{lem:propertiesOfTwoNorms}
 \begin{enumerate}
  \item\label{item:FrobNormIsSmallerThanTraceNorm} For each matrix $A\in
       M_{m}$
       we have $\FrNorm{A}\le\trNorm{A}\le m \FrNorm{A}$;
       therefore the two norms induce the same topology on the set $M_{m}$.
  \item Both norms $\FrNorm{\place}$ and $\trNorm{\place}$ are complete.
  \item The subset $\DM_{m}\subseteq M_{m}$ is closed with respect to both
       norms $\FrNorm{\place}$ and $\trNorm{\place}$.
 \end{enumerate}
\end{mylemma}
\begin{myproof}
 1.  
 Let $\lambda_{1},\dotsc,\lambda_{m}$ be the (nonnegative) eigenvalues 
 of  positive $A^{\dagger}A$. Then the inequality is
 reduced to
 \begin{displaymath}
  \sqrt{\lambda_{1}+\cdots+\lambda_{m}}
 \;\le\;
    \sqrt{\lambda_{1}}+\cdots+\sqrt{\lambda_{m}}
 \;\le\;
  m\cdot   \sqrt{\lambda_{1}+\cdots+\lambda_{m}}
 \end{displaymath}
 which is obvious.

2. $\FrNorm{\place}$ is complete because so is
 $\C$. Then one uses~\ref{item:FrobNormIsSmallerThanTraceNorm}.



3.
 Let $(\rho_{k})_{k\in\nat}$ 
 be a Cauchy sequence in $\DM_{m}$. We show that
 $\lim_{k}\rho_{k}$ belongs to $\DM_{m}$. It is positive because the mapping 
 $\bra{v}\place\ket{v}:\DM_{m}\to \C$ is continuous with respect to
 $\FrNorm{\place}$ (hence also to $\trNorm{\place}$). Similarly, 
 continuity of $\trace(\place)$ yields that $\trace(\lim_{k}\rho_{k})\le 1$.
\myqed
\end{myproof}
We shall henceforth assume the topology on $M_{m}$ that is induced by
either of the norms. It is with respect to this topology that we speak,
for example, continuity of the function $\trace(\place):\DM_{m}\to\C$.
On the one hand, the Frobenius norm $\FrNorm{\place}$ is useful since many
functions---such as $\trace(\place)$---are obviously continuous with
respect to it.
On the other hand,
the trace norm $\trNorm{\place}$ is important for us due to the
following property.

\begin{mylemma}\label{lem:convergenceViaTraceNorm}
 Let $(\rho_{n})_{n\in\nat}$ be a sequence in $\DM_{m}$ 
that is increasing with respect to the L\"{o}wner order in
 Definition~\ref{definition:LoewnerPartialOrder}.
Then $(\rho_{n})_{n\in\nat}$ is Cauchy and hence has a limit
 in $\DM_{m}$.
\end{mylemma}
\begin{myproof}
  For any $n,n'\in\nat$ with $n\le n'$, we have
 \begin{equation}\label{eq:traceNormOfChainAsTrace}
  \trNorm{\rho_{n'}-\rho_{n}}
  \stackrel{(*)}{=}
  \trace(\rho_{n'}-\rho_{n})
  =
  \trace(\rho_{n'})-\trace(\rho_{n})\enspace;
 \end{equation}
 where $(*)$ holds since $\rho_{n'}-\rho_{n}$ is positive
  (see~(\ref{eq:traceNormForPositiveMatrix})).  Now observe that the
  sequence $(\trace(\rho_{n}))_{n\in\nat}$ is an increasing sequence in
  $[0,1]$ hence is Cauchy. Combined
  with~(\ref{eq:traceNormOfChainAsTrace}), we conclude that the
  sequence $(\rho_{n})_{n\in\nat}$ in $\DM_{m}$ is Cauchy with respect to
  $\trNorm{\place}$. By Lemma~\ref{lem:propertiesOfTwoNorms}, it has a
  limit $\lim_{n}\rho_{n}$ in $\DM_{m}$.
 \myqed
\end{myproof}
\auxproof{
Note that in the proof we say $\|A+B\|=\|A\|+\|B\|$, which is quite unusual.
}

\begin{mylemma*}[Lemma~\ref{lem:LoewnerPartialOrderIsOrderAndCPO}, repeated]\label{lem:LoewnerPartialOrderIsOrderAndCPOInDetail}
 The relation $\Lle$  in Definition~\ref{definition:LoewnerPartialOrder} is
 indeed a partial order. Moreover it is an $\omega$-CPO: 
 any increasing $\omega$-chain $\rho_{0}\Lle
 \rho_{1}\Lle\cdots$ in $\DM_{m}$ has the least upper bound.
\end{mylemma*}
\begin{myproof}
 Reflexivity holds because $0$ is a positive matrix; transitivity is
 because a sum of positive matrices is again positive. Anti-symmetry 
 is because, if a positive matrix $A$ is such that $-A$ is also
 positive, all the eigenvalues of $A$ are $0$ hence $A$ itself is the
 zero matrix.

 That $\Lle$ is an $\omega$-CPO is proved
 in~\cite[Proposition~3.6]{Selinger04} via the translation into quadratic
 forms. Here we present a proof using norms.
 By Lemma~\ref{lem:convergenceViaTraceNorm},
 an increasing $\omega$-chain  $(\rho_{n})_{n\in\nat}$ in
 $\DM_{m}$ has a limit $\lim_{n}\rho_{n}$ in $\DM_{m}$.
 We claim that $\lim_{n}\rho_{n}$ is the least upper bound. 

To show that
 $\rho_{k}\Lle\lim_{n}\rho_{n}$, 
\auxproof{recall Definition~\ref{definition:positiveMatrix} and}%
consider
 \begin{equation}\label{eq:tmp1}
  \bra{v}(\textstyle\lim_{n}\rho_{n}) - \rho_{k}\ket{v}
 =
  \textstyle\lim_{n} \bra{v}\rho_{n} - \rho_{k}\ket{v}\enspace;
 \end{equation}
 the equality is due to the continuity of $\bra{v}\place\ket{v}:\DM_{m}\to
 \C$. The value $ \bra{v}\rho_{n} - \rho_{k}\ket{v}$ is a nonnegative real
 for almost all $n$, therefore~(\ref{eq:tmp1}) itself is a
 nonnegative real. This proves  $\rho_{k}\Lle\lim_{n}\rho_{n}$.
 One can similarly prove that $\lim_{n}\rho_{n}$ is the least among the 
 upper bounds of $(\rho_{n})_{n\in\nat}$. \myqed
\end{myproof}

\begin{myproposition*}[Proposition~\ref{proposition:QOisCPO}, repeated]
 The order $\Lle$ on $\QO_{m,n}$ (Definition~\ref{definition:pointwiseorderOnQO}) is an $\omega$-CPO.
\end{myproposition*}
\begin{myproof}
 Let $(\Eop_{k})_{k\in\nat}$ be an increasing chain in $\QO_{m,n}$. We define
 $\Eop$ to be its ``pointwise supremum'':
for
 each $\rho\in \DM_{m}$,
\begin{equation}\label{eq:upperLimitInQOAsALimit}
 \Eop(\rho)\;:=\;
  \sup_{k\to\infty} \Eop_{k}(\rho)
 \;\stackrel{(*)}{=}\;
  \lim_{k\to\infty} \Eop_{k}(\rho)
\end{equation} 
where the supremum is taken in the $\omega$-CPO $\DM_{n}$ (Lemma~\ref{lem:LoewnerPartialOrderIsOrderAndCPO}). In the proof
of   Lemma~\ref{lem:LoewnerPartialOrderIsOrderAndCPO} we  exhibited
that the supremum is indeed the limit ($(*)$ above).
We claim that this $\Eop$ is the supremum of the chain
 $(\Eop_{k})_{k\in\nat}$.

 We check that~(\ref{eq:upperLimitInQOAsALimit}) indeed defines
 a QO $\Eop$. 
 In Definition~\ref{definition:quantumOperation}, the trace condition follows 
 from the continuity of $\trace(\place):\DM_{n}\to \mathbb{R}$. 
 For convex linearity
 we have to show
 \begin{displaymath}
    \lim_{k\to\infty}\bigl((c_{k}(x))_{m,n}(\textstyle\sum_{j}p_{j}\rho_{j})\bigr)
 =
   \textstyle\sum_{j}p_{j}
    \bigl(\displaystyle\lim_{k\to\infty}(c_{k}(x))_{m,n}(\rho_{j})\bigr)\enspace.
 \end{displaymath}
 This follows from the linearity of the limit operation $\lim_{k\to\infty}$, which is 
 straightforward 
 since $\lim_{k\to\infty}$ is with respect to the trace norm $\trNorm{\place}$.
 To prove complete positivity of $\Eop$, 
 one can use Choi's characterization of complete positive maps
 (see~\cite[Theorem~6.5]{Selinger04}).  The operations involved in the
 characterization are all continuous, hence one can conclude complete
 positivity of $\Eop$ from that of $\Eop_{k}$.

\auxproof{A direct proof:
 Lemma~\ref{lem:densityMatrixAsEnsemble}, it suffices to show the following:
 \begin{displaymath}
 \begin{array}{ll}
   \text{for any
   $\ket{u}\in\C^{k}\otimes\C^{m}$,}
  \\
   \bigl(\,\IM_{k}\otimes
   (c(x))_{m,n}\,\bigr)\,\bigl(\,\ket{u}\bra{u}\,\bigr)
   \quad\text{is positive. }
 \end{array} 
 \end{displaymath}
 We have
 \begin{align*}
  & \bigl(\,\IM_{k}\otimes
   (c(x))_{m,n}\,\bigr)\,\bigl(\,\ket{u}\bra{u}\,\bigr)
 \\
 &=
 \bigl(\,\IM_{k}\otimes
   (c(x))_{m,n}\,\bigr)\,\bigl(\,\sum_{i,j,i',j'}a_{i,j}a^{*}_{i,j}
   \ket{v_{i}}\ket{w_{j}}\bra{w_{j'}}\bra{v_{i'}}\,\bigr)
 \tag*{\text{where $(\ket{v_{i}})_{i}$ and $(\ket{w_{j}})_{j}$ are ONBs}}
 \\
 &=
\sum_{i,j,i',j'}a_{i,j}a^{*}_{i,j}
 \bigl(\,\ket{v_{i}} \bra{v_{i'}} \otimes
  (c(x))_{m,n}(\ket{w_{j}}\bra{w_{j'}})
\,\bigr)
 \\
 &=
\sum_{i,j,i',j'}a_{i,j}a^{*}_{i,j}
 \Bigl(\,\ket{v_{i}} \bra{v_{i'}} \otimes
  \bigl(\,\lim_{l}(c_{l}(x))_{m,n}(\ket{w_{j}}\bra{w_{j'}})\,\bigr)
\,\Bigr)
 \tag*{\text{by def.\ of $(c(x))_{m,n}$}}
 \\
 &=
\lim_{l}\sum_{i,j,i',j'}a_{i,j}a^{*}_{i,j}
 \Bigl(\,\ket{v_{i}} \bra{v_{i'}} \otimes
  (c_{l}(x))_{m,n}(\ket{w_{j}}\bra{w_{j'}})
\,\Bigr)
 \tag*{\text{by continuity}}
 \\
  &=\lim_{l} \bigl(\,\IM_{k}\otimes
   (c_{l}(x))_{m,n}\,\bigr)\,\bigl(\,\ket{u}\bra{u}\,\bigr)\enspace.
 \end{align*}
This is positive due to the complete positivity of $c_{l}$. 
}%

 It remains to show that $\Eop$ is indeed the least upper bound.
 This is obvious since $\Lle$ on $\QO_{m,n}$ is a
 pointwise extension of $\Lle$ on density matrices.
 \myqed 
\end{myproof}

\auxproof{\subsection{Another CPO Structure $\QOle$ of $\QO_{m,n}$}
\label{appendix:AnotherCPOStructure}
\begin{mydefinition}[Order $\QOle$ on $\QO_{m,n}$]
\label{definition:anotherOrderOnQX}
Given $\Eop,\Fop\in\QO_{m,n}$, we define
 \begin{displaymath}
  \Eop\QOle \Fop
  \quad\defiff\quad
  \Fop-\Eop
  \text{ is a QO.}
 \end{displaymath}
\end{mydefinition}

\begin{mylemma}\label{lem:QOOrderIsIndeedOrderOnQX}
 The relation $\QOle$ in Definition~\ref{definition:anotherOrderOnQX} is indeed
 a partial order, satisfying reflexivity, transitivity and anti-symmetry.
\end{mylemma}
\begin{myproof}
 Reflexivity holds because the zero map, mapping any state $\rho\in \DM_{m}$ to
 $0\in \DM_{n}$, is a QO. Anti-symmetry is because a QO $\Eop$ such that
 $-\Eop$ is also a QO is necessarily $0$.

To prove transitivity, assume
 $\Eop\QOle \Eop'$ and $\Eop'\QOle \Eop''$. 
Let 
 $\Fop:=\Eop'-\Eop$
 and
 $\Fop':=\Eop''-\Eop'$;
we are done if we show that $\Fop + \Fop'$ is a QO.
We prove its trace condition (Definition~\ref{definition:quantumOperation});
 the other axioms are obvious. Given any $\rho\in \DM_{m}$,
\begin{displaymath}
 \trace\bigl[\,(\Fop +\Fop')(\rho)\,\bigr]
=
 \trace\bigl[\, (\Eop''-\Eop)(\rho)\,\bigr]
\le
 \trace\bigl[\, \Eop''(\rho)\,\bigr]
\le 1\enspace,
\end{displaymath}
since $\Eop$ and $\Eop''$ are QOs.
This concludes the proof. \myqed
\end{myproof}

\begin{myremark}
 The order $\QOle$ in Definition~\ref{definition:anotherOrderOnQX} 
 is finer than the (pointwise) L\"{o}wner order $\Lle$
 (Definition~\ref{definition:pointwiseorderOnQX}).
Obviously $\Eop\QOle \Fop$ implies $\Eop\Lle \Fop$; 
however the converse does not hold in general. To see this, 
define $\Eop,\Fop\in\QO_{2,2}$ by
 \begin{align*}
  \Eop
\left(
 \begin{array}{cc}
  a&b \\ c&d
 \end{array}
\right)
 &:=
  \frac{1}{2}
\left(
 \begin{array}{cc}
  d&0 \\ 0&a
 \end{array}
\right)\enspace;
 \\
 \quad
  \Fop
\left(
 \begin{array}{cc}
  a&b \\ c&d
 \end{array}
\right)
 & :=
  \frac{1}{2}
\left(
 \begin{array}{cc}
  a+d&c \\ b&a+d
 \end{array}
\right)\enspace.
 \end{align*}
Let us first see that these are indeed QOs. One can directly check the
 conditions of Definition~\ref{definition:quantumOperation}: convex linearity is
 obvious; so is the trace property. Both $\Eop$ and $\Fop$ are positive:
 direct calculations show that $\Eop(\ket{v}\bra{v})$ and
 $\Fop(\ket{v}\bra{v})$ are both positive for each $\ket{v}\in\C^{2}$.
 Hence, by diagonalizing a density matrix,
\auxproof{Lemma~\ref{lem:densityMatrixAsEnsemble}, }
 $\Eop(\rho)$ and
 $\Fop(\rho)$ are both positive for each $\rho\in \DM_{2}$.  To check that
 $\Eop$ and $\Fop$ are \emph{completely} positive, one can use the
 characterization in Choi's theorem~\cite{Choi75} (see
 also~\cite[Theorem~6.5]{Selinger04}) which can be verified by direct
 calculations.  

 Alternatively, one can follow the proof of
 \auxproof{Proposition~\ref{proposition:oprSumRepr}}%
 the fact that a QO has an operator-sum representation (found e.g.\
 in~\cite[\S{}8.2.4]{NielsenC00}) to obtain concrete operator-sum
 representations of $\Eop$ and $\Fop$.  This leads to
 \begin{displaymath}
\begin{array}{ll}
   \Eop=\sum_{i=1}^{2}E_{i}(\place)E_{i}^{\dagger}\enspace,
  \quad
  &\Fop=\sum_{i=1}^{3}F_{i}(\place)F_{i}^{\dagger}\enspace,
 \\
  E_{1}=
\left(
 \begin{array}{cc}
  0&1/\sqrt{2} \\ 0&0
 \end{array}
\right)\enspace,
 &
  F_{1}=
\left(
 \begin{array}{cc}
  0&1/\sqrt{2} \\ 1/\sqrt{2}&0
 \end{array}
\right)\enspace,
\\
  E_{2}=
\left(
 \begin{array}{cc}
  0&0 \\ 1/\sqrt{2}&0
 \end{array}
\right)\enspace;\;
&
  F_{2}=
\left(
 \begin{array}{cc}
  1/\sqrt{2}&0 \\ 0&0
 \end{array}
\right)\enspace,
 \\
 &
  F_{3}=
\left(
 \begin{array}{cc}
 0&0 \\ 0&  1/\sqrt{2}
 \end{array}
\right)\enspace.
\end{array} 
\end{displaymath}
One can easily check that these matrices indeed gives representations
for $\Eop$ and $\Fop$. Hence 
 we
 see that $\Eop$ and $\Fop$ are QOs.

Anyway we have shown that $\Eop$ and $\Fop$ are QOs. %
\auxproof{See completePos.eps.}%
The map
\begin{displaymath}
   (\Fop-\Eop)
\left(
 \begin{array}{cc}
  a&b \\ c&d
 \end{array}
\right)
 =
  \frac{1}{2}
\left(
 \begin{array}{cc}
  a&c \\ b&d
 \end{array}
\right)
\end{displaymath}
 is the well-known ``transpose'' example (multiplied by $1/2$): it is a
 map which is positive but fails to be completely positive. See
 e.g.~\cite[Box.~8.2]{NielsenC00}.
\end{myremark}

\begin{myproposition}\label{proposition:QOOrderIsCPO}
 The order $\QOle$ on $\QO_{m,n}$ is an $\omega$-CPO.
\end{myproposition}
\begin{myproof}
 Most of the proof of Proposition~\ref{proposition:QOisCPO} carries over; the
 only gap is the last part, where we have to show that $\Eop$ defined
 in~(\ref{eq:upperLimitInQOAsALimit}) is the least upper bound
 also with respect to $\QOle$.

  The mapping $\Eop\in\QO_{m,n}$ is an upper bound; one can prove that
  $\Eop-\Eop_{k}$ is a QO, in the same way as we proved that $\Eop$ is a
  QO in the proof of Proposition~\ref{proposition:QOisCPO} (exploiting
  continuity). Assume $\Fop\in \QO_{m,n}$ is another upper bound.  We
  can show that $\Fop-\Eop$ is a QO, again in the same way. \myqed
\end{myproof}

}

\section{Proofs for \S{}\ref{subsection:HoqSyntacticProperties}}
\label{section:appendixProofSyntProp}

\subsection{Proof of Lemma~\ref{lem:bottomUpDefOfEvalContext}}
\begin{myproof}
We let 
\begin{itemize}
 \item 
 the set of evaluation contexts that is defined in
 Definition~\ref{definition:valueAndEvCtxt} denoted by $\EV$, and
 \item 
 that which is defined in Lemma~\ref{lem:bottomUpDefOfEvalContext} 
 denoted by $\EValt$.
\end{itemize}
We are set out to show  $\EV=\EValt$. We rely on the following facts:
\begin{enumerate}
 \item\label{item:EVisSubstClosed} if $E, E'\in\EV$ then $E[E']\in \EV$; and
 \item\label{item:EVAltisSubstClosed} if $D,D'\in\EValt$ then $D[D']\in \EValt$.
\end{enumerate}
The former is proved by induction on the construction of $E'$; 
the latter is  by induction on the construction of $D$.

One direction $\EV\subseteq\EValt$ is proved easily by induction. We
 present only one case. For $E\equiv E'[[\place]M]\in \EV$, by the
 induction hypothesis we have $E'\in\EValt$; moreover $[\place]M\in
 \EValt$. Therefore by the fact~\ref{item:EVAltisSubstClosed} above,
 $E\equiv E'[[\place]M]$ belongs to $\EValt$.

 The other direction $\EValt\subseteq\EV$  is similar; we present only
 one case. For $D\equiv D'M\in \EValt$, by the induction hypothesis
 we have $D'\in \EV$; moreover $[\place]M\in \EV$. 
Therefore by the fact~\ref{item:EVisSubstClosed} above,
 $D\equiv D'M=[D']M$ 
belongs to $\EV$.
\myqed
\end{myproof}

\subsection{Proof of Lemma~\ref{lem:propertiesOfTheSubtypeRelation}}
\begin{myproof}
\ref{item:subtpIsPreorder}. 
Reflexivity $A\subtp A$ is easy by induction
 on the construction of $A$. 
 Transitivity
 \begin{displaymath}
  A \subtp A'
  \quad\text{and}\quad
    A' \subtp A''
  \qquad\Longrightarrow\qquad
  A \subtp A''
 \end{displaymath}
 is shown by induction on the derivation. We present one case; the other
 cases are similar. Assume that $A\subtp A'$ is derived by the
 $(\boxtimes)$  rule:
 \begin{equation}\label{eq:subtpTrans1}
\vcenter{     \infer[(\boxtimes)]{A\equiv\bang ^{n}(B\boxtimes C)\subtp
   \bang ^{n'}(B'\boxtimes C')\equiv A'}{B\subtp B' & C\subtp C'
   &{n=0 \Rightarrow n'=0}}
}\enspace. \end{equation}
 The form $A'\equiv \bang
 ^{n'}(B'\boxtimes C')$ requires the relation $A'\subtp A''$ to be
 derived also by the $(\boxtimes)$  rule:
 \begin{equation}\label{eq:subtpTrans2}
\vcenter{     \infer[(\boxtimes)]{A'\equiv\bang ^{n'}(B'\boxtimes C')\subtp
   \bang ^{n''}(B''\boxtimes C'')\equiv A''}{B'\subtp B'' & C'\subtp C''
   &{n'=0 \Rightarrow n''=0}}
}\enspace. 
 \end{equation}
 Now we apply the induction hypothesis to $B\subtp B'$ and $B'\subtp
 B''$ in~(\ref{eq:subtpTrans1}--\ref{eq:subtpTrans2}), and
 obtain that $B\subtp B''$ is derivable. Similarly $C\subtp C''$ is
 derivable. That $n=0\Rightarrow n''=0$ follows immediately from
 (\ref{eq:subtpTrans1}--\ref{eq:subtpTrans2}), too. Using
 the  $(\boxtimes)$ rule we derive $A\subtp A''$. 

\ref{item:bangIsMonotone}.
By cases on the rule that derives $A\subtp B$. We present the
 case  $(\limp)$:
\begin{displaymath}
\vcenter{    \infer[(\limp)]{A\equiv\bang ^{n}(A_{1}\limp A_{2})\subtp
   \bang ^{m}(B_{1}\limp B_{2})\equiv B}{B_{1}\subtp A_{1} & A_{2}\subtp B_{2}
   &{n=0 \Rightarrow m=0}} 
}\enspace.
\end{displaymath}
Since $n+1=0\Rightarrow m+1=0$ is trivially true, using 
$B_{1}\subtp A_{1}$ and $A_{2}\subtp B_{2}$ we derive
$\bang ^{n+1}(A_{1}\limp A_{2})\subtp
   \bang ^{m+1}(B_{1}\limp B_{2})$.
The other cases are similar.

\ref{item:bangSubtp1}.
By cases on the outermost type constructor in $A$ (ignoring $\bang$). Assume
 it is $\limp$, with $A\equiv\bang^{k}(B\limp C)$. Then we have $B\subtp B$
 and $C\subtp C$ due to the item~\ref{item:subtpIsPreorder}.; and 
 $n=0 \Rightarrow m=0$ implies
 $n+k=0 \Rightarrow m+k=0$. Therefore
 \begin{displaymath}
\vcenter{    \infer[(\limp)]{\bang ^{n+k}(B\limp C)\subtp
   \bang ^{m+k}(B\limp C)}{B\subtp B & C\subtp C
   &{n+k=0 \Rightarrow m+k=0}} 
} 
 \end{displaymath}
derives $\bang ^{n}A\subtp \bang ^{m}A$, as
   required. The other cases are similar.

\ref{item:bangSubtp2}.
Straightforward, by cases on the rule that derives $\bang^{n}A\subtp\bang^{m} B$.

\ref{item:subtpDirectedSupInf}.
 We show existence of directed sups and infs by simultaneous induction,
 on the complexity of upper/lower bounds.

 Assume       $A_{1}\subtp \nqbit$ and
      $A_{2}\subtp \nqbit$. Then 
 $A_{1}\equiv \bang^{n_{1}}\nqbit$ and
 $A_{2}\equiv \bang^{n_{2}}\nqbit$ for some $n_{1},n_{2}\in\nat$. We define
 \begin{displaymath}
  A_{0}
  \;:\equiv\;
  \begin{cases}
   \bang\nqbit&\text{if $n_{1}\neq 0$ and $n_{2}\neq 0$,}
   \\
   \nqbit & \text{otherwise.}
  \end{cases}
 \end{displaymath}
 This $A_{0}$ is clearly a supremum of $A_{1}$ and $A_{2}$.

 Assume       $\nqbit\subtp A_{1}$ and
      $\nqbit\subtp A_{2}$. Then $A_{1}$ and $A_{2}$ must both be
 $\nqbit$, and $A_{0}:\equiv \nqbit$ is the infimum.

 In the cases where the given upper (or lower) bound is
 $\bang^{n+1}\nqbit$, $\top$ or $\bang^{n+1}\top$, we can similarly compute a supremum (or a
 infimum). 

 Assume $A_{1}\subtp B\limp C$ and
      $A_{2}\subtp B\limp C$. Then the rules
 in~(\ref{eq:HoqSubtypeRelation}) force that 
 $A_{1}\equiv \bang^{n_{1}}(B_{1}\limp C_{1})$ and
 $A_{2}\equiv \bang^{n_{2}}(B_{2}\limp C_{2})$, with
 $B\subtp B_{1}$,  $B\subtp B_{2}$,
 $C_{1}\subtp C$ and  $C_{2}\subtp C$.
 Since the complexity of $B$ or $C$ is smaller than that of $B\limp C$ we
 can use the induction hypothesis, obtaining
 $B_{0}$ as a infimum of $B_{1},B_{2}$ and $C_{0}$ as a supremum of
 $C_{1},C_{2}$. Now
 \begin{displaymath}
  A_{0}
  \;:\equiv\;
  \begin{cases}
   \bang (B_{0}\limp C_{0})&\text{if $n_{1}\neq 0$ and $n_{2}\neq 0$,}
   \\
   B_{0}\limp C_{0} & \text{otherwise.}
  \end{cases}
 \end{displaymath}
 is easily shown to be a supremum of $A_{1},A_{2}$.
 The other cases are similar.
\myqed
\end{myproof}

\subsection{Proof of Lemma~\ref{lem:monotonicity}}
\begin{myproof}
\auxproof{see scannedNotes/monotonicityHoq.eps}
 By induction on the derivation of $ \Delta\vdash M:A$. The proof is
 mostly straightforward; here we only present one case.

 Assume that the derivation of $ \Delta\vdash M:A$
looks as follows, with
the
 $(\mbox{$\limp$}.\text{I}_{2})$ rule the one applied last, and
 $\Delta=(\bang\Delta_{0},\Gamma)$, $M\equiv \lambda x^{B}. N$, 
 $A\equiv \bang^{n}(B'\limp C)$.
\begin{displaymath}
   \infer[(\mbox{$\limp$}.\text{I}_{2})]
  {\bang\Delta_{0},\Gamma\vdash\lambda x^{B}. N:\bang^{n}(B'\limp C)}
  {
   {x:B, \bang\Delta_{0},\Gamma\vdash N:C}
   &\quad
   {\FV(N)\subseteq |\Delta_{0}|\cup\{x\}}
   &\quad
   B'\subtp B
  }
\end{displaymath}
Since $\Delta'\subtp\Delta=(\bang\Delta_{0},\Gamma)$, we have
 $\Delta'=(\Delta'_{0},\Gamma')$ with $\Delta'_{0}\subtp\bang\Delta_{0}$
 and $\Gamma'\subtp\Gamma$. Furthermore, by
 Lemma~\ref{lem:propertiesOfTheSubtypeRelation}.\ref{item:bangSubtp2},
 $\Delta'_{0}$ must be of the form $\Delta'_{0}=\bang\Delta''_{0}$ with some
 $\Delta''_{0}$. Thus $\Delta'=(\bang\Delta''_{0},\Gamma')$.
 Similarly, from the assumption that $A\equiv\bang^{n}(B'\limp C)\subtp
 A'$ we have $A'\equiv \bang^{m}(B''\limp C'')$ with $B''\subtp B'$,
 $C\subtp C''$ and $n=0\Rightarrow m=0$.

 Now we have
 \begin{math}
  (x:B,\bang\Delta''_{0},\Gamma')
  \subtp
  (x:B,\bang\Delta_{0},\Gamma)
 \end{math}. Using the induction hypothesis we obtain 
 $\Vdash x:B,\bang\Delta''_{0},\Gamma'\vdash N:C''$.
 Since $\FV(N)\subseteq|\Delta_{0}|\cup\{x\}=|\Delta''_{0}|\cup\{x\}$,
 the $(\mbox{$\limp$}.\text{I}_{2})$ rule can be applied.
\begin{displaymath}
   \infer[(\mbox{$\limp$}.\text{I}_{2})]
  {\bang\Delta''_{0},\Gamma'\vdash\lambda x^{B}. N:\bang^{m}(B''\limp C'')}
  {
   {x:B, \bang\Delta''_{0},\Gamma'\vdash N:C''}
   &\quad
   {\FV(N)\subseteq |\Delta''_{0}|\cup\{x\}}
   &\quad
   {B''\subtp B}
  }
\end{displaymath} 
To obtain $B''\subtp B$ we used transitivity
 (Lemma~\ref{lem:propertiesOfTheSubtypeRelation}.\ref{item:subtpIsPreorder}). 
This derives $\Delta'\vdash \lambda x^{B}. N:A'$. \myqed
\end{myproof}


\subsection{Proof of Lemma~\ref{lem:promotionOnlyForValues}}
\begin{myproof}
\ref{item:bangTypeThenFVAreBangType}.
 By induction on the construction of a value $V$.

 If $V\equiv x$, a variable, then the type judgment must be derived by the
 $(\text{Ax.}1)$ rule. Then the claim follows immediately from
 Lemma~\ref{lem:propertiesOfTheSubtypeRelation}.\ref{item:bangSubtp2}. 

 If $V$ is some constant 
 (i.e. $\new, \meas^{n+1}_{i},
  U,
    \cmp_{m,n}, or
    \new_{\rho}$),
or if $V\equiv *$, $\FV(V)$ is empty.

 If $V\equiv\lambda x^{B}. M$, the type judgment $\Delta\vdash V:\bang A$ must be derived by the 
$(\mbox{$\limp$}.\text{I}_{2})$ rule:
\begin{displaymath}
   \infer[(\mbox{$\limp$}.\text{I}_{2})]
  {\bang\Delta_{0},\Gamma\vdash\lambda x^{B}. M:\bang^{n}(B'\limp C)}
  {
   {x:B, \bang\Delta_{0},\Gamma\vdash M:C}
   &\quad
   {\FV(M)\subseteq |\Delta_{0}|\cup\{x\}}
   &\quad
   B'\subtp B
  }
\end{displaymath}  
 with $\Delta=(\bang\Delta_{0},\Gamma)$ and $A\equiv\bang^{n-1}(B\limp
 C)$. 
 Now we have
 \begin{math}
  \FV(V)=\FV(M)\setminus\{x\} \subseteq 
  \bigl(|\Delta_{0}|\cup\{x\}\bigr)\setminus\{x\}
 =|\Delta_{0}|
 \end{math},
 from which the claim follows.

 If $V\equiv\tuple{V_{1},V_{2}}$, the type judgment $\Delta\vdash
 V:\bang A$ must be derived as follows:
 \begin{displaymath}
  \infer[(\mbox{$\boxtimes$}.\text{I}), (\dagger)]
  {\bang\Delta_{0},\Gamma_{1},\Gamma_{2}\vdash
  \tuple{V_{1},V_{2}}:\bang^{n}(A_{1}\boxtimes A_{2})}
  {
  \bang\Delta_{0},\Gamma_{1}\vdash V_{1}:\bang^{n}A_{1}
  &\quad
  \bang\Delta_{0},\Gamma_{2}\vdash V_{2}:\bang^{n}A_{2}
  }
 \end{displaymath}
 with $\Delta=(\bang\Delta_{0},\Gamma_{1},\Gamma_{2})$.
 By the induction hypothesis, we have 
 \begin{displaymath}
  (\bang\Delta_{0},\Gamma_{1})\rest{\FV(V_{1})} 
  \;=\; \bang\Delta_{1}\enspace,
  \qquad
  (\bang\Delta_{0},\Gamma_{2})\rest{\FV(V_{2})} 
  \;=\; \bang\Delta_{2}
 \end{displaymath}
 for some $\Delta_{1}$ and $\Delta_{2}$ (here $
 (\bang\Delta_{0},\Gamma_{1})\rest{\FV(V_{1})} $ denotes the suitable
 restriction of a context). The claim follows immediately. The cases
 where $V\equiv   \injl^{B}\, V'$ or
$V\equiv \injr^{A}\, V'$ are similar. 

\ref{item:promotionOnlyForValues}. 
By induction on the construction of a value $V$.

 If $V\equiv x$, a variable, then the claim follows easily from
 Lemma~\ref{lem:propertiesOfTheSubtypeRelation}.\ref{item:bangSubtp2}. 
 The cases where $V$ is a constant or $V\equiv *$ are similarly easy.

 In case  $V\equiv\lambda x^{B}. M$: if $A$ is of the form
 $A\equiv\bang^{n}(B'\limp C)$ with $n\ge 1$, the type judgment 
$\bang\Delta,\Gamma\vdash V:A$ is derived by the
 $(\mbox{$\limp$}.\text{I}_{2})$ rule and it also derives
 $\bang\Delta,\Gamma\vdash V:\bang A$. If $A$ is of the form 
 $A\equiv B'\limp C$, the derivation of $\bang\Delta,\Gamma\vdash V:A$ looks as follows.
 \begin{equation}\label{eq:lemma:promotionOnlyForValues1}
 \vcenter{ \infer[(\mbox{$\limp$}.\text{I}_{1})]
  {\bang\Delta,\Gamma\vdash\lambda x^{B}. M:B'\limp C}
  {
   x:B, \bang\Delta,\Gamma\vdash M:C
   &\quad
  B'\subtp B
  }}
 \end{equation}
 Now the assumption 
 $\FV(V)\subseteq|\Delta|$ yields $\FV(M)\subseteq|\Delta|\cup\{x\}$; 
 this can be used in
 \begin{displaymath}
  \vcenter{\infer[(\mbox{$\limp$}.\text{I}_{2})]
  {\bang\Delta,\Gamma\vdash\lambda x^{B}. M:\bang(B'\limp C)}
  {
   x:B, \bang\Delta,\Gamma\vdash M:C
   &\quad
   \FV(M)\subseteq|\Delta|\cup\{x\}
   &\quad
  B'\subtp B
  }}\enspace.
 \end{displaymath}
  Thus we have derived $\bang\Delta,\Gamma\vdash V:\bang A$.

 Finally, the cases where
$V\equiv   \injl^{B}\, V'$ or
$V\equiv \injr^{A}\, V'$ are easy using the induction hypothesis. This
 concludes the proof.
\myqed
\end{myproof}

\subsection{Proof of Lemma~\ref{lem:substitutionLemma}}
\begin{myproof}
  \auxproof{See scannedNotes/substSubjRedProgress.eps.}
The first two rules are 
 straightforward, by induction on the derivation of $\bang\Delta,\Gamma_{2},x:A \vdash
 N:B$. Here we make  essential use of the 
monotonicity rule
(Lemma~\ref{lem:monotonicity}) and the
weakening rule
 (Lemma~\ref{lem:lemmasInTyping}.\ref{item:weakeningInHoq}).
The third rule $(\text{Subst}_{3})$ follows from $(\text{Subst}_{2})$ 
via
Lemma~\ref{lem:lemmasInTyping}.\ref{item:removingUnusedContextInHoq}
and~\ref{lem:promotionOnlyForValues}.\ref{item:bangTypeThenFVAreBangType}.
The last rule $(\text{Subst}_{4})$ for evaluation contexts is proved 
by induction on the construction of $E$, where we employ the ``bottom-up''
 definition (Lemma~\ref{lem:bottomUpDefOfEvalContext}) in place of
 Definition~\ref{definition:valueAndEvCtxt}.
\myqed
\end{myproof}

\subsection{Proof of Lemma~\ref{lem:subjectReductionLemma}}
\begin{myproof}
 \auxproof{See scannedNotes/substSubjRedProgress.eps.}
By induction on the construction of an evaluation context
 $E$. The step cases where $E\not\equiv [\place]$ are easy, since the
 local character of the typing rules of $\Hoq$ (for a rule to be
 applied, the terms in the assumptions can be anything). For the base
 case (i.e.\ $E\equiv [\place]$) we prove by cases according to
 Definition~\ref{definition:operationalSemantics}.

 In the case where $M\longrightarrow_{p}N$ is by the $(\limp)$ rule
 of Definition~\ref{definition:operationalSemantics}, we have $p=1$,  
 $M\equiv (\lambda x^{A'}.M')V$ and $N\equiv M'[V/x]$. By the assumption 
 we have $\Vdash \Delta\vdash (\lambda x^{A'}.M')V:A$; inspection of the
 typing rules shows that its derivation must look like the following.
 \begin{equation}
 \vcenter{ 
 \infer[(\mbox{$\limp$}.\text{E})] 
  {\bang\Delta',\Gamma_{1},\Gamma_{2}\vdash (\lambda x^{A'}.M')V:A}
  {
  \infer[(\mbox{$\limp$}.\text{I}_{1})]{\bang\Delta',\Gamma_{1}\vdash
  \lambda x^{A'}.M':B\limp A}{
    \infer*{x:A',\bang\Delta',\Gamma_{1}\vdash M':A}{}
    &
    {B\subtp A'}
   }
  &\quad
  \infer*[]{\bang\Delta',\Gamma_{2}\vdash V:C}{}
  &\quad
  C\subtp B
  }}
 \label{diagram:lemmasubjectReductionLemma1}
 \end{equation}
Here $\Delta=(\Delta',\Gamma_{1},\Gamma_{2})$.
 We have $C\subtp B\subtp A'$, thus $C\subtp A'$
 (Lemma~\ref{lem:propertiesOfTheSubtypeRelation}.\ref{item:subtpIsPreorder}). Using 
 the 
monotonicity rule
(Lemma~\ref{lem:monotonicity}) we have $\Vdash
 \bang\Delta',\Gamma_{2}\vdash V:A'$. Combining this with the top-left judgment
in~(\ref{diagram:lemmasubjectReductionLemma1}) via 
the $(\text{Subst}_{3})$ rule in Lemma~\ref{lem:substitutionLemma}, we
 obtain $\Vdash \bang\Delta',\Gamma_{1},\Gamma_{2}\vdash M'[V/x]:A$,
 which is our goal.

The cases of the $(\boxtimes), (\top)$, and $(+_{i})$ rules are
 similar, where we rely on the substitution rules
in Lemma~\ref{lem:substitutionLemma}.

We consider the case of the $(\text{rec})$ rule, where $p=1$, 
$M\equiv \letreccl{f^{B\limp C} x=M'}{N'}$
and 
\begin{align*}
 N&\equiv
 N'\bigl[
 (\lambda x^{B}.\, \letreccl{f^{B\limp C} x=M'}{M'})/f
\bigr]
\\&
 \equiv
 N'\bigl[
 (\lambda z^{B}.\, \letreccl{f^{B\limp C} x=M'}{M'[z/x]})\,/\,f
\bigr]\enspace.
\end{align*}
Here $z$ is a fresh variable and we used the $\alpha$-equivalence.
By the assumption 
 we have $\Vdash \Delta\vdash
\letreccl{f^{B\limp C} x=M'}{N'}
:A$; inspection of the
 typing rules shows that its derivation must look like the following.
 \begin{equation}
 \vcenter{ 
 \infer[(\text{rec})] 
  {\bang\Delta',\Gamma\vdash
    \letreccl{f^{B\limp C} x=M'}{N'}
   :A}
  {
  \infer*{\bang\Delta',f:\bang(B\limp C), x:B \vdash
   M':C}{}
  &\quad
  \infer*[]{\bang\Delta',\Gamma,f:\bang(B\limp C)\vdash N':A}{}
  }}
 \label{diagram:lemmasubjectReductionLemma2}
 \end{equation}
Here $\Delta=(\Delta',\Gamma)$. Now by $\alpha$-converging 
$\bang\Delta',f:\bang(B\limp C), x:B \vdash
   M':C$---the top-left
 judgment in~(\ref{diagram:lemmasubjectReductionLemma2})---we have
 \begin{math}
  \Vdash \bang\Delta',f:\bang(B\limp C), z:B \vdash
   M'[z/x]:C
 \end{math}. Applying the $(\text{rec})$ rule to the last two judgments,
 we obtain
 \begin{math}
  \Vdash \bang\Delta', z:B \vdash
  \letreccl{f^{B\limp C} x=M'}{M'[z/x]}:C
 \end{math}. By the $(\mbox{$\limp$}.\text{I}_{2})$ rule this leads to
 \begin{math}
  \Vdash \bang\Delta' \vdash
  \lambda z^{B}.\, \letreccl{f^{B\limp C} x=M'}{M'[z/x]}:\bang(B\limp C)
 \end{math}. The last judgment is combined with the second assumption
 in~(\ref{diagram:lemmasubjectReductionLemma2}) via the substitution
 rule $(\text{Subst}_{2})$ in Lemma~\ref{lem:substitutionLemma}, and yields
 \begin{math}
 \Vdash \bang\Delta',\Gamma\vdash N'\bigl[\,
  \lambda z^{B}.\, \letreccl{f^{B\limp C} x=M'}{M'[z/x]}\,/\,f
\,\bigr]:A
 \end{math}. This is our goal.

In the other cases, the reduction $M\longrightarrow_{p}N$ is derived by
 one of the rules in Definition~\ref{definition:operationalSemantics} that
 deal with quantum constants (such as $\new$ and $U$). We do only one case,
 of the rule $(\meas_{1})$. The other cases are similar.

By inspecting the typing rules it is easy to see that the type $A$ of
 the term $\meas^{n+1}_{i}(\new_{\rho})$ must be
 $A\equiv\bang\bit\boxtimes n\text{-}\qbit$. Therefore it suffices to
 show that the term $\ttrue\equiv\injl^{\top}(*)$ can indeed have the
 type $\bang\bit\equiv\bang(\top+\top)$. This is shown as follows.
\begin{displaymath}
 \vcenter{
 \infer[(+.\text{I}_{1})]
   {\Delta\vdash \injl^{\top}(*):\bang(\top+\top)}
   {\infer[(\mbox{$\top$}.\text{I})]{\Delta\vdash *:\bang\top}{}}
}
\end{displaymath}
This concludes the proof. \myqed
\end{myproof}

\subsection{Proof of Lemma~\ref{lem:progress}}
\begin{myproof}
 \auxproof{See scannedNotes/substSubjRedProgress.eps.}
By induction on the construction of the (closed) term $M$. We only present the
 case where $M\equiv NL$. If $N$  is not a value,
by the induction hypothesis $N$ has a reduction
 $N\longrightarrow_{p}N'$; this yields $M\equiv
 NL\longrightarrow_{p}N'L$. It is similar when $N$ is a value but $L$ is
 not.

Now assume that both $N$ and $L$ are values. By the assumption that $M\equiv NL$ is
 typable, we must have $\Vdash\,\vdash N: B\limp A$ for some $B$. A
 value $N$
 of the type $B\limp A$ must be either of the following forms: $\lambda
 x^{B'}.\, N'$, $\new$, $\meas^{n}_{i}$, $U$ or $\cmp$. 

 If $N\equiv\lambda
 x^{B'}.\, N'$, since $L$ is a value we have
 $M\equiv NL\longrightarrow_{1} N'[L/x]$. If $N\equiv\new$, it is easy
 to see that a closed
 value $L$ of type $\bit$ must be either $\ttrue$ or
 $\ffalse$. Therefore the reduction $(\new_{1})$ or $(\new_{2})$ in
 Definition~\ref{definition:operationalSemantics} is
 enabled from $M\equiv NL$. The other cases are similar.

\myqed
\end{myproof}


\section{The Quantum Branching Monad $\Q$}
\label{appendix:quantumBranchingMonad}

The following characterization is standard.
 See e.g.~\cite{NielsenC00}.
\begin{mylemma}\label{lem:reformulatedTraceCondForQuantumMonad}
 The trace condition~(\ref{eq:traceCondForQuantumMonad}) holds if
 and only if: for each $m\in\nat$,
\begin{equation}\label{eq:traceCondReformulated}
 \sum_{x\in X}\sum_{n\in\nat}M\Bigl(\,\bigl(\,c(x)\,\bigr)_{m,n}\,\Bigr)
 \;\Lle\;
 \IM_{m}\enspace.
\end{equation}
 Here $\Lle$ is the L\"{o}wner partial order
 (Definition~\ref{definition:LoewnerPartialOrder});
 $M\bigl((c(x))_{m,n}\bigr)$ is the
 matrix
 from Definition~\ref{definition:theMatrixMofE}. Note that 
 $M\bigl((c(x))_{m,n}\bigr)$
 is an $m\times
 m$ matrix regardless of the choice of $n$, hence the sum
 in~(\ref{eq:traceCondReformulated}) makes sense.
\end{mylemma}
\begin{myproof}
 We define a matrix $A$ by
 \begin{equation}\label{eq:proofInReformulatedTraceCondForQuantumMonad}
  A\;:=\; \IM_{m} - 
 \sum_{x\in X}\sum_{n\in\nat}M\Bigl(\,\bigl(\,c(x)\,\bigr)_{m,n}\,\Bigr)\enspace.
 \end{equation}
 To prove the `if' part, 
 assume that $A$ is positive. We have, for each $\rho\in \DM_{m}$,
 \begin{align*}
  & \trace(A\rho) + \sum_{x\in X}\sum_{n\in\nat}\trace\Bigl((c(x))_{m,n}(\rho)\Bigr)  
  \\
  &= 
   \trace(A\rho) + \sum_{x\in X}\sum_{n\in\nat}\trace
   \Bigl(\sum_{i\in I_{x,m,n}}
   E^{(i)}_{x,m,n}\cdot\rho\cdot(E^{(i)}_{x,m,n})^{\dagger}
  \Bigr)
  \\
  \tag*{\text{where $\{E_{x,m,n}^{(i)}\}_{i\in I_{x,m,n}}$ is an 
 operator-sum representation of
  $(c(x))_{m,n}$}}
  \\
  &= 
   \trace(A\rho) + \sum_{x\in X}\sum_{n\in\nat}\trace
   \Bigl(\sum_{i\in I_{x,m,n}}
   (E^{(i)}_{x,m,n})^{\dagger}\cdot E^{(i)}_{x,m,n}\cdot\rho
  \Bigr)
  \\
  &= 
   \trace\Bigl(\,
   \bigl(\,A+\sum_{x\in X}\sum_{n\in\nat}M\bigl((c(x))_{m,n}\bigr)\,\bigr)(\rho)
   \,\Bigr)
  \\
  &= 
   \trace(\rho) 
  \le 1
  \tag*{\text{by~(\ref{eq:proofInReformulatedTraceCondForQuantumMonad}).}}
\end{align*}
 Hence it suffices to show that $\trace(A\rho)\ge 0$. 
 It is a standard fact that any density matrix
 \begin{math}
  \rho\in\DM_{m}
 \end{math} can be written as
  \begin{displaymath}
 \sum_{i\in I}\lambda_{i}\ket{v_{i}}\bra{v_{i}}\enspace,
 \end{displaymath}
  with $\ket{v_{i}}\in \C^{m}$, $\abs{\ket{v_{i}}}=1$, $\lambda_{i}\ge 0$
 and $\sum_{i}\lambda_{i}\le 1$. Therefore it suffices to show that
$\trace(A\ket{v}\bra{v})\ge 0$ if
 $\abs{\ket{v}}=1$. Now:
 \begin{displaymath}
  \trace(A\rho)
 =
  \trace(A\ket{v}\bra{v})
 \stackrel{(*)}{=}
  \bra{v}A\ket{v}
 \ge 0\enspace,
 \end{displaymath}
 where $(*)$ is 
because $\trace(BC)=\trace(CB)$ for any $B,C$;
and the last inequality holds
 because $A$ is positive.

 For the `only if' part, we must show that the matrix $A$
 in~(\ref{eq:proofInReformulatedTraceCondForQuantumMonad}) is
 positive. For that purpose it suffices to prove: for any
 $\ket{v}\in\C^{m}$ with length $1$, $\bra{v}A\ket{v}\ge 0$.
 \begin{align*}
  \bra{v}A\ket{v}
 &=
  \bigl\langle v\bigl|\,
  \IM_{m}-\sum_{x,n}\sum_{i}(E^{(i)}_{x,m,n})^{\dagger}E^{(i)}_{x,m,n}
  \,\bigr|v\bigr\rangle
 \\
  \tag*{\text{
where $\{E_{x,m,n}^{(i)}\}_{i\in I_{x,m,n}}$ is an 
 operator-sum representation of
  $(c(x))_{m,n}$}}
 \\
 &=
 \inpr{v}{v}
 -\sum_{x,n}\sum_{i}\bra{v}\,
 (E^{(i)}_{x,m,n})^{\dagger}E^{(i)}_{x,m,n}
 \,\ket{v}
 \\
 &=
 1
-\sum_{x,n}\sum_{i}
 \trace\Bigl(\,
E^{(i)}_{x,m,n}\ket{v}\bra{v} (E^{(i)}_{x,m,n})^{\dagger}
  \,\Bigr)
  \tag*{\text{using 
  $\trace(BC)=\trace(CB)$ and  $\inpr{v}{v}=\abs{v}^{2}=1$}}
 \\
 &=
 1
-\sum_{x,n}
 \trace\Bigl(\,
 (c(x))_{m,n}\bigl(\,\ket{v}\bra{v}\,\bigr)
  \,\Bigr)
 \quad\ge 0
 \tag*{by~(\ref{eq:traceCondForQuantumMonad}).}
 \end{align*}
This concludes the proof. \myqed
\end{myproof}
%

\begin{myproposition}\label{proposition:QIsAFunctor}
 The construction $\Q$ in Definition~\ref{definition:quantumBranchingMonad} is 
 indeed a functor.
\end{myproposition}
\begin{myproof}
 First we check that, given a function $f:X\to Y$ and $c\in\Q X$, the
 data $(\Q f)(c)$ defined in~(\ref{eq:QuantumMonadOnArrows})
 indeed satisfies the trace
 condition. This is easy by direct calculations.
\auxproof{For each $m\in\nat$,
 \begin{align*}
 & 
 \sum_{y\in Y}\sum_{n\in\nat} M\bigl(\,  \bigl(\,(\Q f)(c)(y)\,\bigr)_{m,n}\,\bigr)
 \\
 &
 =
 \sum_{y\in Y}\sum_{n\in\nat} M\bigl(\,\sum_{x\in f^{-1}(\{y\})}
  (c(x))_{m,n}\,\bigr)
 \\
 &=
 \sum_{y\in Y}\sum_{n\in\nat} \sum_{x\in f^{-1}(\{y\})}
  M\bigl(\,
  (c(x))_{m,n}\,\bigr)
 \tag*{by Lemma~\ref{lem:MPreservesSums}}
 \\
 &=
 \sum_{x\in X}\sum_{n\in\nat}
  M\bigl(\,
  (c(x))_{m,n}\,\bigr)
 \tag*{\text{(*)}}
 \\ 
 &\Lle
 \IM_{m}
 \tag*{since $c\in \Q X$.}
 \end{align*}
Here $(*)$ holds because, due to $f$ being a function, we have
 \begin{displaymath}
  \coprod_{y\in Y}\{x\mid f(x)= y\} 
 \;=\; X\enspace.
 \end{displaymath}
}
It remains to be shown that: $\Q(\id)=\id$ and $\Q(g\co f)=\Q g\co \Q f$.
These are easy consequences of the facts that $\id^{-1}=\id$ and $(g\co f)^{-1}=
 f^{-1}\co g^{-1}$, respectively. \myqed
\end{myproof}

\begin{mylemma}\label{lem:multiplicationOfQIsIndeedWellDefined}
 The sum in the definition~(\ref{eq:multOfQuantumMonad}) of $\mu$
 is well-defined.
\end{mylemma}
\begin{myproof}
 First we show that, for fixed $\gamma\in\Q\Q X$,  $m\in\nat$
 and $\rho\in \DM_{m}$, there are only countably many pairs $(c,k)\in \Q
 X\times \nat$ such that
 \begin{displaymath}
  (\gamma(c))_{m,k}(\rho)\neq 0\enspace,
  \quad\text{equivalently (because the matrix is positive),}\;
  \trace\Bigl(\,(\gamma(c))_{m,k}(\rho)\,\Bigr)\neq 0\enspace.
 \end{displaymath}
 To see this, observe that the trace
 condition~(\ref{eq:traceCondForQuantumMonad}) for $\gamma\in
 \Q\Q X$ means $\sum_{c,k}\trace\bigl((\gamma(c))_{m,k}(\rho)\bigr)\le 1$.
 It is a standard fact
 that a
 discrete distribution with sum $\le 1$ has at most a countable support;
 from this our claim above   follows.

Therefore we can enumerate all such pairs as $((c_{l},k_{l}))_{l\in\nat}$. Then~(\ref{eq:multOfQuantumMonad}) 
 amounts  to
\begin{displaymath}
   \bigl(\,\mu_{X}(\gamma)(x)\,\bigr)_{m,n}(\rho)
 \;=\;
 \sum_{l\in\nat}
\Bigl(\, \bigl(c_{l}(x)\bigr)_{k_{l},n}  \co
 \bigl(\gamma(c_{l})\bigr)_{m,k_{l}}\,\Bigr)(\rho)\;.
\end{displaymath}
The right-hand side is the limit of a sequence (over $l\in\nat$) in $\DM_{n}$ 
that satisfies the assumption of
 Lemma~\ref{lem:convergenceViaTraceNorm}.
Thus it is well-defined.
\myqed
\end{myproof}

\auxproof{
***The following (using~(\ref{eq:traceCondReformulated})) has
turned out to be more complicated than
using~(\ref{eq:traceCondForQuantumMonad})***

In proving that $\Q$ is indeed a monad
(Proposition~\ref{proposition:QIsAMonad}) we use one lemma. It can
be seen as an analogue of the following property of 2-step probabilistic
branching:

\quad\begin{minipage}[b]{.3\textwidth}
  \noindent Let $p_{i}, q_{i,j}\in [0,1]$, $\sum_{i\in I}p_{i}\le 1$ and $\sum_{j\in
 J_{i}} q_{i,j}\le 1$ for each $i\in I$. Then we have 
 \begin{math}
 \sum_{i\in I}\sum_{j\in J_{i}}q_{i,j}\cdot p_{i}\le 1 
 \end{math}.
\end{minipage}
\includegraphics[width=.1\textwidth]{twoLayerProbBranching.eps}

\begin{mylemma}\label{lem:twoStepQuantumBranching}
 Let $\{\Eop_{i}\}_{i\in I}$ be a family  of QOs with
 $\Eop_{i}\in \QO_{m,k_{i}}$. For each $i\in I$, let
  $\{\Fop_{i,j}\}_{j\in J_{i}}$ be again a family of QOs with
 $\Fop_{i,j}\in \QO_{k_{i}, n_{i,j}}$.
 Assume
 \begin{displaymath}
  \sum_{i\in I}M(\Eop_{i})\Lle \IM_{m}\enspace,
  \quad\text{and}\quad
  \sum_{j\in J_{i}}M(\Fop_{i,j})\Lle \IM_{k_{i}} 
  \;\text{for each $i$.}
 \end{displaymath}
 Then we have
 \begin{displaymath}
  \sum_{i\in I}\sum_{j\in J_{i}} M(\Fop_{i,j}\co \Eop_{i}) \Lle
  \IM_{m}\enspace.
 \end{displaymath}
  Recall that $\Fop_{i,j}\co \Eop_{i}$ is a sequential composition of
 QOs and belongs to $\QO_{m,n_{i,j}}$.
\end{mylemma}
\begin{myproof}
 \marginpar{Perhaps easier using the trace condition~(\ref{eq:traceCondForQuantumMonad})?}
 Let $\{E_{i}^{(h)}\}_{h\in H_{i}}$ be an operator-sum representation of
 $\Eop_{i}$;
 and $\{F_{i,j}^{(l)}\}_{l\in L_{i,j}}$ be that of $\Fop_{i,j}$.
 Then the composition $\Fop_{i,j}\co \Eop_{i}$ can be represented as
 \begin{equation}\label{eq:oprSumReprOfComposedQO}
  \Fop_{i,j}\co \Eop_{i} 
  =
  \sum_{h\in H_{i}}\sum_{l\in L_{i,j}} F_{i,j}^{(l)}E_{i}^{(h)}
  (\place)
  \bigl(E_{i}^{(h)}\bigr)^{\dagger} \bigl(F_{i,j}^{(l)}\bigr)^{\dagger}\enspace.
 \end{equation}
 Now we have
 \begin{align*}
 & \IM_{m} -   \sum_{i\in I}\sum_{j\in J_{i}} M(\Fop_{i,j}\co \Eop_{i})
 \\
 &= 
   \IM_{m} -   \sum_{i\in I}\sum_{j\in J_{i}} \sum_{h\in H_{i}}\sum_{l\in
   L_{i,j}}\bigl(E_{i}^{(h)}\bigr)^{\dagger}
   \bigl(F_{i,j}^{(l)}\bigr)^{\dagger}F_{i,j}^{(l)}E_{i}^{(h)}
  \tag*{\text{by~(\ref{eq:oprSumReprOfComposedQO})}}
 \\
 &=
   \IM_{m} -   \sum_{i\in I}\sum_{h\in H_{i}}
   \bigl(E_{i}^{(h)}\bigr)^{\dagger}
   \Bigl(\, \sum_{j\in J_{i}} 
    M(\Fop_{i,j})
   \,\Bigr)
   E_{i}^{(h)}
 \\
 &=
   \IM_{m} -   \sum_{i\in I}\sum_{h\in H_{i}}
   \bigl(E_{i}^{(h)}\bigr)^{\dagger}
   \Bigl(\, 
    \IM_{k_{i}} - B_{i}^{\dagger} B_{i}
   \,\Bigr)
   E_{i}^{(h)}
 \tag*{by $\sum_{j} 
    M(\Fop_{i,j})\Lle \IM_{k_{i}}$ and
  Lemma~\ref{lem:charOfPositiveMatrix}.\ref{item:positiveMatrixCharThree}.}
 \\
 &=
   \IM_{m} -   \sum_{i\in I}\sum_{h\in H_{i}}
   \bigl(E_{i}^{(h)}\bigr)^{\dagger}
   E_{i}^{(h)}
   +
   \sum_{i\in I}\sum_{h\in H_{i}}
   \bigl(E_{i}^{(h)}\bigr)^{\dagger}
   B_{i}^{\dagger} B_{i}
   E_{i}^{(h)}
 \\
 &=
\Bigl(\,
   \IM_{m} -   
   \sum_{i\in I}
   M(\Eop_{i})
\,\Bigr)   +
   \sum_{i\in I}\sum_{h\in H_{i}}
   \bigl(B_{i} E_{i}^{(h)}\bigr)^{\dagger}
    B_{i}
   E_{i}^{(h)}
 \enspace.
 \end{align*}
 The first summand $ \IM_{m} - \sum_{i\in I} M(\Eop_{i}) $ is positive
   by assumption; the matrix $ \bigl(B_{i} E_{i}^{(h)}\bigr)^{\dagger}
   B_{i} E_{i}^{(h)} $ in the second summand is positive by
   Lemma~\ref{lem:charOfPositiveMatrix}.\ref{item:positiveMatrixCharThree}.
   It immediately follows from Definition~\ref{definition:positiveMatrix} that
   a sum of positive matrices is again positive; this concludes the
 proof.
  \myqed
\end{myproof}
}

\begin{myproposition}\label{proposition:QIsAMonad}
 The construction $\Q$ in Definition~\ref{definition:quantumBranchingMonad} is 
 indeed a monad.
\end{myproposition}
\begin{myproof}
 First we verify that the data $\eta_{X}(x)$
 in~(\ref{eq:unitOfQuantumMonad}) and $\mu_{X}(\gamma)$
 in~(\ref{eq:multOfQuantumMonad}) satisfy the trace
 condition~(\ref{eq:traceCondForQuantumMonad})  and hence belong
 indeed to $\Q X$. For the unit $\eta_{X}(x)$ this is obvious.  For the
 multiplication $\mu_{X}(\gamma)$ we shall
 verify~(\ref{eq:traceCondForQuantumMonad}). For any $\rho\in
 \DM_{m}$, we have
\begin{align*}
&
 \sum_{x\in X}\sum_{n\in \nat}
\trace\Bigl(\,\bigl(\,(\mu_{X}(\gamma))(x)\,\bigr)_{m,n}(\rho)\,\Bigr)
\\
&=
 \sum_{x\in X}\sum_{n\in \nat}
 \sum_{c\in \Q X}\sum_{k\in\nat}
\trace\Bigl(\,
(c(x))_{k,n}
\Bigl(\,(\gamma(c))_{m,k}(\rho)\,\Bigr)
\,\Bigr)
\\
&=
 \sum_{x\in X}\sum_{n\in \nat}
 \sum_{c\in \Q X}\sum_{k\in\nat}
\trace\Bigl(\,
(\gamma(c))_{m,k}(\rho)
\,\Bigr)
\cdot
\trace\Bigl(\,
(c(x))_{k,n}
\Bigl(\,
\frac{(\gamma(c))_{m,k}(\rho)}
{
\trace\Bigl(\,
(\gamma(c))_{m,k}(\rho)
\,\Bigr)
}
\,\Bigr)
\,\Bigr)
\tag*{\text{since $(c(x))_{k,n}$ and $\trace$ are linear}}
\\
&=
 \sum_{c\in \Q X}\sum_{k\in\nat}
\trace\Bigl(\,
(\gamma(c))_{m,k}(\rho)
\,\Bigr)
\cdot
\biggl(\,
 \sum_{x\in X}\sum_{n\in \nat}
\trace\Bigl(\,
(c(x))_{k,n}
\Bigl(\,
\frac{(\gamma(c))_{m,k}(\rho)}
{
\trace\Bigl(\,
(\gamma(c))_{m,k}(\rho)
\,\Bigr)
}
\,\Bigr)
\,\Bigr)
\,\biggr)
\\
&\le
 \sum_{c\in \Q X}\sum_{k\in\nat}
\trace\Bigl(\,
(\gamma(c))_{m,k}(\rho)
\,\Bigr)
\cdot 1
\tag*{\text{by the trace condition for $c\in\Q X$,
 (\textasteriskcentered)}}
\\
&\le
 1
\tag*{\text{by the trace condition for $\gamma\in\Q\Q X$.}}
\end{align*}
Note that in the above (\textasteriskcentered), the matrix
\begin{displaymath}
 \frac{(\gamma(c))_{m,k}(\rho)}
{
\trace\Bigl(\,
(\gamma(c))_{m,k}(\rho)
\,\Bigr)
}
\end{displaymath}
has its trace $1$ hence is a density matrix.

 Next we verify that the maps $\eta_{X}$ and $\mu_{X}$
 in~(\ref{eq:unitOfQuantumMonad}--\ref{eq:multOfQuantumMonad})
 are natural in $X$. For $\eta_{X}$ it is obvious. For $\mu_{X}$, given
 $\gamma\in\Q\Q X$ and $f:X\to Y$:
 \begin{align*}
 & \Bigl(\,(\mu_{Y}\co \Q\Q f)(\gamma)(y)\,\Bigr)_{m,n}
 \\
 &=
  \sum_{c'\in \Q Y}\sum_{k\in \nat}\,
  (c'(y))_{k,n}\co \bigl(\,(\Q\Q
  f)(\gamma)(c')\,\bigr)_{m,k}
 \\
 &=
  \sum_{c'\in \Q Y}\sum_{k\in \nat}\,
  (c'(y))_{k,n}\co
  \bigl(\,
  \sum_{c\in (\Q f)^{-1}(\{c'\})}
  (\gamma(c))_{m,k}
\,\bigr)
 \\
 &=
  \sum_{c'\in \Q Y}
  \sum_{c\in (\Q f)^{-1}(\{c'\})}
  \sum_{k\in \nat}\,
  (c'(y))_{k,n}\co
  (\gamma(c))_{m,k}
 \tag*{since $(c'(y))_{k,n}$ is linear}
 \\
 &=
  \sum_{c\in \Q X}
  \sum_{k\in \nat}\,
  \bigl(\,(\Q f)(c)(y)\,\bigr)_{k,n}\co
  (\gamma(c))_{m,k}
 \\
 &=
  \sum_{c\in \Q X}
  \sum_{k\in \nat}\,
  \bigl(\,\sum_{x\in f^{-1}(\{y\})}(c(x))_{k,n}\,\bigr)
  \co
  (\gamma(c))_{m,k}
 \\
 &=
 \sum_{x\in f^{-1}(\{y\})}
  \sum_{c\in \Q X}
  \sum_{k\in \nat}\,
  (c(x))_{k,n}
  \co
  (\gamma(c))_{m,k}
 \\
 &=
 \sum_{x\in f^{-1}(\{y\})}
  (\mu_{X}(\gamma)(x))_{m,n}
 =
 \Bigl(\,(\Q f\co \mu_{X})(\gamma)(y)\,\Bigr)_{m,n}\enspace.
 \end{align*}
 This proves the naturality of $\mu$.

 Finally we verify that $\eta$ and $\mu$ indeed satisfy the monad laws,
 that is, that the following diagrams commute.
 \begin{equation}\label{diagram:monadLaws}
\vcenter{  \xymatrix@R=1em@C-.7em{
   {\Q X}
      \ar[r]^-{\eta_{\Q X}}
      \ar@{=}[rd]
  &
   {\Q\Q X}
      \ar[d]^-{\mu_{X}}
  &
   {\Q X}
      \ar[l]_-{\Q\eta_{X}}
      \ar@{=}[ld]
  &
   {\Q\Q\Q X}
      \ar[r]^-{\Q\mu_{X}}
      \ar[d]_-{\mu_{\Q X}}
  &
   {\Q\Q X}
      \ar[d]^-{\mu_{X}}
  \\
  &
   {\Q X}
  &
  &
   {\Q\Q X}
      \ar[r]_-{\mu_{X}}
  &
   {\Q X}
}
} \end{equation}
The leftmost triangle is obvious; for the other triangle, we first
 observe
 \begin{equation}\label{eq:QOfEtaExplicitly}
  \Bigl(\,(\Q\eta_{X})(c)(c')\,\Bigr)_{m,n}
 \;=\;
  \begin{cases}
   (c(x'))_{m,n} & \text{if $c'=\eta_{X}(x')$ for some $x'\in X$,}
  \\
   0 & \text{otherwise.}
  \end{cases}
 \end{equation}
This is used in the following calculation.
 \begin{align*}
  \Bigl(\,\bigl(\,\mu_{X}\co (\Q \eta_{X})\, \bigr)(c)(x)\,\Bigr)_{m,n}
  &=
  \sum_{c'\in \Q X}\sum_{k\in \nat}\,(c'(x))_{k,n}\co
  \Bigl(\,(\Q\eta_{X})(c)(c')\,\Bigr)_{m,k}
 \\ 
  &=
  \sum_{x'\in X}\sum_{k\in \nat}\,\bigl(\,(\eta_{X}(x'))(x)\,\bigr)_{k,n}\co
  (c(x'))_{m,k}
  \tag*{\text{by~(\ref{eq:QOfEtaExplicitly})}}
 \\ 
  &=
  \IM_{n}\co (c(x))_{m,n} = (c(x))_{m,n}\enspace.
 \end{align*}
 This proves the commutativity of the triangle in the middle
 of~(\ref{diagram:monadLaws}).  For the square on the right, given
 $\Gamma\in \Q\Q\Q X$:
\begin{align*}
 \Bigl(\,(\mu_{X}\co \Q \mu_{X})(\Gamma)(x)\,\Bigr)_{m,n}
 &=
 \sum_{c\in \Q X}
 \sum_{k\in \nat}
 (c(x))_{k,n}
 \co
 \Bigl(\,(\Q\mu_{X})(\Gamma)(c)\,\Bigr)_{m,k}
 \\
 &=
 \sum_{c\in \Q X}
 \sum_{k\in \nat}
 (c(x))_{k,n}
 \co
 \Bigl(\,
  \sum_{\gamma\in \mu_{X}^{-1}(\{c\})} (\Gamma(\gamma))_{m,k}
 \,\Bigr)
 \\
 &=
 \sum_{c\in \Q X}
 \sum_{k\in \nat}
  \sum_{\gamma\in \mu_{X}^{-1}(\{c\})}
 (c(x))_{k,n}
 \co
 (\Gamma(\gamma))_{m,k}
 \\
 &=
  \sum_{\gamma\in \Q\Q X}
 \sum_{k\in \nat}\,
 (\mu_{X}(\gamma)(x))_{k,n}
 \co
 (\Gamma(\gamma))_{m,k}
 \\
 &=
  \sum_{\gamma\in \Q\Q X}
 \sum_{k\in \nat}
 \sum_{c\in \Q X}
 \sum_{l\in \nat}\,
 (c(x))_{l,n}
 \co
 (\gamma(c))_{k,l}
 \co
 (\Gamma(\gamma))_{m,k}\enspace;
\\
 \Bigl(\,(\mu_{X}\co  \mu_{\Q X})(\Gamma)(x)\,\Bigr)_{m,n}
&=
 \sum_{c\in \Q X}
 \sum_{l\in \nat}
 (c(x))_{l,n}
 \co
 (\mu_{\Q X}(\Gamma)(c))_{m,l}
\\
&=
 \sum_{c\in \Q X}
 \sum_{l\in \nat}
 (c(x))_{l,n}
 \co
 \bigl(\,\sum_{\gamma\in\Q\Q X}\sum_{k\in \nat}(\gamma(c))_{k,l}\co
 (\Gamma(\gamma))_{m,k}\,\bigr)
 \\
 &=
  \sum_{\gamma\in \Q\Q X}
 \sum_{k\in \nat}
 \sum_{c\in \Q X}
 \sum_{l\in \nat}\,
 (c(x))_{l,n}
 \co
 (\gamma(c))_{k,l}
 \co
 (\Gamma(\gamma))_{m,k}\enspace.
\end{align*}
This concludes the proof.
\myqed
\end{myproof}

\subsection{The Kleisli Category $\Kleisli{\Q}$}\label{subsection:KleisliQ}
\begin{mylemma*}[Lemma~\ref{lem:compositionOfQKleisliArrowsInDetail}, repeated]
 Given two successive arrows $f:X\relto Y$ and $g:Y\relto U$ in
 $\Kleisli{\Q}$, their composition $g\Kco f:X\relto U$ is concretely
 given as follows.
 \begin{displaymath}
\Bigl(\,( g \Kco f)(x)(u)\,\Bigr)_{m,n}
 \;=\;
  \sum_{y\in Y}\sum_{k\in\nat}
   \bigl(g(y)(u)\bigr)_{k,n}\co
  \bigl(f(x)(y)\bigr)_{m,k}\enspace.
\end{displaymath}
\end{mylemma*}
\begin{myproof}
Given $x\in X$, $ u\in U$ and $\rho\in \DM_{m}$:
 \begin{align*}
\Bigl(\,( g \Kco f)(x)(u)\,\Bigr)_{m,n}
 &
 =
  \Bigl(\,(\mu_{U}\co \Q g \co f)(x)(u)\,\Bigr)_{m,n}
 \\
 &
 =
  \Bigl(\,\mu_{U} \bigl(\,(\Q g) (f(x))\,\bigr)(u)\,\Bigr)_{m,n}
 \\
 &
 =
  \sum_{c\in\Q U}\sum_{k\in \nat}
   (c(u))_{k,n}\co \Bigl(\,\bigl(\,(\Q g)
  (f(x))\,\bigr)(c)\,\Bigr)_{m,k}
 \tag*{by def.\ of $\mu$}
 \\
 &
 =
  \sum_{c\in\Q U}\sum_{k\in \nat}
   (c(u))_{k,n}\co
  \Bigl(\,
    \sum_{y\in g^{-1}(\{c\})}
  (f(x)(y))_{m,k}
  \,\Bigr)
 \tag*{by def.\ of $\Q g$}
 \\
 &
 =
  \sum_{c\in\Q U}\sum_{k\in \nat}
    \sum_{y\in g^{-1}(\{c\})}
   (c(u))_{k,n}\co
  (f(x)(y))_{m,k}
 \\
 &
 =
  \sum_{y\in Y}\sum_{k\in\nat}
   (g(y)(u))_{k,n}\co
  (f(x)(y))_{m,k}\enspace.
 \tag*{$(*)$}
 \end{align*}
Here the equality $(*)$ holds because, due to $g:Y\to \Q U$ being a function, we
have
 \begin{math}
 Y= \coprod_{c\in \Q U}\{y\mid g(y)=c\} 
 \end{math}.
\myqed
\end{myproof}

Note that $\Kleisli{\Q}$ has finite coproducts, carried over from
$\Sets$ by the Kleisli inclusion functor.
\begin{mytheorem}
\label{theorem:KleisliOfQIsTracedInDetail}
 The monad $\Q$ on $\Sets$ satisfies the
 following conditions (from~\cite[Requirements~4.7]{Jacobs10trace}); and therefore 
 by~\cite[Proposition~4.8]{Jacobs10trace}, the category $\Kleisli{\Q}$ is partially
 additive.
\begin{enumerate}
 \item $\Kleisli{\Q}$ is $\omega$-$\mathbf{CPO}$ enriched.
 \item $\Kleisli{\Q}$ has monotone cotupling.
 \item For each $X,Y\in\Kleisli{\Q}$, the least element $\bot_{X,Y}\in
       \Kleisli{Q}(X,Y)$
       in the homset is preserved by both pre- and
       post-composition: that is, $f\Kco \bot =\bot$ and $\bot\Kco
       g=\bot$.

       We note that, under this condition, there exist ``projection'' maps
       $p_{j}:\coprod_{i\in I}X_{i}\relto X_{j}$ such that 
       \begin{displaymath}
	p_{j}\Kco \kappa_{i} \;=\;
	\begin{cases}
	 \id & \text{if $i=j$,}
	 \\
	 \bot & \text{otherwise,}
	\end{cases}
       \end{displaymath}
       where $\kappa_{j}:X_{j}\relto\coprod_{i\in I}X_{i}$ denotes a coprojection.
 \item The ``bicartesian'' maps
       \begin{displaymath}
	\bc_{(X_{i})_{i\in I}}\;:=\;
        \left(\,
	 \Q(\coprod_{i\in I}X_{i}) \stackrel{\tuple{p^{\flat}_{i}}_{i\in
        I}}{\longrightarrow}
	 \prod_{i\in I}\Q X_{i}
	\,\right)
	\quad\text{where}\quad
	p^{\flat}_{i}:= \mu\co Tp_{i}
       \end{displaymath}
       form a cartesian natural transformation with monic
       components. This means that all the naturality squares
       \begin{displaymath}
	\vcenter{\xymatrix@R=1em{
	 {T(\coprod_{i}X_{i})}
	      \ar@{ >->}[r]^-{\bc}
	      \ar[d]_{T(\coprod_{i}f_{i})}
	&
	 {\prod_{i}TX_{i}}
	      \ar[d]^{\prod_{i}Tf_{i}}
	\\
	 {T(\coprod_{i}Y_{i})}
	      \ar@{ >->}[r]_-{\bc}
	&
	 {\prod_{i}TY_{i}}
	}}
       \end{displaymath}
       are pullback diagrams in $\Sets$, for each $f_{i}:X_{i}\to Y_{i}$
       in $\Sets$.
\end{enumerate}
\end{mytheorem}
 The original
 condition~\cite[Requirements~4.7]{Jacobs10trace} is stated in terms of
 DCPOs instead of $\omega$-CPOs. This difference is not important.
\begin{myproof}
 We use the pointwise extension of the order $\Lle$ in
 Definition~\ref{definition:pointwiseorderOnQX}
 in homsets $\Kleisli{\Q}(X,Y)$. It is an $\omega$-CPO due to
 Proposition~\ref{proposition:QOisCPO}.
 It is easy to see that the bottoms are preserved by pre- and
 post-composition.
 \auxproof{Note 15/12/2010, p.2}
 To see that supremums are preserved too, one uses the following facts.
 \begin{itemize}
  \item  A QO is
 continuous, since its operator-sum representation is.
  \item The fact at the beginning of the proof of
	Lemma~\ref{lem:multiplicationOfQIsIndeedWellDefined} (that the
	support of each of the relevant functions is at most countable).
  \item  The limit operator $\lim_{k\to\infty}$
 (for increasing chains) and the countable sum operator $\sum_{l\in\nat}$
	 are interchangeable: $\lim_{k}\sum_{l}\rho_{k,l}=\sum_{l}\lim_{k}\rho_{k,l}$.
 \end{itemize}
 \auxproof{Note 15/12/2010, pages around 19} 
 Cotupling is monotone since the order in the homsets are pointwise.

 To see $\bc$ is monic, assume $\bc(c)=\bc(d)$. Then
 $p^{\flat}_{i}(c)=p^{\flat}_{i}(d)$ for each $i\in I$. It is easy to
 see that $p^{\flat}_{i}(c)=c\co \kappa_{i}$, therefore
 \begin{displaymath}
  c = [c\co \kappa_{i}]_{i}=[d\co \kappa_{i}]_{i} = d\enspace.
 \end{displaymath}
 It is straightforward to see that the naturality squares are pullbacks.
 \myqed
 \auxproof{Note 15/12/2010, page 12} 
\end{myproof}

\section{Proofs for~\S{}\ref{section:denotationalModel}}
\label{section:proofsForTheFixedPointOperatorSection}

\subsection{Proof of Lemma~\ref{lem:LCAIsCPO}}
\begin{myproof}
The set $A_{\Q}=\Kleisli{\Q}(\nat,\nat)$ is an $\omega$-CPO
due to the  $\omega\text{-}\mathbf{CPO}$ enriched structure of
the category $\Kleisli{\Q}$ (see Theorem~\ref{theorem:KleisliOfQIsTraced}).
Therefore the order $\sqsubseteq$ on $A_{\Q}$ is  essentially  the L\"{o}wner partial order
(Definition~\ref{definition:LoewnerPartialOrder}).

To show the item~\ref{item:applBangConti}, 
we use the fact that 
   composition $\Kco$ of arrows and the trace operator $\trace$ are both
continuous in the Kleisli category  $\Kleisli{\Q}$. Indeed, the former is part of the fact that $\Kleisli{\Q}$
 is $\omega\text{-}\mathbf{CPO}$ enriched
 (Theorem~\ref{theorem:KleisliOfQIsTracedInDetail}). The proof for the
 latter 
is not hard either, exploiting the explicit presentation of $\trace$ by
 Girard's execution formula (see~\cite[Chap.~3]{Haghverdi00PhD}). In the
proof the following Fubini-like result is essential: if
 $(x_{n,m})_{n,m\in\nat}$ is increasing both in $n$ and $m$, then
 \begin{math}
  \sup_{n}\sup_{m} x_{n,m}=\sup_{n} x_{n,n}
 \end{math}.
The item~\ref{item:applBangConti} then follows immediately from the 
definitions of $\cdot$ and $\bang$
in~(\ref{eq:applicativeStrOfGoILCAInStringDiagrams}) and (\ref{eq:bangStrOfGoILCAInStringDiagrams}). 

The item~\ref{item:applLeftStrict} is proved using
the presentation of $\trace$ by
 Girard's execution formula and the fact that composition $\Kco$
in $\Kleisli{\Q}$ is (left and right) strict.
\auxproof{see scannedNotes/AQCpo.eps} \myqed
\end{myproof}

\subsection{Proof of Lemma~\ref{lem:closurePropertyOfAdmissibility}}
\begin{myproof}
 For inductiveness of $U\dtimes V$, assume
 $x_{0}\sqsubseteq
 x_{1}\sqsubseteq \cdots$,
 $x'_{0}\sqsubseteq
 x'_{1}\sqsubseteq \cdots$ and $(x_{i},x'_{i})\in U\dtimes V$ for each $i$. 
 By the definition of $U\dtimes V$, we find 
 $k_{i},l_{i},u_{i},k'_{i},l'_{i},u'_{i}
$ 
 such that
 $x_{i}=\cmbt{\dot{P}} k_{i} (\cmbt{\dot{P}} l_{i} u_{i})$,
 $x'_{i}=\cmbt{\dot{P}} k'_{i} (\cmbt{\dot{P}} l'_{i} u'_{i})$,
 $(k_{i}u_{i},k'_{i}u'_{i})\in U$ and
 $(l_{i}u_{i},l'_{i}u'_{i})\in V$. 
 Since $k_{i}=\cmbt{\dot{P}_{l}} x_{i}$,  by continuity of $\cdot$
 we have that $(k_{i})_{i}$ is an increasing chain. So are
 $
(l_{i})_{i},
(u_{i})_{i},
(k'_{i})_{i},
(l'_{i})_{i},
(u'_{i})_{i}
$; therefore
 $
(k_{i}u_{i})_{i},
(l_{i}u_{i})_{i},
(k'_{i}u'_{i})_{i},
(l'_{i}u'_{i})_{i}
$ are increasing, too. By the admissibility of $U$ and $V$ we have
\begin{displaymath}
 \bigl(\,
 \sup_{i}k_{i}u_{i}
 \,,\,
 \sup_{i}k'_{i}u'_{i}
\,\bigr)\in U
 \quad\text{and}\quad
 \bigl(\,
 \sup_{i}l_{i}u_{i}
 \,,\,
 \sup_{i}l'_{i}u'_{i}
\,\bigr)\in V\enspace.
\end{displaymath}
Again by continuity of $\cdot$ we have
\begin{displaymath}
 \sup_{i}x_{i}
=\cmbt{\dot{P}}(\sup_{i}k_{i})
 \bigl(\cmbt{\dot{P}}(\sup_{i} l_{i})(\sup_{i}u_{i})\bigr)
 \quad\text{and}\quad
 \sup_{i}x'_{i}
=\cmbt{\dot{P}}(\sup_{i}k'_{i})
 \bigl(\cmbt{\dot{P}}(\sup_{i} l'_{i})(\sup_{i}u'_{i})\bigr)
\enspace;
\end{displaymath} 
since $\sup_{i}k_{i}u_{i} 
=(\sup_{i}k_{i})(\sup_{i}u_{i})$ (and so on) we conclude that
$(\sup_{i}x_{i},\sup_{i}x'_{i})\in U\dtimes V$.


 Strictness is the reason we use $\dtimes$ instead of $\times$. 
We have
 \begin{math}
  \cmbt{\dot{P}}
  \bot
  (
  \cmbt{\dot{P}}
  \bot
  \bot
  ) =\bot
 \end{math}: this is because $\cmbt{\dot{P}}xy = j\Kco (x+y)\Kco k$
 (see~(\ref{eq:dotPinStringDiagrams})) and that $\Kco$ is (left
 and right) strict. This shows  $(\bot,\bot)\in U\dtimes V$.

 Inductiveness of $X\limp U$ is easily shown by similar arguments.
 Finally, strictness of $X\limp U$ is because for each $(x,x')\in X$,
 $(\bot x,\bot x')=(\bot,\bot)\in U$. Here the left strictness of
 $\cdot$ is crucial.
 \myqed
\end{myproof}

\subsection{Proof of Lemma~\ref{lem:resultTypeIsAdmissible}}
\begin{myproof}
The PER  $\sem{\zeroqbit}$ (Definition~\ref{definition:semQbitAndSemBit})
is admissible. Indeed,
 $(\bot,\bot)=(\cmbt{Q}_{0},\cmbt{Q}_{0})\in\sem{\zeroqbit}$ and an
 increasing
 chain in $\sem{\zeroqbit}$ is precisely an increasing chain in $[0,1]$.
Therefore Lemma~\ref{lem:closurePropertyOfAdmissibility} shows that
the functor 
$\Fpbt=\sem{\bit}\limp(\sem{\zeroqbit}\dtimes\place)$
preserves admissibility. Since $\Bt=\{(\bot,\bot)\}$ is admissible,
each object $\Fpbt ^{i}B$ in the
 final sequence is admissible.

We prove strictness of $R$. By
 Definition~\ref{definition:sequenceAndNThDereliction}  we  have
$\bot=(\bot)_{i\in\nat}$; hence $\cmbt{\dot{P}}(\bot)_{i}\bot=\bot$
by~(\ref{eq:dotPinStringDiagrams}).
Thus it suffices to show that for any $j$ and any $i$ such that $j\le i$, 
$(c_{i,j}\bot,\bot)\in \Fpbt^{j}\Bt$. This is by cases: we distinguish
 $j=0$ and $j>0$.

If $j=0$, $[c_{i,j}]:\Fpbt^{i}\Bt\to\Bt$ is the unique map to
 $\Bt=\{(\bot,\bot)\}$ and hence $c_{i,j}\bot=\bot$. Therefore
$(c_{i,j}\bot,\bot)\in \Fpbt^{j}\Bt$ for any $i$.

Assume $j>0$. Let us first note the functor $\sem{\bit}\limp \place$'s
 action on arrows: it carries
\begin{equation}\label{eq:expFunctorActionOnArrows}
[c]:X\stackrel{}{\longrightarrow} Y
 \quad\text{to}\quad
[\lambda tb. c(tb)]:
 \sem{\bit}\limp X
 \longrightarrow
 \sem{\bit}\limp Y\enspace.
\end{equation}
The functor $\sem{\zeroqbit}\times \place$ carries
\begin{math}
 [c]:X\stackrel{}{\to} Y
\end{math}
to
\begin{equation}\label{eq:timesFunctorActionOnArrows}
\bigl[
\lambda v. v(\lambda k_{1}w. w(\lambda k_{2}u. 
\cmbt{P}
k_{1}
(
\cmbt{P} (\lambda z. c(k_{2}z))
u
)
))
\bigr]
:
 \sem{\zeroqbit}\times X
 \longrightarrow
 \sem{\zeroqbit}\times Y
 \enspace;
\end{equation}
after suitable insertion of the conversion combinators
$\cmbt{C_{P\mapsto\dot{P}}}$ and   $\cmbt{C_{\dot{P}\mapsto P}}$,
it
describes the functor 
$\sem{\zeroqbit}\dtimes \place$'s action on arrows.

Our aim now is to show
\begin{displaymath}
 (c_{i,j}\bot,\bot)
\;\in\; \Fpbt^{j}\Bt 
 \;=\;\sem{\bit}\limp(\sem{\zeroqbit}\dtimes \Fpbt^{j-1}\Bt)\enspace;
\end{displaymath}
by~(\ref{eq:homInPER}) it suffices to show
\begin{equation}\label{eq:lemmaresultTypeIsAdmissibleInDetail1}
 \bigl(\,
 c_{i,j}\bot b,\,\bot b'
\,\bigr)
\;\in\; 
\sem{\zeroqbit}\dtimes \Fpbt^{j-1}\Bt
\quad\text{for each $(b,b')\in \sem{\bit}$.}
\end{equation}
By~(\ref{eq:expFunctorActionOnArrows}) we have
\begin{displaymath}
 \bigl(\,c_{i,j}\,,\,\lambda tb. d_{i-1,j-1}(tb)\,\bigr)
\;\in\; 
 \Fpbt^{i}\Bt
\limp
 \Fpbt^{j}\Bt\enspace,
\end{displaymath}
where $d_{i-1,j-1}$ is the realizer of the arrow 
\begin{displaymath}
 \sem{\zeroqbit}\dtimes[c_{i-1,j-1}]\;:\;\sem{\zeroqbit}\dtimes\Fpbt^{i-1}\Bt\longrightarrow\sem{\zeroqbit}\dtimes\Fpbt^{j-1}\Bt
\end{displaymath}
described as in~(\ref{eq:timesFunctorActionOnArrows}).
Therefore
\begin{equation}\label{eq:lemmaresultTypeIsAdmissibleInDetail2}
 \bigl(\,c_{i,j}\bot b\,,\,d_{i-1,j-1}(\bot b)\,\bigr)
\;\in\; 
\sem{\zeroqbit}\dtimes \Fpbt^{j-1}\Bt\enspace.
\end{equation}
Now using that $\cdot$ is left strict,
\begin{align*}
 d_{i-1,j-1}(\bot b)
 =
d_{i-1,j-1}\bot
=
\bot
=
\bot b'\enspace,
\end{align*}
where for the second equality we also used the concrete
 description~(\ref{eq:timesFunctorActionOnArrows}) of
 $d_{i-1,j-1}$. Therefore~(\ref{eq:lemmaresultTypeIsAdmissibleInDetail2})
 proves~(\ref{eq:lemmaresultTypeIsAdmissibleInDetail1}).

 Inductiveness of $R$ is proved much like the proof of Lemma~\ref{lem:closurePropertyOfAdmissibility}.
 \auxproof{scannedNotes/finalCoalgebraAdmissible.eps}
\myqed
\end{myproof}

\section{Well-Definedness of
  Interpretation of Well-Typed Terms}\label{section:wellDefinednessOfInterpretation}
\auxproof{
Records on principal typing:
\begin{itemize}
 \item First trial: a derivable judgment $\Delta\vdash M:A$ is such that
       $|\Delta|=\FV(M)$. But then rules such as
       $(\mbox{$\limp$}.\text{I})$ become not applicable.
 \item Second trial (June 2013): weakening allowed. But then
       Hoshino-san's counter example:
       \begin{displaymath}
	\infer{x,y,z\vdash xy:A}{x,z\vdash x:A\limp A & y\vdash y:A}
	\text{  vs. }
       \infer{x,y,z\vdash xy:A}{x\vdash x:A\limp A & y,z\vdash y:A}
       \end{displaymath}
       (see scannedNotes/welldfd.eps)
 \item Third solution (July 2013): go back to the first trial, allowing 
       some free variables to be absent
\end{itemize}
}
Towards our goal of proving Lemma~\ref{lem:interpretationIsWellDfd},
 we introduce another set of typing rules, and  we
call them the \emph{principal typing rules}. The system is a restriction of the
$\Hoq$ typing rules (Table~\ref{table:typingRules}).

\begin{mydefinition}[Principal typing in $\Hoq$]\label{definition:principalTypingRules}
 The \emph{principal typing rules} of $\Hoq$ are in
 Table~\ref{table:principalTypingRules}. 

\begin{table}[tbp] \normalsize
\begin{displaymath}
\begin{array}{ll}
   \vcenter{\infer[(\text{Ax.}1)_{\mathrm{P}}]{
    x:A\vdash x:A}{}}
 &
   \vcenter{\infer[(\text{Ax.}2)_{\mathrm{P}}]{
   \vdash c:\bang \DType(c)}{}
}
 \\[+1em]
 \multicolumn{2}{l}{ \infer[(\mbox{$\limp$}.\text{I}_{1})_{\mathrm{P}}]
  {\Delta\vdash\lambda x^{A}. M:A\limp B}
  {
   [x:A], \Delta\vdash M:B
   &\quad
   \Delta
\not\equiv\bang\Delta'
   \;\text{for any $\Delta'$}
  }}
 \\[+1em]
 \multicolumn{2}{l}{  \infer[(\mbox{$\limp$}.\text{I}_{2})_{\mathrm{P}}]
   {\Delta\vdash\lambda x^{A}. M:\bang(A\limp B)}
   {
    {[x:A], \Delta\vdash M:B}
    &\quad
     \Delta
     \equiv\bang\Delta'
     \;\text{for some $\Delta'$}
   }
 }
  \\[+1em]
  \multicolumn{2}{l}{\infer[(\mbox{$\limp$}.\text{E})_{\mathrm{P}}]
  {\bang\Delta,\Gamma_{1},\Gamma_{2}\vdash MN:B}
  {
  \bang\Delta,\Gamma_{1}\vdash M:\bang^{n}(A\limp B)
  &\quad
  \bang\Delta,\Gamma_{2}\vdash N:C
  &\quad
  C\subtp A
  }}
  \\[+1em]
  \multicolumn{2}{l}{\infer[(\mbox{$\boxtimes$}.\text{I})_{\mathrm{P}}]
  {\bang\Delta,\Gamma_{1},\Gamma_{2}\vdash
  \tuple{M_{1},M_{2}}:\bang^{m}(\bang^{m} A_{1}\boxtimes \bang^{m} A_{2})}
  {
  \bang\Delta,\Gamma_{1}\vdash M_{1}:\bang^{n
     }A_{1}
  &
  \bang\Delta,\Gamma_{2}\vdash M_{2}:\bang^{n
     }A_{2}
  &
  {\footnotesize
  \begin{array}[b]{l}
   \bigl(m=0\Leftrightarrow  n=0 
    \bigr)
   \\  \land
  \bigl(m=1\Leftrightarrow n\ge 1  
  \bigr)
  \end{array}  
  }
  &
  {\footnotesize
  \begin{array}[b]{l}
   \text{At least one of  $A_{1}$ and $A_{2}$ is}
    \\
   \text{not of the form $\bang B$}
  \end{array}
  }
}}
 \\[+1em]
  \multicolumn{2}{l}{
  \infer[(\mbox{$\boxtimes$}.\text{E})_{\mathrm{P}}]
  {\bang\Delta,\Gamma_{1},\Gamma_{2}\vdash
  \letcl{\tuple{x_{1}^{\bang^{n
        }A_{1}},x_{2}^{\bang^{n
        }A_{2}}}=M}{N}:A}
  {
    \begin{array}[b]{l}
     {\bang\Delta,\Gamma_{1}\vdash M:\bang^{m}(C_{1}\boxtimes
    C_{2})}
    \\
    {\bang\Delta,\Gamma_{2},[x_{1}:\bang^{n
     }A_{1}],[x_{2}:\bang^{n
     }A_{2}]\vdash
    N:A}
    \end{array}    
    &\quad
    {
    \begin{array}[b]{l}
     m=0\Rightarrow n=0 
     \\
     C_{1}\subtp A_{1}\quad
    C_{2}\subtp A_{2}
    \end{array}
    }
  }}
  \\[+1em]
  \infer[(\mbox{$\top$}.\text{I})_{\mathrm{P}}]
  {\Delta\vdash *:\bang\top}
  {\phantom{\top}
  }
  &
  \infer[(\mbox{$\top$}.\text{E})_{\mathrm{P}}]
  {\bang\Delta,\Gamma_{1},\Gamma_{2}\vdash
  \letcl{*=M}{N}:A}
  {
  \bang\Delta,\Gamma_{1}\vdash M:\bang^{n}\top
  &\quad
  \bang\Delta,\Gamma_{2}\vdash N:A
  }
 \\[+1em]
   \multicolumn{2}{l}{
  \vcenter{\infer[(+.\text{I}_{1})_{\mathrm{P}}]
  {\Delta\vdash\injl^{A_{2}} M:\bang^{m}(\bang^{m}A_{1}+A_{2})}
  {\Delta\vdash M:\bang^{n}A_{1}
  &
  {\footnotesize
  \begin{array}[b]{l}
   \bigl(m=0\Leftrightarrow  n=0 
    \bigr)
   \\  \land
  \bigl(m=1\Leftrightarrow n\ge 1  
  \bigr)
  \end{array}  
  }
  &
  {\footnotesize
  \begin{array}[b]{l}
   \text{$A_{1}$ is}
    \\
   \text{not of the form $\bang B$}
  \end{array}
  }
  }}}
 \\[+1em]
   \multicolumn{2}{l}{
  \vcenter{\infer[(+.\text{I}_{2})_{\mathrm{P}}]
  {\Delta\vdash\injr^{A_{1}} N:\bang^{m}(A_{1}+\bang^{m} A_{2})}
  {\Delta\vdash N:\bang^{n}A_{2}
  &
  {\footnotesize
  \begin{array}[b]{l}
   \bigl(m=0\Leftrightarrow  n=0 
    \bigr)
   \\  \land
  \bigl(m=1\Leftrightarrow n\ge 1  
  \bigr)
  \end{array}  
  }
  &
  {\footnotesize
  \begin{array}[b]{l}
   \text{$A_{2}$ is}
    \\
   \text{not of the form $\bang B$}
  \end{array}
  }
}}}
  \\[+2em]
 \multicolumn{2}{l}{  \infer[(+.\text{E})_{\mathrm{P}}]
  {
     \bang\Delta,\bang\Delta_{1},\bang\Delta_{2},\Gamma,\Gamma',
     \Gamma'_{1},\Gamma'_{2}
     \vdash
     \matchcl{P}{(x_{1}^{\bang^{n
         }A_{1}}\mapsto
      M_{1}\mid x_{2}^{\bang^{n
      }A_{2}}\mapsto M_{2})}:B
  }
  {
   \begin{array}[b]{l}
    \bang\Delta,\bang\Delta_{1},\bang\Delta_{2},\Gamma\vdash P:\bang^{m}(C_{1}+C_{2})
    \\
    \bang\Delta,\bang\Delta_{1},\Gamma',\Gamma'_{1},[x_{1}:\bang^{n
        }A_{1}]\vdash M_{1}:B
    \\
    \bang\Delta,\bang\Delta_{2},\Gamma',\Gamma'_{2},[x_{2}:\bang^{n
        }A_{2}]\vdash M_{2}:B
   \end{array}
  &\quad
    \begin{array}[b]{l}
     m=0\Rightarrow n=0 
     \\
     C_{1}\subtp A_{1}\quad
    C_{2}\subtp A_{2}
    \\
  x_{1}\not\in|\Gamma,\Delta_{2},\Gamma_{2}|,
  x_{2}\not\in|\Gamma,\Delta_{1},\Gamma_{1}|
    \end{array}
  }}
 \\[+1em]
 \multicolumn{2}{l}{
  \infer[(\text{rec})_{\mathrm{P}}]
  {\bang\Delta,\Gamma\vdash
  \letreccl{f^{A\limp B}x=M}{N}:C}
  {
  \begin{array}[b]{l}
   {\bang\Delta,[f:\bang(A\limp B)],[x:A]
          \vdash M:B''}
   \\
   {\bang\Delta,\Gamma,[f:\bang(A\limp B)]\vdash N:C}
  \end{array}  
  &\quad
  {B''\subtp B}
}}
\end{array}
\end{displaymath}
 \caption{Principal typing rules for $\Hoq$}
 \label{table:principalTypingRules}
\end{table}

\auxproof{
\begin{table}[tbp] \normalsize
(** This is obsolete!! **)
\begin{displaymath}
\begin{array}{ll}
   \vcenter{\infer[(\text{Ax.}1)_{\mathrm{P}}]{
   \Delta, x:A\vdash x:A}{}}
 &
   \vcenter{\infer[(\text{Ax.}2)_{\mathrm{P}}]{
   \Delta\vdash c:\bang \DType(c)}{}
}
 \\[+1em]
 \multicolumn{2}{l}{ \infer[(\mbox{$\limp$}.\text{I}_{1})_{\mathrm{P}}]
  {\Delta\vdash\lambda x^{A}. M:A\limp B}
  {
   x:A, \Delta\vdash M:B
   &\quad
   \Delta\rest{\FV(M)\setminus\{x\}}\not\equiv\bang\Delta'
   \;\text{for any $\Delta'$}
  }}
 \\[+1em]
 \multicolumn{2}{l}{  \infer[(\mbox{$\limp$}.\text{I}_{2})_{\mathrm{P}}]
   {\Delta\vdash\lambda x^{A}. M:\bang(A\limp B)}
   {
    {x:A, \Delta\vdash M:B}
    &\quad
     \Delta\rest{\FV(M)\setminus\{x\}}\equiv\bang\Delta'
     \;\text{for some $\Delta'$}
   }
 }
  \\[+1em]
  \multicolumn{2}{l}{\infer[(\mbox{$\limp$}.\text{E})_{\mathrm{P}}]
  {\bang\Delta,\Gamma_{1},\Gamma_{2}\vdash MN:B}
  {
  \bang\Delta,\Gamma_{1}\vdash M:\bang^{n}(A\limp B)
  &\quad
  \bang\Delta,\Gamma_{2}\vdash N:C
  &\quad
  C\subtp A
  }}
  \\[+1em]
  \multicolumn{2}{l}{\infer[(\mbox{$\boxtimes$}.\text{I})_{\mathrm{P}}]
  {\bang\Delta,\Gamma_{1},\Gamma_{2}\vdash
  \tuple{M_{1},M_{2}}:\bang^{m}(\bang^{m} A_{1}\boxtimes \bang^{m} A_{2})}
  {
  \bang\Delta,\Gamma_{1}\vdash M_{1}:\bang^{n
     }A_{1}
  &
  \bang\Delta,\Gamma_{2}\vdash M_{2}:\bang^{n
     }A_{2}
  &
  {\footnotesize
  \begin{array}[b]{l}
   \bigl(m=0\Leftrightarrow  n=0 
    \bigr)
   \\  \land
  \bigl(m=1\Leftrightarrow n\ge 1  
  \bigr)
  \end{array}  
  }
  &
  {\footnotesize
  \begin{array}[b]{l}
   \text{$A_{1}$ or $A_{2}$ is}
    \\
   \text{not of the form $\bang B$}
  \end{array}
  }
}}
 \\[+1em]
  \multicolumn{2}{l}{
  \infer[(\mbox{$\boxtimes$}.\text{E})_{\mathrm{P}}]
  {\bang\Delta,\Gamma_{1},\Gamma_{2}\vdash
  \letcl{\tuple{x_{1}^{\bang^{n
        }A_{1}},x_{2}^{\bang^{n
        }A_{2}}}=M}{N}:A}
  {
    \begin{array}[b]{l}
     {\bang\Delta,\Gamma_{1}\vdash M:\bang^{m}(C_{1}\boxtimes
    C_{2})}
    \\
    {\bang\Delta,\Gamma_{2},x_{1}:\bang^{n
     }A_{1},x_{2}:\bang^{n
     }A_{2}\vdash
    N:A}
    \end{array}    
    &\quad
    {
    \begin{array}[b]{l}
     m=0\Rightarrow n=0 
     \\
     C_{1}\subtp A_{1}\quad
    C_{2}\subtp A_{2}
    \end{array}
    }
  }}
  \\[+1em]
  \infer[(\mbox{$\top$}.\text{I})_{\mathrm{P}}]
  {\Delta\vdash *:\bang\top}
  {\phantom{\top}
  }
  &
  \infer[(\mbox{$\top$}.\text{E})_{\mathrm{P}}]
  {\bang\Delta,\Gamma_{1},\Gamma_{2}\vdash
  \letcl{*=M}{N}:A}
  {
  \bang\Delta,\Gamma_{1}\vdash M:\bang^{n}\top
  &\quad
  \bang\Delta,\Gamma_{2}\vdash N:A
  }
 \\[+1em]
   \multicolumn{2}{l}{
  \vcenter{\infer[(+.\text{I}_{1})_{\mathrm{P}}]
  {\Delta\vdash\injl^{A_{2}} M:\bang^{m}(\bang^{m}A_{1}+A_{2})}
  {\Delta\vdash M:\bang^{n}A_{1}
  &
  {\footnotesize
  \begin{array}[b]{l}
   \bigl(m=0\Leftrightarrow  n=0 
    \bigr)
   \\  \land
  \bigl(m=1\Leftrightarrow n\ge 1  
  \bigr)
  \end{array}  
  }
  &
  {\footnotesize
  \begin{array}[b]{l}
   \text{$A_{1}$ is}
    \\
   \text{not of the form $\bang B$}
  \end{array}
  }
  }}}
 \\[+1em]
   \multicolumn{2}{l}{
  \vcenter{\infer[(+.\text{I}_{2})_{\mathrm{P}}]
  {\Delta\vdash\injr^{A_{1}} N:\bang^{m}(A_{1}+\bang^{m} A_{2})}
  {\Delta\vdash N:\bang^{n}A_{2}
  &
  {\footnotesize
  \begin{array}[b]{l}
   \bigl(m=0\Leftrightarrow  n=0 
    \bigr)
   \\  \land
  \bigl(m=1\Leftrightarrow n\ge 1  
  \bigr)
  \end{array}  
  }
  &
  {\footnotesize
  \begin{array}[b]{l}
   \text{$A_{2}$ is}
    \\
   \text{not of the form $\bang B$}
  \end{array}
  }
}}}
  \\[+1em]
 \multicolumn{2}{l}{  \infer[(+.\text{E})_{\mathrm{P}}]
  {
     \bang\Delta,\Gamma,\Gamma'
     \vdash
     \matchcl{P}{(x_{1}^{\bang^{n
         }A_{1}}\mapsto
      M_{1}\mid x_{2}^{\bang^{n
      }A_{2}}\mapsto M_{2})}:B
  }
  {
   \bang\Delta,\Gamma\vdash P:\bang^{m}(C_{1}+C_{2})
  &\quad
   \begin{array}[b]{l}
    \bang\Delta,\Gamma',x_{1}:\bang^{n
        }A_{1}\vdash M_{1}:B
    \\
    \bang\Delta,\Gamma',x_{2}:\bang^{n
        }A_{2}\vdash M_{2}:B
   \end{array}
  &\quad
    \begin{array}[b]{l}
     m=0\Rightarrow n=0 
     \\
     C_{1}\subtp A_{1}\quad
    C_{2}\subtp A_{2}
    \end{array}
  }}
 \\[+1em]
 \multicolumn{2}{l}{
  \infer[(\text{rec})_{\mathrm{P}}]
  {\bang\Delta,\Gamma\vdash
  \letreccl{f^{A\limp B}x=M}{N}:C}
  {
  \begin{array}[b]{l}
   {\bang\Delta,f:\bang(A\limp B),x:A
          \vdash M:B''}
   \\
   {\bang\Delta,\Gamma,f:\bang(A\limp B)\vdash N:C}
  \end{array}  
  &\quad
  {B''\subtp B}
}}
\end{array}
\end{displaymath}
 \caption{Principal typing rules for $\Hoq$, second trial (obsolete)}
 \label{table:principalTypingRules}
\end{table}
}
In the rules, a square-bracketed entry like $[x:A]$ in a context
means that it can be absent. The contexts in the
 $(+.\text{E})_{\mathrm{P}}$
rule are complicated: 
they form the following partition of the free variables of the term 
in the conclusion. Notice that $\Gamma'$ need not be of the form $\bang\Gamma''$.
 \begin{displaymath}
\scalebox{1} 
{
\begin{pspicture}(0,-2.2643652)(5.3792577,2.2643652)
\pscircle[linewidth=0.04,dimen=outer](1.6941992,-0.918916){1.3}
\pscircle[linewidth=0.04,dimen=outer](3.2941992,-0.918916){1.3}
\pscircle[linewidth=0.04,dimen=outer](2.4941993,0.481084){1.3}
\usefont{T1}{ptm}{m}{n}
\rput(0.31507814,-2.113916){$\FV(P)$}
\usefont{T1}{ptm}{m}{n}
\rput(4.863828,-2.113916){$\FV(M_2)$}
\usefont{T1}{ptm}{m}{n}
\rput(2.649326,2.086084){$\FV(M_1)$}
\usefont{T1}{ptm}{m}{n}
\rput(1.3574121,-1.313916){$\Gamma$}
\usefont{T1}{ptm}{m}{n}
\rput(2.5806642,-0.513916){$\Delta$}
\usefont{T1}{ptm}{m}{n}
\rput(2.4636524,-1.313916){$\Delta_2$}
\usefont{T1}{ptm}{m}{n}
\rput(1.8491504,-0.11391602){$\Delta_1$}
\usefont{T1}{ptm}{m}{n}
\rput(2.8258984,0.88608396){$\Gamma_1$}
\usefont{T1}{ptm}{m}{n}
\rput(4.0404005,-1.113916){$\Gamma_2$}
\usefont{T1}{ptm}{m}{n}
\rput(3.1826563,-0.11391602){$\Gamma'$}
\end{pspicture} 
}
\end{displaymath}

 We shall write $\Pi\VdashP \Delta\vdash M:A$ if a derivation tree
 $\Pi$, according to these rules, derives the type judgment. 
We write $\VdashP \Delta\vdash M:A$ if there
 exists such $\Pi$, that is, the type judgment is derivable.
\end{mydefinition}

\begin{mylemma}\label{lem:fromPrincipalTypingToTyping}
 $\VdashP\Delta\vdash M:A$ implies
 $\Vdash\Delta\vdash M:A$.
\end{mylemma}
\begin{myproof}
 By induction on the principal type derivation of $\VdashP\Delta\vdash
 M:A$. We only present some cases.

 When the last rule applied is $(\mbox{$\limp$}.\text{E})_{\mathrm{P}}$,
 that is
 \begin{displaymath}
  \infer[(\mbox{$\limp$}.\text{E})_{\mathrm{P}}]
  {\bang\Delta',\Gamma_{1},\Gamma_{2}\vdash M'N':A}
  {
  \bang\Delta',\Gamma_{1}\vdash M':\bang^{n}(A'\limp A)
  &\quad
  \bang\Delta',\Gamma_{2}\vdash N':C
  &\quad
  C\subtp A'
  }
 \end{displaymath}
 with $\Delta=(\bang\Delta',\Gamma_{1},\Gamma_{2})$ and $M\equiv
 M'N'$, by the induction hypothesis we have
 \begin{displaymath}
  \Vdash \bang\Delta',\Gamma_{1}\vdash M': \bang^{n}(A'\limp A)
  \quad\text{and}\quad
  \Vdash \bang\Delta',\Gamma_{2}\vdash N':C\enspace.
 \end{displaymath} 
 Applying
Corollary~\ref{corollary:derelictionAdmsble} to the
 former yields
  \begin{math}
  \Vdash \bang\Delta',\Gamma_{1}\vdash M':A'\limp A\enspace.
 \end{math} Then, together with $C\subtp A'$, we can use the
 $(\mbox{$\limp$}.\text{E})$ rule (of the original type system) to
 derive $\bang\Delta',\Gamma_{1},\Gamma_{2}\vdash M'N':A$.
 The case $(\mbox{$\boxtimes$}.\text{I})_{\mathrm{P}}$ is similar using
  Corollary~\ref{corollary:derelictionAdmsble}. For the
 case
$(\mbox{$\boxtimes$}.\text{E})_{\mathrm{P}}$ we additionally use
 Lemma~\ref{lem:monotonicity} to show that
 \begin{displaymath}
    \Vdash  
    \bang\Delta,\Gamma_{2},x_{1}:\bang^{n
     }A_{1},x_{2}:\bang^{n
     }A_{2}\vdash
    N:A
    \quad\text{implies}\quad
    \Vdash  
    \bang\Delta,\Gamma_{2},x_{1}:\bang^{n
     }C_{1},x_{2}:\bang^{n
     }C_{2}\vdash
    N:A
    \enspace.
 \end{displaymath}
 The case $(+.\text{E})_{\mathrm{P}}$ and $(\text{rec})_{\mathrm{P}}$
 are similar. \myqed
 \auxproof{For the latter, 
 Lemma~\ref{lem:monotonicity} is used to show that, under
 the assumptions that $C_{1}\subtp A_{1}$ and $C_{2}\subtp A_{2}$,
 \begin{align*}
    &\Vdash  
    {\bang\Delta,f:\bang^{n}(A\limp B),x:A\vdash M:B''}
    \quad\text{implies}\quad
    \Vdash  
    {\bang\Delta,f:\bang^{n}(A\limp B''),x:A\vdash M:B''}
    \enspace,
  \\
&\Vdash  
   {\bang\Delta,\Gamma,f:\bang(A\limp B)\vdash N:C}
    \quad\text{implies}\quad
    \Vdash  
   {\bang\Delta,\Gamma,f:\bang(A\limp B'')\vdash N:C}
    \enspace.
 \end{align*}
}
\end{myproof}

\begin{mylemma}\label{lem:uniquePrincipalTyping}
\begin{enumerate}
 \item $\VdashP \Delta\vdash M:A$ implies $|\Delta|=\FV(M)$.
 \item 
 Principal typing is unique in the following sense:  
 \begin{quote}\normalsize
 $\VdashP \Delta\vdash M:A$ and
 $\VdashP \Delta\vdash M:A'$ 
imply
 $A\equiv A'$.
 \end{quote}
 \item 
  Derivation in principal typing is unique: if $\Pi\VdashP\Delta\vdash
       M:A$
 and $\Pi'\VdashP\Delta\vdash M:A$, then $\Pi\equiv\Pi'$.
\end{enumerate}
\end{mylemma}
\begin{myproof}
1. Straightforward by induction.

2.
 By induction on the construction of $M$. We only present one case; the
 other cases are similar.
\auxproof{see scannedNotes/welldfd.eps}

 Assume $M$ is of the form 
\begin{math}
\letcl{\tuple{x_{1}^{\bang^{n}A_{1}},x_{2}^{\bang^{n}A_{2}}}=M'}{N}  
\end{math}. Then its principal type derivation must end with the 
$(\mbox{$\boxtimes$}.\text{E})_{\mathrm{P}}$ rule, as below.
\begin{equation}\label{eq:lemma:uniquePrincipalTyping1}
 \infer[(\mbox{$\boxtimes$}.\text{E})_{\mathrm{P}}]
  {\bang\Delta,\Gamma_{1},\Gamma_{2}\vdash
  \letcl{\tuple{x_{1}^{\bang^{n}A_{1}},x_{2}^{\bang^{n}A_{2}}}=M'}{N}:A}
  {
    \begin{array}[b]{l}
     {\bang\Delta,\Gamma_{1}\vdash M':\bang^{m}(C_{1}\boxtimes
    C_{2})}
    \\
    {\bang\Delta,\Gamma_{2},[x_{1}:\bang^{n}A_{1}],[x_{2}:\bang^{n}A_{2}]\vdash
    N:A}
    \end{array}    
    &\quad
    {
    \begin{array}[b]{l}
     m=0\Rightarrow n=0 
     \\
     C_{1}\subtp A_{1}\quad
    C_{2}\subtp A_{2}
    \end{array}
    }
  }
\end{equation}
The context $\bang\Delta,\Gamma_{2},
[x_{1}:\bang^{n_{1}}A_{1}],[x_{2}:\bang^{n_{2}}A_{2}]
$ is
determined by the given context $\bang\Delta,\Gamma_{1},\Gamma_{2}$---in
particular we can read off the types $\bang^{n}A_{i}$ of the
variables $x_{i}$ from the explicit type labels in $M$. Therefore by the
induction hypothesis, the principal type $A$ of $N$ is determined; hence
 so is
 the principal type of $M$, too.

3. Straightforward by induction on the construction of a term $M$. 
In many cases (including $M\equiv NL$, where the rule
 $(\mbox{$\limp$}.\text{E})_{\mathrm{P}}$ is involved) 
the items 1--2.\ play an essential role. \myqed
\end{myproof}

\begin{mydefinition}[Interpretation of principal type judgments]
\label{definition:interpretationOfPrincipalTypeJudgments}
 For each derivation $\Pi\VdashP \Delta\vdash M:A$ by the rules in
Table~\ref{table:principalTypingRules}, we assign an arrow 
 \begin{displaymath}
  \semP{\Pi}\;:\;\sem{\Delta}\longrightarrow T\sem{A} 
 \end{displaymath}
 in the way that is a 
 straightforward adaptation of 
Definition~\ref{definition:interpretationOfTypeJudgments}. 

\end{mydefinition}

\begin{mylemma}\label{lem:typableThenPrincipalTypable}
 Assume $\Pi\Vdash\Delta\vdash M:A$ (in the original rules in
 Table~\ref{table:typingRules}). Then there exist a type $A^{\circ}$ 
 and a derivation $\Pi^{\circ}$ (in the principal typing rules in
 Table~\ref{table:principalTypingRules}) such that:
 \begin{enumerate}
  \item $\Pi^{\circ}\VdashP\Delta\rest{\FV(M)}\vdash M:A^{\circ}$,
  \item $A^{\circ}\subtp A$, and
  \item the following diagram commutes.
	\begin{equation}\label{diagram:lemmatypableThenPrincipalTypable1}
	 \vcenter{\xymatrix@R=1em@C+1em{
	  {\sem{\Delta}}
	     \ar[r]^-{\weak}
             \ar@/_/[rrd]_-{\sem{\Pi}}
         &
	  {\sem{\Delta\rest{\FV(M)}}}
	     \ar[r]^-{\semP{\Pi^{\circ}}}
         &
          {T\sem{A^{\circ}}}
             \ar[d]^{T\sem{A^{\circ}\subtp A}}
         \\
	 &&
	  {T\sem{A}}
	 }}
	\end{equation}
 \end{enumerate}
\end{mylemma}
\begin{myproof}
 The diagram
 in~(\ref{diagram:lemmatypableThenPrincipalTypable1}) can be refined
 into the following one; we shall prove that the triangle therein commutes.
	\begin{equation}\label{diagram:lemmatypableThenPrincipalTypable2}
	 \vcenter{\xymatrix@R=1em@C+2em{
	  {\sem{\Delta}}
	     \ar[r]^-{\weak}
         &
	  {\sem{\Delta|_{\FV(M)}}}
	     \ar[r]^-{\semP{\Pi^{\circ}}}
             \ar[rd]_-{\fvsem{\Pi}}
         &
          {T\sem{A^{\circ}}}
             \ar[d]^{T\sem{A^{\circ}\subtp A}}
         \\
	 &&
	  {T\sem{A}}
	 }}
	\end{equation}
Here $\fvsem{\Pi}$ is as in Definition~\ref{definition:interpretationOfTypeJudgments}.
The proof is by induction on $\Pi$. We present one case; the others
 are similar.   

Assume $\Pi$ is in the following form, 
with the last rule applied being $(\mbox{$\limp$}.\text{E})$.
\begin{displaymath}
 \Pi\;\equiv\;
 \left[
 \vcenter{ 
 \infer[(\mbox{$\limp$}.\text{E})] 
  {\bang\Delta,\Gamma_{1},\Gamma_{2}\vdash MN:B}
  {
  \infer*[\Pi_{1}]{\bang\Delta,\Gamma_{1}\vdash M:A\limp B}{}
  &\quad
  \infer*[\Pi_{2}]{\bang\Delta,\Gamma_{2}\vdash N:C}{}
  &\quad
  C\subtp A
  }}
\right]
\end{displaymath}
By the induction hypothesis, there exist types $D,E$ and derivations 
$\Pi_{1}^{\circ}, \Pi_{2}^{\circ}$ such that
\begin{displaymath}
 \begin{array}{ll}
\Pi_{1}^{\circ}\VdashP\bang\Delta,\Gamma_{1}\vdash M:D\enspace,
  & 
\Pi_{2}^{\circ}\VdashP\bang\Delta,\Gamma_{2}\vdash N:E\enspace;
\\
  D\subtp A\limp B\enspace,
 &
  E\subtp C\enspace;
\\
\fvsem{\Pi_{1}}=T\sem{D\subtp A\limp B}\co
       \semP{\Pi_{1}^{\circ}}\enspace,\;\text{and}\qquad
 &
\fvsem{\Pi_{2}}=T\sem{E\subtp C}\co \semP{\Pi_{2}^{\circ}}\enspace.
 \end{array}
\end{displaymath}
Since $D\subtp A\limp B$, the type $D$ must be of the form $D\equiv
 \bang^{m}(A'\limp B')$ with $A\subtp A'$ and $B'\subtp B$. Now consider
the following derivation $\Pi^{\circ}$.
\begin{displaymath}
 \Pi^{\circ}\;:\equiv\;
 \left[
 \vcenter{ 
 \infer[(\mbox{$\limp$}.\text{E})_{\mathrm{P}}] 
  {\bang\Delta,\Gamma_{1},\Gamma_{2}\vdash MN:B'}
  {
  \infer*[\Pi_{1}^{\circ}]{\bang\Delta,\Gamma_{1}\vdash M:\bang^{m}(A'\limp B')}{}
  &\quad
  \infer*[\Pi_{2}^{\circ}]{\bang\Delta,\Gamma_{2}\vdash N:E}{}
  &\quad
  E\subtp A'
  }}
\right]
\end{displaymath}
Here the side condition $E\subtp A'$ holds since $E\subtp C\subtp
 A\subtp A'$. Thus we obtain
 $\Pi^{\circ}\VdashP\bang\Delta,\Gamma_{1},\Gamma_{2}\vdash MN:B'$ with
 $B'\subtp B$. 

 It remains to show that $\fvsem{\Pi}=(T\sem{B'\subtp B})\co
 \semP{\Pi^{\circ}}$. This is, however, an immediate consequence of
 \begin{itemize}
  \item the induction hypothesis,
  \item  \begin{math}
  \sem{A\subtp C} = \sem{B\subtp C}\co \sem{A\subtp B}
 \end{math}
	 (Lemma~\ref{lem:interpSubtpRelTransitive}),
  \item bifunctoriality of $\boxtimes$ and $\limp$, and
  \item naturality of $\str,\str',\ev$ and $\mu$.
 \end{itemize}
This can be checked by straightforward diagram chasing.  \myqed
\end{myproof}

We are ready to prove Lemma~\ref{lem:interpretationIsWellDfd}.

\begin{myproof} (Of Lemma~\ref{lem:interpretationIsWellDfd})
Assume $\Pi\Vdash\Delta\vdash M:A$ and
$\Pi'\Vdash\Delta\vdash M:A$. We apply
 Lemma~\ref{lem:typableThenPrincipalTypable} to obtain
 $A^{\circ},\Pi^{\circ},
 (A')^{\circ}, (\Pi')^{\circ}$  such that
\begin{displaymath}
\begin{array}{ll}
 \Pi^{\circ}\VdashP\Delta\rest{\FV(M)}\vdash M:A^{\circ}\enspace,
 \qquad
 &(\Pi')^{\circ}\VdashP\Delta\rest{\FV(M)}\vdash M:(A')^{\circ}\enspace,
 \\
  \sem{\Pi} 
 = T\sem{A^{\circ}\subtp A}\co \semP{\Pi^{\circ}}\co \weak
 \quad\text{and}\quad
  &
  \sem{\Pi'} 
 = T\sem{(A')^{\circ}\subtp A}\co \semP{(\Pi')^{\circ}}\co \weak\enspace.
\end{array}
\end{displaymath}
Applying
 Lemma~\ref{lem:uniquePrincipalTyping} to the first line  we have $A^{\circ}\equiv
 (A')^{\circ}$, and moreover $\Pi^{\circ}\equiv (\Pi')^{\circ}$. This is
 used in the second line to conclude $\sem{\Pi}=\sem{\Pi'}$. \myqed
\end{myproof}



\section{Proofs for \S{}\ref{section:adequacy}}
\label{section:proofsForAdequacySection}

\begin{mylemma}\label{lem:07121732}
  Let $M$ be a term such that
 $\Gamma \vdash M:A$ is derivable. Assume a subtype relation $A \subtp
 B$.
 Then  we
  have
  \begin{displaymath}
    T\llbracket A \subtp B \rrbracket \co 
    \llbracket \Gamma \vdash M : A\rrbracket 
   \quad = \quad\llbracket \Gamma \vdash M :B \rrbracket\enspace. 
  \end{displaymath}
\end{mylemma}
\begin{myproof}
  By induction on the term $M$. We only present two cases. When $M
  \equiv x$, the composition $T\llbracket A \subtp B \rrbracket \co
  \llbracket \Gamma \vdash M : A\rrbracket$ is equal to
  \begin{equation}\label{eq:07121627}
    \llbracket \Gamma \rrbracket
    \xrightarrow{\mathsf{weak}}
    \llbracket A' \rrbracket
    \xrightarrow{\llbracket A'\subtp A \rrbracket}
    \llbracket A \rrbracket
    \xrightarrow{\llbracket A\subtp B \rrbracket}
    \llbracket B \rrbracket
    \xrightarrow{\eta_{\llbracket B \rrbracket}^{T}}
    T\llbracket B \rrbracket.
  \end{equation}
  Since $\llbracket A\subtp B \rrbracket \co \llbracket A'\subtp A
  \rrbracket$ is equal to $\llbracket A'\subtp B \rrbracket$ by
  Lemma~\ref{lem:interpSubtpRelTransitive}, the arrow
  (\ref{eq:07121627}) is equal to $\llbracket \Gamma \vdash x : B
  \rrbracket$. 

 When $M \equiv M_{1}\,M_{2}$, the term environment
  $\Gamma$ is of the form $\bang \Delta,\Gamma_{1},\Gamma_{2}$, and there
  is a type $C$ such that $\bang \Delta,\Gamma_{1} \vdash M_{1}:C \limp
  A$ and $\bang \Delta,\Gamma_{2} \vdash M_{2}:C$ are derivable. The
  interpretation $\llbracket \Gamma \vdash M_{1}\,M_{2}:A\rrbracket$
  is
  \begin{multline*}
    \mu_{\llbracket A \rrbracket} \co \mu_{T\llbracket A \rrbracket}
    \co TT\mathsf{ev}_{\llbracket C \rrbracket,T\llbracket A
      \rrbracket} \co T\mathrm{str}_{\llbracket C \rrbracket
      \limp T\llbracket A \rrbracket, \llbracket C \rrbracket}
    \co \mathrm{str}'_{\llbracket C \rrbracket \limp T\llbracket
      A \rrbracket, T\llbracket C \rrbracket} \\ %
    \co \bigl(\,\llbracket \bang \Delta,\Gamma_{1} \vdash M_{1}: C \limp A
    \rrbracket \boxtimes \llbracket \bang \Delta,\Gamma_{2} \vdash M_{2}:C
    \rrbracket\,\bigr) \co c\enspace,
  \end{multline*}
  where $c: \llbracket \bang \Delta,\Gamma_{1},\Gamma_{2}\rrbracket
  \to \llbracket \bang \Delta,\Gamma_{1} \rrbracket \boxtimes \llbracket
  \bang \Delta,\Gamma_{2}\rrbracket$ is a suitable permutation
  followed by contractions. By naturality of $\mu$,
  $\mathrm{str}$, $\mathrm{str}'$ and $\mathsf{ev}$, the composition
  $T\llbracket A \subtp B \rrbracket \co \llbracket \Gamma \vdash
  M_{1}\,M_{2}:A\rrbracket$ is equal to
  \begin{multline*}
    \mu_{\llbracket B \rrbracket} \co \mu_{T\llbracket B \rrbracket}
    \co TT\mathsf{ev}_{\llbracket C \rrbracket,T\llbracket B
      \rrbracket} \co T\mathrm{str}_{\llbracket C \rrbracket
      \limp T\llbracket B \rrbracket, \llbracket C \rrbracket}
    \co \mathrm{str}'_{\llbracket C \rrbracket \limp T\llbracket
      B \rrbracket, T\llbracket C \rrbracket} \\
    \co  
    \Bigl(\bigl(\,T(\llbracket C \rrbracket \limp T\llbracket A \subtp B
    \rrbracket) \co
\llbracket \bang \Delta,\Gamma_{1} \vdash M_{1}: C \limp A
    \rrbracket\,\bigr) \boxtimes \llbracket \bang \Delta,\Gamma_{2} \vdash M_{2}:C
    \rrbracket\Bigr) \co c\enspace,
  \end{multline*}
  where the arrow
  \begin{displaymath}
   T\bigl(\,\llbracket C \rrbracket \limp T\llbracket A \subtp B
  \rrbracket\,\bigr)
   \quad:\quad
   T(\sem{C}\limp T\sem{A})
   \;\longrightarrow\;
T(\sem{C}\limp T\sem{B})
  \end{displaymath}
  is obtained by applying suitable functors to the arrow $\sem{A\subtp
 B}:\sem{A}\to\sem{B}$.

  Since $T\bigl(\,\llbracket C \rrbracket \limp T\llbracket A \subtp B
  \rrbracket\,\bigr) \co \llbracket \bang \Delta,\Gamma_{1} \vdash M_{1}: C
  \limp A \rrbracket$ is equal to $\llbracket \bang \Delta,\Gamma_{1}
  \vdash M_{1}:C \limp B \rrbracket$, we see that $T\llbracket
  A\subtp B\rrbracket \co \llbracket \Gamma \vdash
  M_{1}\,M_{2}:A\rrbracket$ is equal to $\llbracket \Gamma \vdash
  M_{1}\,M_{2}:B\rrbracket$. 

 We can prove the other cases in the same way.
  \myqed
\end{myproof}

As is usual with the categorical interpretation of call-by-value languages in 
Kleisli categories, the interpretation of a value $\Gamma \vdash V:A$ in
$\Hoq$ factorizes through the monad unit $\eta^{T}$ and is the form
 $\eta^{T}_{\llbracket A \rrbracket} \co f $
for some arrow $f:\llbracket \Gamma \rrbracket \to \llbracket A
\rrbracket$ (see Definition~\ref{definition:interpretationOfTypeJudgments}).
We write $\llbracket \Gamma \vdash V : A \rrbracket_{\mathsf{v}}$ for
the arrow such that $\llbracket \Gamma \vdash V :A \rrbracket =
\eta^{T}_{\llbracket A \rrbracket} \co \llbracket \Gamma \vdash 
V : A\rrbracket_{\mathsf{v}}$ given in
Definition~\ref{definition:interpretationOfTypeJudgments}.  It is easy to
see that $\llbracket \vdash V:\bang A \rrbracket_{\mathsf{v}}$ is of the
form $\bang f \co \varphi'$ for some $f:\punit \to \llbracket A
\rrbracket$. Here $\varphi':\punit\iso\bang\punit$ is from Theorem~\ref{theorem:PERQIsLinear}.

\begin{mylemma}\label{lem:07121953}
 Assume that
 $\Gamma,x:A \vdash M:B$ and
 $\vdash V:A$ are derivable for a term $M$ and a closed value $V$.
 Then
  the composition $\llbracket \Gamma , x:A \vdash M : B\rrbracket
  \co (\mathrm{id}_{\llbracket \Gamma \rrbracket} \boxtimes
  \llbracket \vdash V : A\rrbracket_{\mathsf{v}})$ is equal to
  $\llbracket \Gamma \vdash M[V/x] : B\rrbracket$. 
\end{mylemma}
In Lemma~\ref{lem:07121953}, we assume that $x$
  is the largest in $|\Gamma|\cup \{x\}$ with respect to the linear
  order $\prec$
  in Definition~\ref{definition:interpOfContexts}.  It
  is straightforward to generalize the statement to an arbitrary
  variable $x$ in a term context.

\begin{myproof}
  By induction on the term $M$. We only present two cases. When $M
  \equiv x$, the composition $\llbracket \Gamma , x:A \vdash x:B
  \rrbracket \co (\mathrm{id}_{\llbracket \Gamma \rrbracket}
  \boxtimes \llbracket \vdash V :A\rrbracket_{\mathsf{v}})$ is equal
  to $\eta^{T}_{\llbracket B \rrbracket} \co \llbracket
  A\subtp B\rrbracket \co \llbracket \Gamma \vdash
  V:A\rrbracket_{\mathsf{v}}$, which is equal to $\llbracket \Gamma
  \vdash V:B \rrbracket$ by Lemma~\ref{lem:07121732}.  

 When $M \equiv
  M_{1}\,M_{2}$, the term environment $\Gamma,x:A$ is of the form
  $\bang \Delta,\Gamma_{1},\Gamma_{2}$ and there is a type $C$ such that
  $\bang \Delta,\Gamma_{1} \vdash M_{1}: C \limp B$ and
  $\bang \Delta,\Gamma_{2} \vdash M_{2} : C$ are derivable. The
  interpretation $\llbracket \Gamma,x:A \vdash M_{1}\,M_{2}:B\rrbracket$
  is
  \begin{multline*}
    \mu_{\llbracket B \rrbracket} \co \mu_{T\llbracket B \rrbracket}
    \co TT\mathsf{ev}_{\llbracket C \rrbracket,T\llbracket B
      \rrbracket} \co T\mathrm{str}_{\llbracket C \rrbracket
      \limp T\llbracket B \rrbracket, \llbracket C \rrbracket}
    \co \mathrm{str}'_{\llbracket C \rrbracket \limp T\llbracket
      B \rrbracket, T\llbracket C \rrbracket} \\ %
    \co (\llbracket \bang \Delta,\Gamma_{1} \vdash M_{1}: C \limp B
    \rrbracket \boxtimes \llbracket \bang \Delta,\Gamma_{2} \vdash M_{2}:C
    \rrbracket) \co c 
  \end{multline*}
  where $c : \llbracket \bang \Delta,\Gamma_{1},\Gamma_{2}\rrbracket \to
  \llbracket \bang \Delta,\Gamma_{1} \rrbracket \boxtimes \llbracket
  \bang \Delta,\Gamma_{2}\rrbracket$ is a suitable permutation
  followed by contractions. By the induction hypothesis 
  and naturality of the permutation and
  contractions, the composition $\llbracket \Gamma ,x:A\vdash
  M_{1}\,M_{2}:B\rrbracket \co (\mathrm{id}_{\llbracket \Gamma
    \rrbracket} \boxtimes \llbracket \vdash
  V:A\rrbracket_{\mathsf{v}})$ is equal to
  \begin{multline*}
    \mu_{\llbracket B \rrbracket} \co \mu_{T\llbracket B \rrbracket}
    \co TT\mathsf{ev}_{\llbracket C \rrbracket,T\llbracket B
      \rrbracket} \co T\mathrm{str}_{\llbracket C \rrbracket
      \limp T\llbracket B \rrbracket, \llbracket C \rrbracket}
    \co \mathrm{str}'_{\llbracket C \rrbracket \limp T\llbracket
      B \rrbracket, T\llbracket C \rrbracket} \\ %
    \co (\llbracket \bang \Delta',\Gamma'_{1} \vdash M_{1}[V/x]: C
    \limp B \rrbracket \boxtimes \llbracket \bang \Delta',\Gamma'_{2}
    \vdash M_{2}[V/x]:C \rrbracket)  \co c'
  \end{multline*}
  where $\bang \Delta',\Gamma'_{1},\Gamma'_{2}$ is the term environment
  obtained by removing $x:A$ from $\bang \Delta,\Gamma_{1},\Gamma_{2}$, and
  $c' : \llbracket \bang \Delta',\Gamma'_{1},\Gamma'_{2} \rrbracket \to
  \llbracket \bang \Delta',\Gamma'_{1} \rrbracket \boxtimes \llbracket
  \bang \Delta,\Gamma'_{2}\rrbracket$ is a suitable permutation
  followed by contractions. 
  Hence, $\llbracket \Gamma , x:A \vdash
  M_{1}\,M_{2} : B\rrbracket \co (\mathrm{id}_{\llbracket \Gamma
    \rrbracket} \boxtimes \llbracket \vdash V :
  A\rrbracket_{\mathsf{v}})$ is equal to $\llbracket \Gamma \vdash
  (M_{1}\,M_{2})[V/x] : B\rrbracket$. 

 We can prove the other cases in the
  same way. \myqed
\end{myproof}

\begin{mylemma}\label{lem:07121942}
  If $\vdash E[M]:A$ is derivable, then there exist a type $B$ such
  that $\vdash M:B$ and $x:B \vdash E[x]:A$.
\end{mylemma}
\begin{myproof}
  By induction on the evaluation context $E$.  \myqed
\end{myproof}

\begin{mylemma}[Lemma~\ref{lem:07121946}, repeated]\label{alem:07121946}
  Let $E$ be  an evaluation context, and $x$ be a variable that does
 not occur in $E$. Assume that $x:A \vdash E[x]:B$ is derivable.
  Then for any term $M$ such that $\Vdash\Gamma \vdash M:A$, the interpretation
  $\llbracket \Gamma \vdash E[M]:B\rrbracket:\sem{\Gamma}\to T\sem{B}$ is calculated by
 \begin{displaymath}
  \llbracket \Gamma \vdash E[M]:B\rrbracket
  \quad=\quad
  \mu^{T}_{\llbracket B \rrbracket} \co T\llbracket x:A \vdash
  E[x]:B\rrbracket \co \llbracket \Gamma \vdash M:A
  \rrbracket\enspace.
 \end{displaymath}
\end{mylemma}
\begin{myproof}
  By induction on the evaluation context $E$, where we use the
 characterization in Lemma~\ref{lem:bottomUpDefOfEvalContext}.  We only present two
  cases. When $E\equiv [\place]$, since $x:A \vdash x:B$ is derivable,
 we must have
  $A \subtp  B$. By Lemma~\ref{lem:07121732}, $\llbracket
  \Gamma \vdash E[M]:B\rrbracket$ is equal to $T\llbracket A \subtp  B
  \rrbracket \co \llbracket \Gamma \vdash M:A\rrbracket$, which is
  nothing but $\mu_{\llbracket B \rrbracket} \co T\llbracket x:A \vdash
  E[x]:B\rrbracket \co \llbracket \Gamma \vdash M:A \rrbracket$.

  When $E\equiv E'\,N$, there exists a type $C$ such that $x:A \vdash
  E'[x]: C \limp B$ and $\vdash N : C$ are derivable.  The
  interpretation $\llbracket \Gamma \vdash \bigl(E'[M]\bigr)\,N:B\rrbracket$ is
  given, by Definition~\ref{definition:interpretationOfTypeJudgments}, by
  \begin{multline} \label{eq:07121928} \mu_{\llbracket B \rrbracket}
    \co \mu_{T\llbracket B \rrbracket} \co
    TT\mathsf{ev}_{\llbracket C \rrbracket,T\llbracket B \rrbracket}
    \co T\mathrm{str}_{\llbracket C \rrbracket \limp T\llbracket
      B \rrbracket, \llbracket C \rrbracket} \co
    \mathrm{str}'_{\llbracket C \rrbracket \limp T\llbracket B
      \rrbracket, T\llbracket C \rrbracket} \\ %
    \co (\llbracket \Gamma \vdash E'[M]: C \limp B \rrbracket
    \boxtimes \llbracket \vdash N:C \rrbracket)\enspace.
  \end{multline}
  By the induction hypothesis we have
  \begin{displaymath}
   \llbracket \Gamma \vdash E'[M]:C \limp
  B \rrbracket
  \quad=\quad
    \mu_{\llbracket C \limp B \rrbracket}
    \co T\llbracket x:A \vdash E'[x]:C \limp B\rrbracket
    \co \llbracket \Gamma \vdash M :A\rrbracket\enspace.
  \end{displaymath}
  Using this we see that (\ref{eq:07121928}) is equal to
 \begin{align*}
  &\mu_{\llbracket B
    \rrbracket} \co Tf \co \llbracket \Gamma \vdash M:A
  \rrbracket\enspace, \quad\text{where}\quad
 \\
 &
  f\;:=\;
  \left[
 \begin{array}{l}
    \mu_{\llbracket B \rrbracket} \co \mu_{T\llbracket B \rrbracket}
    \co TT\mathsf{ev}_{\llbracket C \rrbracket,T\llbracket B
      \rrbracket} \co T\mathrm{str}_{\llbracket C \rrbracket
      \limp T \llbracket B
      \rrbracket, \llbracket C \rrbracket} \co
   \\
    \mathrm{str}'_{\llbracket C \rrbracket \limp T \llbracket
      B \rrbracket, T\llbracket C
      \rrbracket}  \co \bigl(\,\llbracket x:A \vdash E'[x]:C \limp
    B\rrbracket \boxtimes \llbracket \vdash N : C\rrbracket\,\bigr)
 \end{array}
\right]
 \quad:\quad
 \llbracket A \rrbracket \longto T\llbracket B
  \rrbracket\enspace.
 \end{align*}
 Here we notice that $f$ coincides with  the interpretation of $x:A
 \vdash E'[x]\,N:B$. This concludes the case when $E\equiv E'\,N$.

We can prove the
  other cases in the same way.  \myqed
\end{myproof}

\begin{mylemma}[Lemma~\ref{lem:07122017}, repeated]\label{alem:07122017}
  For a closed term $M$ such that $\Vdash\,\vdash M:A$, if there is a reduction $M \to_{1}
  N$ that is not due to a measurement rule (($\meas_{1}$--$\meas_{4}$)
 in Definition~\ref{definition:operationalSemantics}), then 
\begin{displaymath}
 \llbracket \vdash M
  :A\rrbracket
 \quad=\quad
 \llbracket \vdash N
  :A\rrbracket\enspace.
\end{displaymath}
Note
  that $\vdash N:A$ is derivable by
  Lemma~\ref{lem:subjectReductionLemma}.
\end{mylemma}
\begin{myproof}
  Assume $M\equiv E\bigl[(\lambda x^{C}.\,L)\,V\bigr]$ and $N \equiv E\bigl[L[V/x]\bigr]$. By
  Lemma~\ref{lem:07121942}, there exists a type $B$ such that
  $y:B\vdash E[y]:A$ and $\vdash (\lambda x^{C}.\,L)\,V : B$ are
  derivable. By Lemma~\ref{lem:subjectReductionLemma}, we also have
  $\vdash L[V/x]:B$. Since 
\begin{align*}
 \llbracket \vdash
   E[(\lambda x^{C}.\,L)\,V]:A\rrbracket
 \quad&=\quad
\mu_{\llbracket A
     \rrbracket} \co T\llbracket x:B \vdash E[x]:A \rrbracket \co
   \llbracket \vdash (\lambda x^{C}.\,L)\,V : B \rrbracket
 \qquad\text{and}
 \\
\llbracket \vdash E[L[V/x]]:A\rrbracket
 \quad&=\quad
\mu_{\llbracket A \rrbracket} \co T\llbracket x:B \vdash E[x]:A
  \rrbracket \co \llbracket \vdash L[V/x] :B \rrbracket
\end{align*}
 by Lemma~\ref{lem:07121946}, 
it is enough to show that $\llbracket
  \vdash (\lambda x^{C}.\,L)\,V : B\rrbracket$ is equal to $\llbracket
  \vdash L[V/x]:B \rrbracket$. By unfolding the definition of the
  interpretation (Definition~\ref{definition:interpretationOfTypeJudgments}),
 we obtain 
\begin{displaymath}
 \llbracket \vdash (\lambda x^{C}.\,L)\,V :
  B\rrbracket
 \quad = \quad
\llbracket x:A \vdash L: B \rrbracket \co \llbracket
  \vdash V:A \rrbracket_{\mathsf{v}}\enspace;
\end{displaymath}
this coincides with $\llbracket
  \vdash L[V/x] : B\rrbracket$ by Lemma~\ref{lem:07121953}. 

 For the other
  reduction rules, we can prove the statement in the same way.  \myqed
\end{myproof}

\begin{mylemma}[Lemma~\ref{lem:07141800}\,(\ref{enu:07130423}), repeated]
  \label{lem:07130423} 
  If $(t,V)$ is in $R_{A}$, then $(\eta_{\llbracket A \rrbracket}^{T}
  \co t,V)$ is in $R_{A}^{\top\top}$.
\end{mylemma}
\begin{myproof}
  For $(k,E) \in R_{A}^{\top}$, since $\mu_{\llbracket \bit 
    \rrbracket} \co Tk \co \eta_{A}^{T} \co t = k \co t$, we
  have
  \begin{displaymath}
    \mu_{\llbracket \bit  \rrbracket} \co Tk \co
    \eta_{A}^{T} \co t \;\lessdot\; E[V]\enspace.    
  \end{displaymath}
  Therefore $(\eta_{\llbracket A \rrbracket}^{T} \co t,V)$ is in
  $R_{A}^{\top\top}$.
\end{myproof}

\begin{mylemma}[Lemma~\ref{lem:07141800}\,(\ref{enu:07130422}), repeated]
  \label{lem:07130422}
  If $(t,V) \in R_{A}$ and $A \subtp  A'$, then $(\llbracket A \subtp  A'
  \rrbracket \co t,V) \in R_{A'}$.
\end{mylemma}
\begin{myproof}
  By induction on $A$. When $A$ is $\top$ or
  $\nqbit$, the type $A'$ is equal to $A$, and the
  statement is straightforward. 

  When $A$ is $B \boxtimes C$, by the
  definition of subtyping relation, $A'$ must be of the form $B'
  \boxtimes C'$ for some $B' :> B$ and $C' :> C$. For $(t \boxtimes
  s,\langle V,W \rangle) \in R_{B \boxtimes C}$, the composition
  $\llbracket B \boxtimes C \subtp  B' \boxtimes C' \rrbracket \co (t
  \boxtimes s)$ is equal to $(\llbracket B \subtp  B' \rrbracket \co t)
  \boxtimes (\llbracket C \subtp  C' \rrbracket \co s)$. By the induction
  hypothesis, $(\llbracket B \subtp  B' \rrbracket \co t,V)$ is in
  $R_{B'}$, and $(\llbracket C \subtp  C' \rrbracket \co s,W)$ is in
  $R_{C'}$, and therefore $(\llbracket B \boxtimes C \subtp  B' \boxtimes
  C' \rrbracket \co (t \boxtimes s),\langle V,W\rangle)$ is in
  $R_{B' \boxtimes C'}$. We can similarly show the statement for $A \equiv B
  + C$. 

 When $A$ is $B \limp C$, the type $A'$ is of the form $B'
  \limp C'$ for some $B' \subtp  B$ and $C \subtp  C'$.  For $(t,V) \in R_{B
    \limp C}$ and $(s,W) \in R_{B'}$, by naturality of
  $\mathsf{ev}$ we have
  \begin{multline}\label{eq:07140403}
    \mathsf{ev}_{\llbracket B' \rrbracket,\llbracket C'\rrbracket}
    \co ((\llbracket B \limp C \subtp  B' \limp C' \rrbracket
    \co t) \boxtimes s) \\
    = T\llbracket C \subtp C' \rrbracket \co \mathsf{ev}_{\llbracket B
      \rrbracket,\llbracket C\rrbracket} \co ( t \boxtimes
    (\llbracket B' \subtp B \rrbracket \co s)).
  \end{multline}
  By the induction hypothesis, $(\llbracket B' \subtp  B \rrbracket \co
  s, W)$ is in $R_{B}$. Since $(t,V)$ is in $R_{B \limp C}$, we
  see that $\bigl(\mathsf{ev}_{\llbracket B \rrbracket,\llbracket C
    \rrbracket} \co (t \boxtimes (\llbracket B' \subtp  B \rrbracket
  \co s)), V\,W\bigr)$ is in $R_{C}^{\top\top}$. By the induction
  hypothesis, $(k \co \llbracket C \subtp  C' \rrbracket,E)$ is 
 easily seen to be in
  $R_{C}^{\top}$ for any $(k,E) \in R_{C'}^{\top}$.  Therefore,
  \begin{displaymath}
    \mu_{\llbracket \bit  \rrbracket} \co
    Tk \co T\llbracket C \subtp  C' \rrbracket \co
    \mathsf{ev}_{\llbracket B \rrbracket,\llbracket C \rrbracket} 
    \co (t \boxtimes (\llbracket B' \subtp  B \rrbracket \co
    s)) \;\lessdot\; E[V\,W]  
  \end{displaymath}
  for any $(k,E) \in R_{C'}^{\top}$. By the definition of
  $R_{C}^{\top\top}$ and (\ref{eq:07140403}), we obtain
  \begin{displaymath}
    \bigl(\,\mathsf{ev}_{\llbracket B' \rrbracket,\llbracket C' \rrbracket}
    \co ((\llbracket B \limp C \subtp  B' \limp C' \rrbracket
    \co t) \boxtimes s) ,\; V\,W\,\bigr)
    \quad\in R_{C'}^{\top\top}
  \end{displaymath}
  for any $(s,W) \in R_{B'}$, which implies that $(\llbracket B
  \limp C \subtp  B' \limp C' \rrbracket \co t, V)$ is in $R_{B'
    \limp C'}$.  

 When $A$ is $\bang B$, the type $A'$ is of the form
  $\bang ^{n}B'$ for some $n \geq 0$ and $B \subtp  B'$.  If $(\bang t \co
  \varphi',V)$ is in $R_{A}$, then $\llbracket A \subtp  A' \rrbracket
  \co \bang t \co \varphi' = \bang ^{n}(\llbracket B \subtp  B'\rrbracket \co
  t) \co \bang ^{n-1}\varphi' \co \cdots \co \bang \varphi' \co
  \varphi'$. By the induction hypothesis, $(\llbracket B \subtp 
  B'\rrbracket \co t,V)$ is in $R_{B'}$, and therefore by~(\ref{eq:defOfLogicalRelationR}), $(\llbracket
  A \subtp  A'\rrbracket \co \bang t \co \varphi' ,V)$ is in $R_{A'}$. \myqed
\end{myproof}

\begin{mylemma}[Lemma~\ref{lem:bot}\,(\ref{enu:bot}), repeated]\label{alem:bot}
  For any type $A$ and $M \in \ClTerm(A)$, we have $([\bot],M)
  \in R_{A}^{\top\top}$.
\end{mylemma}
\begin{myproof}
  Let $(k,E) \in R_{A}^{\top}$. We claim 
 \begin{equation}\label{eq:lemBotEq}
    \mu_{\llbracket \bit 
    \rrbracket} \co Tk \co [\bot] \;=\; [\bot]\enspace,
 \end{equation}  
 where $[\bot]$ denotes the arrow $\punit\to T\sem{A}$ in $\PER_{\Q}$
 that is realized by $\bot\in A_{\Q}$ (cf.\
 Lemma~\ref{lem:resultTypeIsAdmissible},~\ref{lem:LCAIsCPO}.\ref{item:applLeftStrict}).
We have
 \begin{align*}
   \mu &=[\lambda k^{(((\llbracket A \rrbracket \limp R) \limp R) \limp R) \limp R}
         y^{\llbracket A \rrbracket \limp R}.
         k(\lambda h^{(\llbracket A \rrbracket \limp R) \limp R}.hy)] \\
   Tk  &=[\lambda v^{(\llbracket A \rrbracket \limp R) \limp R}x^{((\sem{\bit} \limp R) \limp R) \limp R}.
         v(\lambda a^{\llbracket A \rrbracket}.x(c_{k}a))]
 \end{align*}
 where $c_{k}$ is a choice of a realizer of $k$. We put type
 annotations to explain intentions of these realizers. 
 The arrow  $\mu_{\llbracket \bit \rrbracket} \co Tk:T\sem{A}\to T\sem{\bit}$ is
 realized by
 \begin{align*}
   \lambda v.(\lambda ky.k(\lambda h.hy)) ((\lambda vx.v(\lambda a.x(c_{k}a)))v) 
   &= \lambda vy.(\lambda x.v(\lambda a.x(c_{k}a)))(\lambda h.hy)  \\
   &\overset{(1)}{=} \lambda vy.v(\lambda a.(\lambda h.hy)(c_{k}a))  \\
   &\overset{(2)}{=} \lambda vy.v(\lambda a.c_{k}ay) .
 \end{align*}
 We note that (1) and (2) follow from Remark~\ref{rem:int}: the LCA
 $A_{\mathcal{Q}}$ is a model of the untyped linear lambda calculus
 modulo beta-reductions.
 Therefore the left-hand side of~(\ref{eq:lemBotEq}) is
 \begin{displaymath}
   \bigl[\,
   \lambda x.\, \lambda y.\, \bot (\lambda a.\, c_{k}ay)
   \,\bigr]
 \end{displaymath}
 that is nothing but $[\bot]$ due to the left strictness of
 application of $A_{\Q}$.

 From~(\ref{eq:lemBotEq}) it easily follows that
 $\prb(\tree(\mu_{\llbracket \bit \rrbracket} \co Tk \co [\bot])) =
 (0,0)$.
 Therefore we always have
 $\mu_{\llbracket \bit \rrbracket} \co Tk \co [\bot] \lessdot
 E[M]$. \myqed
\end{myproof}
\begin{mylemma}[Lemma~\ref{lem:bot}\,(\ref{enu:lab}),
  repeated]\label{alem:sup}
  For any type $A$ and $M \in \ClTerm(A)$, if there exists a sequence
  of realizers $a_{1} \sqsubseteq a_{2} \sqsubseteq \cdots$ of arrows
  in $\PER_{\Q}(\punit,T\llbracket A \rrbracket)$ such that
  $([a_{n}],M) \in R_{A}^{\top\top}$, then we have
  $([\bigvee_{n \geq 1}a_{n}],M) \in R_{A}^{\top\top}$.
\end{mylemma}
\begin{myproof}
  Let $(k,E)$ be an element in $R_{A}^{\top}$.  Since the application
  of $A_{\Q}$ is continuous, the value assigned to any edge of the
  tree
  \begin{displaymath}
    \tree
    \Bigl(\mu_{\llbracket \bit  \rrbracket} \co Tk \co
    \bigl[\bigvee_{n \geq 1} a_{n}\bigr]\Bigr)
  \end{displaymath}
  is the least upper bound of the value on the corresponding edge of
  the trees
  \begin{displaymath}
    \tree (\mu_{\llbracket \bit  \rrbracket} \co Tk \co
    [a_{n}]).
  \end{displaymath}
  Therefore, if we have
  $\mu_{\llbracket \bit \rrbracket} \co Tk \co [a_{n}] \lessdot E[M]$
  for every $n \geq 1$, then it follows that
  $\mu_{\llbracket \bit \rrbracket} \co Tk \co [\bigvee_{n \geq
    1}a_{n}] \lessdot E[M]$. \myqed
\end{myproof}

\begin{mylemma}\label{lem:07130433}
  Let $M \to_{1} N$ be a reduction that is not due to a measurement
  rule (($\meas_{1}$--$\meas_{4}$) in
  Definition~\ref{definition:operationalSemantics}). Then
  \begin{displaymath}
    (t,M)\in R_{A}^{\top\top}
    \quad\Longleftrightarrow\quad
    (t,N)\in R_{A}^{\top\top}\enspace.
  \end{displaymath}
\end{mylemma}
\begin{myproof}
  Let $(k,E)$ be an element in $R_{A}^{\top}$.  Assume $(t,M)$ is in
  $R_{A}^{\top\top}$; then we have
  $\mu_{\llbracket \bit \rrbracket} \co Tk \co t \lessdot E[M]$. By
  the definition of big-step semantics, $E[M] \oprred (p,q)$ if and
  only if $E[N] \oprred (p,q)$. Therefore,
  $\mu_{\llbracket \bit \rrbracket} \co Tk \co t \lessdot E[N]$.  The
  other direction is similar.  \myqed
\end{myproof}

\begin{mylemma}\label{lem:new}
  For a type $A$ such that
  $\bang (\bit \limp \mathtt{qbit}) \subtp A$, the pair
  $(\llbracket \vdash \new:A\rrbracket, \new)$ is in
  $R_{A}^{\top\top}$.
\end{mylemma}
\begin{myproof}
  When $A = \bit \limp \mathtt{qbit}$, since
  $(\llbracket \new_{\rho} \rrbracket_{\mathrm{const}}, \new_{\rho})$
  is in $R_{\mathtt{qbit}}$ for any $\rho \in \DM_{2}$, both
  $(\llbracket \vdash \new_{\mid 0 \rangle \langle 0 \mid}
  :\mathtt{qbit} \rrbracket,\new\,\ttrue)$
  and
  $(\llbracket \vdash \new_{\mid 1 \rangle \langle 1 \mid}
  :\mathtt{qbit} \rrbracket,\new\,\ffalse)$
  are in $R_{\mathtt{qbit}}^{\top\top}$ by Lemma~\ref{lem:07130423}
  and Lemma~\ref{lem:07130433}. Therefore,
  $(\llbracket \new \rrbracket_{\mathrm{const}},\new)$ is in
  $R_{\bit \limp \mathtt{qbit}}$, and by Lemma~\ref{lem:07130423},
  $(\llbracket \vdash \new:A \rrbracket,\new)$ is in
  $R_{\bit \limp \mathtt{qbit}}^{\top\top}$. When
  $A = \bang (\bit \limp \mathtt{qbit})$, we have
  $\eta_{\llbracket A \rrbracket}^{T} \co \bang \llbracket \new
  \rrbracket_{\mathrm{const}} \co \varphi' = \llbracket \vdash \new
  :A\rrbracket$.
  Since $(\llbracket \new \rrbracket_{\mathrm{const}},\new)$ is in
  $R_{\bit \limp \mathtt{qbit}}$, the pair
  $(\bang \llbracket \new \rrbracket_{\mathrm{const}} \co
  \varphi',\new)$
  is in $R_{\bang (\bit \limp \mathtt{qbit})}$. Therefore, by
  Lemma~\ref{lem:07130423},
  $(\llbracket \vdash \new:A\rrbracket,\new)$ is in
  $R_{A}^{\top\top}$.  When $A$ satisfies
  $\bang (\bit \limp \mathtt{qbit}) \subtp A$, the statement follows
  from Lemma~\ref{lem:07130422}, Lemma~\ref{lem:07130423} and that
  $(\bang \llbracket \new \rrbracket_{\mathrm{const}} \co
  \varphi',\new)$ is in $R_{\bang (\bit \limp \mathtt{qbit})}$. \myqed
\end{myproof}

\begin{mylemma}\label{lem:U}
  For a type $A$ such that
  $\bang (\nqbit \limp n\text{-} \mathtt{qbit}) \subtp A$, the pair
  $(\llbracket \vdash U:A\rrbracket, U)$ is in $R_{A}^{\top\top}$.
\end{mylemma}
\begin{myproof}
  Similar to the proof of Lemma~\ref{lem:new}. \myqed
\end{myproof}

\begin{mylemma}\label{lem:cmp}
  For a type $A$ such that
  $\bang (\nqbit \boxtimes m\text{-}\mathtt{qbit}\limp (n+m)\text{-}
  \mathtt{qbit}) \subtp A$,
  the pair
  $(\llbracket \vdash \mathtt{cmp}_{n,m}:A\rrbracket,
  \mathtt{cmp}_{n,m})$ is in $R_{A}^{\top\top}$.
\end{mylemma}
\begin{myproof}
  Similar to the proof of Lemma~\ref{lem:new}.  \myqed
\end{myproof}

\begin{mylemma}\label{lem:meas1}
  For a type $A$ such that
  $\bang \bigl((n+1)\text{-}\mathtt{qbit} \limp \bang \bit \boxtimes
  \nqbit\bigr) \subtp A$, the pair $(\llbracket \vdash \mathtt{meas}_{i}^{n+1}:A\rrbracket, \mathtt{meas}_{i}^{n+1})$ is in $R_{A}^{\top\top}$.
\end{mylemma}
\begin{myproof}
  First we shall prove that, when  $A \equiv (n+1)\text{-}\mathtt{qbit} \limp \bang \bit 
  \boxtimes \nqbit$, we have $(\llbracket
  \mathtt{meas}_{i}^{n+1}\rrbracket_{\mathrm{const}},
  \mathtt{meas}_{i}^{n+1})$ is in $R_{A}$.  It is enough to show that
  for any $\rho \in \DM_{2^{n+1}}$, the pair $(\llbracket \vdash
  \mathtt{meas}_{i}^{n+1}\,\new_{\rho}: \bang \bit 
  \boxtimes \nqbit\rrbracket,\,
  \mathtt{meas}_{i}^{n+1}\,\new_{\rho})$ is in
  $R_{\bang \bit  \boxtimes \nqbit}^{\top\top}$.
  Let $(k,E)$ be an element in $R_{\bang \bit  \boxtimes
    \nqbit}^{\top}$.  We define
  $k_{0},k_{1}:\llbracket \nqbit \rrbracket\to
  T\llbracket \bit \rrbracket$ to be the following arrows.
  \begin{align*}
    k_{0} \;=\; k \co ((\bang \kappa_{\ell} \co \varphi')\boxtimes
    \mathrm{id}_{\llbracket \nqbit \rrbracket})  &&
    k_{1} \;=\; k \co ((\bang \kappa_{r} \co \varphi') \boxtimes 
    \mathrm{id}_{\llbracket \nqbit \rrbracket})
  \end{align*}
  Then $\prb(\tree(\mu_{\llbracket \bit  \rrbracket} \co
  Tk \co \llbracket \vdash
  \mathtt{meas}_{i}^{n+1}\,\new_{\rho}: \bang \bit 
  \boxtimes \nqbit\rrbracket))$ is equal to
  \begin{displaymath}
    (0,0) +   \prb(
    \tree(k_{0} \co \llbracket \new_{
      \langle 0_{i}\mid \rho \mid 0_{i}\rangle} \rrbracket_{\mathrm{const}})) +
    \prb(\tree(k_{1} \co \llbracket \new_{
      \langle 1_{i}\mid \rho \mid 1_{i}\rangle} \rrbracket_{\mathrm{const}}))\enspace;
  \end{displaymath}
  this is seen much like in the proof of Theorem~\ref{thm:sound}.
  Since
  \begin{align*}
    k_{0} \co \llbracket \new_{ \langle 0_{i}\mid \rho
      \mid 0_{i}\rangle} \rrbracket_{\mathrm{const}} &\;\lessdot\;
    E[\langle \ttrue,
    \new_{ \langle 0_{i}\mid \rho \mid 0_{i}\rangle}\rangle]
   \quad\text{and}\\
    k_{1} \co \llbracket \new_{ \langle 1_{i}\mid \rho
      \mid 1_{i}\rangle} \rrbracket_{\mathrm{const}} &\;\lessdot\;
    E[\langle \ffalse,
    \new_{ \langle 1_{i}\mid \rho \mid 1_{i}\rangle}\rangle] \enspace,
  \end{align*}
  it follows that $\mu_{\llbracket \bit \rrbracket} \co Tk
  \co \llbracket \vdash
  \mathtt{meas}_{i}^{n+1}\,\new_{\rho}: \bang \bit 
  \boxtimes \nqbit\rrbracket \lessdot
  E[\mathtt{meas}_{i}^{n+1}\,\new_{\rho}]$, and therefore,
  $\bigl(\llbracket \vdash \mathtt{meas}_{i}^{n+1}\,\new_{\rho}:
  \bang \bit  \boxtimes \nqbit\rrbracket,\,
  \mathtt{meas}_{i}^{n+1}\,\new_{\rho}\bigr)$ is in
  $R_{\bang \bit  \boxtimes
    \nqbit}^{\top\top}$. 
 
   When $A \equiv
  \bang \bigl((n+1)\text{-}\mathtt{qbit} \limp \bang \bit  \boxtimes
  \nqbit\bigr)$, since $(\llbracket \vdash
  \mathtt{meas}_{i}^{n+1}\rrbracket_{\mathrm{const}}
  ,\mathtt{meas}_{i}^{n+1})$ is in $R_{(n+1)\text{-}\mathtt{qbit}
    \limp \bang \bit  \boxtimes \nqbit}$, the
  pair $(\bang \llbracket
  \mathtt{meas}_{i}^{n+1}\rrbracket_{\mathrm{const}} \co \varphi'
  ,\mathtt{meas}_{i}^{n+1})$ is in $R_{A}$.  Therefore, by
  Lemma~\ref{lem:07130423}, $(\llbracket \vdash
  \mathtt{meas}_{i}^{n+1}:A\rrbracket ,\mathtt{meas}_{i}^{n+1})$ is in
  $R_{A}^{\top\top}$.  

  Finally, when $A$ satisfies
  $\bang ((n+1)\text{-}\mathtt{qbit} \limp \bang \bit  \boxtimes
  \mathtt{qbit}) \subtp  A$, the statement follows from
  Lemma~\ref{lem:07130422}, Lemma~\ref{lem:07130423} and that
  $(\bang \llbracket \mathtt{meas}_{i}^{n+1}\rrbracket_{\mathrm{const}}
  \co \varphi' ,\mathtt{meas}_{i}^{n+1})$ is in
  $R_{\bang ((n+1)\text{-}\mathtt{qbit} \limp \bang \bit  \boxtimes
    \mathtt{qbit})}$. \myqed
\end{myproof}

\begin{mylemma}\label{lem:meas2}
  For a type $A$ such that $\bang (\mathtt{qbit} \limp \bang \bit )
  \subtp  A$, the pair $(\llbracket \vdash
  \mathtt{meas}_{1}^{1}:A\rrbracket, \mathtt{meas}_{1}^{1})$ is in
  $R_{A}^{\top\top}$.
\end{mylemma}
\begin{myproof}
  Similar to the proof of Lemma~\ref{lem:meas1}. 
 \myqed
\end{myproof}


\newpage
\bibliographystyle{elsarticle-num}
\bibliography{./myrefs}

\newpage

\auxproof{
\subsubsection{Linear Algebra}

We list the standard results that are used in the rest of the paper,
with
citations where the proofs are found. (To-do! Use~\cite{NielsenC00})
\begin{mylemma}
 A matrix $A\in M_{m}$ is diagonalizable if and only if it is
 \emph{normal}, that is, $A^{\dagger}A=AA^{\dagger}$. \myqed
\end{mylemma}
\begin{mydefinition}[Positive matrix]
\label{definition:positiveMatrix}
An $m\times m$ matrix $A$ is said to be \emph{positive} (or
 \emph{positive semidefinite}) if for any $\ket{v}\in\C^{m}$ we have
\begin{displaymath}
 \bra{v}A\ket{v} \ge 0\enspace.
\end{displaymath} 
\end{mydefinition}

\begin{mylemma}
\label{lem:charOfPositiveMatrix}
For an $m\times m$ matrix $A$, the following are equivalent.
\begin{enumerate}\setlength\itemsep{0em}
 \item $A$ is positive.
 \item $A$ is Hermitian and
its eigenvalues are all nonnegative.
 \item\label{item:positiveMatrixCharThree} $A=B^{\dagger}B$ for some
      $m\times m$ matrix $B$. \myqed
\end{enumerate}
\end{mylemma}
\auxproof{ \noindent$[2.\Rightarrow 1.]$ 
A Hermitian $A$ is diagonalizable: let $\lambda_{i}$ ($i\in[1,m]$) be
 its eigenvalues; and $\ket{v_{i}}$ be corresponding eigenvectors such
 that $(\ket{v_{i}})_{i\in[1,m]}$ is an ONB. Any vector in $\C^{m}$ can
 be written as a linear combination $\sum_{i}a_{i}\ket{v_{i}}$. Now
 \begin{align*}
 & \bigl\langle\textstyle\sum_{j}a_{j}v_{j}\bigl|A\bigr|\textstyle\sum_{i}a_{i}v_{i}\bigr\rangle
 \\
 &=
 \textstyle\sum_{i} \textstyle\sum_{j}a_{i}a_{j}^{*}\bigl\langle v_{j}\bigl|A\bigr|v_{i}\bigr\rangle
  &&\text{linearity of $\bra{\place}\place\rangle$}
 \\
 &=
 \textstyle\sum_{i}
  \textstyle\sum_{j}a_{i}a_{j}^{*}\lambda_{i}\bigl\langle
  v_{j}\bigl|\bigr.v_{i}\bigr\rangle
  && \text{by } Av_{i}= \lambda_{i}v_{i}
 \\
 &=
 \textstyle\sum_{i}a_{i}a_{i}^{*}\lambda_{i}
  && \text{$(\ket{v_{i}})_{i}$ is an ONB}
 \\
 &=
 \textstyle\sum_{i}\lambda_{i}|a_{i}|^{2}
 \ge 0
  && \text{by }\lambda_{i}\ge 0.
 \end{align*}

 \noindent $[2.\Rightarrow 3.]$ 
 By assumption, the matrix $A$ can be diagonalized as follows:
 \begin{displaymath}
  A = U^{\dagger}
  \left[
  \begin{array}{ccc}
   \lambda_{1}& & \\
   &\ddots & \\
   & & \lambda_{m}
  \end{array}
\right] U
 \end{displaymath}
 with unitary $U$ and $\lambda_{i}\ge 0$. Now let $B$ be the matrix
 \begin{displaymath}
  B:=
U^{\dagger}
  \left[
  \begin{array}{ccc}
   \sqrt{\lambda_{1}}& & \\
   &\ddots & \\
   & & \sqrt{\lambda_{m}}
  \end{array}
\right] U\enspace;
 \end{displaymath}
 then obviously $B^{\dagger} = B$ and $A = B^{2}=B^{\dagger}B$, since 
 for unitary $U$ we have $UU^{\dagger} = \IM$.

 \noindent $[3.\Rightarrow 1.]$ 
 \begin{displaymath}
  \bra{v}A\ket{v} 
 = 
  \bra{v}B^{\dagger}B\ket{v} 
 = 
  \bra{Bv}Bv\rangle
 = 
  \abs{Bv}^{2} \ge 0\enspace.
 \end{displaymath}

 \noindent $[1.\Rightarrow 2.]$  We have, for each $v\in\C^{m}$, 
 \begin{align*}
  \inpr{v}{Av} &= \inpr{v}{Av}^{*} 
 && \text{$\inpr{v}{Av}$ is real}
 \\
 &= \inpr{Av}{v}
 &&\text{by }  \inpr{u}{v} = \inpr{v}{u}^{*}.
 \end{align*} 
 Thus by the definition of $A^{\dagger}$, we have $A=A^{\dagger}$ hence Hermitian.

 In particular $A$ is diagonalizable: let $\lambda_{i}$ and
 $\ket{v_{i}}$ be eigenvalues and eigenvectors of $A$ such that
 $(\ket{v_{i}})_{i}$ is an ONB. Now
 \begin{displaymath}
  \lambda_{i}=\lambda_{i}\abs{\ket{v_{i}}}^{2}
  = \bra{v_{i}} A \ket{v_{i}} \ge 0\enspace.
 \end{displaymath}
 This concludes the proof. \myqed
}

\begin{mylemma}\label{lem:tracePres}
\begin{enumerate}
 \item For a $1\times 1$ matrix $A$, $\trace(A)=A$.
 \item $\trace(\sum_{i}A_{i})=\sum_{i}\trace(A_{i})$.
 \item 
  Let $A,B$ be $m\times n$ and $n\times m$ matrices, respectively.
 We have $\trace(AB)=\trace(BA)$. \myqed
\end{enumerate}
\end{mylemma}
}

\auxproof{
\begin{mylemma}\label{lem:densityMatrixAsEnsemble}
 Let $\rho$ be an $m\times m$ matrix. It is a density matrix if and
 only if it can be written as
 \begin{displaymath}
  \rho = \sum_{i\in I}\lambda_{i}\ket{v_{i}}\bra{v_{i}}\enspace,
 \end{displaymath}
 with $\ket{v_{i}}\in \C^{m}$, $\abs{\ket{v_{i}}}=1$, $\lambda_{i}\ge 0$
 and $\sum_{i}\lambda_{i}\le 1$.
\end{mylemma}
\begin{myproof}
The `if' part is straightforward.
 For the `only if' part, use diagonalization. 
A detailed proof is found
 in~\cite[Theorem~2.5]{NielsenC00}. 
\myqed
\end{myproof}

We will also use the following standard fact. For the proof, see
e.g.~\cite[Remark~2.1]{Selinger04}.
\begin{mylemma}\label{lem:densityMatricesSpanAllMatrices}
Any matrix $A\in M_{m}$ can be written as a linear combination of 
density matrices. \myqed
\end{mylemma}

 First, we have
       \begin{displaymath}
	A = 
	\frac{1}{2}(A^{\dagger} + A) +i\cdot \frac{i}{2}(A^{\dagger}-A)\enspace;
       \end{displaymath}
here both $\frac{1}{2}(A^{\dagger} + A)$ and $\frac{i}{2}(A^{\dagger}-A)$ are
       Hermitian. Further, let $B$ be an arbitrary Hermitian matrix.  It
       is written as a combination of positive matrices, namely
       \begin{displaymath}
	B=(B + \lambda I) - \lambda I\enspace,
       \end{displaymath}
       where $-\lambda$ is the most negative eigenvalue of $B$.
       Finally, given a positive matrix $C$, $C/\trace(c)$ is a density
 matrix.
 \myqed
}

\auxproof{
\section{Material that can be used: QPL 2011 abstract}
\paragraph{Quantum Programming Language (QPL)}
 Study of \emph{high-level quantum programming languages} and their
 \emph{mathematically formulated semantics} has great potential
 benefits.  It may aid discovery of new quantum algorithms via intuitive
 presentation. More importantly, it allows us to transfer various
 verification techniques for classical computation (program logics, type
 systems, bisimulation, etc.) into quantum computation. Such
 verification methods are also desired in the realm of \emph{quantum
 communication}: quantum communication is closer to practical deployment
 than quantum computation, but our experience with classical communication
 protocols makes us wary of its correctness issue.

\paragraph{Quantum Data, Classical Control}
 There have been several QPLs proposed in the
 literature. Some are imperative like QCL or qGCL; others are functional.
 Those which are more widely accepted share one design philosophy:
 \emph{quantum data, classical control}. This means that, while qubits
 (``quantum data'') play an essential role in quantum computation, there
 is no quantum superposition of execution threads.  Thus one can follow
 the control structure of a quantum program much like a classical
 program, although he might encounter \emph{quantum primitives} that
 operate on quantum data (namely: preparation, unitary transformation,
 measurement).

\paragraph{Functional QPL}
We find the style of functional programming particularly useful in this
application domain. Besides being ``real men's programming style,''
its proximity to mathematics provides a \emph{mathematical support} to
language design as well as semantic study. Indeed, the current work
fully exploits the genericity of categorically formulated semantic
techniques for (classical) functional programming. On language design, 
\emph{linear $\lambda$-calculus} provides a useful prototype of a
functional QPL. The ``no-cloning'' property of
quantum data is nicely enforced by the linear type system, while
duplicable classical data is represented via the $\bang$ modality.
}

\end{document}